\documentclass[prd,aps,notitlepage,nofootinbib,floatfix,twocolumn,superscriptaddress,twoside]{revtex4-1}

\pdfoutput=1
\usepackage{natbib}
\usepackage{amssymb}
\usepackage{amsmath}    
\usepackage{graphicx}   
\usepackage{verbatim}   
\usepackage{color}      
\usepackage{hyperref}   

\usepackage{array}
\usepackage{enumerate}
\usepackage{multirow}

 \usepackage[normalem]{ulem}
 \usepackage{mathtools}

\newcommand{\aap}{{Astron.~Astrophys.}}
\newcommand{\apjl}{{Astrophys.~J.~Lett.}}
\newcommand{\apjs}{{Astrophys.~J.~Supp.}}
\newcommand{\aj}{{Astron.~J.}}
\newcommand{\mnras}{{Mon.~Not.~R.~Astron.~Soc.}}
\newcommand{\jcap}{{J.~Cosmol.~Astropart.~Phys.}}

\newcommand{\Planck}{{\slshape Planck~}}
\newcommand{\Planckc}{{\slshape Planck}}

\begin{document}

\title{Matter bispectrum of large-scale structure: Three-dimensional comparison between theoretical models and numerical simulations}
\author{Andrei Lazanu}
\email{A.Lazanu@damtp.cam.ac.uk}
\affiliation{Centre for Theoretical Cosmology, DAMTP, University of Cambridge, CB3 0WA, United Kingdom}
\author{Tommaso Giannantonio}
\email{T.Giannantonio@ast.cam.ac.uk}
\affiliation{Kavli Institute for Cosmology Cambridge, Institute of Astronomy, University of Cambridge, Madingley Road, Cambridge CB3 0HA, United Kingdom}
\affiliation{Centre for Theoretical Cosmology, DAMTP, University of Cambridge, CB3 0WA, United Kingdom}
\author{Marcel Schmittfull}
\email{M.Schmittfull@berkeley.edu}
\affiliation{Berkeley Center for Cosmological Physics, Department of Physics and Lawrence Berkeley National Laboratory, University of California, Berkeley, California 94720, USA}
\author{E.P.S. Shellard}
\email{E.P.S.Shellard@damtp.cam.ac.uk}
\affiliation{Centre for Theoretical Cosmology, DAMTP, University of Cambridge, CB3 0WA, United Kingdom}

\date{\today}

\begin{abstract}
We study the matter bispectrum of the large-scale structure by comparing different perturbative and phenomenological models with measurements from $N$-body simulations obtained with a modal bispectrum estimator. Using shape and amplitude correlators, we directly compare simulated data with theoretical models over the full three-dimensional domain of the bispectrum, for different redshifts and scales. We review and investigate the main perturbative methods in the literature that predict the one-loop bispectrum: standard perturbation theory, effective field theory, resummed Lagrangian and renormalised perturbation theory, calculating the latter also at two loops for some triangle configurations. We find that effective field theory (EFT) succeeds in extending the range of validity furthest into the mildly nonlinear regime, albeit at the price of free extra parameters requiring calibration on simulations: EFT is found to be accurate to 5\% up to a scale of $k_{\max}^* \simeq 0.4 h/Mpc$ at $z = 1$, compared with $k_{\max}^* \simeq 0.2 h/Mpc$ at $z = 1$ for most other one-loop perturbative methods. For the more phenomenological halo model, we confirm that despite its validity in the deeply nonlinear regime it has a deficit of power on intermediate scales, which worsens at higher redshifts (the maximum deficit in the amplitude correlator is $\sim 20\%$ at $z =1$, and up to $40\%$ at $z=2$); this issue is ameliorated, but not solved, by combined halo-perturbative models. We show from simulations that in this transition region there is a strong squeezed bispectrum component that is significantly underestimated in the halo model at earlier redshifts. We thus propose a phenomenological method for alleviating this deficit, which we develop into a simple phenomenological "three-shape" benchmark model based on the three fundamental shapes we have obtained from studying the halo model. When calibrated on the simulations, this three-shape benchmark model accurately describes the bispectrum on all scales and redshifts considered, providing a prototype bispectrum \textsc{Halofit}-like methodology that could be used to describe and test parameter dependencies.
\end{abstract}

\maketitle

\section{Introduction}

The $\Lambda$CDM model has so far been successful in describing the properties of the Universe, as recently confirmed by the  latest \Planck satellite results \cite{2015arXiv150201582P,Planck2015_Params}. The initial conditions of this model are based on the assumption that all the structure in the Universe was generated by quantum fluctuations at primordial times, during an inflationary phase \cite{Guth1981,Mukhanov1981}. The physics of inflation has been extensively studied in recent years and many scenarios have been proposed \cite{Hawking1982295, 1475-7516-2003-10-013, 0264-9381-19-13-305, baumannbook, Martin2014}; distinguishing between the numerous existing models 
 is one of the ultimate goals of cosmology. This problem can be tackled observationally by studying the properties of the perturbations at later times: the cosmic microwave background (CMB) and the large-scale structure of the Universe (LSS). 

CMB anisotropies have provided in the past two decades  a wealth of cosmological information, which has been exploited with increasing efficiency by subsequent observational campaigns, up to the exquisite accuracy of the latest results from the \Planck satellite \cite{Planck2015_Params}.
The CMB has also provided some of the strongest constraints on inflation. On the one hand, the shape of the CMB two-point statistics (power spectrum) is directly related to the power spectrum of perturbations at the end of inflation, whose parameters and features can thus be accurately constrained  \cite{Planck2015_Inflation}. On the other hand, many inflationary models predict a significant non-Gaussian component in the distribution of primordial perturbations \cite{Chen2010}: higher-order statistics of the CMB anisotropies, such as the three-point correlation function (bispectrum) have provided strict constraints on such models \cite{Planck2015_PNG}.

 Nevertheless, the CMB can primarily supply only two-dimensional data from the surface of last scattering, which in temperature has been already almost fully exploited to the limit of cosmic variance by \Planckc. 
The LSS, traced by current and upcoming galaxy surveys, contains much more information than the CMB due to its three-dimensional nature, and it can thus provide
further complementary insight on cosmology across cosmic time.
 In principle, there is roughly a 1000-fold increase in the number of modes available compared to the CMB ~\cite{Carrasco2012}, but this information is more challenging to extract due to the more limited theoretical understanding of the LSS physics in the low-redshift universe, where additional complexity is added by nonlinear structure formation, the relationship between dark and visible matter (galaxy bias), and redshift-space effects \cite{2014arXiv1412.4671A}.
Indeed, the modelling of galaxy clustering is first based on a description of dark matter clustering; the clustering of collapsed dark matter haloes is then defined by introducing halo bias, while the connection to observable galaxies
can be made by using halo occupation distribution \cite{Kwan2013} or halo-abundance matching \cite{RodriguezTorres2015} methods, calibrated on $N$-body simulations.
 In this paper, we only describe the clustering of dark matter in real space, leaving the connection to galaxy observables, including bias and redshift-space effects, to subsequent work.


Galaxy surveys like SDSS \cite{York2000} and BOSS \cite{Eisenstein2011,1538-3881-145-1-10} have dramatically increased our understanding of the Universe. On-going and future surveys, like DES \cite{DES2005,Diehl2014}, LSST \cite{2008arXiv0805.2366I}, \textit{Euclid} \cite{Laureijs2011}, DESI \cite{2014SPIE.9147E..0SF}, \textit{WFIRST} \cite{2015arXiv150303757S} and the proposed SPHEREX mission \cite{2014arXiv1412.4872D} are expected to increase the precision of the measurements even further.

%

To date, most cosmological implications from  large-scale structure data have been drawn from the power spectrum of galaxies. At linear level, the matter power spectrum encodes all the information available if the primordial random fluctuations are Gaussian.
The power spectrum is also sensitive to some classes of primordial non-Gaussianity (PNG) via the scale-dependent galaxy bias \cite{Dalal2008,Matarrese2008}, which has been widely used to obtain competitive PNG constraints \cite{Slosar2008,Xia2010,Mana2013,Ross2013,Giannantonio2014,Giannantonio2014b}.

 However, in order to fully exploit the LSS information and to test all types of PNG, it is important to also study higher-order statistics, such as the
bispectrum \cite{Verde2000, Scoccimarro2004, Bartolo2005, Jeong2009, Nishimichi2010, Liguori2010, Baldauf2011}. 
Even for Gaussian initial conditions, where the primordial bispectrum is zero, nonlinear coupling between Fourier modes produces a non-zero bispectrum due to gravitational collapse \cite{Fry1994}. 
 This gravitational bispectrum must be well understood in order to be able to separate the primordial component and to constrain the physics of inflation.  At the same time, it can provide additional cosmological information \cite{Sefusatti2006}, for example on the growth of structure \cite{Matarrese1997,Scoccimarro1998,Shirata2007}, and bias parameters \cite{Matarrese1997,Verde1998,Scoccimarro1999,Sefusatti2007,Guo2009,Pollack2012,PhysRevD.90.123522}.

Modelling the evolution of matter density perturbations beyond linear scales is a complex problem. On relatively large scales, in the quasi-linear regime, significant progress has been made using perturbative methods.
Arguably, the most common procedure is Eulerian standard perturbation theory (SPT) \cite{Fry1984,1986ApJ...311....6G,1994ApJ...431..495J,Bernardeau20021}, where the growth of structure is described by a set of differential equations in terms of the present-time density perturbations, expanded to the desired order. Alternatively, in Lagrangian perturbation theory (LPT) \cite{Buchert1989,Moutarde1991,Buchert1992,Buchert1994,Bouchet1995,Ehlers1997,Tatekawa2004} the fluid equations are written in terms of the initial density perturbations via a displacement field, which reduces to the Zel'dovich approximation at linear order \cite{White2014}.
Both methods have advantages and shortcomings \cite{PhysRevD.80.043531,Valageas2013a,Tassev2014,Vlah2015}; in particular, SPT has a narrow range of validity at low redshift, and its series expansion shows poor convergence properties. LPT has the additional drawback that its perturbative approach can not predict clustering beyond shell crossing.
For these reasons, the recent years have seen a proliferation of further developments: SPT has been re-formulated in the language of field theory by Ref.~\cite{NYAS:NYAS13,PhysRevD.73.063519, PhysRevD.73.063520, PhysRevD.77.023533, Bernardeau2008, PhysRevD.85.123519}, re-organising the series expansion in terms of vertices and propagators, and improving its convergence properties (renormalised perturbation theory, RPT); this has been later simplified to the \textsc{MPTbreeze} scheme \cite{Crocce2012}. Related developments include the large-$N$ expansion \cite{Valageas2007}, the closure theory \cite{Taruya2008}, and renormalisation group approaches \cite{McDonald2007, Matarrese2007}. A resummation technique in Lagrangian space (RLPT) was developed by Ref.~\cite{Matsubara2008}; subsequent extensions were developed by Refs.~\cite{Matsubara2008,Okamura2011,Carlson2013,Rampf2012,Rampf2012b,Valageas2013,Matsubara2014,Sugiyama2014}.
Most recently, the effective field theory of LSS (EFTofLSS) has been developed by Refs.~\cite{1475-7516-2012-07-051, Carrasco2012,Hertzberg2014,Carrasco2014,Carrasco2014b,Pajer2013,Carroll2014,Porto2014,Senatore2015}, based on the idea that the contribution of small-scale physics to the quasi-linear perturbations can be encapsulated into an set of additional, unknown source terms in the equations of motion, whose value can be fixed by comparison with $N$-body simulations.

 In the fully nonlinear regime, perturbation theories necessarily break down and numerical $N$-body simulations have to be used to calibrate phenomenological models of gravitational clustering, such as the halo model \cite{Seljak11102000, MaFry2000, Peacock2000}. This formalism is based on the approximation that all matter in the Universe is in the form of spherical haloes with a universal density profile and without sub-structure, and it can be used to describe the matter power spectrum and bispectrum relatively accurately (typically better than $10\%$ at $k < 1 \, h/$Mpc at $z=0$) \cite{Cooray20021}.  
It is however difficult to significantly improve the halo model accuracy beyond the limits set by its underlying assumptions,
 especially on intermediate scales. For this reason, Refs.~\cite{valageas1,valageas2,Valageas2013} combined a revised version of the halo model, valid on small scales, with perturbative recipes that are more accurate on quasi-linear scales. 
Ref. \cite{Mead:2015yca} also proposed a halo model extension that improves its accuracy at the cost of 12 extra parameters.

A more drastic approach was introduced by Refs.~\cite{Mohammed2014,Seljak2015}, 
where the physically-motivated small-scale one-halo term was replaced with a series expansion in the even powers of $k$, with free parameters to be calibrated on $N$-body simulations.
It is possible to extend these ideas even further into the direction of phenomenology at the cost of a reduced physical understanding: the \textsc{Halofit} method \cite{Smith01062003,2012ApJ...761..152T} achieves a higher accuracy matter power spectrum by combining halo model-inspired templates with numerous heuristic parameters fit to $N$-body simulations while, in the ultimate numerical and agnostic approach, matter clustering is directly calculated by interpolating over a grid of $N$-body simulations spanning a range of different cosmologies \cite{Heitmann2010}. No bispectrum counterpart exists to date for these numerical methods.

  At the same time, there has been progress in $N$-body simulations studies and bispectrum estimators
\cite{Sefusatti2006, Liguori2010, 2010MNRAS.406.1014S, 2012MNRAS.425.2903S, Schmittfull2013, Schmittfull2015}. 
In contrast with the standard brute-force method of measuring the bispectrum for all possible triangular configurations, Refs.~\cite{Fergusson2012,Schmittfull2013} applied to the LSS the modal decomposition of the bispectrum introduced for CMB studies by Refs.~\cite{Fergusson2010,Liguori2010}, thus developing a significantly faster and more efficient estimator.
A simplified version tailored to estimating the projection of the simulation bispectrum on the tree-level prediction was presented in Ref.~\cite{Schmittfull2015}.

Relatively few measurements of the bispectrum from galaxy surveys exist \cite{Jing1998, Feldman2001, Scoccimarro2001, Verde2002,Jing2004,Wang2004,Marin2011,Marin2013}. The state of the art results have recently been obtained by Ref.~\cite{GilMarin2015a, GilMarin2015b}
from the BOSS luminous red galaxies. These data have been used to improve the power spectrum constraints on galaxy bias and structure growth; however, to date no primordial non-Gaussianity constraints exist from the LSS bispectrum.

In this paper, we make the first comprehensive comparison of models describing the matter bispectrum as a function of scale and redshift. We review a selection of different models from the literature and we analyse their accuracy on different scales by comparing their predictions with direct estimates of the bispectrum from $N$-body simulations. We compare two classes of models: methods based on perturbative approaches and phenomenological halo models. The perturbative models considered are: tree-level, nonlinear tree-level, SPT, RPT, RLPT, and EFT (all at one loop).
In the nonlinear regime, we investigate the standard halo model and a modified halo model combined with EFT, based on the method by Ref.~\cite{valageas2}. We base our analysis on a full three-dimensional comparison of the shapes and amplitudes of the bispectra, which allows us to compare all the triangular configurations in the bispectra at once, rather then confronting individual slices in specific configuration limits, as usual with previous work.  This approach is relevant to observational forecasts of predicted signal-to-noise where the full statistical significance requires summation over all triangle configurations.

We then develop a simple phenomenological model based on the three fundamental shapes of the halo model components, which provides a good global fit to the simulations. We quantitatively compare the simulations with each of the theoretical models considered and we discuss their advantages and limitations. We also show how to numerically calculate the two-loop bispectrum in the \textsc{MPTbreeze} formalism in an infrared-safe manner and we present the results for several scaled triangular configurations.

The plan of this paper is as follows. After a brief introduction to the bispectrum and its three-dimensional estimators in Sec.~\ref{sec:bisp}, we review the theoretical models we consider in Sec.~\ref{sec:pert} (perturbation theory) and Sec.~\ref{sec:halo} (nonlinear and phenomenological models).
 We then describe the $N$-body simulations in Sec.~\ref{sec:sims}. In Sec.~\ref{sec:benchmark} we discuss the measured bispectrum shapes, and use this to introduce the phenomenological two-halo boost and three-shape benchmark model. 
 We next present the results of the comparison between the different theoretical models and simulations in Sec.~\ref{sec:comparison}, before concluding in Sec.~\ref{sec:concl}.  Several appendices provide details of the considered models.

\section{Bispectrum introduction}
\label {sec:bisp}
The statistical analysis of random fields, such as the matter density perturbation $\delta \equiv \left(\rho - \bar \rho \right) / \bar \rho $, where $\rho$ is the matter density of mean $\bar \rho$, involves measuring its $N$-point correlation functions in real space, or its $N$-spectra in Fourier space. We consider here the power spectrum and bispectrum, which are defined as:
\begin{align}
\label{ps}
\langle	\delta (\textbf{k}_1) \delta (\textbf{k}_2) \rangle &= (2\pi)^3 \delta_D (\textbf{k}_1 + \textbf{k}_2) P(k) \\
\langle	\delta (\textbf{k}_1) \delta (\textbf{k}_2) \delta (\textbf{k}_3) \rangle &= (2 \pi)^3 \delta_D (\textbf{k}_1 + \textbf{k}_2 + \textbf{k}_3) B(k_1,k_2,k_3) \, ,
\label{bis}
\end{align}
where $\delta_{D}$ is the Dirac delta function. For statistically homogeneous and isotropic cosmologies, to which we restrict our attention here, the bispectrum only depends on the wavenumbers $k_1,k_2,k_3$. While the power spectrum is a 1D quantity, as it is simply a real function of the wavenumber $k$, the bispectrum is a more complex 3D quantity, as it is a real function of $k_1,k_2,k_3$. The bispectrum therefore contains more information, but it is also more cumbersome to study, and it thus requires relatively more advanced techniques to be measured and exploited. We introduce in this section our method for analysing the full 3D matter bispectrum, and for comparing its observations with theoretical models.

\subsection{Shape and amplitude correlators}
\begin{figure}[tb]
\begin{center}
\includegraphics[width=3.5in]{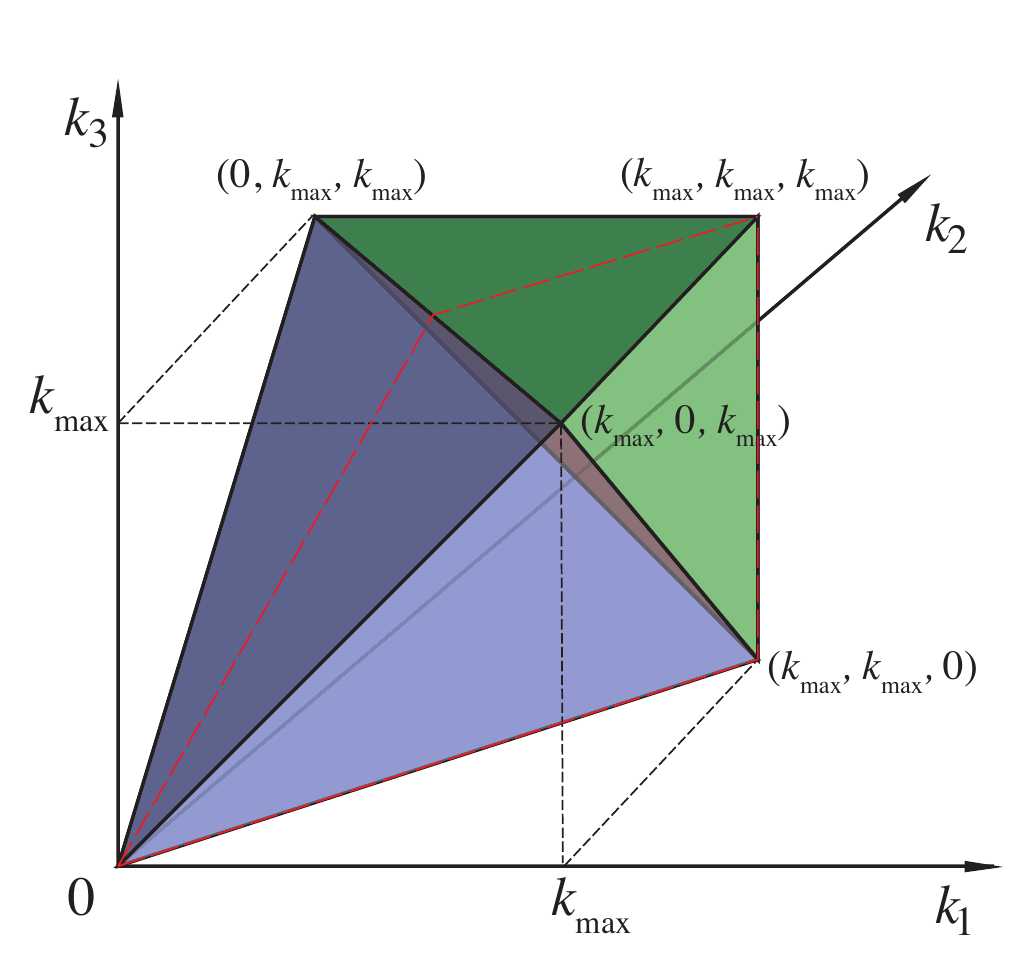} 
\caption{The tetrapyd bispectrum domain consists of a tetrahedral region (blue)  defined by the wavevector triangle condition in Eq.~(\ref{bis}), together with a pyramidal region (green) bounded by the resolution limit $k_{\rm max}$.  For the autocorrelator bispectrum this has a sixfold symmetry, so to illustrate the internal structure of the bispectrum (equilateral limit) we will split the tetrapyd across the vertical plane given by the red-dashed lines, removing the front half as shown in Fig.~\ref{tetrapyd_split}. }
\label{tetrapyd}
\end{center}
\end{figure}

\begin{figure}[tb]
\begin{center}
\includegraphics[width=3.5in]{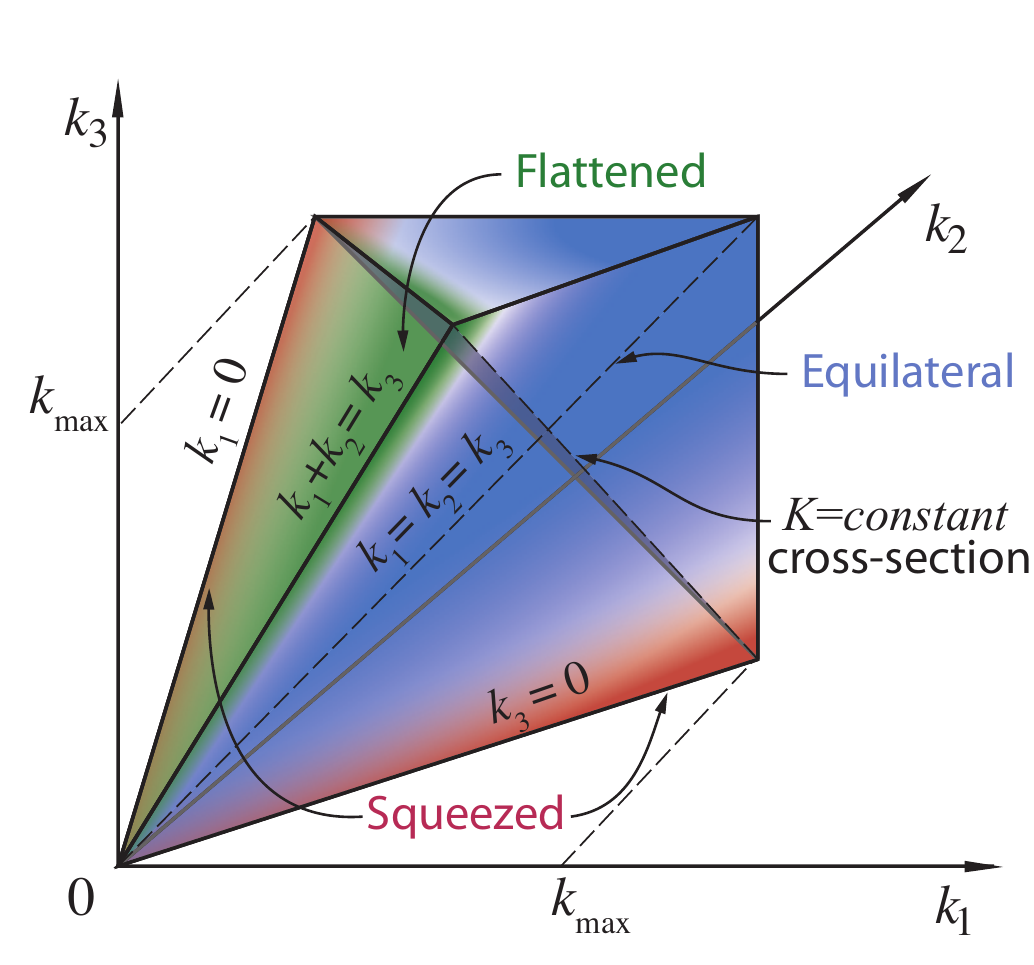}
\caption{The split 3D tetrapyd region used to illustrate the SN-weighted bispectrum showing only the back half with $k_1<k_2$.  Colour-coded regions show the location of  the `squeezed' (red), `flattened' (green) and `equilateral' or `constant' (blue) shape signals.  In the bispectrum ansatz Eq.~(\ref{separable}) the shape $S(k_1,k_2,k_3)$ is defined on the $K\equiv k_1+k_2+k_3=\hbox{const.}$ cross-sectional planes, while the scale-dependence $f(K)$ is given along the dashed diagonal  $k_1=k_2=k_3$.}
\label{tetrapyd_split}
\end{center}
\end{figure}

In order to compare the observed or simulated bispectra with the corresponding theoretical predictions, we define the signal-to-noise weighted scalar product between two bispectrum shapes \textit{i} and \textit{j}  \cite{1475-7516-2004-08-009, Fergusson2012}:
\begin{equation}
\label{shapeprod}
\langle B_i, B_j \rangle \equiv \frac{V}{\pi}\int_{\mathcal{V}_B}dV_k\, \frac{k_1k_2k_3 \,B_i(k_1,k_2,k_3)\,B_j(k_1,k_2,k_3)}{ P(k_1)P(k_2)P(k_3)} \ ,
\end{equation}
where the integration domain $\mathcal{V}_B$ is the tetrahedral region of volume $V$ satisfying the triangle condition on the wavenumbers $k_1$, $k_2$ and $k_3$ (such that $\textbf{k}_1 + \textbf{k}_2 + \textbf{k}_3=0$), together with a chosen resolution limit $k_1, k_2, k_3< k_{\max}$.    The bispectrum domain is the union of a tetrahedron with a triangular pyramid on top (denoted the `tetrapyd') and is illustrated in Fig.~\ref{tetrapyd}.   The inner product Eq.~(\ref{shapeprod}) provides a natural definition for the signal-to-noise (SN) weighted bispectrum, 
\begin{equation}
\label{SNweight}
B^{\rm SN}_i(k_1,k_2,k_3)    \equiv  \sqrt{ \frac{k_1k_2k_3}{ P(k_1)P(k_2)P(k_3)} }\;B_i(k_1,k_2,k_3)\,,
\end{equation}
where we use the measured (or \textsc{Halofit}) power spectrum $P_{\text{\rm NL}}(k)$ for wavenumbers in the quasilinear and nonlinear regimes (rather than the linear power spectrum $P_{\rm lin}$).   
The  SN-weighted bispectrum  $B^{\rm SN}_i$ is the relevant quantity observationally if the matter bispectrum could be measured directly, providing optimal forecasts for an ideal survey (i.e.\ one without experimental noise or systematics).   To develop an intuitive understanding of the distinct gravitational bispectrum contributions, we will plot the SN-weighted bispectrum in three dimensions on half the tetrapyd domain as shown in Fig.~\ref{tetrapyd_split}.  Although the full tetrapyd has a sixfold symmetry for the isotropic bispectrum of Eq.~(\ref{bis}), leaving this redundancy allows us to view $B^{\rm SN}_i$ from equilateral, flattened and squeezed limits simultaneously. (Future work will include bispectrum cross-correlators, e.g., the matter-matter-halo bispectrum where Fig.~\ref{tetrapyd_split} shows the complete domain, as for recent CMB polarisation results \cite{Planck2015_PNG}.)

 Based on the scalar product of Eq.~(\ref{shapeprod}),  we define the \textit {shape correlator} (or cosine) by:
\begin{equation}
\label{shapecor}
\mathcal{S}\left(B_i,B_j\right) \equiv \frac{\langle B_i, B_j \rangle}{\sqrt{\langle B_i, B_i \rangle \langle B_j, B_j \rangle}} \, ,
\end{equation}
which is restricted to $-1 \le \mathcal{S} \le 1$. In the following, we will typically calculate the shape correlators between theoretical and simulated bispectra, to which the shapes $i,j$ will  correspond respectively.  

In order to measure how well the magnitude of the theoretical bispectra $i$ fit the (simulated) data $j$, we define the \textit{amplitude correlator} as:
\begin{equation}
\label{ampcor1}
\mathcal{A} \left( B_i, B_j \right) \equiv  \sqrt{ \frac{\langle B_i, B_i \rangle}{\langle B_j, B_j \rangle}} \, .
\end{equation}
We can thus introduce a single quantity that combines the shape and amplitude information, the \textit{total correlator}, defined as:
\begin{multline}
\mathcal{T} \left(B_i,B_j\right) \equiv 1-\sqrt\frac{\langle B_j-B_i, B_j-B_i \rangle}{\langle B_j, B_j \rangle} \\
=1-\sqrt{1 - 2 \mathcal{S}\left(B_i,B_j\right)\mathcal{A}\left(B_i,B_j\right)+\mathcal{A}^2\left(B_i,B_j\right)} \, .
\label{totalcor}
\end{multline}
This total correlator offers an excellent  means by which to determine the overall goodness of fit as we essentially measure the magnitude of the residual $B_i-B_j$ relative to the measured bispectrum $|B_j|$. If $B_i=B_j$, this is zero and the total correlator is $\mathcal{T}=1$. If $B_i$ and $B_j$ are misaligned ($\mathcal{S} < 1$) or differ in amplitude ($\mathcal{A} \ne 1$), the residual $B_i-B_j$ is non-zero and the total correlator $\mathcal{T} < 1$. For increasing relative bispectrum residual, the total correlator always decreases. 
 (Note that this is a more stringent test than the shape correlator of Eq.~\ref{shapecor} alone because $\mathcal{S}$ appears under a square root in Eq.~\ref{totalcor}).

It is possible to relate the total correlator $\mathcal{T}$ to the $\chi^2$ goodness of fit determined between the theoretical bispectrum $B_i$ and the estimated (or simulated) bispectrum $B_j$, as \cite{Baldauf:2012hs}:
\begin{align}
 \chi^2&=\sum_{k_1,k_2,k_3} \frac{\left[B_j(k_1,k_2,k_3)-B_i(k_1,k_2,k_3) \right]^2}{\text{var}(B_i)} \nonumber \\
 &= \langle B_j-B_i,\,B_j-B_i \rangle \, ,
\end{align}
so that $\chi^2$ and the total correlator $\mathcal{T}$ are simply linked by:
\begin{equation}
\chi^2 =  \left[1-\mathcal{T}(B_i,B_j) \right]^2  \langle B_j,B_j\rangle \, .
\end{equation}
As we are using a small number of simulations of limited resolution, in the following we will consider the total correlator $\mathcal{T}$ together with its uncertainty as a measurement of the goodness of fit of each model.
In principle, the use of $\chi^2/ \text{d.o.f.}$ may be more suitable than $\mathcal{T}$ to distinguish overfitting ($\chi^2/ \text{d.o.f.} < 1$) from poor model performance ($\chi^2/ \text{d.o.f.} > 1$). However, our focus here is to determine the $k_{\max}$ at which the model starts to become a poor description of our present simulations, which corresponds to the $k_{\max}$ where $\mathcal{T}$ becomes significantly smaller than unity (given the 
estimated errors between simulations).

The three correlators here, $\mathcal{S}$, $\mathcal{A}$ and $\mathcal{T}$, are all cumulative functions of $k_{\max}$, which is the resolution cut-off used in the  scalar product of Eq.~(\ref{shapeprod}).
 We therefore obtain an overall integrated measure of how well a particular theory matches simulations (or observations) up to  $k_{\max}$.

\subsection{Three canonical shape functions} 
\label{sec:shapefn}
As we shall see in subsequent sections of this paper, we are able to obtain an accurate global description of the nonlinear gravitational  bispectrum from a sum over a limited number of simple bispectrum shapes, provided that we have the flexibility to modify an overall scale-dependent amplitude.  For this reason, we consider the following non-trivial bispectrum ansatz:
\begin{equation}
B(k_1,k_2,k_3) = f(K) \, S(k_1,k_2,k_3) \, , 
\label{separable}
\end{equation}
where $ K \equiv k_1+k_2+k_3$, and the `shape function' $S$ is taken, in turn, to be a separable function of the form \begin{equation}
S(k_1,k_2,k_3) = A(k_1)\, B(k_2)\, C(k_3)+ \hbox{perms}\,.
\label{separable_shape}
\end{equation}  
This separation between transverse $K= \hbox{const.}$ slices and the $K$-dependent diagonal  is illustrated in Fig.~\ref{tetrapyd_split}.  

The separable ansatz (Eq.~\ref{separable}) is motivated in part by comparison with primordial non-Gaussian models, for which we define the shape function $S$ by taking out an overall scaling $(k_1k_2k_3)^{-2}$ after which $S$ is (almost) scale-invariant, that is, independent of the summed wavenumber $K$ along the tetrapyd diagonal.  For this reason, most primordial bispectra depend only on the two degrees of freedom transverse to the diagonal and can be completely defined by the shape $S$ on the triangular surface $K= \hbox{const}$.   

At late times, this simple separation of variables (Eq.~\ref{separable}) may not apply accurately because of the scale-dependent transfer functions, which means that perturbations with different wavenumbers $k_i$ receive different amplifications.  Nevertheless, this is encoded in the turnover of the late-time linear matter power spectrum $P_{\mathrm{lin}}(k)$, which can still be used to create a separable (though scale-dependent) `shape function', e.g.\ as we will see  for the tree-level gravitational bispectrum.  For this reason, the separable description (Eq.~\ref{separable}) can still prove very useful if physically well-motivated shapes $S(k_1,k_2,k_3)$ are chosen and an overall scaling dependence $f(K)$ is allowed.   

\begin{figure}[tb]
\begin{center}
\includegraphics[width=2.4in]{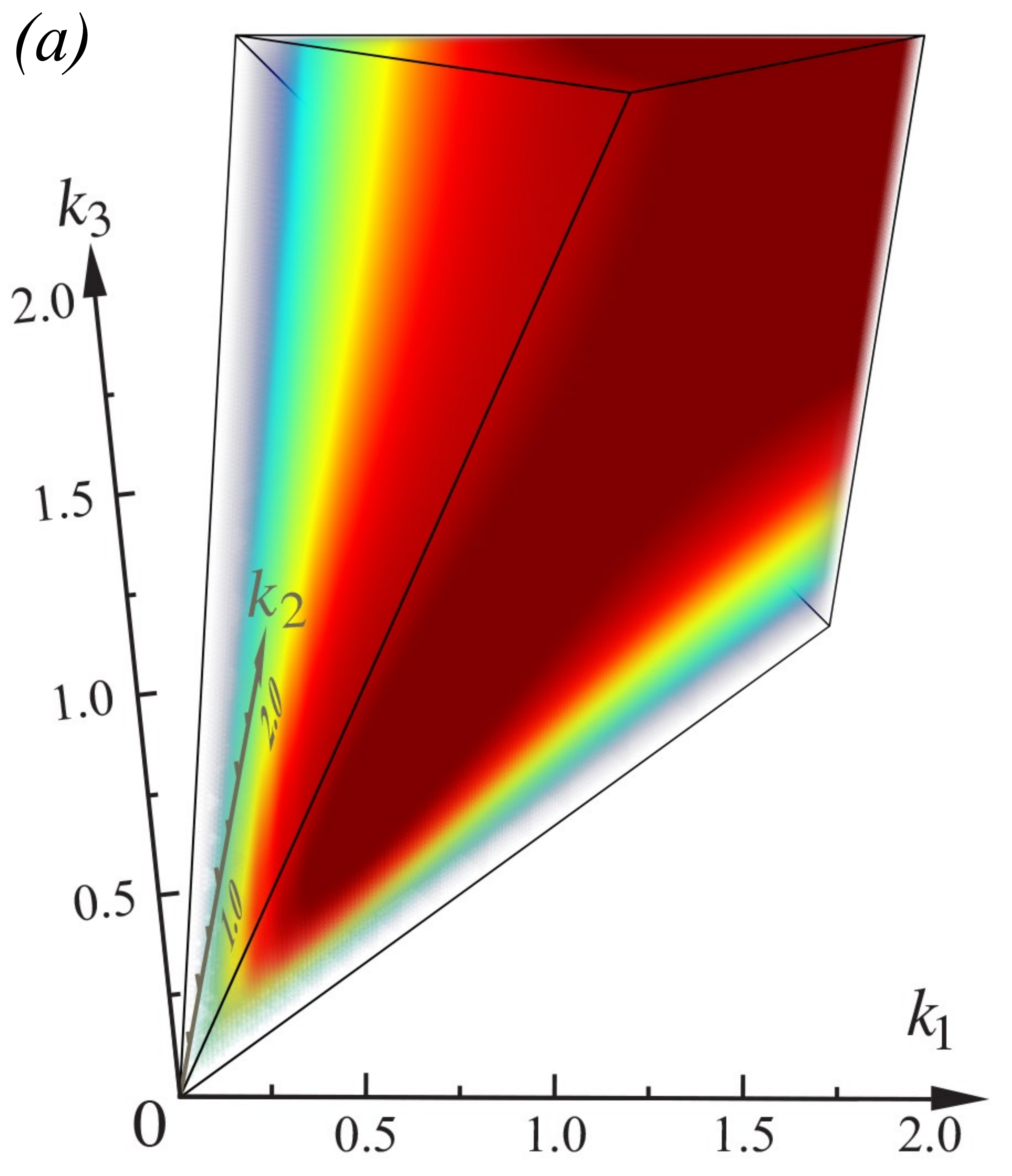}\\
\medskip
\includegraphics[width=2.4in]{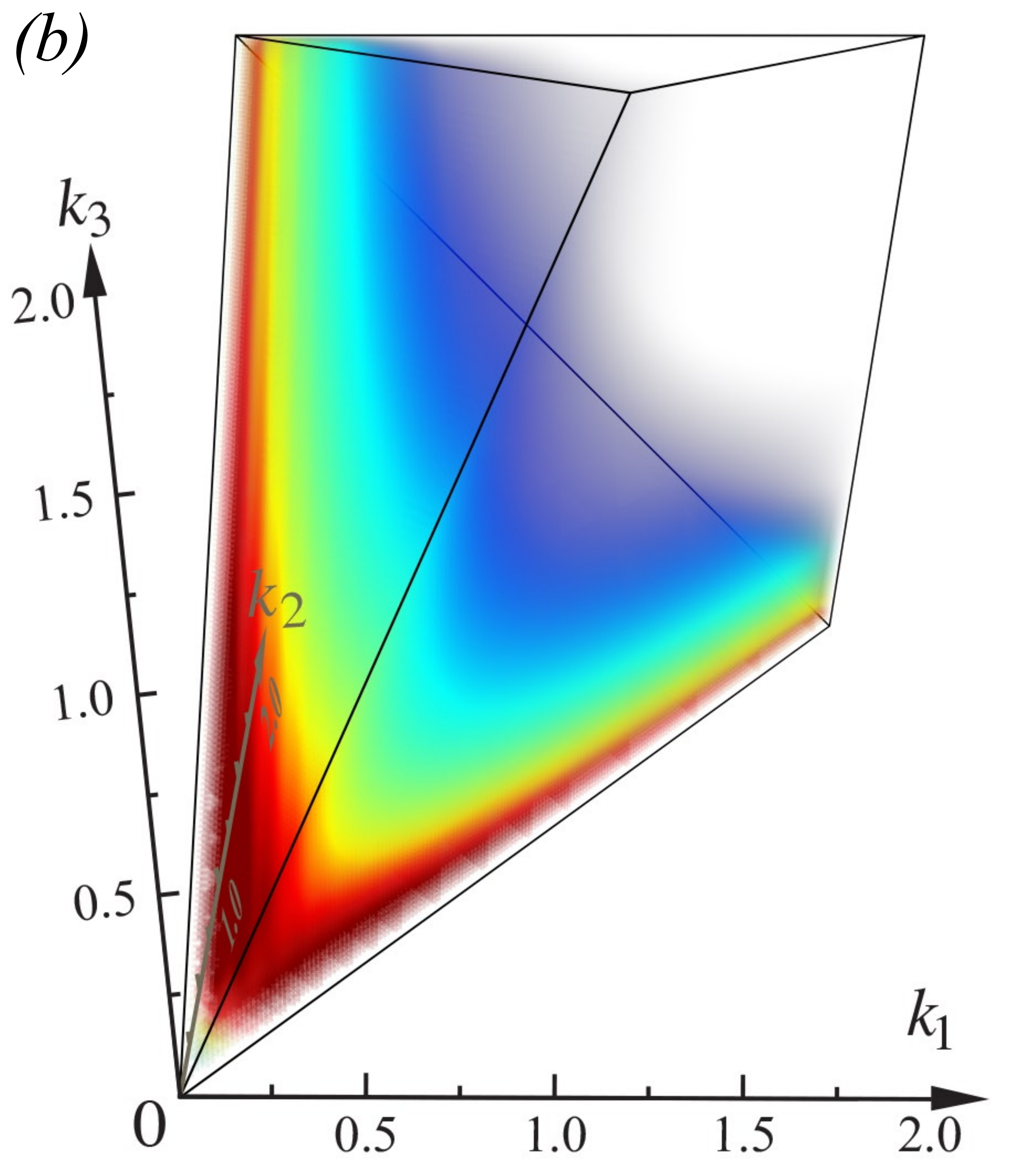}\\           
\includegraphics[width=2.4in]{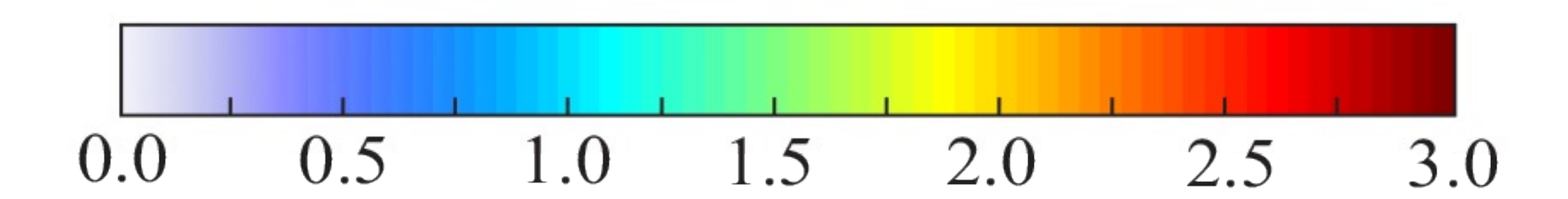}
\caption{ (a) The SN-weighted `constant' bispectrum of Eq.~(\ref{constantsh}) with a broadly equilateral signal shown together with (b) the `squeezed' or local model (Eq.~\ref{squeezed}) with high signal at the edges near $k_i \approx 0$ (shown at redshift $z=0$). Note that the plotted `constant' bispectrum does not have a constant cross-sectional shape because of the non-uniform signal-to-noise weighting (Eq.~\ref{SNweight}) particularly near the edges; here $S^\text{const.}$ in Eq.~(\ref{constantsh}) is multiplied by $f(K) = K^{3}$ (the colour scale  is normalised).  }
\label{bispectrum_shapes}
\end{center}
\end{figure}

\begin{figure}[tb]
\begin{center}
\includegraphics[width=2.4in]{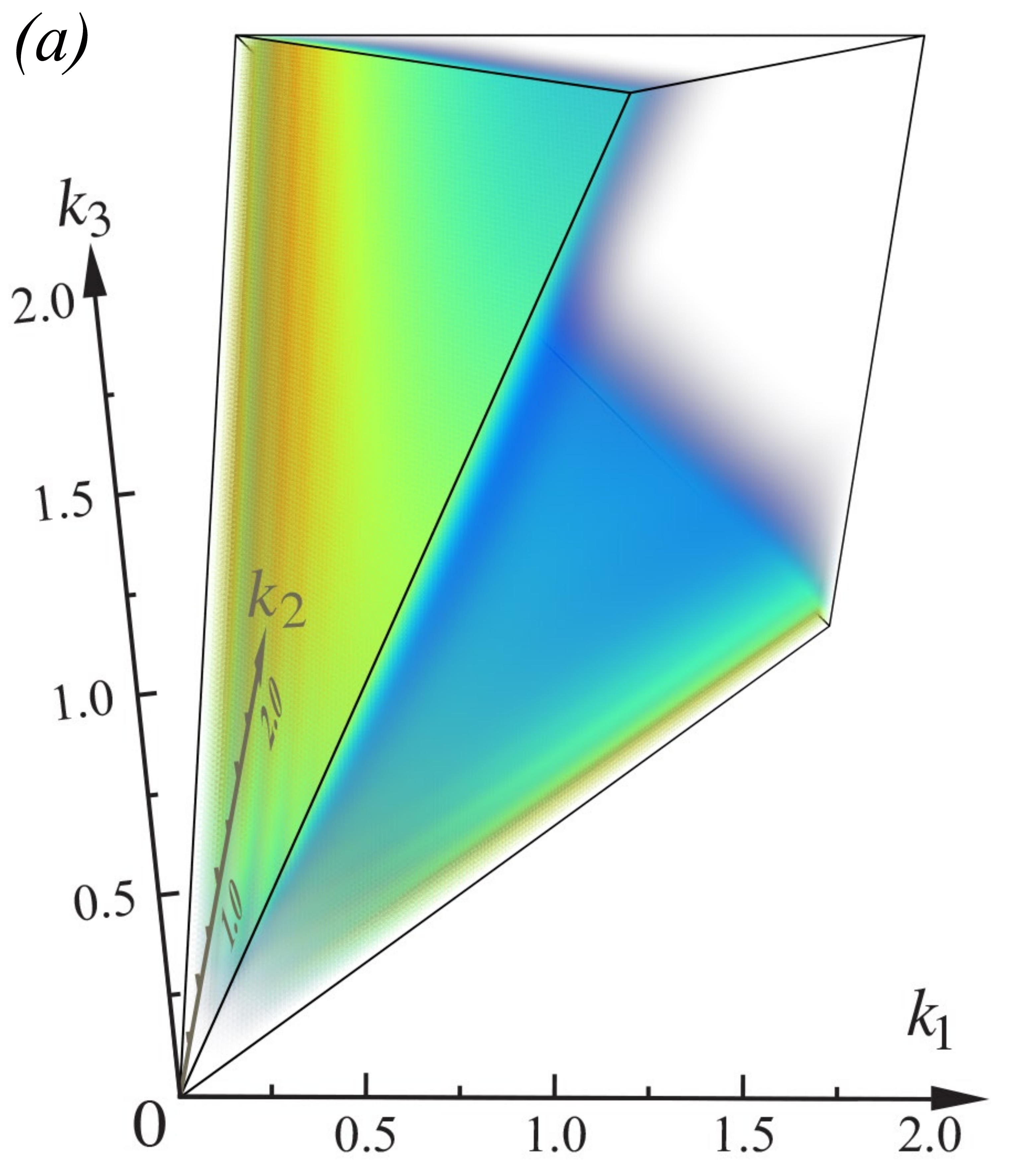}\\
\includegraphics[width=2.4in]{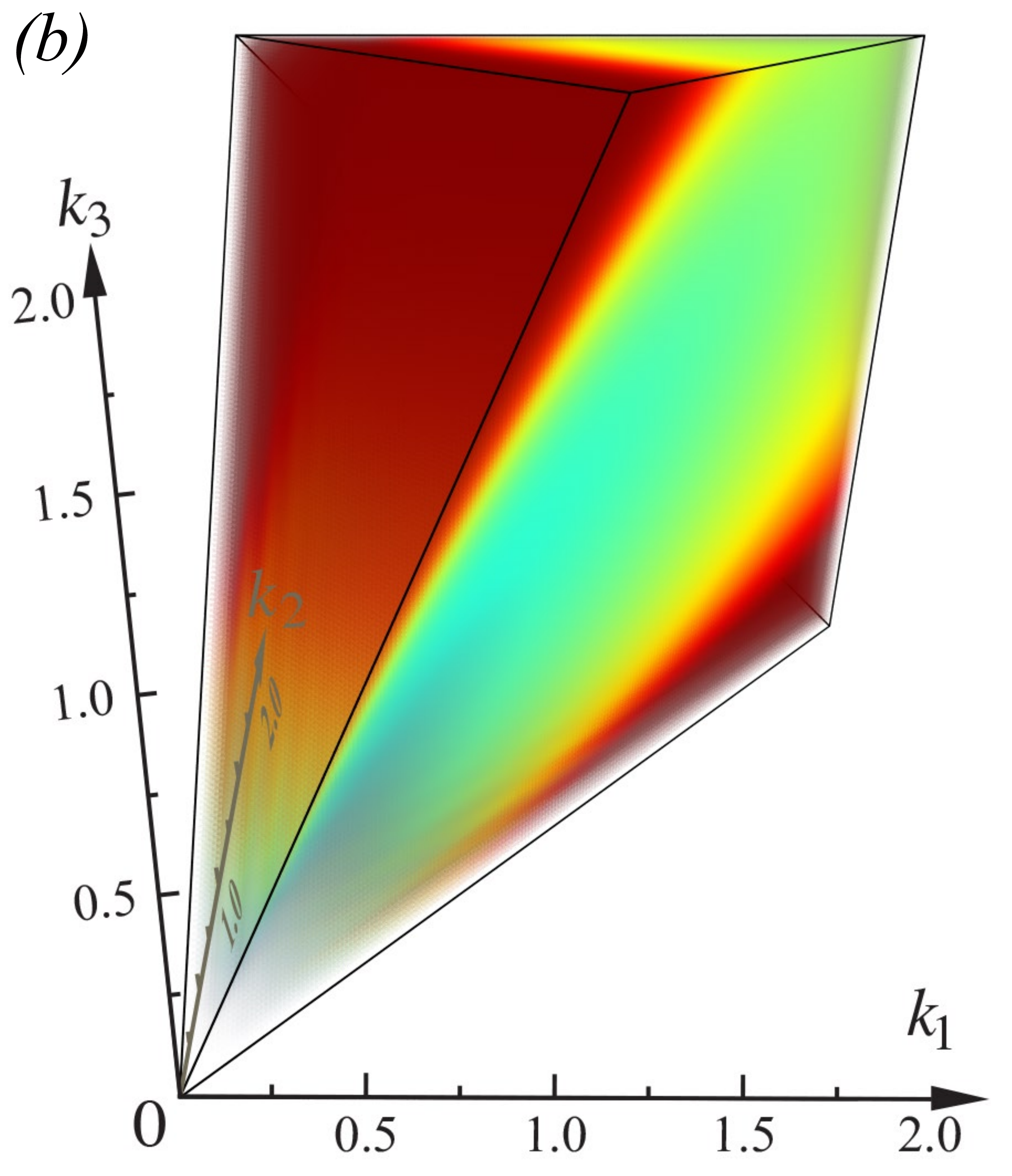}\\
\includegraphics[width=2.4in]{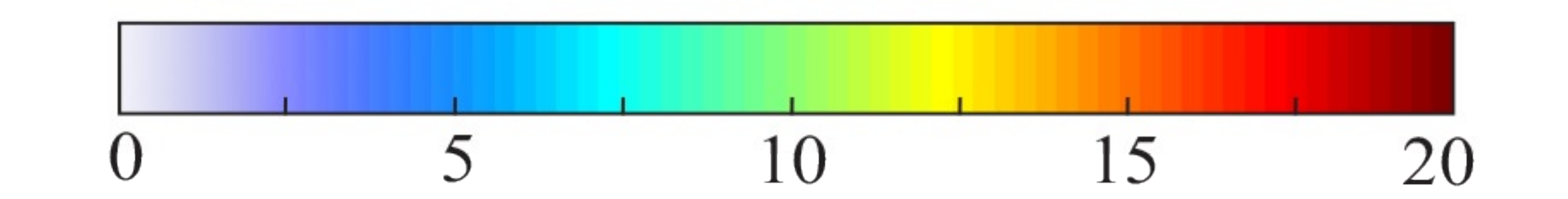}
\caption{Flattened shapes: (a) The SN-weighted tree-level bispectrum of Eq.~(\ref{stree}) compared with (b) the nonlinear tree-level model (Eq.~\ref{streeNL}), both shown at redshift $z=2$. Note that this flattened shape is dominated by signal  on the outer tetrapyd face (front left) where $k_1+k_2\approx k_3$ (see Fig.~\ref{tetrapyd_split} for the geometry). The nonlinear tree-level amplitude is substantially higher than the tree-level, but they share an excellent  binned shape correlation (Eq.~\ref{shapecorbin}), which always remains above 99\%. }
\label{bispectrum_tree}
\end{center}
\end{figure}

The three basic separable bispectrum shape functions $S(k_1,k_2,k_3)$ we shall employ are the constant shape \cite{PhysRevD.80.043510}, the  squeezed (or local) shape \cite{1475-7516-2006-05-004, 1475-7516-2009-05-018, 2010AdAst2010E..72C} and the tree-level (or flattened) shape from standard perturbation theory discussed earlier.  These three functions are essentially weighting functions for specific triangular configurations, that is, constant treats all triangles equally across the tetrapyd, squeezed favours those along the edges, and flattened those near the faces, as illustrated in Fig.~\ref{tetrapyd_split} (qualitatively encompassing the commonly discussed equilateral, local and orthogonal shapes respectively).  The constant shape is simply given by 
\begin{equation}
S^{\text{const}}(k_1,k_2,k_3)=1\,(\text{Mpc}/h)^6.
\label{constantsh}
\end{equation}
Physically, the constant bispectrum is produced by a random set of point sources, together with an appropriate scaling dependence $f(K)$.   It is our first approximation to the bispectrum of the nonlinear virialised end products of gravitational collapse assumed in halo models, with a SN-weighted version illustrated in Fig.~\ref{bispectrum_shapes}(a).    

The second shape is `squeezed'  and we shall define it as
\begin{multline}
S^{\text{squeez}} \left(k_1,k_2,k_3\right) = {\textstyle{\frac{1}{3}}} \left[P_{\text{lin}}(k_1)P_{\text{lin}}(k_2) \right. \\ \left. + P_{\text{lin}}(k_2)P_{\text{lin}}(k_3)+P_{\text{lin}}(k_3)P_{\text{lin}}(k_1) \right] \, ,
\label{squeezed}
\end{multline}
which incorporates the scale dependence of the transfer functions within the linear power spectrum $P_{\text{lin}}(k)$.  It is illustrated in Fig.~\ref{bispectrum_shapes}(b).  This squeezed shape is motivated by `local' non-Gaussianity in which perturbation fields are simply squared, and where the leading contribution has a large wavelength mode affecting nonlinearity on small scales (i.e., for `squeezed' triangles with $k_1 \ll k_2,k_3$).  However, Eq.~(\ref{squeezed}) regularises  the related scale-invariant primordial local shape,
\begin{equation}
S^{\text{local}} \left(k_1,k_2,k_3\right) = \frac{1}{3}\left(\frac{k_1^2}{k_2k_3}+\frac{k_2^2}{k_3k_1}+\frac{k_3^2}{k_1k_2}\right)\,,
\label{squeezed1}
\end{equation}
which behaves poorly because it diverges for very squeezed triangles.

Finally, the third flattened shape is the tree-level gravitational bispectrum given by \cite{Fry1984}
\begin{multline}
S^{\text{tree}} (k_1,k_2,k_3)=2P_{\text{lin}}(k_1)P_{\text{lin}}(k_2)F_2^{\left(s\right)} (\textbf{k}_1,\textbf{k}_2) \\ + 2 \text{ perms.} \, ,
\label{stree}
\end{multline}
where the kernel $F_2^{(s)}$ can be expressed as:
\begin{equation}
F_2^{(s)}(\textbf{q}_1,\textbf{q}_2)=\frac{5}{7}+\frac{1}{2}\frac{\textbf{q}_1 \cdot \textbf{q}_2}{q_1q_2}\left(\frac{q_1}{q_2}+\frac{q_2}{q_1}\right)+\frac{2}{7}\frac{(\textbf{q}_1 \cdot \textbf{q}_2)^2}{q_1^2q_2^2}\, ,
\label{f2s}
\end{equation}
which, although not immediately apparent, is also a separable shape of the form of Eq.~(\ref{separable_shape}).  Eq.~(\ref{f2s}) represents the leading-order gravitational non-Gaussianity generated by nonlinear terms in the equations of motion.   As we shall see, the scaling dependence $f(K)$ in Eq.~(\ref{separable}) allows us to approximately incorporate higher-order perturbative corrections.   However, the actual gravitational bispectrum is more closely approximated if the tree-level shape (Eq.~\ref{stree}) is modified by employing the nonlinear power spectrum \cite{Scoccimarro21082001}, given by the \textsc{Halofit} method \citep{2003MNRAS.341.1311S, 2012ApJ...761..152T}:
\begin{multline}
S^{\text{treeNL}} (k_1,k_2,k_3)=2P_{\text{NL}}(k_1)P_{\text{NL}}(k_2)F_2^{\left(s\right)} (\textbf{k}_1,\textbf{k}_2) \\ + 2 \text{ perms.} \, .
\label{streeNL}
\end{multline}
For this reason, we will generally employ this improved flattened bispectrum as the third shape in our subsequent modelling.  Both the tree-level (Eq.~\ref{stree}) and the nonlinear tree-level (Eq.~\ref{streeNL}) shapes are illustrated in Fig.~\ref{bispectrum_tree}.

\subsection{Scale-dependent or `sliced' correlators}

Having given the key shapes $S^i$ that we will use to describe gravitational non-Gaussianity using the separable ansatz Eq.~(\ref{separable}), we must also define a scale-dependent correlator that can be used to test the accuracy of this approximation.   
To determine this we need a more `localised', binned (or `sliced') correlator, which only integrates over the transverse degrees of freedom on the $K=\hbox{constant}$ surfaces,  modifying Eq.~(\ref{shapeprod}) to have the restricted domain of integration,
\begin{multline}
\label{shapeprod2}
\langle B_i, B_j \rangle^S_K \equiv \\ \frac{V}{\pi}\int_{\Delta\mathcal{V}_B}dV_k\, \frac{k_1k_2k_3 B_i(k_1,k_2,k_3)B_j(k_1,k_2,k_3)}{ P_\delta(k_1)P_\delta(k_2)P_\delta(k_3)} \, ,
\end{multline}such that the integral is now evaluated in a specific thin slice of the tetrahedron with 
$$K<k_1+k_2+k_3<K+\Delta K\,,$$ 
and where the index $S$ denotes slice. Substituting the localised inner product definitions in the correlators (Eqs.~\ref{shapecor}-\ref{totalcor}), this allows us to define the \emph{sliced} correlators $\mathcal{S}^S$, $\mathcal{A}^S$ and $\mathcal{T}^S$; for example, the binned shape correlator becomes 
\begin{equation}
\label{shapecorbin}
\mathcal{S}^S(K) \equiv \frac{\langle B_i, B_j \rangle^S_K}{\sqrt{\langle B_i, B_i \rangle^S_K \langle B_j, B_j \rangle^S_K}} \, .
\end{equation}
Importantly, if we find a good binned shape correlation $S^S(K)\approx 1$ between our target model (or simulation) and the canonical shapes above (Eqs.~\ref{constantsh}, \ref{squeezed}, \ref{stree}), then we can use the binned amplitude correlation  $\mathcal{T}^S$ to determine  the overall scale-dependence $f(K)$ in our separable ansatz of Eq.~(\ref{separable}).   Later in Sec.~\ref{sec:benchmark} we will combine these in a ``three-shape benchmark'' model and establish that it can achieve an excellent fit to simulations, thus dramatically reducing the number of degrees of freedom required to accurately describe the matter bispectrum.

\section{Perturbation theory for large-scale structure}
\label{sec:pert}
The amplitude of the matter density fluctuations in the Universe $\delta$ is small at early times and on large scales, so that $\delta \ll 1$ and linear dynamics suffices for an accurate modelling in this regime. At later times and on smaller scales, perturbations grow under gravity, making linear theory increasingly inaccurate. Various methods exist that can extend the range of validity of the model, accurately describing the large-scale structure to smaller scales and later times than linear theory.

A first possible approach, which we consider in this section, is to extend linear theory perturbatively, by expanding the evolution equations to higher order. This leads directly to standard (Eulerian) perturbation theory (SPT), which we review in Sec.~\ref{sec:SPT}; we next summarise more recent developments, which extend the range of validity by improving the SPT expansion convergence, removing divergences, and adding counterterms. The methods we consider are effective field theory (EFT) in Sec.~\ref{sec:EFT}, renormalised (Eulerian) perturbation theory (RPT) in Sec.~\ref{sec:RPT}, and resummed Lagrangian perturbation theory (RLPT) in Sec.~\ref{sec:RLPT}. For each method, we provide a more complete review in the Appendices~\ref{sec:SPT_full}, \ref{sec:EFT_full}, \ref{sec:RPT_full}, \ref{sec:RLPT_full} respectively.
We discuss possible extensions to two loops in Sec.~\ref{sec:twoloops} and Appendix \ref{app:twoloops}.
We finally discuss the shapes of the perturbation theory bispectra in Sec.~\ref{sec:shapespert}.
  
\subsection{Standard perturbation theory}
\label{sec:SPT}

Eulerian standard perturbation theory is derived by expanding
 the evolution equations for the dark matter density and velocity fields 
as a series of the linearly evolved density field $\delta_1$. 
In analogy with field theory, the resulting expansion for the power spectrum and bispectrum can be grouped to loop orders according to the number of $\delta_1$'s involved.
 We present here in the following the expressions for the SPT matter power spectrum and bispectrum, whose derivation is summarised in Appendix~\ref{sec:SPT_full}; see also Ref.~\cite{Bernardeau20021} for a comprehensive review.

The tree-level (zero-loop) power spectrum is simply given by the linear power spectrum:
\begin{equation}
\label{p0loop}
P_{\text{tree}}^{\mathrm{SPT}} \left(k,z\right)=P_{11}\left(k,z\right)=D^2\left(z\right)P_{\text{lin}}\left(k\right) \, ,
\end{equation}
where $D (z)$ is the linear growth function normalised to one today.
This can be evaluated numerically by evolving the primordial fluctuations through the Boltzmann equations through codes such as \textsc{Camb} \cite{Lewis:1999bs}. The one-loop contribution can be obtained from two diagrams and has the following form  \cite{1994ApJ...431..495J}:
\begin{equation}
P_{\text{1-loop}}^{\mathrm{SPT}} \left(k,z\right)=P_{13}\left(k,z\right)+P_{22}\left(k,z\right) \, ,
\end{equation}
where the two contributions have the following expressions:
\begin{align}
\label{p13}
P_{13}\left(k,z\right) &= D^4\left(z\right)\int \frac{d^3q}{\left(2\pi\right)^3} \, 6 \, P_{\text{lin}}\left(k\right)P_{\text{lin}}\left(q\right)F_3^{\left(s\right)}\left(\textbf{k},\textbf{q},-\textbf{q}\right) \\
P_{22}\left(k,z\right) &= D^4\left(z\right)\int \frac{d^3q}{\left(2\pi\right)^3} \, 2 \, P_{\text{lin}}\left(q\right)P_{\text{lin}}\left(|\textbf{k}-\textbf{q}|\right)  \nonumber \\
&~~~~ \times\left[F_2^{\left(s\right)}\left(\textbf{q},\textbf{k}-\textbf{q}\right)\right]^2 \, ,
\label{p22}
\end{align}
and where the kernels $F_n^{(s)}$ are defined in Appendix~\ref{sec:SPT_full}.

The tree-level bispectrum has the following expression \cite{Fry1984}:
\begin{multline}
B_{\text{tree}}^{\mathrm{SPT}} (k_1,k_2,k_3, z)= 2 D^4(z) P_{\text{lin}} (k_1) P_{\text{lin}} (k_2) F_2^{(s)}(\textbf{k}_1,\textbf{k}_2)  \\
+ \text{2 perms.} \, 
\label{btree}
\end{multline}
In order to improve the accuracy of the tree-level bispectrum, Ref.~\cite{Scoccimarro21082001} proposed simply replacing the linear power spectrum in the tree-level formula with the nonlinear power spectrum  estimated e.g. with the \textsc{Halofit} method \citep{2003MNRAS.341.1311S, 2012ApJ...761..152T}. This heuristically extends the range of validity of the model, and is what we call `nonlinear tree level' bispectrum (see Eq.~\ref{streeNL}).
Ref.~\cite{Scoccimarro21082001} showed that a further improvement can be achieved if, in addition to using the nonlinear power spectrum, the $F_2$ kernel is replaced with a modified version $F_2^{\text{eff}}$, which includes six free parameters that are fit to $N$-body simulations. Later, this method was extended by Ref.~\cite{GilMarin2012} by adding three extra parameters, as described in Appendix~\ref{sec:9params}, and re-calibrated on more precise $N$-body data over an extended range. This is what we indicate as `nine-parameter fit' model in the discussion below.

For the one-loop bispectrum, there are four diagrams that can be drawn \cite{0004-637X-496-2-586}:
\begin{equation}
B_{\text{1-loop}}^{\mathrm{SPT}} =B_{222}+B_{321}^{(I)}+B_{321}^{(II)}+B_{411} \, .
\end{equation}
These have the following expressions:
\begin{align}
\label{b222}
B_{222}\left(k_1,k_2,k_3, z\right)&=8D^6\left(z\right)\int_{\textbf{q}}P_{\text{lin}}\left(q\right)P_{\text{lin}}\left(|\textbf{k}_2-\textbf{q}|\right)  \nonumber \\
\times P_{\text{lin}}&\left(|\textbf{k}_3+\textbf{q}|\right)F_2^{\left(s\right)}\left(-\textbf{q},\textbf{k}_3+\textbf{q}\right) \nonumber \\
\times F_2^{\left(s\right)}&\left(\textbf{k}_3+\textbf{q},\textbf{k}_2-\textbf{q}\right)F_2^{\left(s\right)}\left(\textbf{k}_2-\textbf{q},\textbf{q}\right)
\end{align}
\begin{align}
\label{b321i}
B_{321}^{(I)}\left(k_1,k_2,k_3, z\right) &= 6D^6\left(z\right)P_{\text{lin}}\left(k_3\right)\int_{\textbf{q}}P_{\text{lin}}\left(|\textbf{k}_2-\textbf{q}|\right)  \nonumber \\
\times P_{\text{lin}}\left(q\right)F_3^{\left(s\right)}&\left(-\textbf{q},-\textbf{k}_2+\textbf{q},-\textbf{k}_3\right)F_2^{\left(s\right)}\left(\textbf{k}_2-\textbf{q},\textbf{q}\right) \nonumber \\
&+ \text{5 perms.}
\end{align}
\begin{align}
\label{b321ii}
B_{321}^{(II)}(k_1,k_2,k_3, z) &= 6 D^6(z) P_{\text{lin}}(k_2) P_{\text{lin}}(k_3)F_2^{(s)}(\textbf{k}_2,\textbf{k}_3)  \nonumber \\
\times \int_{\textbf{q}}P_{\text{lin}}\left(q\right)F_3^{\left(s\right)}&\left(\textbf{k}_3,\textbf{q},-\textbf{q}\right) + \text{5 perms.}
\end{align}
\begin{align}
\label{b411}
B_{411}\left(k_1,k_2,k_3, z\right)&=12D^6\left(z\right)P_{\text{lin}}\left(k_2\right)P_{\text{lin}}\left(k_3\right)  \nonumber \\
\times \int_{\textbf{q}}P_{\text{lin}}\left(q\right)F_4^{\left(s\right)}&\left(\textbf{q},-\textbf{q},-\textbf{k}_2,-\textbf{k}_3\right) + \text{2 perms.} \, ,
\end{align}
where $\int_{\textbf{q}} \equiv \int \frac{d^3q}{\left(2\pi\right)^3}$. The numerical integration of the expressions above is non-trivial, and we discuss the necessary procedures in Appendix~\ref{sec:SPT_full}.

It is known \cite{Bernardeau20021} that SPT only succeeds in extending the range of validity of linear theory by a small amount at low redshift, while it overpredicts the power seen in $N$-body simulations on smaller scales. This is because the SPT loop corrections are integrated over all $k$ modes, including scales that are not in the linear regime, which are actually suppressed in reality compared to SPT \cite{Seljak2015}.
Furthermore, the convergence of the SPT expansion is problematic, as it relies on the near cancellation of large positive and negative terms, so that increasing the loop order does not necessarily improve the accuracy of the expansion, especially at low redshift.

\subsection{Effective field theory}
\label{sec:EFT}
Some of the problems of SPT mentioned in the previous section are addressed by  the effective field theory of LSS (EFTofLSS, or simply EFT).
 At nonlinear level, the Fourier modes do not evolve independently any more, and hence small-scale fluctuations can influence much larger scales. 
The basic assumption of EFT is to introduce additional free parameters that describe the effect of non-perturbative small-scale physics onto the larger observable scales.
The SPT expansion can only be expected to work when the density contrast is small, $\delta \ll 1$, so that its range of validity at low redshift becomes increasingly limited. Nevertheless, even when this condition is not satisfied, the gravitational potential is still small and can be used to produce a valid perturbative expansion.
 Based on this fact, EFTofLSS has been developed in Refs.~\cite{1475-7516-2012-07-051, Carrasco2012}.

This method consists of adding to the equations of motion an effective stress-energy tensor $\tau_{\mu \nu}$, induced by short wavelength modes. This has the effect of adding corrections to the fluid equations, with terms corresponding to the speed of sound, viscosity and stochastic pressure. 
As we describe in more detail in Appendix~\ref{sec:EFT_full}, the EFT method leads to additional contributions to the SPT matter power spectrum and bispectrum, with free parameters to be calibrated with $N$-body simulations.

At one loop, one term is added to the SPT matter power spectrum \cite{1475-7516-2014-07-057}:
\begin{equation}
P^{\text{EFT}} (k, z) = P^{\text{SPT}} (k, z) + P_{c_s} (k, z) \, ,
\end{equation}
where
\begin{equation} \label{eq:eftP}
P_{c_s} (k, z) = -2\left(2\pi\right)c_{s(1)}^2\frac{k^2}{k_{\text{NL}}^2}D^{2+\zeta}(z)P_{\text{lin}}\left(k\right) \, .
\end{equation}
Here the parameters $c_{s(1)}$ and $\zeta$ are fit to $N$-body simulations, and $k_{\text{NL}}$ is defined as the scale where the perturbative ansatz ($\delta \ll 1$) breaks down.

Likewise, one term is added to the one-loop SPT  matter bispectrum:
\begin{equation}
B^{\text{EFT}} (k_1,k_2,k_3, z) = B^{\text{SPT}} (k_1,k_2,k_3, z) + B_{c_s} (k_1,k_2,k_3, z) \, ,
\end{equation}
where
\begin{multline}  \label{eq:eftB}
B_{c_s}\left(k_1,k_2,k_3, z\right)=  \\
[2P_{\text{lin}}\left(k_1\right)P_{\text{lin}}\left(k_2\right) \tilde{F}_2^{\left(s\right)}\left(\textbf{k}_1,\textbf{k}_2\right) +\text{2 perms.}]D(z)^{4+\zeta} \\
-[2\bar{c}_1k_1^2P_{\text{lin}}\left(k_1\right)P_{\text{lin}}\left(k_2\right) F_2^{\left(s\right)}\left(\textbf{k}_1,\textbf{k}_2\right) + \text{5 perms.}]D(z)^{4+\zeta} \, .
\end{multline}
Here, $\bar{c}_1=2\pi\frac{c_{s(1)}^2}{k_{\text{NL}}}$, and the sound speed parameter $c_{s(1)}$ is fixed at the power spectrum level only, so that the bispectrum includes no extra free parameters.  $\tilde{F}$ is given by Eq.~(\ref{eq:TildeF2}).

The additional EFT terms effectively subtract the excess power that is present in the SPT results, so that an accurate modelling can be achieved over an extended range of scales.

\subsection{Renormalised perturbation theory}
\label{sec:RPT}
The renormalised perturbation theory (RPT)  model has been developed in Refs.~\cite{PhysRevD.73.063519, PhysRevD.73.063520, PhysRevD.77.023533, Bernardeau2008, Crocce2012, PhysRevD.85.123519}. This method uses the formalism of the SPT and re-organises the infinite expansion differently using an idea from Ref.~\cite{NYAS:NYAS13}. 
As described in more detail in Appendix~\ref{sec:RPT_full}, this approach is based on the study of the nonlinear propagator connecting the initial with the evolved fields describing density and velocity perturbations. In this way, the perturbative expansion can be written as a series of the nonlinear propagator. This infinite series can be re-summed, yielding the RPT expressions for power spectrum and bispectrum at any number of loops. Compared to SPT, this method has the advantage that all the contributions involved are positive and the resummation of the propagator terms gives a well-defined perturbative expansion in the nonlinear regime. However, the expressions involved are complicated and the solutions are computationally demanding, requiring to solve numerically a set of integro-differential equations. Moreover, more than one loop is required to obtain an accurate result, even on mildly nonlinear scales. 

In order to solve these problems, Refs.~\cite{Bernardeau2008, Crocce2012} proposed a method that simplifies the calculation dramatically. The scheme is called \textsc{MPTbreeze} and in this formalism only the late-time propagator is calculated and hence no time integrations are required. As described in Appendix~\ref{sec:RPT_full},
the \textsc{MPTbreeze} power spectrum contributions can be expressed in terms of their SPT counterparts as follows:
\begin{align} \label{eq:RPT_P}
P_{\text{tree}}^{\text{MPTbreeze}}\left(k,z\right)&=P_{11}\left(k,z\right)\exp\left[2f(k)D^2(z)\right] \\
P_{\text{1-loop}}^{\text{MPTbreeze}}\left(k,z\right)&=P_{22}\left(k,z\right)\exp\left[2f(k)D^2(z)\right]  \, ,
\end{align}
where the function $f(k)$ is given in Eq.~(\ref{eq:fk}).

The bispectrum contributions can be treated in a similar manner \cite{PhysRevD.85.123519}, and the result up to one loop is given in terms of the SPT one-loop contributions (Eqs.~\ref{b222}, \ref{b321i}):
\begin{align} \label{eq:RPT_B}
B^{\text{MPTbreeze}}&\left(k_1,k_2,k_3,z\right)= \nonumber \\
&\left(B_{\text{tree}}^{\mathrm{SPT}} +B_{222}+B_{321}^I\right)\left(k_1,k_2,k_3,z\right) \times \nonumber \\
&\exp\left[\left(f(k_1)+f(k_2)+f(k_3)\right)D^2(z)\right] \, .
\end{align}

The main advantages of RPT and its \textsc{MPTbreeze} variant are that the expansion series becomes positive definite, so that no cancellation occurs and each successive term improves the range of validity of the theory; and the exponential prefactor term, which effectively suppresses the theory outside its range of validity, thus avoiding some of the SPT problems.

\subsection{Resummed Lagrangian perturbation theory}
\label{sec:RLPT}

Alternatively, perturbation theory can be derived as a function of the Lagrangian coordinates of the initial conditions.
As the observable statistical quantities (power spectra and bispectra) are always defined in the evolved (Eulerian) coordinates, Lagrangian perturbation theory (LPT) has to deal with the evolution of the displacement field $\mathbf{\Psi}$, which relates the two coordinate systems.

By expanding the evolved density and velocity perturbations as a series of $\mathbf{\Psi}$, it is possible to calculate perturbative predictions for power spectrum and bispectrum at any chosen order, although the calculations are complex \cite{Buchert1989,Moutarde1991,Buchert1992,Buchert1994,Bouchet1995,Ehlers1997,Tatekawa2004}.
A general drawback of LPT is that this method can not describe accurately the physics of shell crossing, as particles continue to stream according to their initial velocity; thus dark matter haloes never collapse, and LPT presents a power deficit on small scales.

 More recently Ref.~\cite{Matsubara2008} used the cumulant expansion theorem to obtain a simpler resummed expression for the polyspectra, called resummed Lagrangian perturbation theory (RLPT). This method yields a resummed series expansion similar to, but simpler than, RPT.

We summarise the LPT and RLPT methods in Appendix~\ref{sec:RLPT_full}; the final results are
the RLPT power spectrum \cite{Matsubara2008}
\begin{multline} \label{eq:RLPT_P2}
P^{\mathrm{RLPT}}(k) = \exp \left[ - \frac{k^2}{6 \pi^2} \int dp \, P_{\mathrm{lin}} (p) \right] \\ \times \left[P_{\mathrm{lin}} (k) + P^{\mathrm{SPT}}_{\text{1-loop}} (k)  + \frac{k^2}{6 \pi^2} P_{\mathrm{lin}} (k) \int dp \, P_{\mathrm{lin}} (p)  \right] \, ,
\end{multline}
where $P^{\mathrm{SPT}}_{\text{1-loop}} $ is the one-loop SPT term (without the tree-level term); and the bispectrum \cite{Rampf2012b}
\begin{multline} \label{eq:RLPT_B2}
B^{\mathrm{RLPT}}(k_1, k_2, k_3) = \exp \left[ -\frac{k_1^2 + k_2^2 + k_3^2}{12 \pi^2} \int dp \, P _{\mathrm{lin}} (p) \right] \\
\times \left[  B^{\mathrm{SPT}}_{\mathrm{tree}} + B^{\mathrm{SPT}}_{\text{1-loop}} + \frac{k_1^2 + k_2^2 + k_3^2}{12 \pi^2} B^{\mathrm{SPT}}_{\mathrm{tree}}  \int dp \, P _{\mathrm{lin}} (p)  \right] \, .
\end{multline} 
From Eqs.~(\ref{eq:RLPT_P2}, \ref{eq:RLPT_B2}) it is evident that the RLPT power spectrum and bispectrum reduce back to SPT if the exponential prefactor is expanded to first order.
Furthermore, this prefactor is similar to the RPT results: in both cases, the theory decays rapidly to zero outside its range of validity. Thus this method is not expected to yield realistic predictions in the fully nonlinear regime where the exponential cut-off dominates, but only on quasi-linear scales.

\begin{figure*}[tb]
\begin{center}
\includegraphics[width=0.32\linewidth]{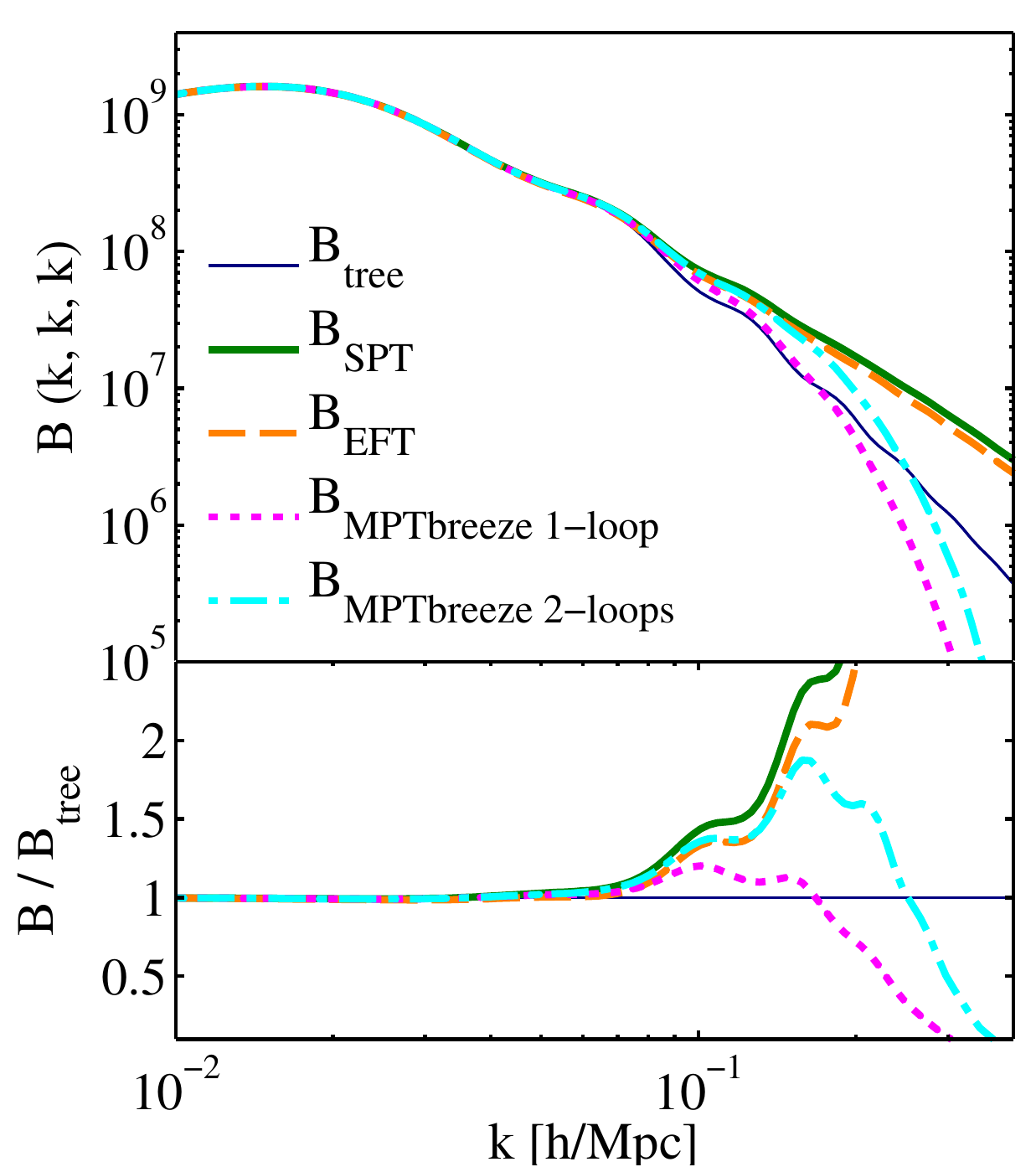} 
\includegraphics[width=0.32\linewidth]{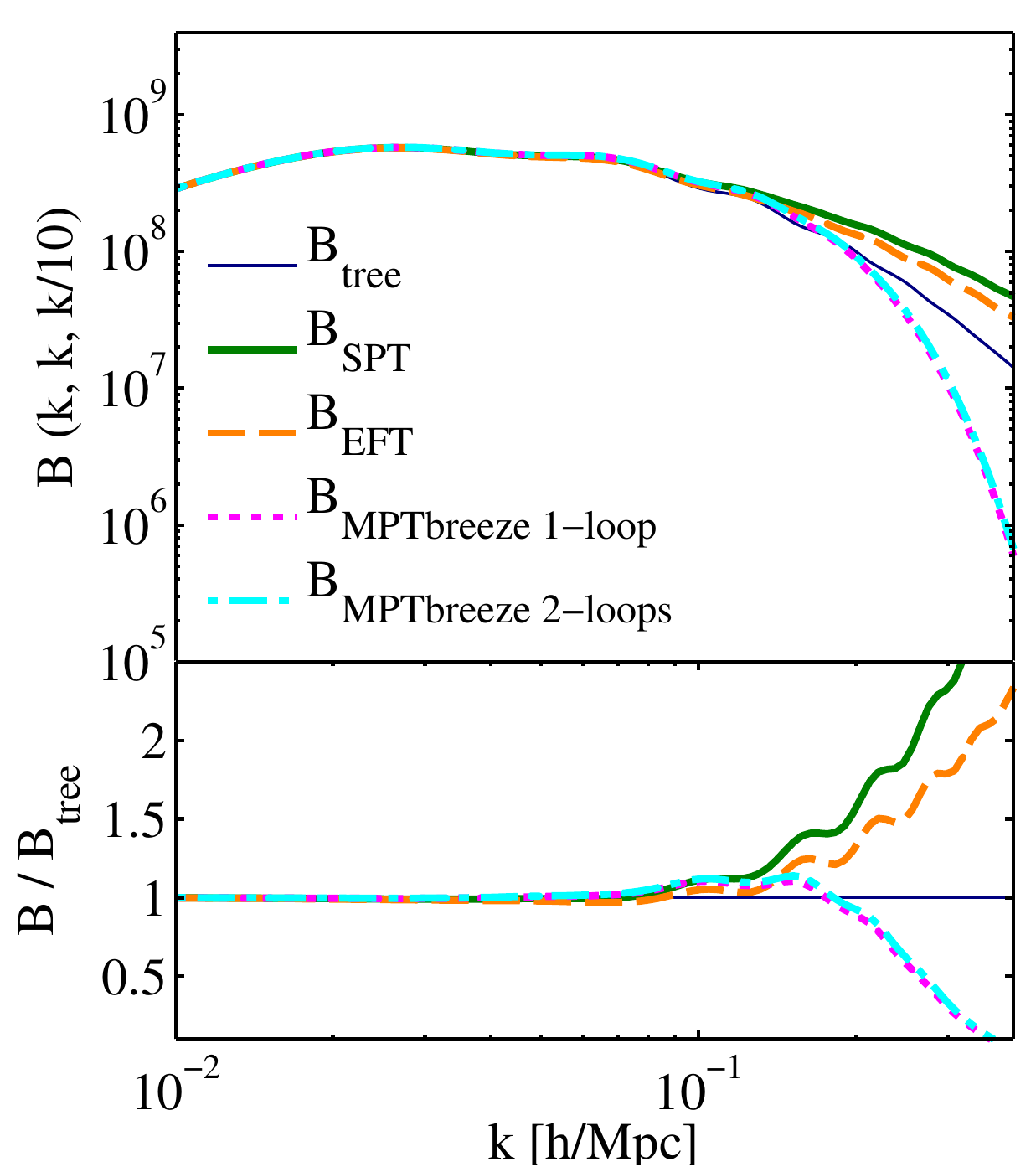} 
\includegraphics[width=0.32\linewidth]{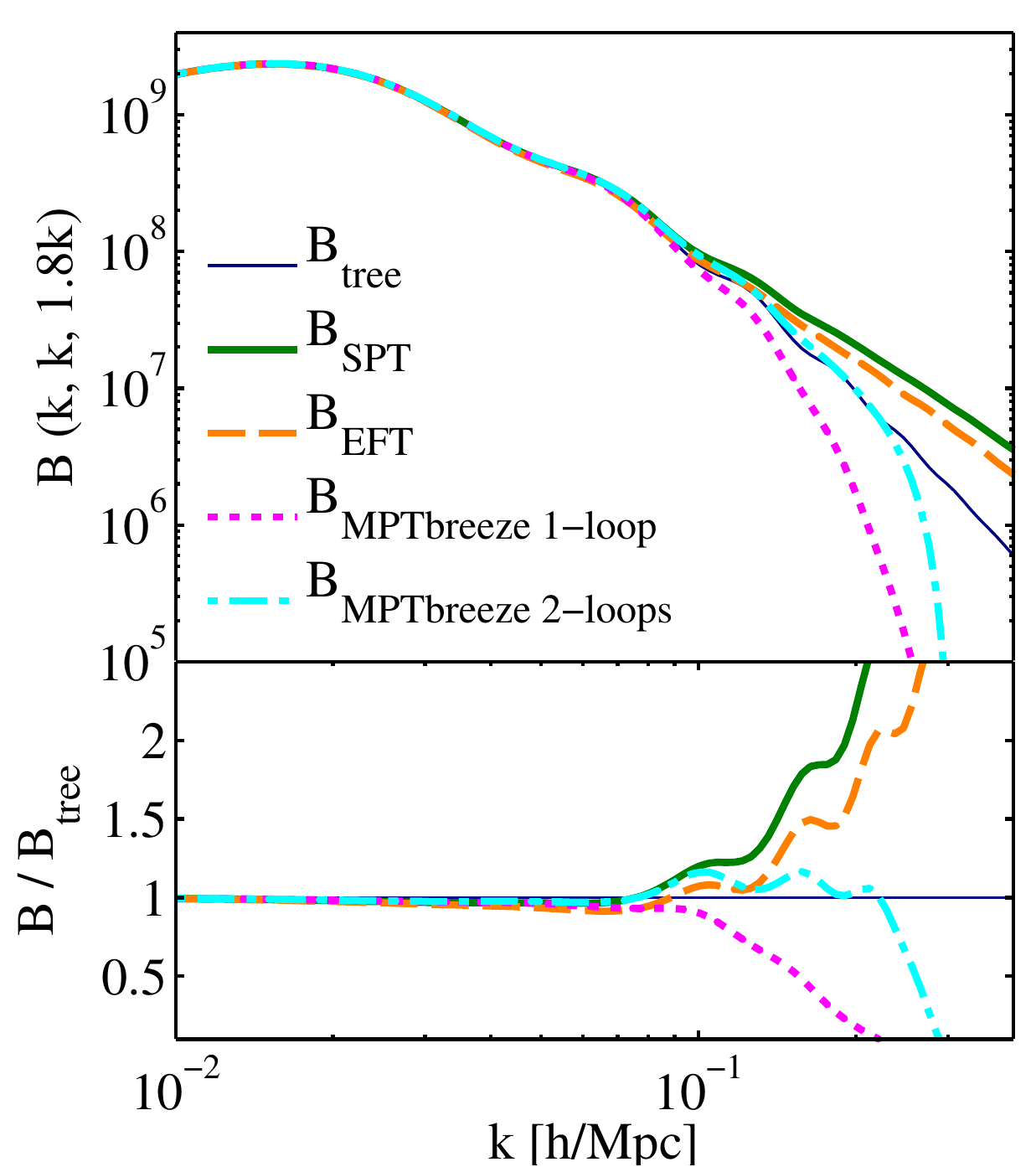}
\caption{Equilateral (left), squeezed (middle) and flattened bispectra (right) from perturbation theories at $z=0$. We show the theoretical predictions of the tree-level bispectrum, SPT, EFT and the one- and two-loop \textsc{MPTbreeze} bispectra. For this last model, we observe that the wavenumber at which the theory starts decaying increases significantly when adding the two-loop terms in the case of the equilateral and flattened configurations, closely following the EFT model down to smaller scales, while for the squeezed configuration the improvement is negligible.  Bispectra are plotted in units of $(\text{Mpc}/h)^6$ throughout the paper.}
\label{MPT2}
\end{center}
\end{figure*}

\subsection{Going to two-loops and estimating perturbation residuals}
\label{sec:twoloops}
So far, most of the LSS perturbation theory work has considered up to two loops in the power spectrum \cite{PhysRevD.80.043531, 1475-7516-2014-07-057} and up to one loop in the bispectrum. The reason has been mainly computational, but there is also a theoretical constraint: perturbation theories are expected only to work close to the linear regime, as they rely on perturbing around small density fluctuations. Even in the EFTofLSS approach, which allows to significantly extend the range of validity of the nonlinear power spectrum over SPT \cite{1475-7516-2014-07-057}, it is not possible to  push the model further to scales associated with  dark matter haloes: in the fully 
nonlinear regime only phenomenological halo models and fits to $N$-body simulations can be used.  Nevertheless, it is interesting to investigate how far into intermediate nonlinear scales perturbation theory can be extrapolated.

Since the bispectrum is a  three-dimensional quantity, its expansion at two loops requires the computation of challenging six-dimensional integrals. Moreover, the integrals involved have divergences that cancel between different terms, so these must be identified and eliminated before numerical computation to ensure convergence. Fortunately, in the \textsc{MPTbreeze} formalism, the number of terms that appear is reduced; as described  in Appendix~\ref{app:twoloops}, we have therefore calculated analytically the terms involved, and we have eliminated the divergences based on the ideas developed in Refs.~\cite{1475-7516-2014-07-056, baldauf, Angulo:2014tfa}. Unfortunately, due to the complexity of the integrals, we have not been able to perform the full three-dimensional bispectrum calculation as in the other one-loop cases. The analytic divergence-free expressions obtained are presented in Appendix~\ref{app:twoloops}, while here we show in Fig.~\ref{MPT2} three triangle configurations: equilateral, squeezed and flattened, also with a comparison between all the tree-level and one-loop perturbative methods at $z = 0$. The EFT bispectrum is expected to be accurate up to higher $k$ than one-loop RPT, as discussed by Ref.~\cite{Angulo:2014tfa} and as shown in Sec.~\ref{sec:comparison} below; therefore, knowing that the RPT approach is a convergent expansion with the precision increasing as the number of loops is increased, we can estimate the range of validity of the one-loop and two-loop \textsc{MPTbreeze} results by comparison with EFT.
In Fig.~\ref{MPT2}, we observe that  the two-loop  \textsc{MPTbreeze} bispectrum closely follows the EFT prediction for an extra $0.04 \, h/\text{Mpc}$ more than the one-loop bispectrum in the equilateral and flattened cases, while the squeezed limit shows a more modest improvement. 
It is therefore clear that extending  \textsc{MPTbreeze} to two loops in the quantitative comparisons of Sec.~\ref{sec:comparison} would significantly improve its range of validity, but we decide not to pursue this for consistency with the other PT methods, and because of the huge analytic and numerical challenges which seem to be entailed.

Controlled perturbative expansions become increasingly accurate as the number of loops is increased, so a criterion for determining where perturbation theory at a given order breaks down is to calculate the next-order contribution and find where they become significant. In Table \ref{check_tspt} we show the value of the wavenumber where the higher order expansion deviates by more than 10\% (20\%) from the lower order. Hence, we compare SPT with tree-level, the \textsc{MPTbreeze} at one loop to the tree-level and the two-loop \textsc{MPTbreeze} bispectrum to its one-loop counterpart. For completeness, we also determine the effect of the counter-term in EFT which corrects SPT. At one-loop we evaluate deviations with the total correlator $\mathcal{T}$, but at two-loop order we determine the worst case amongst the three limiting configurations evaluated.   
\begin{table}
\begin{center}
\caption{Domain of validity for perturbation theory:  wavenumber $k_{\max}^*$ where the two perturbative expansions being compared show relative deviations greater than 10\% (20\%).}
\medskip
\begin{tabular}{  c c c c }
\toprule
\multicolumn{4}{ c }{Perturbation theories}\\
\colrule
  Threshold $10\%$ ($20\%$)          & \multicolumn{3}{c}{$k_{\max}^*\,[h/\text{Mpc}]$} \\ 
\colrule
 Theory                      &  $z = 0 $  &  $z = 1 $  &  $z = 2 $ \\
\colrule
 SPT/Tree-level              &  0.07 (0.08) & 0.08 (0.12)  &  0.12 (0.14)  \\
 EFT/SPT                     &  0.12 (0.41) & 0.41 (0.93)  &  0.77 (1.52)\\
 \hline
 RPT 1-loop/Tree             &  0.08 (0.10) &  0.09 (0.14)  &  0.13 (0.20) \\
 RPT 2-loops/1-loop          &  0.09 (0.11) &  0.13 (0.16)  &  0.19 (0.23)\\ 
\botrule
 \end{tabular}
\label{check_tspt}
\end{center}
\end{table}
Table~\ref{check_tspt}  indicates that the tree-level bispectrum is in fact valid only for small wavenumbers $k \lesssim 0.1 \, h/$Mpc at $z=0$ and $k\lesssim 0.2 \, h/$Mpc at $z=2$, with one-loop contributions apparently offering only a small incremental improvement. However, the comparison of SPT results with the EFT controlled expansion indicates that it may be possible to extrapolate perturbative expansions considerably further.  As we shall see in Sec.~\ref{sec:comparison}, there is an unexpectedly good correspondence between some perturbative bispectra and the results of numerical simulations, going well beyond the thresholds estimated in Table~\ref{check_tspt}.

\begin{figure*}[t]
\begin{center}
\includegraphics[width=\linewidth]{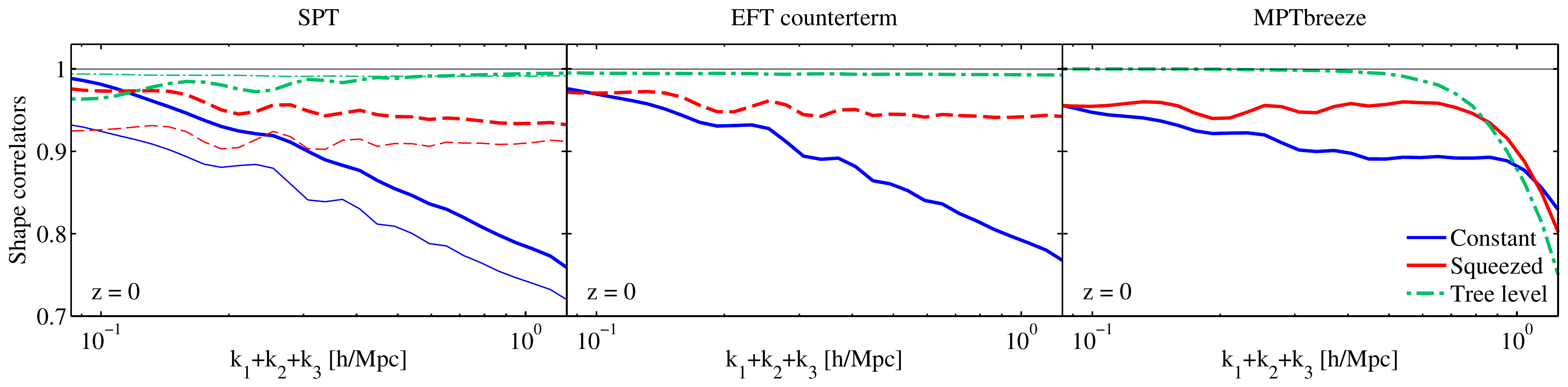}
\caption{Shapes of the perturbation theory bispectra. For each of the theoretical bispectra considered, we show the shape correlators in $k$ slices $\mathcal{S}^S$ (Eq.~\ref{shapeprod2}) with respect to the constant, squeezed and tree-level shapes (Eqs.~\ref{constantsh}-\ref{stree}). The left panel shows the one-loop SPT shape correlators, the central panel shows the EFT counterterm ($-B_{c_s}$) shapes, and the right panel refers to the \textsc{MPTbreeze} one-loop shapes. All panels refer to $z = 0$. In the case of SPT (left), the thick lines represent the sum of the positive terms of the one-loop expansion ($B_{222}$ and $B_{321}^{(I)}$), while the thin lines refer to the sum of the negative terms ($B_{321}^{(II)}$ and $B_{411}$).}
\label{comparatiepert}
\end{center}
\end{figure*}

\subsection{Shapes of perturbative bispectrum models}
\label{sec:shapespert}
We conclude this section by describing the shapes of the various terms appearing in the different perturbative approaches using the binned shape correlator $\mathcal{S}^S$, defined on $K=\hbox{const.}$ slices in Eq.~(\ref{shapeprod2}).  We determine $\mathcal{S}^S$ for each perturbative model against the tree-level, squeezed and constant shapes in Sec.~\ref{sec:shapefn}. The results of this comparison are illustrated in Fig.~\ref{comparatiepert}. In the SPT and EFT bispectra, the tree-level term is always present, and so inevitably the `flat' tree-level shape dominates the large-scale results. For this reason, we restrict our attention to an analysis of the one-loop SPT terms and EFT counterterms separately, in order to achieve a better understanding of the underlying shape corrections. This also simplifies the figures, because in this way there is no mixture of different powers of the growth factor and it is sufficient to test the shapes of these terms at $z=0$. The left panel of Fig.~\ref{comparatiepert} shows the shape correlators in scale-invariant slices of $K\equiv k_1+k_2+k_3=\text{constant}$ for the sum of the positive one-loop terms of SPT (thick lines) and the negative contributions (thin lines). The central panel represents the EFT counterterm for the tree level, $-B_{c_s}$, and the right panel shows the shapes of the \textsc{MPTbreeze} bispectrum.    Figure~\ref{comparatiepert} shows strong correlations with the tree-level shape in the range $0.1 \, h/$Mpc $<k<0.5\, h/$Mpc and beyond, with only the exponential cut-off in the MPTbreeze affecting the correlation. 

Since the tree-level shape correlator is so dominant with respect to the others, we conclude that the perturbative approaches are indistinguishable in shape from the tree-level shape Eq.~(\ref{stree}) in each scale-invariant $K$-bin.  This is for the relevant range of scales probed by this analysis, with the possible exception of some small deviations appearing in the one-loop SPT terms at small $k$.  Overall, Fig.~\ref{comparatiepert} implies these one-loop correction terms are not adding any qualitatively new shape degrees of freedom, thus perturbative methods can be well-approximated in terms of the tree-level shape using the separable ansatz:
\begin{equation}
\label{pertshape}
B_{\text{PT}}(k_1,k_2,k_3) = f(K) \, S^{\text{tree}}(k_1,k_2,k_3) \, ,
\end{equation}
where $K=k_1+k_2+k_3$ and $f(K)$ is an appropriate scale-dependent function defined in Eq.~(\ref{separable}). We will use this result in the construction of the phenomenological benchmark model in Sec.~\ref{sec:sims} below.

\section{Non-perturbative models of large-scale structure}
\label{sec:halo}
\subsection{Halo model basics}
We next extend the clustering modelling deeper into the nonlinear regime using the halo model of the large-scale structure \cite{Seljak11102000,MaFry2000,Peacock2000}. This framework is based on the assumption that all the matter in the Universe is concentrated into discrete regions called haloes.
As summarised in the review by Ref.~\cite{Cooray20021}, the matter power spectrum in this model is described by two contributions:
\begin{equation}
P (k, z) = P_{1h} (k, z) + P_{2h} (k, z) \, ,
\end{equation}
where the one- and two-halo terms describe contributions from dark matter particle pairs that reside in the same or in different haloes respectively, given by:
\begin{align}
\label{p1h}
\MoveEqLeft[3]  P_{1h} (k, z) = \int_0^{\infty} dm  \, n(m, z) \left( \frac {m}{\bar{\rho}} \right)^2 u^2 ( k|m , z) \, , \\
\label{p2h} 
\MoveEqLeft[3]  P_{2h} (k, z) = \int_0^{\infty} dm_1 \, n(m_1, z) \left( \frac{m_1}{\bar{\rho}} \right) u ( k|m_1 , z)  \\
\times{}&   \int_0^{\infty} dm_2  \, n ( m_2 , z) \left( \frac{m_2}{\bar{\rho}} \right) u ( k|m_2 , z) \, \nonumber \\
\times{}& \, P_{h} ( k|m_1,m_2 , z) \, . \nonumber
\end{align}

Here $\bar \rho$ is the mean density of the Universe today, and the one- and two-halo terms can be calculated once the following ingredients are specified: the halo mass function $n(m, z)$, the Fourier transform of the halo profile $u(k|m, z)$, and the halo power spectrum $P_{h} (k|m_1,m_2, z)$, which we describe in Appendix~\ref{sec:haloZutaten} below.

Likewise, the matter bispectrum can be expressed as a sum of three terms:
\begin{multline}
B (k_1,k_2,k_3, z) = B_{1h} (k_1,k_2,k_3, z)  \nonumber \\
+ B_{2h} (k_1,k_2,k_3, z) + B_{3h} (k_1,k_2,k_3, z) \, ,
\end{multline}
where the one-, two-, and three-halo contributions refer to dark matter particle triplets residing in one, two, or three haloes, given by:
\begin{align}
\label{b1h}
\MoveEqLeft[3] B_{1h} (k_1, k_2, k_3, z) = \nonumber \\
{}& \int_0^{\infty} dm \, n(m, z) \left(\frac{m}{\bar{\rho}}\right)^3 \prod_{i=1}^3 u(k_i|m, z) \, , 
\end{align}
\begin{align}
\label{b2h}
\MoveEqLeft[3] B_{2h}(k_1, k_2, k_3, z) = \nonumber \\
{}& \left[ \int_0^{\infty} dm_1 \, n(m_1, z) \left( \frac{m_1}{\bar{\rho}} \right) u(k_1|m_1, z) \right. \nonumber \\
\times{}& \int_0^{\infty} dm_2 \, n(m_2, z) \left( \frac{m_2}{\bar{\rho}} \right)^2 u(k_2|m_2, z) \, u(k_3|m_2, z) \nonumber \\
\times{}& \left.\vphantom{\frac{A}{B}}  P_{h} (k_1|m_1,m_2, z) \right] + 2 \text{ cyc.} \, , \\
\label{b3h}
\MoveEqLeft[3] B_{3h}(k_1, k_2, k_3, z) = \nonumber \\
 {}& \prod_{i=1}^3  \left[ \int_0^{\infty} dm_i \, n(m_i, z) \left(\frac{m_i}{\bar{\rho}}\right) u(k_i|m_i, z) \right]  \nonumber \\
\times{}& \, B_{h} (k_1, k_2, k_3| m_1, m_2, m_3, z) \, .
\end{align}
Here $B_{h} $ is the halo bispectrum, which we describe in Appendix~\ref{sec:haloZutaten} below.

\subsection{Combined halo-PT model}
\label{sec:ihm}
As we show below in Sec.~\ref{sec:comparison}, the halo model provides a good description of $N$-body simulations in the fully nonlinear regime; however, some well-known shortcomings of this formalism are that \cite{Cooray20021, 1475-7516-2012-08-036, PhysRevD.78.023523}: (i) in the transition between linear and nonlinear scales, the halo model description is less accurate, and in the mildly nonlinear regime, perturbative methods are often more successful; (ii) in the linear limit, the nonlinear contributions $P_{1h}, B_{1h}, B_{2h}$ do not vanish, leading to excess power with respect to linear theory  for $k \to 0$; (iii) at higher redshift, as the fraction of matter in virialised structures decreases, the accuracy of the halo model degrades rapidly.

The issues (i) and (ii) are addressed by a combined formalism developed by Valageas and Nishimichi (VN) \cite{valageas1, valageas2}, which we briefly summarise here; we will call this model `halo-PT model' in later sections.

\subsubsection{Power spectrum}

The one- and two-halo power spectrum terms can be combined with perturbation theory as follows \cite{valageas1}:
\begin{align}
  P_{1h}^{\mathrm{VN}} (k) &= \int_0^{\infty} dm \, n (m) \left( \frac{m}{\bar{\rho}} \right)^2 \left[ u^2 (k|m) - W_f^2 (k q_m) \right] \, \label{p1VN} ,\\
  P_{2h}^{\mathrm{VN}} (k) &= F_{2h} \left(1/k\right) P_{\mathrm{PT}} (k) \, \label{p2VN} ,
\end{align}
where $q_m = R_f = [3 m / (4 \pi \bar \rho)]^{1/3}$ is the Lagrangian radius of a halo of mass $m$, $F_{2h}$ describes the probability that two particles at this Fourier space separation are in distinct haloes, and $P_{\mathrm{PT}} (k) $ is the nonlinear matter power spectrum in perturbation theory, e.g. SPT or EFT. With respect to the standard halo model presented in Sec.~\ref{sec:halo} above, the one-halo term is modified by subtracting the filter function $W_f^2 (k q_m)$, which ensures that the one-halo term vanishes in the limit $k \to 0$; the two-halo term is based on a perturbation theory of choice, corrected by the probabilistic prefactor $F_{2h}$ given in Eq.~(\ref{eq:F2}).
The derivation of this model is summarised in Appendix~\ref{appendix:haloPT}.

\subsubsection{Bispectrum}
\label{sec:comb_B}
Using a similar approach, Ref.~\cite{valageas2} derived a combined model for the bispectrum.
In analogy with the power spectrum case, the only term that should contribute to the bispectrum on very large scales is the three-halo term. Hence, that is the only perturbative contribution, while the one- and two-halo terms are non-perturbative.

The one-halo bispectrum term is:
\begin{multline}
B^{\mathrm{VN}}_{1h} (k_1,k_2,k_3) =  \\
\int_0^{\infty} dm \, n (m) \left( \frac{m} {\bar{\rho}} \right)^3   \times  \prod_{i=1}^3 \left[ u (k_i|m) - W_f (k_i q_m) \right] \, .
\end{multline}
This function has the correct behaviour on large scales, as its slope is at least $B^{\mathrm{VN}}_{1h} (k_1,k_2,k_3) \propto k_j^2$ for any $k_j \to 0$.

\begin{figure*}[t]
\begin{center}
\includegraphics[width=\linewidth]{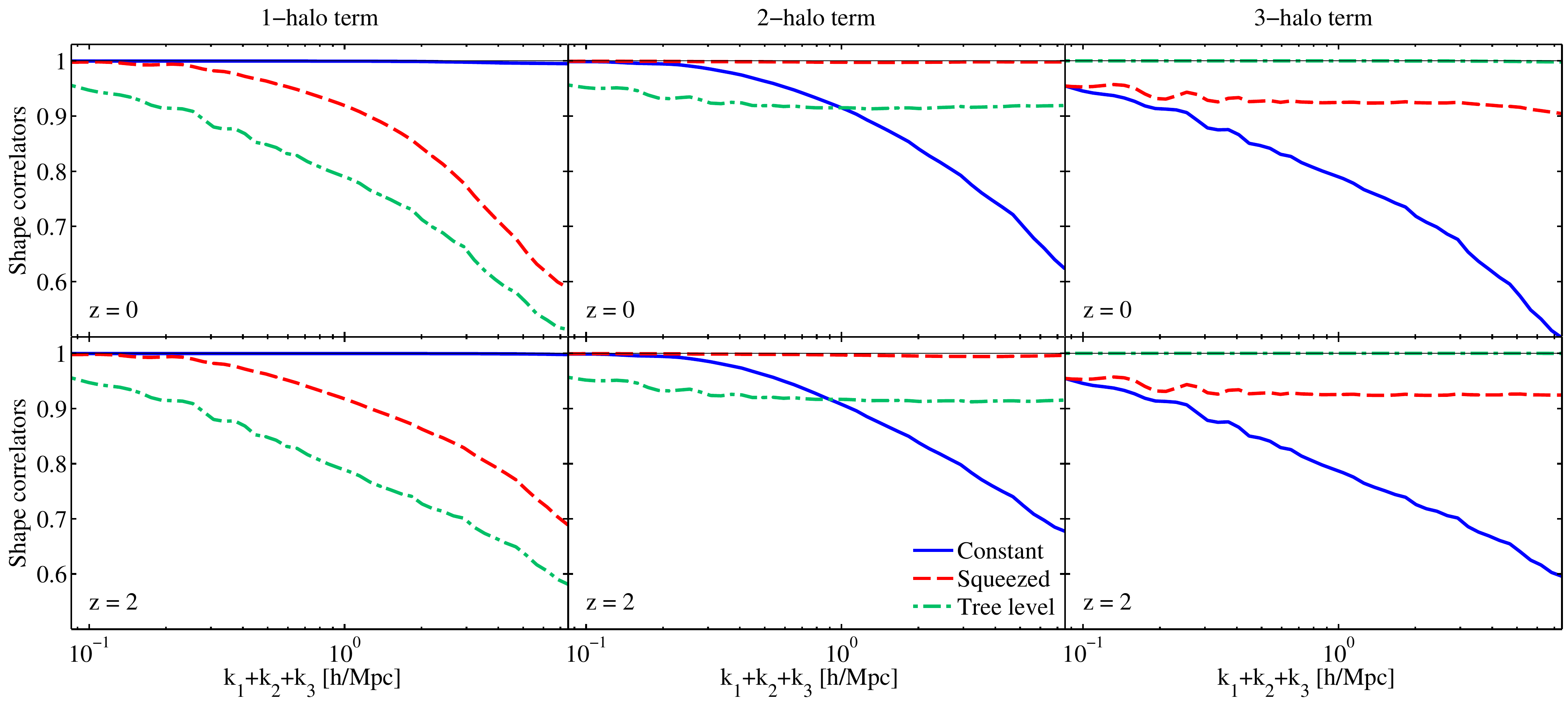}
\caption{Shapes of the halo model bispectrum. We show the correlation of the three components of the halo model with the constant, squeezed and tree-level shapes at redshifts $z = 0$ (upper panels) and $z=2$ (lower panels). The left panels show that the one-halo term has a constant shape (Eq.~\ref{constantsh}), the central panels demonstrate that the two-halo term is nearly fully correlated with the squeezed shape (Eq.~\ref{squeezed}), and  the right panels indicate that the three-halo term has the same shape as the tree-level bispectrum (Eq.~\ref{stree}). These results hold independent of scale and redshift.}
\label{comparatietot}
\end{center}
\end{figure*}

\begin{figure}[tb]
\begin{center}
\includegraphics[width=\linewidth]{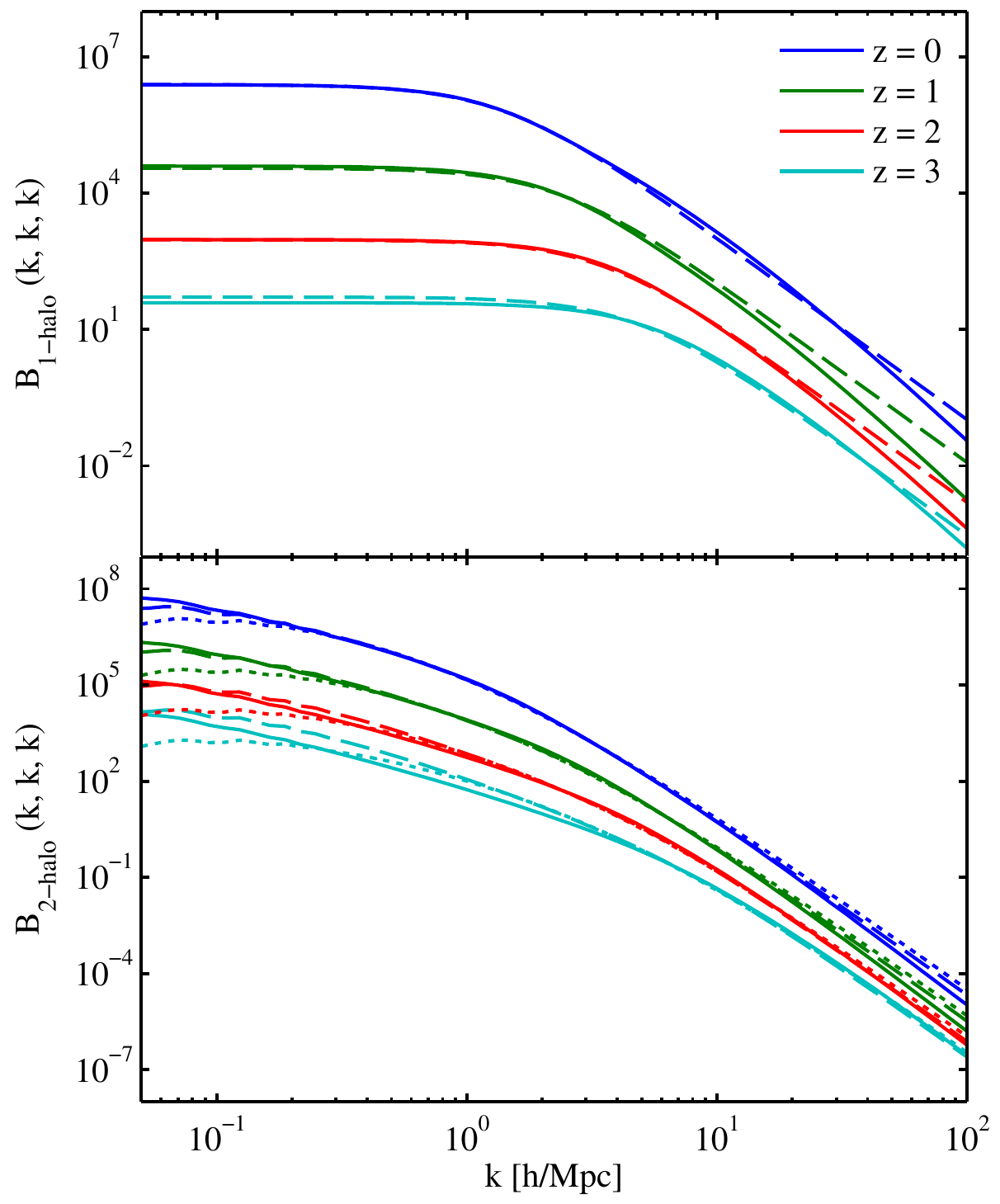} 
\caption{Equilateral one-halo (top panel) and two-halo (bottom panel) bispectra at $z = \{0,1,2,3\}$ (solid lines, from top to bottom), compared with the corresponding fitting function from Eqs.~(\ref{eq:f1h}, \ref{eq:f2h}) (dashed lines). The dotted lines refer to the corrected two-halo fitting function of Eq.~(\ref{2ht}).}
\label{h1eq}
\end{center}
\end{figure}

The full result for the two-halo bispectrum is:
\begin{align}
\MoveEqLeft[3]  B^{\mathrm{VN}}_{2h} (k_1,k_2,k_3) =    \nonumber \\
{}&  \int_0^{\infty}  dm_1 \, n (m_1)   \left( \frac{m_1} {\bar{\rho}} \right) \left[ u (k_1|m_1) - W_f (k_1 \, q_{m_1} ) \right]   \nonumber \\
\times{}& \int_0^{\infty}  dm_2 \, n (m_2)  \left( \frac{m_2} {\bar{\rho}} \right)^2 \left[u (k_2|m_2) - W_f (k_2 \, q_{m_2}) \right]  \nonumber \\
\times{}& \left[ u (k_3|m_2) - W_f (k_3 \, q_{m_3}) \right] P_{hh} (k_1 | m_1, m_2) + 2 \, \text{cyc.}
\end{align}
Here $P_{hh} (k_1 | m_1, m_2) = b(m_1) \, b(m_2) \, P_{\mathrm{lin}} (k_1) $.
This result was however found to be unsatisfactory \cite{valageas2}, because it scales as $B^{\mathrm{VN}}_{2h} \sim k_1^2 \, P(k_1)$ for $k_1 \to 0$, while a scaling $\propto P(k_1)$ is expected; this implies that the approximations made in the derivation of this term are not accurate enough.  Ref.~\cite{valageas2} therefore changes tack and argues for an alternative result that scales more appropriately in the large-scale limit, by replacing the halo with the matter power spectrum, and removing one prefactor:
\begin{align} \label{eq:B2hprime}
\MoveEqLeft[3]  B^{\mathrm{VN}}_{2h'} (k_1,k_2,k_3) =    \nonumber \\
{}&  \int_0^{\infty}  dm_1 \, n (m_1)   \left( \frac{m_1} {\bar{\rho}} \right) \left[ u (k_1|m_1) - W_f (k_1 \, q_{m_1} ) \right]   \nonumber \\
\times{}& \int_0^{\infty}  dm_2 \, n (m_2)  \left( \frac{m_2} {\bar{\rho}} \right)^2 \left[u (k_2|m_2) - W_f (k_2 \, q_{m_2}) \right]  \nonumber \\
\times{}&  P_{\mathrm{lin}} (k_1) + 2 \, \text{cyc.}
\end{align}
The large-scale limit of this result is $B^{\mathrm{VN}}_{2h'} \sim P(k_j)$ for $k_j \to 0$, as desired. Notice however that the rougher approximations assumed while deriving Eq.~(\ref{eq:B2hprime}) make the accurate prediction of this term more uncertain.

Finally, the three-halo bispectrum is obtained with a perturbative approach. Similarly to the two-halo power spectrum, this contribution should match the tree-level bispectrum on very large scales. The probability that the three wavevectors belong to different haloes can be approximated by:
\begin{multline}
  F_{3h} (k_1,k_2,k_3) = \nonumber \\
\int_0^{\nu_{k_1}} d\nu_1 \int_0^{\nu_{k_2}} d\nu_3 \int_0^{\nu_{k_3}} d\nu_3 \, f(\nu_1) \, f(\nu_2) \, f(\nu_3) \, .
\end{multline}
The three-halo bispectrum can then be written as
\begin{equation}
B_{3h}^{\mathrm{VN}} (k_1,k_2,k_3) = F_{3h} (1/k_1, 1/k_2, 1/k_3) \, B_{\mathrm{PT}} (k_1,k_2,k_3) \, \label{b3VN} ,
\end{equation}
where $B_{\mathrm{PT}} (k_1,k_2,k_3)$ is the matter bispectrum in the perturbative method of choice. In practice, the probabilistic prefactor appears to be neglected and set to unity for the bispectrum case \cite{valageas2}.

\subsection{Halo model shapes}
\label{hms}
By analogy with the shape investigation of perturbation theory bispectra we described in Sec.~\ref{sec:shapespert}, we characterise here the shapes of the distinct halo model contributions, each of which has been evaluated numerically for a specific set of cosmological parameters (see Sec.~\ref{sec:sims}).  
In Fig.~\ref{comparatietot} we show the binned shape correlator results $\mathcal{S}^S$ (Eq.~\ref{shapeprod2}), by projecting the three halo model bispectrum components onto the canonical constant, squeezed and tree-level shapes (Eqs.~\ref{constantsh}-\ref{stree}), defined on slices of $K= k_1+k_2+k_3=$ constant, for redshifts $ z = \{0, 2\}$.
The respective panels of Fig.~\ref{comparatietot} showing the one-, two-, and three-halo terms demonstrate that they are maximally correlated with the constant, squeezed, and tree-level shapes respectively, on all scales, and independently of redshift.   This clear observation confirms the accuracy of the separable ansatz (Eq.~\ref{separable}) and the completeness of our canonical three shapes (Eqs.~\ref{constantsh}-\ref{stree}) when characterising the degrees of freedom needed to describe the standard halo bispectrum.   This motivates us to find simple fitting functions $f_i(K)$  for each of the three halo model components.  

\subsubsection{One-halo term}
\begin{figure}[tb]
\begin{center}
\includegraphics[width=2.5in]{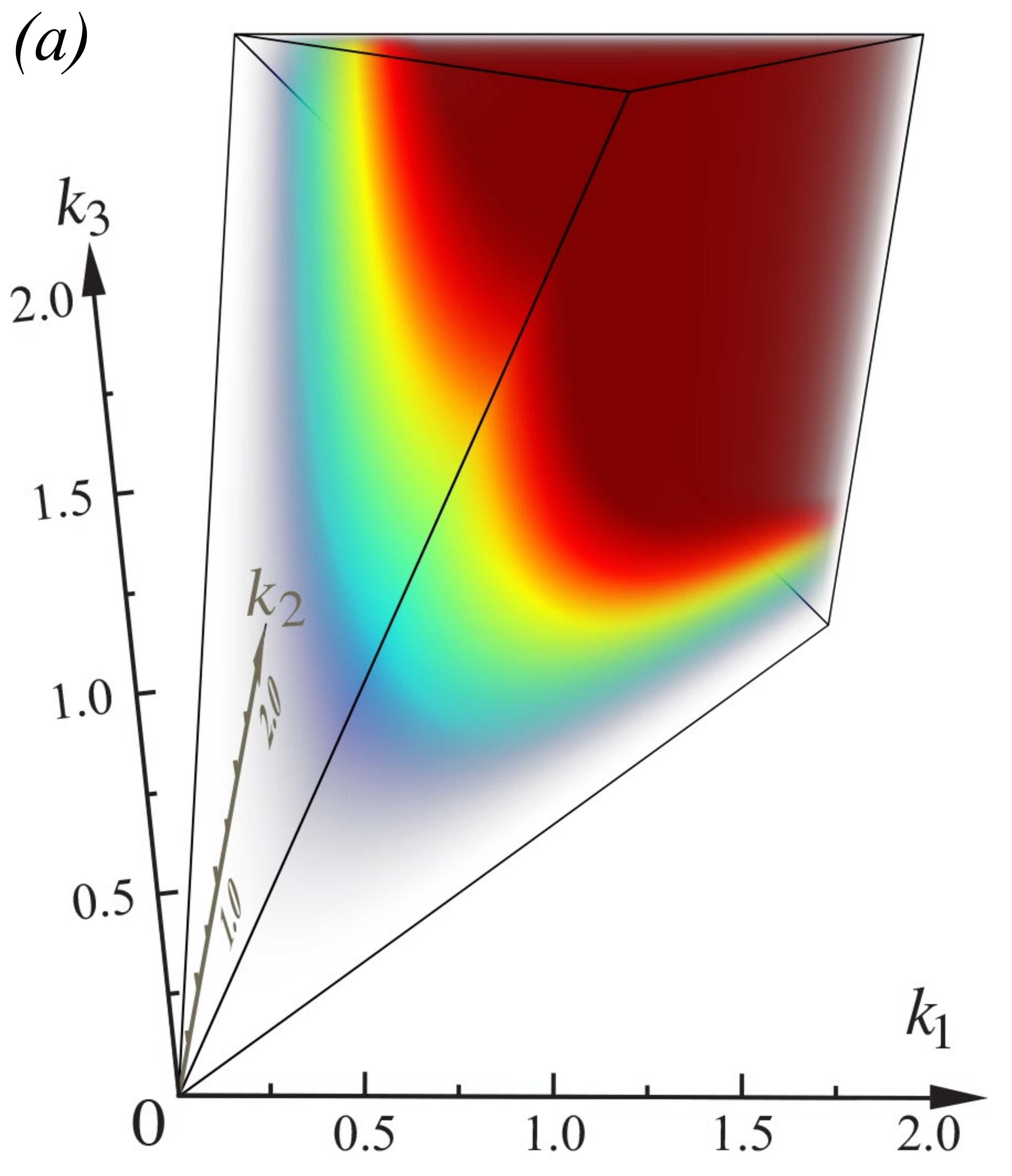}\\
\medskip
\includegraphics[width=2.5in]{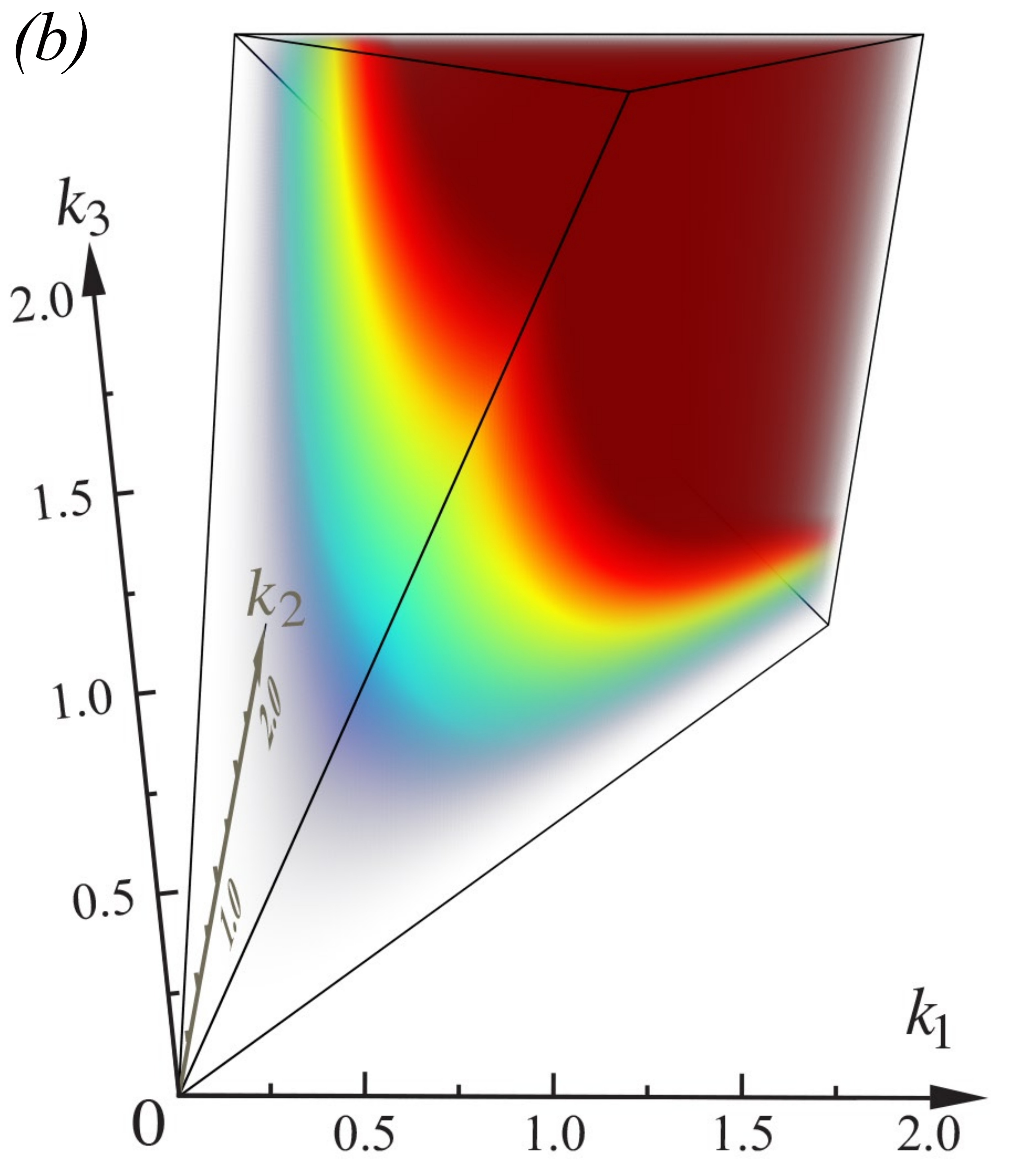}\\
\includegraphics[width=2.5in]{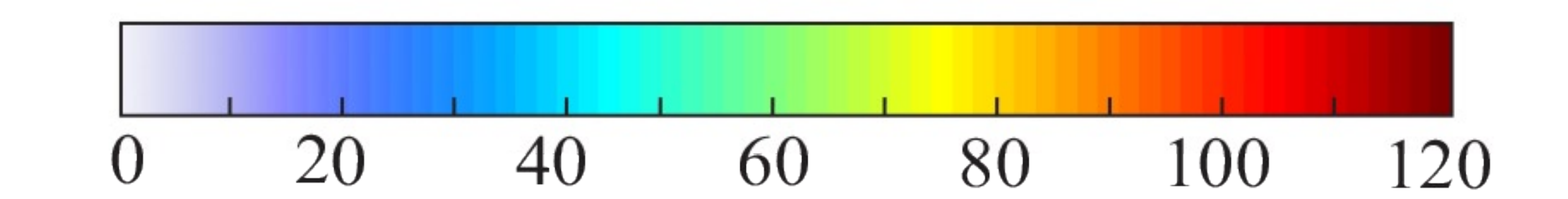}
\caption{The SN-weighted one-halo bispectrum of Eq.~(\ref{b1h}) (upper panel) compared at $z=0$ with the one-halo constant shape ansatz of Eq.~(\ref{b1hbis}) with scale-dependence $f_{1h}(K)$ given by Eqs.~(\ref{eq:f1h}, \ref{fit1b}) (lower panel).  This fit is visually hard to distinguish reflecting the high total correlation achieved over all lengthscales (and redshifts).  The cross-sectional shape does not appear constant because of the SN-weighting (Eq.~\ref{shapeprod}).}
\label{bispectrum_one-halo}
\end{center}
\end{figure}

Given the excellent shape correlation between the one-halo bispectrum (Eq.~\ref{b1h}) and the constant shape (Eq.~\ref{constantsh}) that we observe in Fig.~\ref{comparatietot}, we note that this term can be approximated by:
\begin{equation}
\label{b1hbis}
B_{1h}(k_1,k_2,k_3) = f_{1h}(K) \, S^{\text{const}}(k_1,k_2,k_3) \, ,
\end{equation}
where $K \equiv k_1+k_2+k_3$. Because of the constant cross-sectional form of Eq.~(\ref{b1hbis}), without loss of generality we can focus exclusively on the equilateral case to find a good fit. 
In Fig.~\ref{h1eq} (top panel) we illustrate the equilateral one-halo bispectrum obtained from Eq.~(\ref{b1h}) at $z = \{0,1,2,3\}$, compared with the following square-Lorentzian fitting function we introduce:
\begin{equation}
f_{1h}(K)=\frac{A}{\left[1+bK^2 \right]^2} \, ,
\label{eq:f1h}
\end{equation}
where $A$ and $b$ are functions of redshift $z$ through the perturbation growth factor $D(z)$. We first fit $A, b$ for each redshift separately, and then we obtain two overall redshift-dependent fitting functions, taking account of the growth factors in the following form:
\begin{align}
\label{fit1a}
A&=\frac{2.45 \times 10^6 \, D(z)^8}{0.8 + 0.2 \, D(z)^{-3}} \\
b&=0.054 \, D(z)^{2.2} \,\,  h^{-2}\text{Mpc}^2\,.
\label{fit1b}
\end{align}
We can see in Fig.~\ref{h1eq} (top panel) for the equilateral case, and in Fig.~\ref{bispectrum_one-halo} over the full 3D domain that this is a good approximation of the full one-halo term.

While this phenomenological fit may not be particularly well-motivated physically, it does illustrate that once the one-halo shape has been identified, then a relatively simple combination of growth factors can be used to describe the scale-dependent amplitude for the relevant wavenumber range around $K \sim 1\,h/\text{Mpc}$.   Alternatively, it is sufficient to model the one-halo bispectrum directly by evaluating Eq.~(\ref{b1h}) for equilateral values only $k_1=k_2=k_3$.  More significantly, knowing empirically that ansatzes like Eq.~(\ref{eq:f1h}) are accurate may offer insight which leads to a much simpler mathematical derivation of the individual halo contributions.

\subsubsection{Two-halo term}
As seen in  Fig.~\ref{comparatietot}, the two-halo bispectrum (Eq.~\ref{b2h})  is strongly correlated on all $K=\text{const.}$ slices with the squeezed shape $S^{\text{squeez}}(k_1,k_2,k_3)$ constructed from products of the power spectrum defined in Eq.~(\ref{squeezed}).  This means that we can write:
\begin{equation}
\label{b2hbis}
B_{2h}(k_1,k_2,k_3) = f_{2h}(K) \, S^{\text{squeez}}(k_1,k_2,k_3) \, .
\end{equation} 
In order to obtain a phenomenological fit, we consider again the equilateral configuration, which we show in Fig.~\ref{h1eq} (bottom panel).
From this simple analysis, we find that a useful fitting function valid for the redshift range considered is:
\begin{equation} \label{eq:f2h}
f_{2h}(K) = \frac{155}{1 + 26.2\,h^2\text{Mpc}^{-2} \, D(z)^{-8/3} \, K^{-2}  } \, ,
\end{equation}
where it should be noted that the squeezed shape form already includes a $D^4(z)$ redshift dependence from the linear power spectrum in Eq.~(\ref{squeezed}). 

However, as discussed above in Sec.~\ref{sec:comb_B}, the standard two-halo term causes some large-scale power excess in the full bispectrum, because it does not decay appropriately as $k \to 0$; thus the full bispectrum does not recover the tree-level form on large scales. We can modify our fitting function in order to solve this issue, by considering the functional form:
\begin{equation} 
f_{2h}(K)=\frac{C}{(1+DK^{-1})^3} \, .
\label{2ht}
\end{equation}
This function is chosen to decay more rapidly on very large scales, as in that regime there should be no contribution from the two-halo term. By fitting the full two-halo term at different redshifts and considering the halo-PT VN-model, we obtain:
\begin{align}
\label{2htf1}
C&=240 \\
D&=2.5 \,h\,\text{Mpc}^{-1}\, D(z)^{-4/3} \, .
\label{2htf2}
\end{align}

\begin{figure}[tb]
\begin{center}
\includegraphics[width=2.5in]{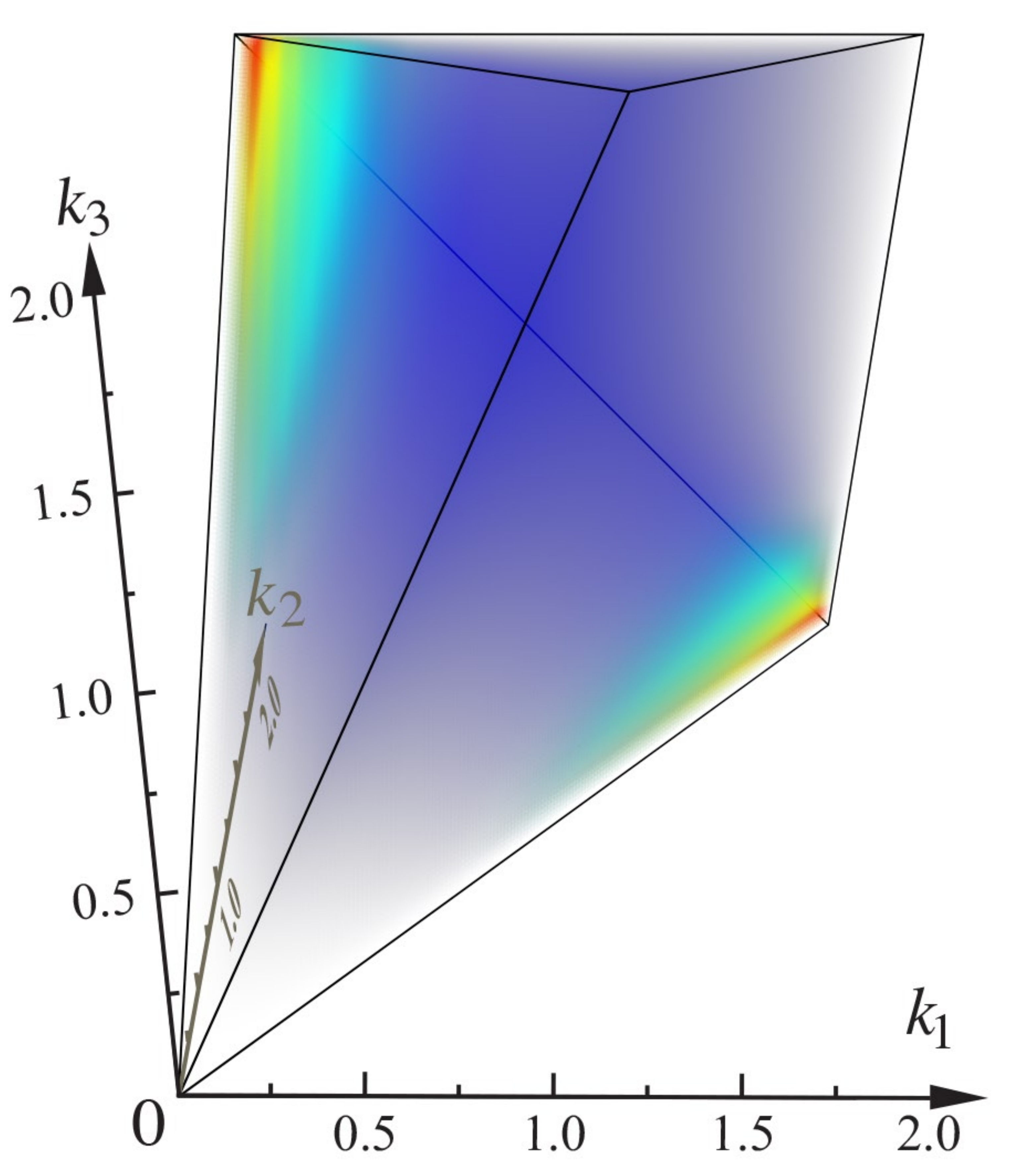}\\
\includegraphics[width=2.5in]{scale120.pdf}
\caption{The SN-weighted two-halo bispectrum of Eq.~(\ref{b2h})  at $z=0$, showing the strongly squeezed signal.  Like the one-halo bispectrum shown in Fig.~\ref{bispectrum_one-halo}, an excellent fit to this model can be obtained with the separable ansatz of Eq.~(\ref{b2hbis}) using the standard `squeezed' shape (Eq.~\ref{squeezed}).}
\label{bispectrum_two-halo}
\end{center}
\end{figure}

We can see in Fig.~\ref{h1eq} (bottom panel) for the equilateral case, and in Fig.~\ref{bispectrum_two-halo} over the full 3D domain that this is a good approximation of the two-halo bispectrum term.

Nevertheless, despite this improvement at $z=0$ as $k\rightarrow 0$ we will show later that the two-halo model does not predict the appropriate growth rates at  redshifts $z>0$ when compared to simulations.  

\subsubsection{Three-halo term}
The three-halo term (Eq.~\ref{b3h}) has a good shape correlation with the tree-level bispectrum (Eq.~\ref{stree}), because it is essentially constructed out of this solution or its one-loop extensions, all of which share the  same highly-correlated flattened shape (as discussed in Sec.~\ref{sec:shapespert}).   Hence, as we have discussed previously, the three-halo  term can be expressed again with a simple fitting function (Eq.~\ref{pertshape}) using the tree-level shape $S^\text{tree}$.  The standard halo model effectively identifies the three-halo term with the tree-level bispectrum so we can take the fitting function to be unity.   Extensions taking a perturbative result with one-loop corrections  can also be described at high accuracy with Eq.~(\ref{pertshape}) but with non-trivial scaling $f(K)$  (e.g. to simplify the halo-PT VN-model which uses $B^\text{EFT}$).   Since the closely-related nonlinear tree-level bispectrum $S^\text{treeNL}$ given in Eq.~(\ref{streeNL}) provides a better approximation to the perturbative models, we can more conveniently use this as our base tree-level ansatz:
\begin{equation}
\label{b3hbis}
B_{3h}(k_1,k_2,k_3)=f_{3h}(K) \, S^{\text{treeNL}}(k_1,k_2,k_3) \, .
\end{equation}
Both tree-level and nonlinear tree-level shapes are plotted in Fig.~\ref{bispectrum_tree}. We will employ Eq.~(\ref{b3hbis}) when developing the phenomenological three-shape model in Sec.~\ref{sec:sims}.

\section{Polyspectra from simulations}
\label{sec:sims}

\subsection{$N$-body simulations}
\label{sec:simset}

We use the $N$-body simulations with Gaussian initial conditions described in detail in Ref.~\cite{Schmittfull2013}. The simulations contain $512^3$ particles that are evolved from an initial redshift of $z=49$ until today using the $N$-body \textsc{Gadget-3} code \cite{Springel200179, Springel21122005} with 2LPT initial conditions \cite{Scoccimarro:1997gr,Crocce:2006ve}. These yield a less than 2\% accuracy in the bispectrum, as shown in Ref. \cite{McCullagh:2015oga}. The simulations are run using a flat $\Lambda$CDM universe with the following WMAP7 \cite{Komatsu2011} parameters: baryon energy density $\Omega_bh^2=0.0226$, dark matter energy density $\Omega_ch^2=0.11$, cosmological constant energy density $\Omega_\Lambda=0.734$, dimensionless Hubble constant $h=0.71$, optical depth $\tau=0.088$, amplitude of primordial perturbations $\Delta^2_\mathcal{R}(k_0)=2.43\times 10^{-9}$ and scalar spectral index $n_s(k_0)=0.963$, where $k_0 = 0.002 \, h \, \mathrm{Mpc}^{-1}$. 
 We use simulations of three different box sizes of 1600, 400 and 100 $\text{Mpc}/h$ respectively; the first one has glass Gaussian initial conditions and the other two have regular grid initial conditions. We denote the simulations using their names from Ref.~\cite{Schmittfull2013}: \textit{G512g}, $G^{512}_{400}$, $G^{512}_{100}$. Given the fixed number of particles, the three box sizes lead to the following wavenumber ranges: $[0.0039, 0.5] \, h/\text{Mpc}$, $[0.016, 2.0] \, h/\text{Mpc}$ and $[0.062, 8.0] \, h/\text{Mpc}$ respectively. For each box size, three independent realisations are available.

\begin{figure*}[t]
\begin{center}
\includegraphics[height=0.34\textheight, trim={0 0 0.25cm 0},clip]{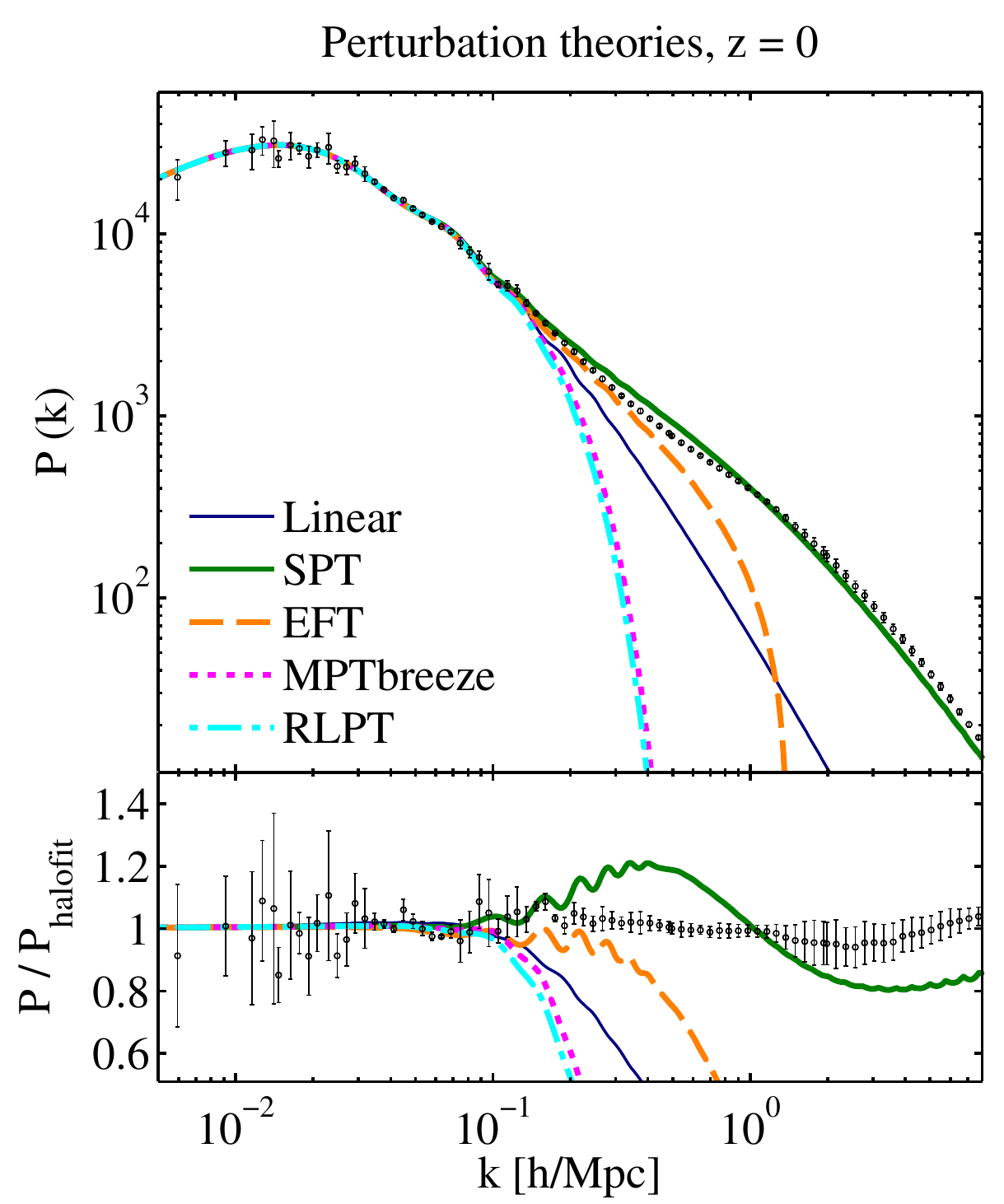} 
\includegraphics[height=0.34\textheight, trim={0 0 0.25cm 0},clip]{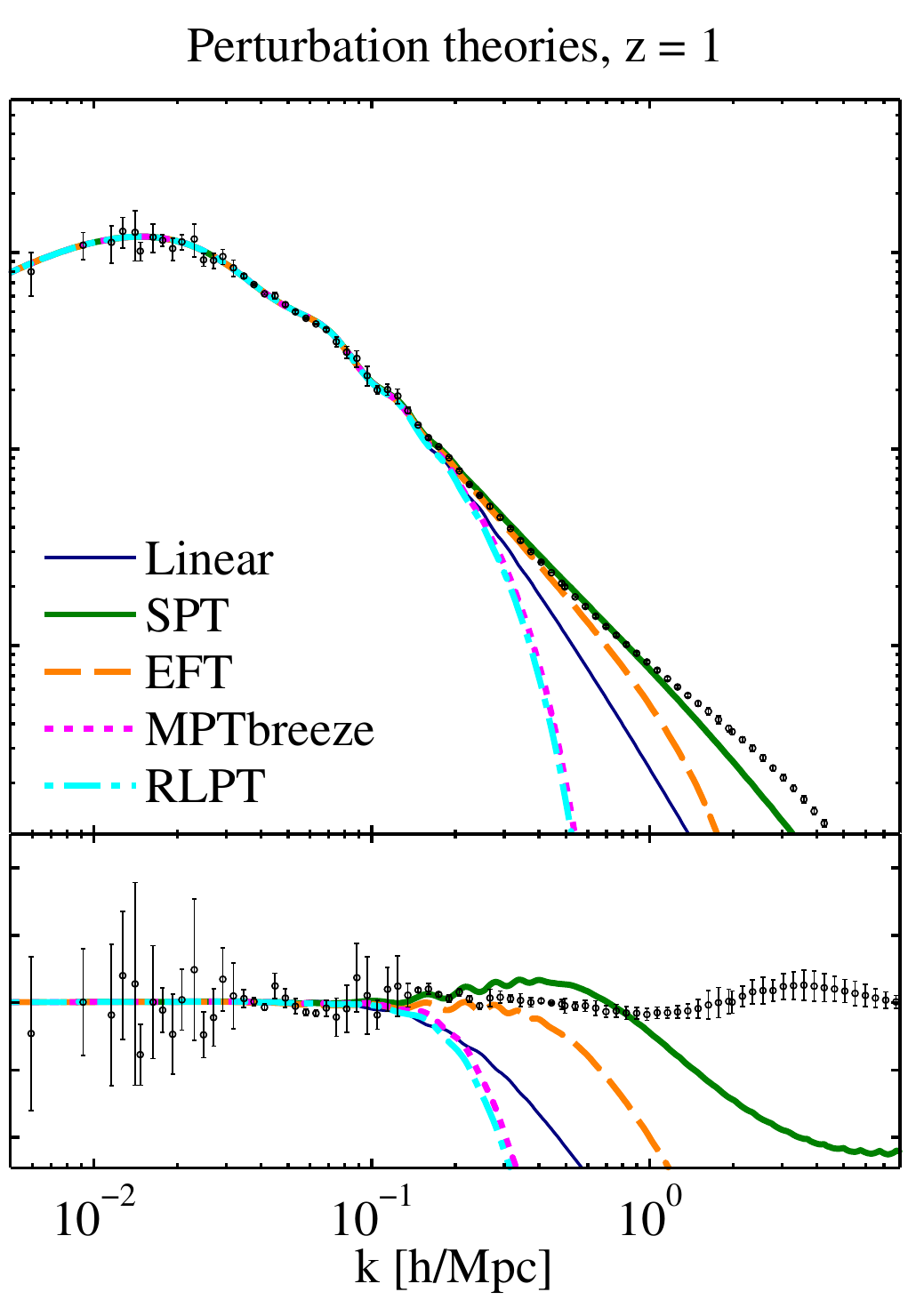}
\includegraphics[height=0.34\textheight, trim={0 0 0.25cm 0},clip]{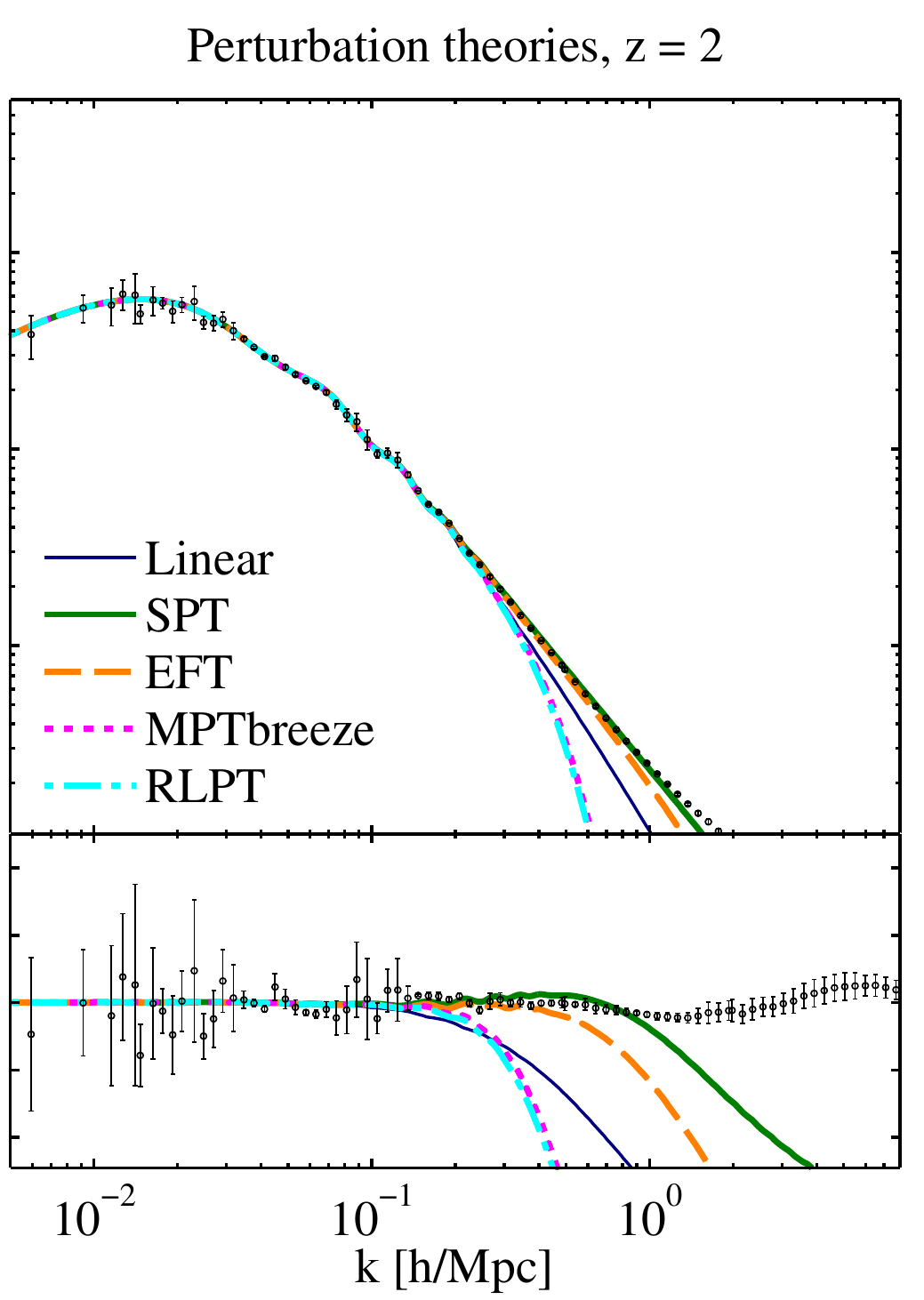}
\includegraphics[height=0.34\textheight, trim={0 0 0.25cm 0},clip]{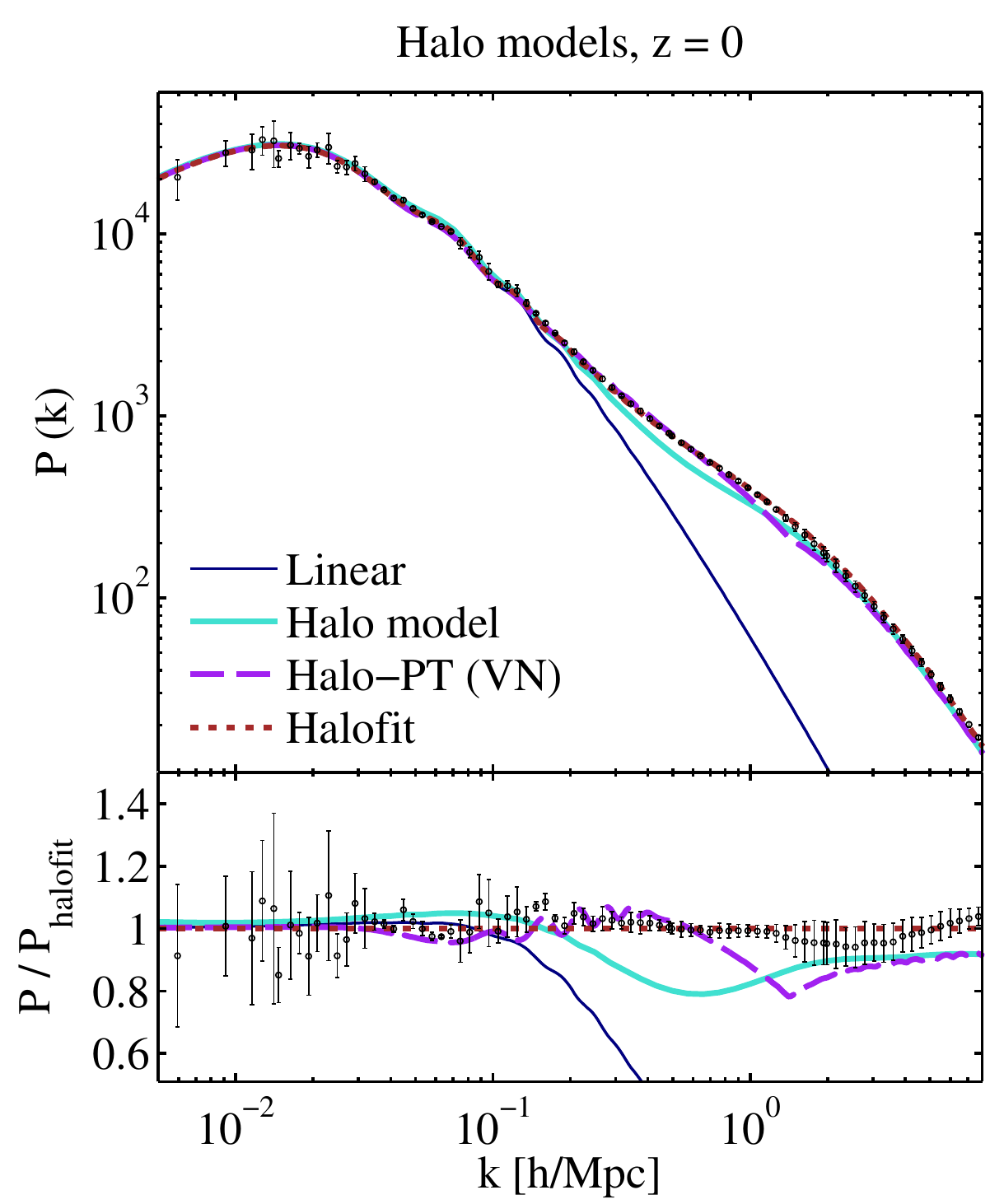}
\includegraphics[height=0.34\textheight, trim={0 0 0.25cm 0},clip]{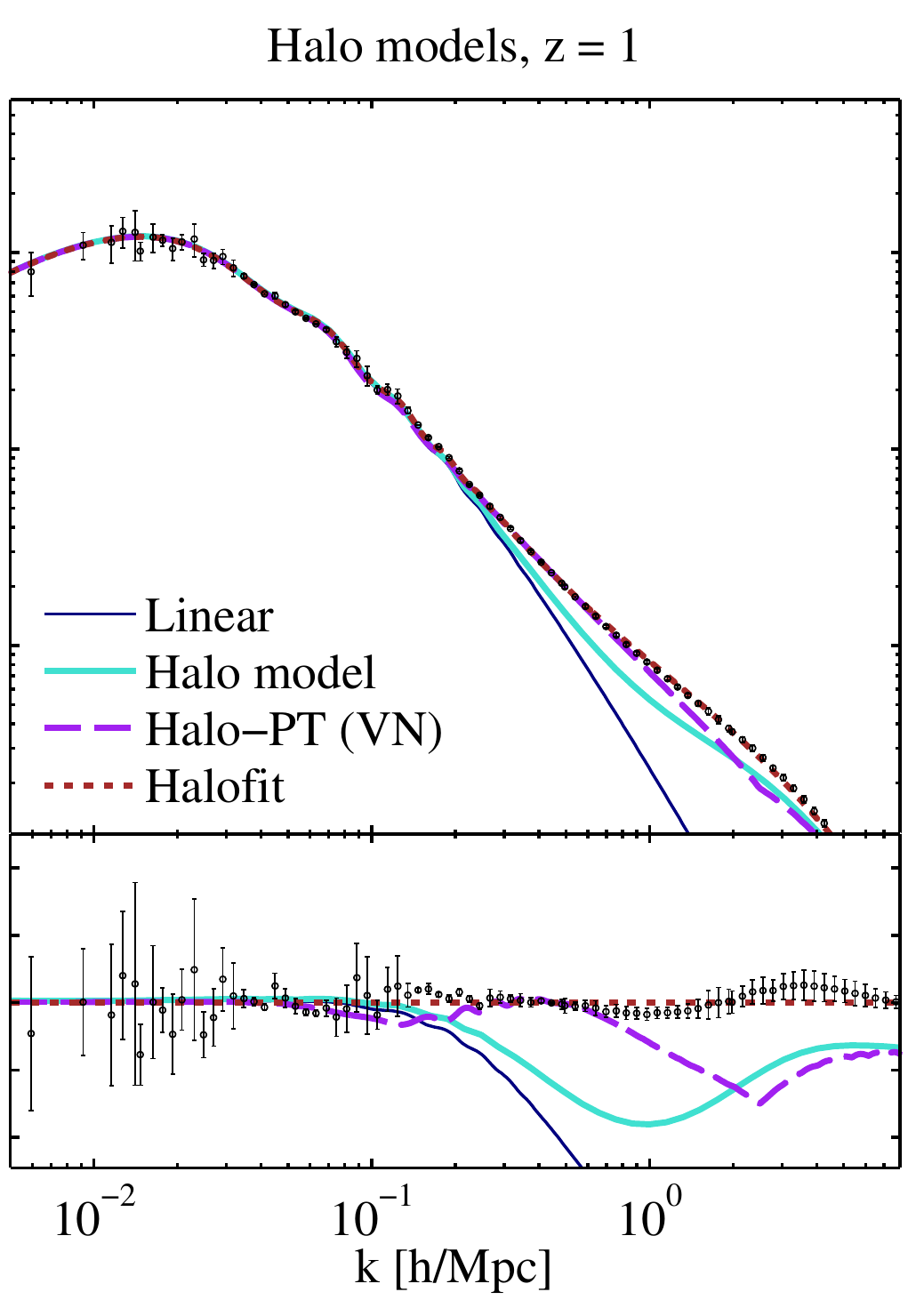}
\includegraphics[height=0.34\textheight, trim={0 0 0.25cm 0},clip]{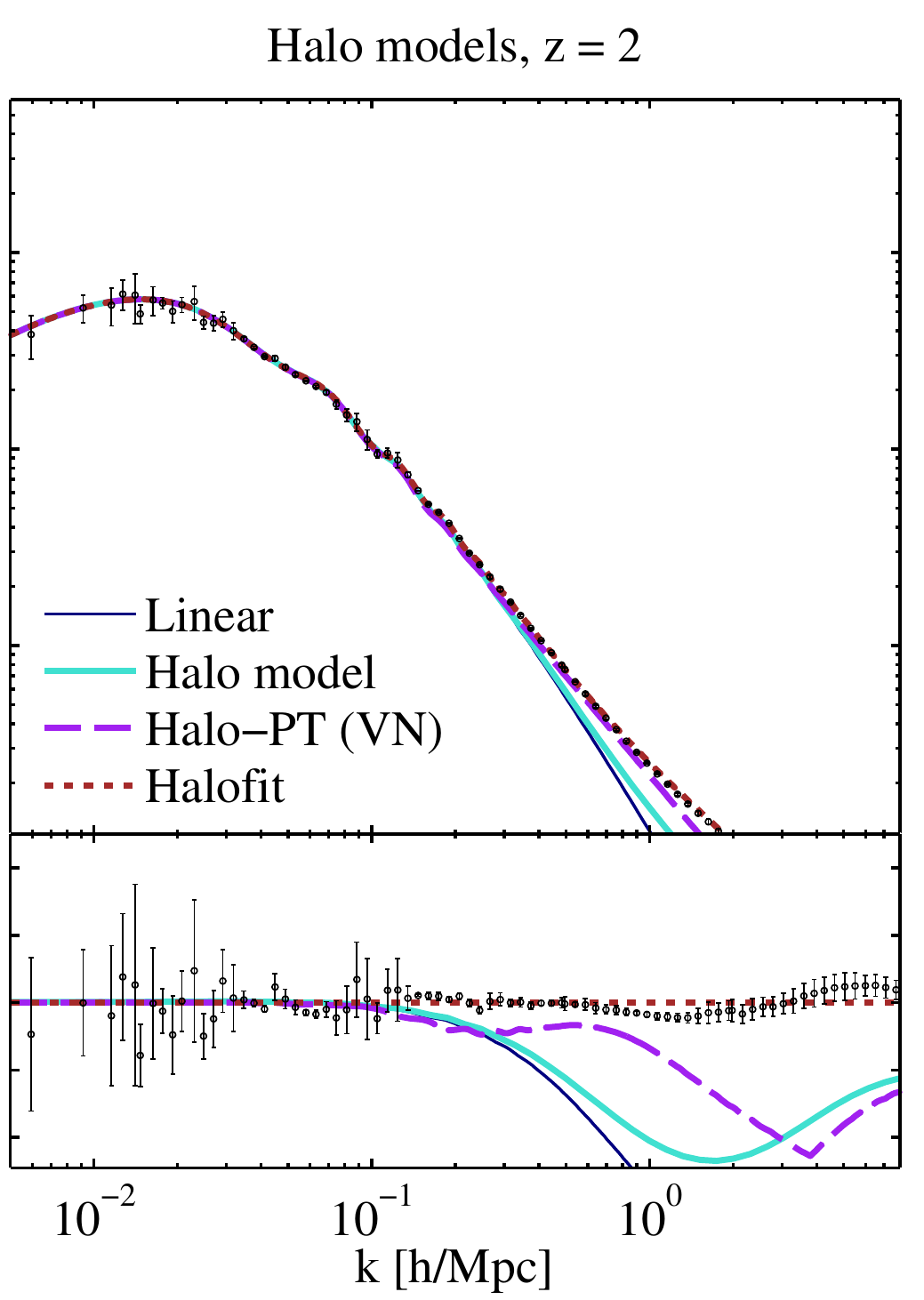}
\caption{Overview of the matter power spectra predicted by the range of theoretical models we consider, compared with data measured from $N$-body simulations. The data points are combined from $N$-body simulations with three different box sizes. The upper and lower rows refer to perturbation theories and halo models respectively. The columns refer to $z = 0$, 1 and 2 from left to right; in each plot, the main upper panel shows the power spectra comparison, while the smaller lower panel shows the residuals with respect to 
 the \textsc{Halofit} prediction.}
\label{ps0}
\end{center}
\end{figure*}

We combine the power spectra and bispectra from the different simulation boxes as follows. As it can be seen in Fig.~6 of Ref.~\cite{Schmittfull2013}, where the matter power spectra from the three simulations considered are compared to the \textsc{Halofit} model, at any redshift $z > 0$ the power spectrum of simulation \textit{G512g} only follows the  \textsc{Halofit} model up to $k_{\max} \simeq 0.2 \, h/$Mpc; however, for $k \gtrsim 0.1 \, h/$Mpc, the simulation $G^{512}_{400}$ matches \textsc{Halofit} more closely. The same behaviour is seen at larger $k$ for the $G^{512}_{400}$ and $G^{512}_{100}$ boxes. Therefore, we combine the power spectra and bispectra from the simulations in order to use each simulation in the range of scales where its results are the closest to \textsc{Halofit}, and we apply a smooth transition between the different boxes. We define a smoothing function $H(k)$ in the range $k \in [k_s,k_e]$ of the form:
\begin{equation}
H(k) = \frac {1 - \sin \left( \pi \frac {k - k_e / 2 - k_s / 2} {k_e - k_s} \right) } {2}  \, .
\label{fse}
\end{equation}
As we have three realisations for each of the simulations, we match each realisation $ i = 1, 2, 3$ from each simulations with the same $i$ realisation in the other simulations, thus obtaining three combined realisations of the power spectra and bispectra over the full $k$ range we consider. 
We have checked that modifying the smoothing function has only a small impact on the overall results.  This procedure allows us to use an overall large simulation data set  covering the entire region of interest in wavevector space with three realisations. However,  larger errors appear in the interior of the domain where the transition between the simulations occurs.

\subsection{Power spectrum}

We estimate the power spectrum of the simulations in each $k$-bin by averaging the squared absolute value of the matter overdensity $|\delta_\mathbf{q}|^2$ over all modes that fall into the shell with distance $k$ from the origin (i.e. over $\mathbf{q}$ with $|\mathbf{q}|-\Delta k/2\le k < |\mathbf{q}|+\Delta k/2$, where $\Delta k$ is the bin width).
We compare in Fig.~\ref{ps0} the power spectrum measured from the simulations with 
the models that we consider: linear theory, the nonlinear power spectrum from \textsc{Halofit}, EFT, \textsc{MPTbreeze} and RLPT at one loop, the halo model and the combined halo-PT model (based on EFT). 
 The lower panels of Fig.~\ref{ps0} show the power spectrum residuals 
with respect to the \textsc{Halofit} model.

Focusing first on the perturbative methods, we note that they increase their range of validity to higher $k$ modes as the redshift is increased, as expected.  We confirm that SPT presents excess power in the quasi-linear regime, departing from the simulations by more than 10\% at $k \simeq 0.15 \, h/$Mpc at $z=0$.
The SPT excess power is however reduced at higher redshifts, as expected given that the one-loop corrections have a higher growth rate compared to the tree level.
 The EFT method can extend the range of validity by subtracting the SPT excess power.
However, the scale range over which EFT is accurate strongly depends on which simulations were used to calibrate the counterterm, and over which range of scales and redshifts. In the present case, the $c_s^2$ counterterm we are using was calibrated by Ref.~\cite{Angulo:2014tfa} with the $G512g$ simulation box we are presenting at $z=0$; therefore, there is no guarantee that this same counterterm will be accurate at higher $k$ over the smaller-box simulations $G_{400}^{512}$, $G_{100}^{512}$, and at $z>0$. Indeed, it is likely that a re-fitting of $c_s^2$ over the combined range of simulations we are using would improve the EFT model accuracy over an extended range of $k$ and $z$.
 The  \textsc{MPTbreeze} and RLPT approaches include an exponential cut-off: this reduces the range in which the model is accurate to 10\% to $k < 0.10 \, h/$Mpc at $z=0$; nonetheless, these models feature an improved accuracy in the mildly nonlinear regime before the cut-off sets in, although the precision of our $N$-body simulations does not allow detailed quantitative statements at the percent level.

 We then consider the halo models: we see that at $z = 0$ this formalism provides a good description of the matter power spectrum on small scales and in the range $k \in [0.01,0.2] \, h/$Mpc, after which we find the well-known power deficit in the transition region between the one- and two-halo terms. The model performs again better at smaller scales ($k \gtrsim 2 \, h/$Mpc at $z=0$), reaching an accuracy of $\sim 10 \%$.
 On very large scales, the halo model amplitude exceeds the simulations, as the one-halo term does not decay to zero as it physically should.
By moving to higher redshifts, we see that at $z = 2$ the halo model provides a worse description of the simulations at intermediate and small scales, as the power deficit in the transition region is exacerbated. This is because the total fraction of dark matter particles that belong to collapsed structures is drastically reduced at this redshift, which undermines the assumptions underlying the halo model approach. On large scales on the other hand, the excess power nearly disappears at high redshift, due to the quick decay of the one-halo term as a function of $z$.

The combined halo-PT model based on EFT succeeds in removing the excess power seen on large scales at $z=0$; as we discuss below, this excess will appear even more evidently in the bispectrum. This model is also partly successful in reducing the power deficit on intermediate scales, thanks to the extra power that is added there from the perturbative term. However, due to the negative counterterm, the EFT power spectrum prediction becomes negative on small scales ($k \gtrsim 1 \, h/$Mpc at $z=0$). After this point, we base the halo-PT model on the SPT prediction: this is the reason of the cusp we see in the halo-PT model residuals in the nonlinear regime.

We finally note the results from the simulations are in good agreement with the nonlinear \textsc{Halofit} power spectrum, as they are within 10\% accuracy over the entire $k$-range considered at all $z$.

\begin{figure*}[tb]
\begin{center}
\includegraphics[width=3in]{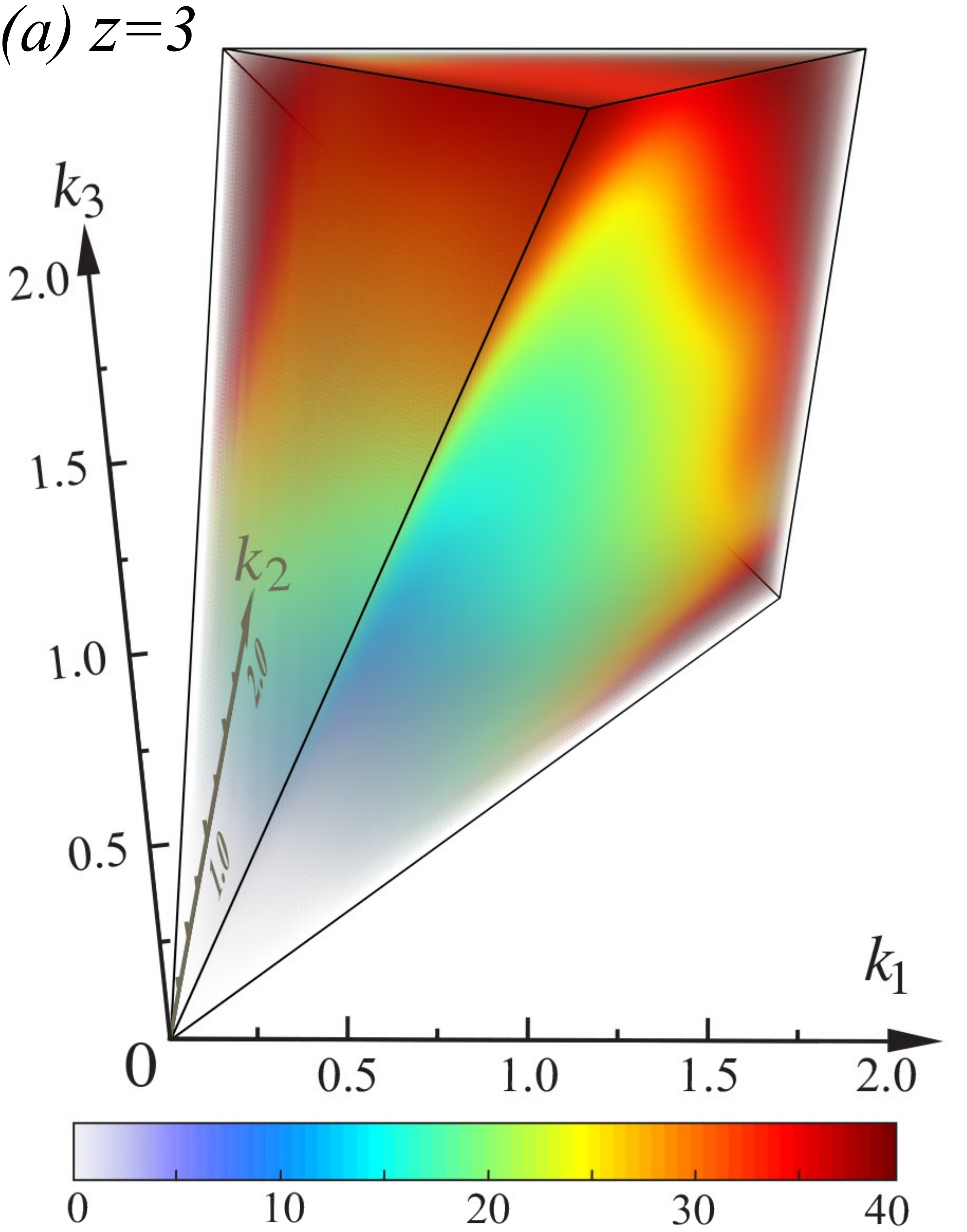}\hfil \includegraphics[width=3in]{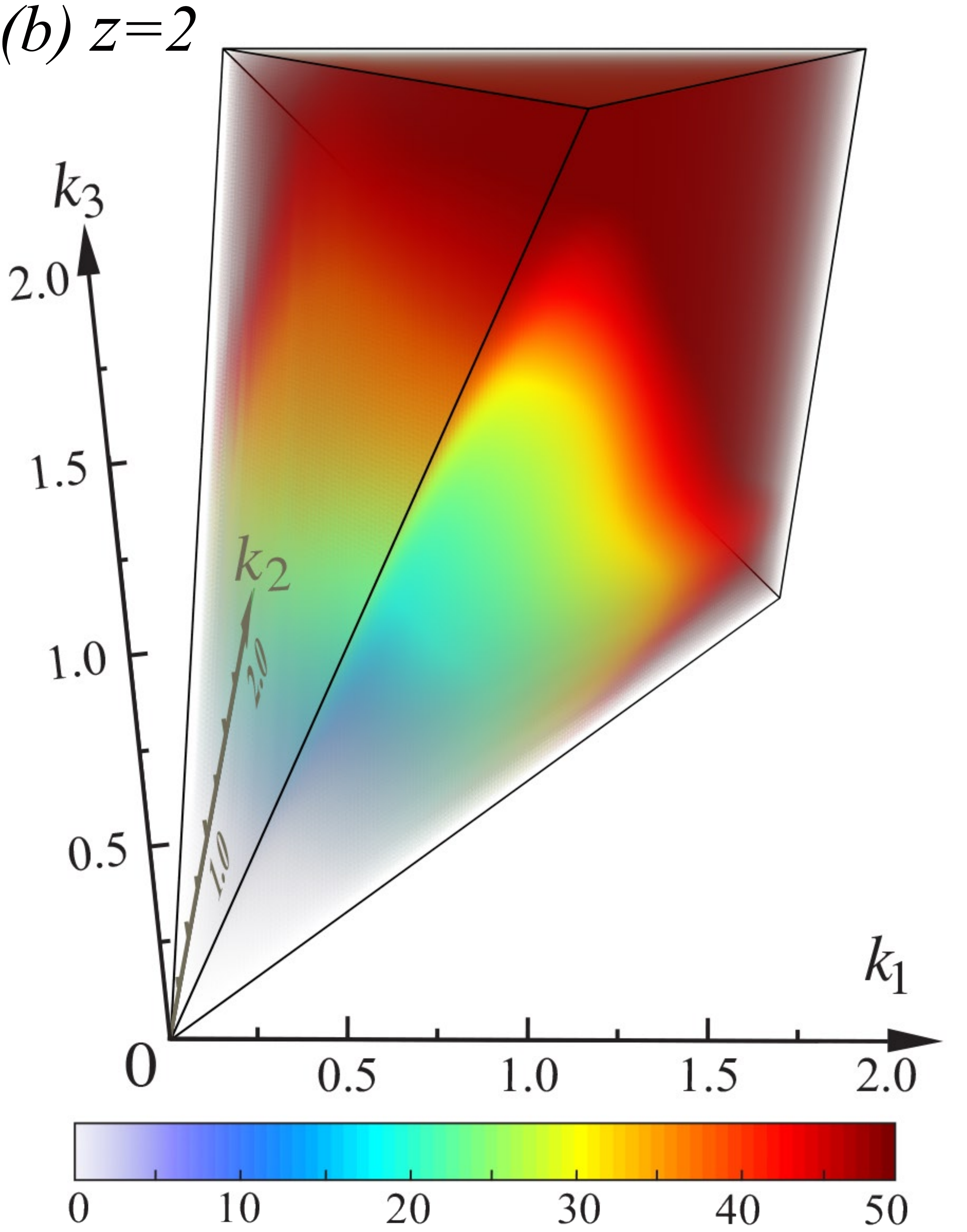}\\
\bigskip\medskip
 \includegraphics[width=3in]{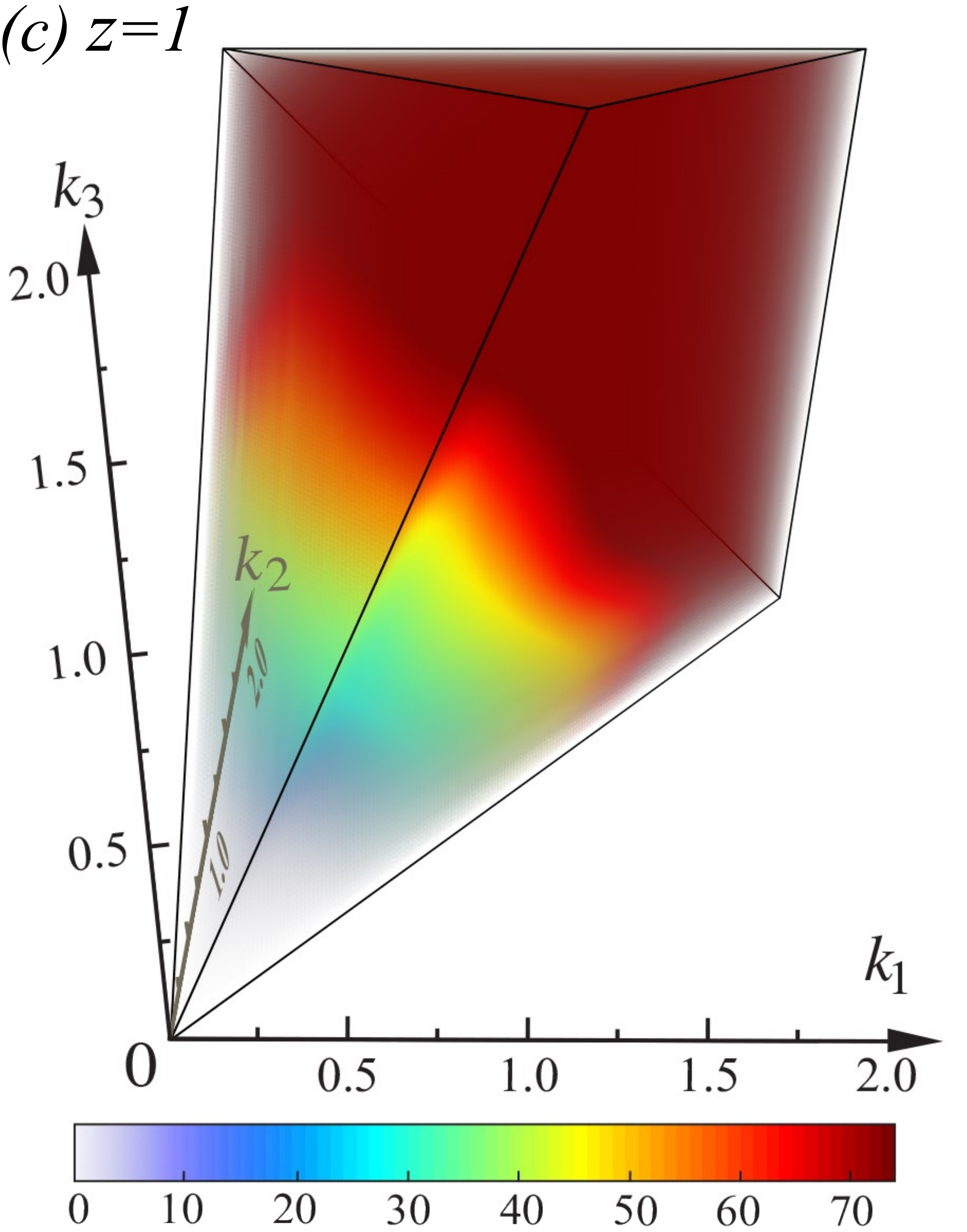} \hfil \includegraphics[width=3in]{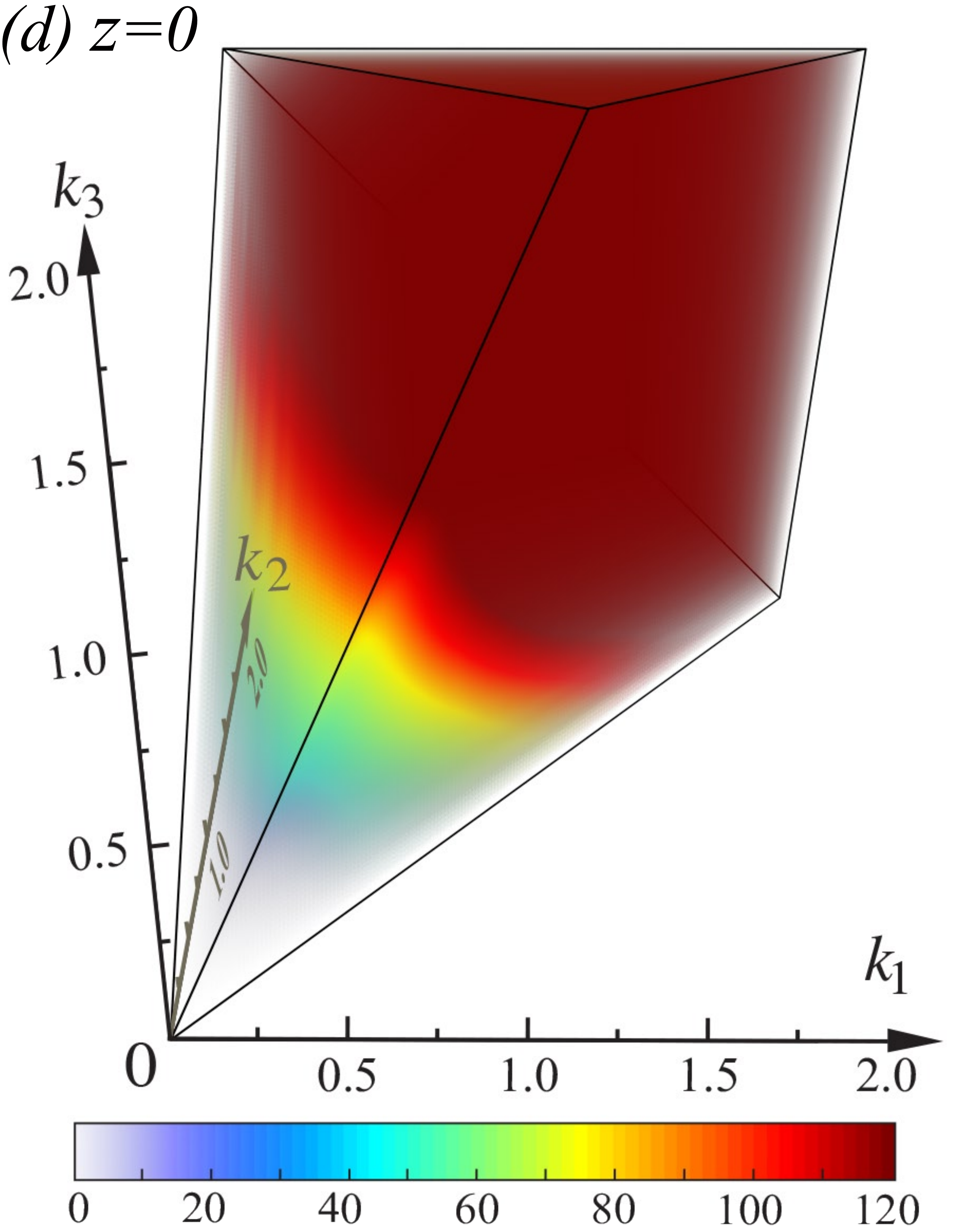} \\
 \medskip
\caption{Evolution of the SN-weighted 3D bispectrum from $N$-body simulations into the nonlinear regime with $k_i\le 2 \, h/$Mpc at redshifts (a) $z=3$, (b) $z=2$, (c) $z=1$, and (d) $z=0$.  The bispectrum colour scheme is scaled with the growth factor $D(z)$ and the tetrahedral geometry of the bispectrum domain is illustrated in Fig.~\ref{tetrapyd_split}.  Note the presence of both a strong flattened and squeezed signal shape at redshifts $z=2,\,3$ (front left face of tetrapyd and lower edge respectively in panels a, b).  At lower redshift this is overtaken by a strong uniform or one-halo signal throughout the interior region for $k\gtrsim 1 \, h/$Mpc  (front right face in panel d). The colour scale is fixed at $z=3$ in (a) to encompass all values up to the maximum.   It is then scaled with the growth rate expected for the tree-level signal to aid physical interpretation and reveal nonlinear growth rates.  This means at small scales in (d) at $z=0$ the colour scale is saturated, which is useful to highlight features at intermediate scales.}
\label{bispectrum_halo}
\end{center}
\end{figure*}

\subsection{Modal bispectrum methodology}
We next follow the modal decomposition method to reconstruct the dark matter bispectrum, using the method developed by Refs.~\cite{Fergusson2010, Fergusson2012, Regan2012}.
In this approach, the full 3D bispectrum $B(k_1,k_2,k_3)$ is expanded on an orthonormal basis defined on the same tetrapyd domain $ Q_n(k_1,k_2,k_3)$, with $n = 0, ..., n_{\max}$. In this way, the full bispectrum information is encoded in the expansion coefficients $\beta_n^Q$, and the bispectrum estimator $\hat B$ can be written as
\begin{equation}
\frac{\hat B (k_1,k_2,k_3) \, \sqrt{k_1k_2k_3}}{\sqrt{P(k_1) P(k_2) P(k_3)}} = \sum_{n=0}^{n_{\max}-1} \beta_n^Q \, Q_n(k_1,k_2,k_3) \, .
\label{eigenfns}
\end{equation}
We note that the left-hand-side is the signal-to-noise weighted bispectrum $B^{\rm SN}_i(k_1,k_2,k_3) $ defined in Eq.~(\ref{SNweight}).
The accuracy of this estimator is regulated by the dimension of the expansion basis, $n_{\max}$; for the smooth bispectra that are typical of the LSS, Ref.~\cite{Schmittfull2013} demonstrated that the choice $n_{\max} \sim 100 $ suffices to achieve a convergence of the total bispectrum signal-to-noise, i.e. considering higher $n_{\max}$ has negligible effect on the matter bispectrum. This highlights the benefits of the modal method: once the basis $Q_n$ is chosen, the entire three-dimensional bispectrum information can be simply compressed in a set of $\sim 100$ numbers.

Ref.~\cite{Fergusson2010} tested several different choices of the basis $Q_n$, 
demonstrating that the modal method successfully reconstructs the bispectrum in all cases. The most suitable choice for $Q_n$ is however built from a set of tetrahedral polynomials $q_p(x)$, which are analogues of the Legendre polynomials on the unit interval.
In more detail, the basis $Q_n$
 can be written as
\begin{equation}
Q_n(x,y,z) = q_{\{r}(x) \, q_s(y) \, q(z)_{t\}} \, ,
\end{equation}
where $n=r+s+t$, $\{rst\}$ means symmetrisation over the three indices, and the order of the permutations is taken as in Ref.~\cite{Fergusson2010}. In turn, the tetrahedral polynomials of order $n$, $q_n(x)$, can be generated by taking the determinant
\begin{equation}
\label{modalfns}
q_n(x) = \frac{1}{\mathcal{N}}
 \begin{vmatrix}
 1/2  & 7/24 & \cdots & w_n \\
 7/24 & 1/5 & \cdots & w_{n+1} \\
  \vdots  & \vdots  & \ddots & \vdots  \\
  w_{n-1} & w_n & \cdots & w_{2n-1} \\
  1       & x & \cdots & x^n 
 \end{vmatrix} \, ,
\end{equation}
where
\begin{equation}
w_n=\frac{n+6}{2(n+3)(n+2)} \, ,
\end{equation}
and the normalisation $\mathcal{N}$ is chosen so that the polynomials $q_n(x)$ are orthonormal with respect to the product:
\begin{equation}
\langle q_n,q_m\rangle=\int_0^1 q_n(x) \, q_m(x) \,\frac{1}{2} \, x \, (4-3x) \, dx=\delta_{nm} \, .
\end{equation}

\subsection{Bispectrum reconstruction from simulations}

A modal reconstruction for the matter bispectrum $B^{\rm SN}_i(k_1,k_2,k_3) $  (Eq.~\ref{eigenfns}) was obtained using the mode functions (Eq.~\ref{modalfns}) for the full array of simulations described in Sec.~\ref{sec:simset}.   This decomposition and its validation were described in detail in Ref.~\cite{Schmittfull2013}:  a relatively small number of modes were sufficient to recover the full bispectrum at the required resolutions, that is, using 120 modes for the \textit{G512g} simulations and 50 modes for the other two simulations.   We focus attention here on the low-redshift regime $z<3$ where the bispectrum is accessible to current and future galaxy surveys and where nonlinearities become important.   To obtain the full bispectrum across the widest range of scales we combined and averaged all the simulation bispectra, interpolating in overlapping regions using the same prescription as that described for the power spectrum.   Error bars for bispectrum correlators were estimated by determining variances from the different simulations. 

In Fig.~\ref{bispectrum_halo} we plot the full three-dimensional matter bispectrum we have obtained across the tetrapyd domain for $0.02 \, h /$Mpc $ \le k\le 2 \, h/$Mpc and at four different redshifts $z= \{0,\,1,\,2,\,3\}$.   The colour scheme is scaled using the growth factor $D(z)$ such that the tree-level bispectrum would appear constant in the perturbative regime.    These plots range from quasi-linear to highly nonlinear regions and several qualitative observations about the nature and evolution of the matter bispectrum are immediately apparent.  

At the higher redshifts $z=2,\,3$ shown in Fig.~\ref{bispectrum_halo}(a,b),  a flattened signal is dominant up to $K\equiv\sum_i k_i \lesssim 4,\, 3.5 \, h/$Mpc respectively (i.e. the tetrahedron region).  This is consistent with the flattened tree-level shape (Eq.~\ref{stree}) which is shown in Fig.~\ref{bispectrum_tree}(a) at $z=2$, but at much lower amplitude on a more sensitive scale.  This means the flattened signal extrapolates with growing amplitude well beyond the perturbative regime at these redshifts (e.g.\ from Table~\ref{check_tspt} $K\lesssim 0.6 \, h/$Mpc at $ z=2$).   We focus further on the perturbative regime with $K\lesssim 1 \, h/$Mpc in Sec.~\ref{sec:testpert}.   For larger $K$, the bispectrum is dominated by a nearly uniform signal associated with halo formation (i.e.\ the top pyramidal region with $K\gtrsim 4 \, h/$Mpc).   Also in Fig.~\ref{bispectrum_halo}(a,b), we note that a significant squeezed signal  is visible for $1 \, h/$Mpc $ \lesssim K\lesssim 4 \, h/$Mpc (on the left and bottom tetrapyd edges), which can be compared with Fig.~\ref{bispectrum_two-halo}.  

At the lower redshifts $z=0,\,1$ in Fig.~\ref{bispectrum_halo}(c,d), the strong halo signal grows to become completely dominant for $K\gtrsim 1 \, h/$Mpc (saturating the colour scheme with $B^{SN}_\text{max}\approx 350$).  At $z=0$, this `constant' halo signal is so large the other contributions seem to be absent (compare with Fig.~\ref{bispectrum_one-halo}).   However, this apparent suppression of flattened and squeezed signals at $z=0$ is only relative, due to the signal-to-noise weighting (Eq.~\ref{SNweight}) with the nonlinear power spectrum $P_{\rm NL}(k)$.  This deeply nonlinear nature of perturbations today  is reflected in the greater difficulty of matching phenomenological models to simulations at low redshift.

\begin{figure}[tb]
\begin{center}
\includegraphics[width=\linewidth]{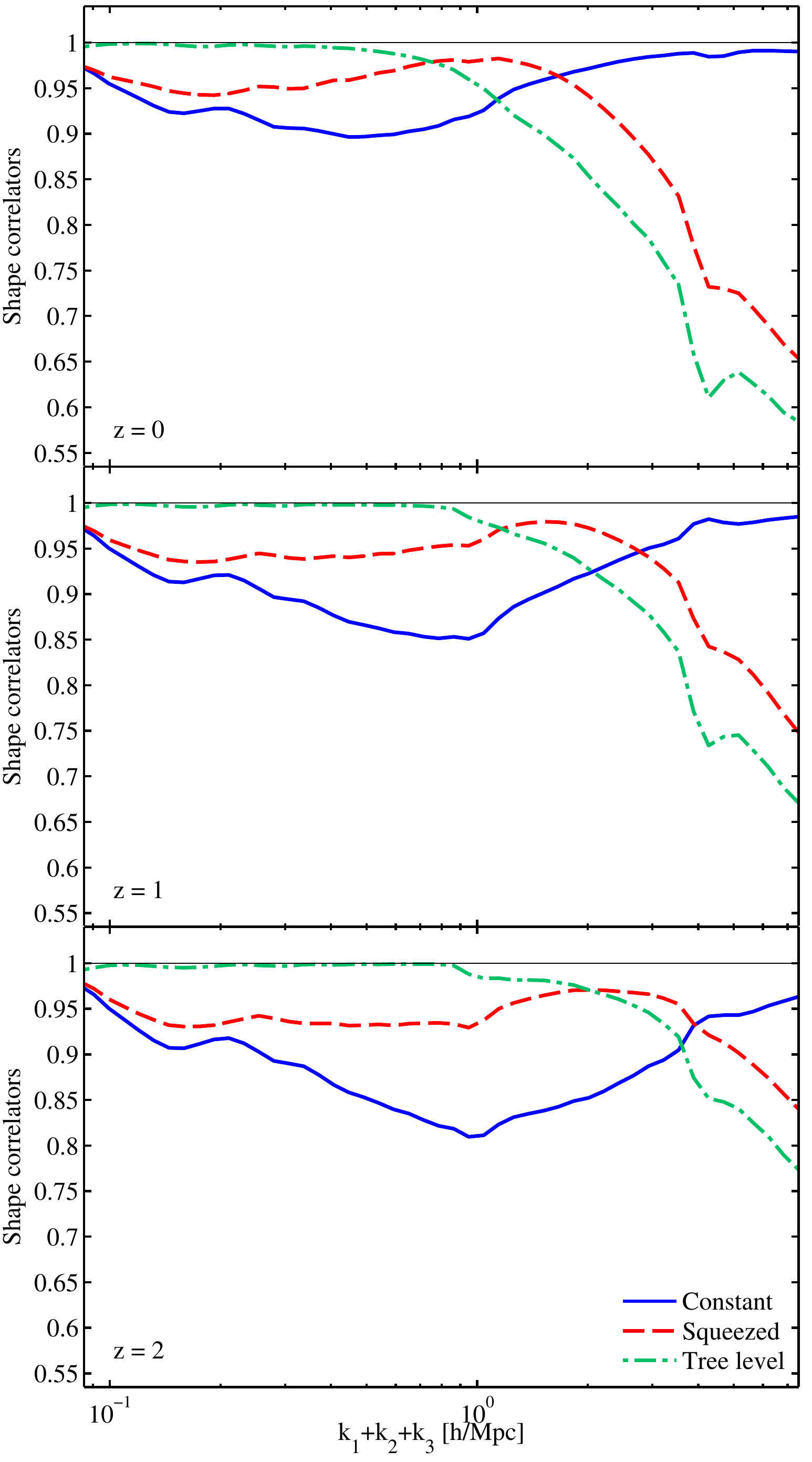} 
\caption{Sliced shape correlations of the measured $N$-body bispectrum with the three canonical shapes: constant  (Eq.~\ref{constantsh}), squeezed (Eq.~\ref{squeezed}) and tree-level  (Eq.~\ref{stree}) shown at redshifts $z = \{0,1,2\}$ (upper to lower panels).   The sliced or binned shape correlator on a given $K=k_1+k_2+k_3$ slice is defined in Eq.~(\ref{shapecorbin}).}
\label{bissh}
\end{center}
\end{figure}

\section{Towards a three-shape bispectrum benchmark model}
\label{sec:benchmark}

In this section we analyse the measured bispectrum to identify the shape degrees of freedom required for its accurate construction.   We study the growth rates of each of these contributions, highlighting differences with the standard halo model particularly for the squeezed shape.  We use these results to guide the development of simple phenomenological bispectrum models: the two-halo boost model and the three-shape benchmark model.

\subsection{Simulation bispectrum  shapes}
We first analyse the shapes of the bispectra measured from $N$-body simulations, in analogy with the investigation of the perturbative and halo model shapes we presented in Figs.~\ref{comparatiepert}, \ref{comparatietot} above.   We calculate the sliced or binned shape correlators $\mathcal{S}^S(K) $ between the $N$-body matter bispectrum and the tree-level (Eq.~\ref{stree}), squeezed (Eq.~\ref{squeezed}), and constant (Eq.~\ref{constantsh}) shapes to determine whether, in combination, these three canonical shapes are sufficient to describe the actual bispectrum. 
The panels of Fig.~\ref{bissh} show a consistent behaviour across the range of redshifts considered. We know that, on large scales, perturbations approach linearity and therefore the tree-level bispectrum is expected to be a good approximation to the $N$-body data. The plots show that this is indeed the case, as on these scales ($K \lesssim 0.5 \, h/$Mpc at $z = 0$) there is a high correlation between the simulated bispectrum and the tree-level shape. The scales up to which the bispectrum is completely dominated by the tree-level shape move significantly to larger values of $K$ as the redshift increases, as expected. On small scales, Fig.~\ref{bissh} shows that deep into the nonlinear regime ($K \gtrsim 3 \, h/$Mpc at $z = 0$) the constant shape dominates, which closely corresponds to the one-halo model discussed in Sec.~\ref{sec:comparison} (and as shown previously in Ref.~\cite{Schmittfull2013}). On intermediate scales, there are several competing contributions of comparable magnitude in the transition between constant and flattened regimes.  Nevertheless, Fig.~\ref{bissh} reveals that at all redshifts there is a range of wavenumbers where the squeezed shape exhibits the highest correlation, which is a new result.   These quantitative shape correlation results  confirm the qualitative picture developed from the evolution  of the 3D bispectrum reconstructions shown in Fig.~\ref{bispectrum_halo}. 

These observations can be interpreted using the halo model formalism for which the basic underlying physical assumptions appear to be corroborated qualitatively. On large scales, the three-halo term is dominant because in this regime the particle triplets over which the bispectrum is estimated should typically be in different haloes, thus reflecting the large-scale quasi-linear bispectrum predicted by perturbation theory. 
As shown in Sec.~\ref{hms}, at small $K$ the tree-level shape is the most important contribution to the observed bispectrum.
On small scales, the three particles are typically in the same nonlinear virialised halo, and hence the one-halo component dominates;  this has a constant shape, which we confirm to be the leading observed bispectrum shape in the high-$K$ limit.
The two-halo term contributes over intermediate lengthscales, where two particles are in one halo and the third particle is elsewhere; this corresponds to the squeezed shape, which indeed we find to be dominating the bispectrum on intermediate scales (though with a larger contribution for $z>0$ than expected in the standard halo model).

\begin{figure}[tb]
\begin{center}
\includegraphics[width=\linewidth]{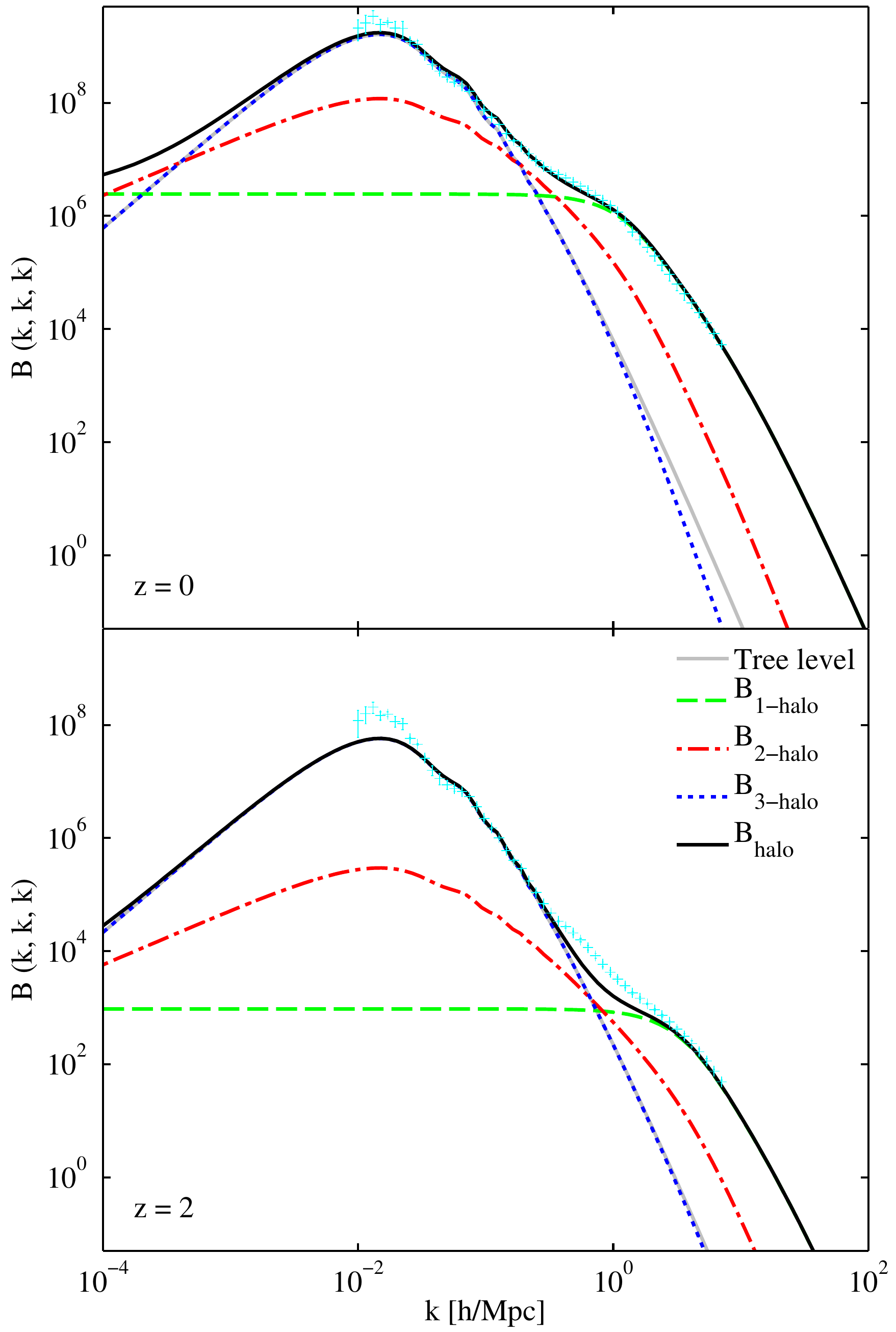} 
\caption{Equilateral configuration of the halo model bispectrum at $z=0$ (top panel) and $z=2$ (bottom panel), showing the contributions of the three components of the halo model, contrasted with the measurements from $N$-body simulations (cyan points).  Note the emerging deficit on intermediate scales at $z=2$.}
\label{tinkereq0}
\end{center}
\end{figure}

As a further illustration, we show in Fig.~\ref{tinkereq0} the equilateral bispectrum ($k_1=k_2=k_3$) of the halo model at $z=0$ and $z=2$ compared with the measured equilateral $N$-body bispectrum. 
Here we can see more clearly the three terms contributing to the halo model and how the two-halo term provides the most significant contribution at intermediate scales at $z=0$.   However, a deficit emerges relative to $N$-body simulations at $z=2$ where the predicted two-halo term no longer dominates over the one- and three-halo terms.

\subsection{Two-halo boost model}

Based on the observation that the halo model has a deficit at intermediate scales, which is found for bispectrum slices in different configurations and becomes more severe as the redshift increases, we have explored simple phenomenological ways of improving the model.
 The two-halo term of the halo model has its highest and most important contribution where the deficit is worst. 
 
\begin{figure}[tb]
\begin{center}
\includegraphics[width=3in]{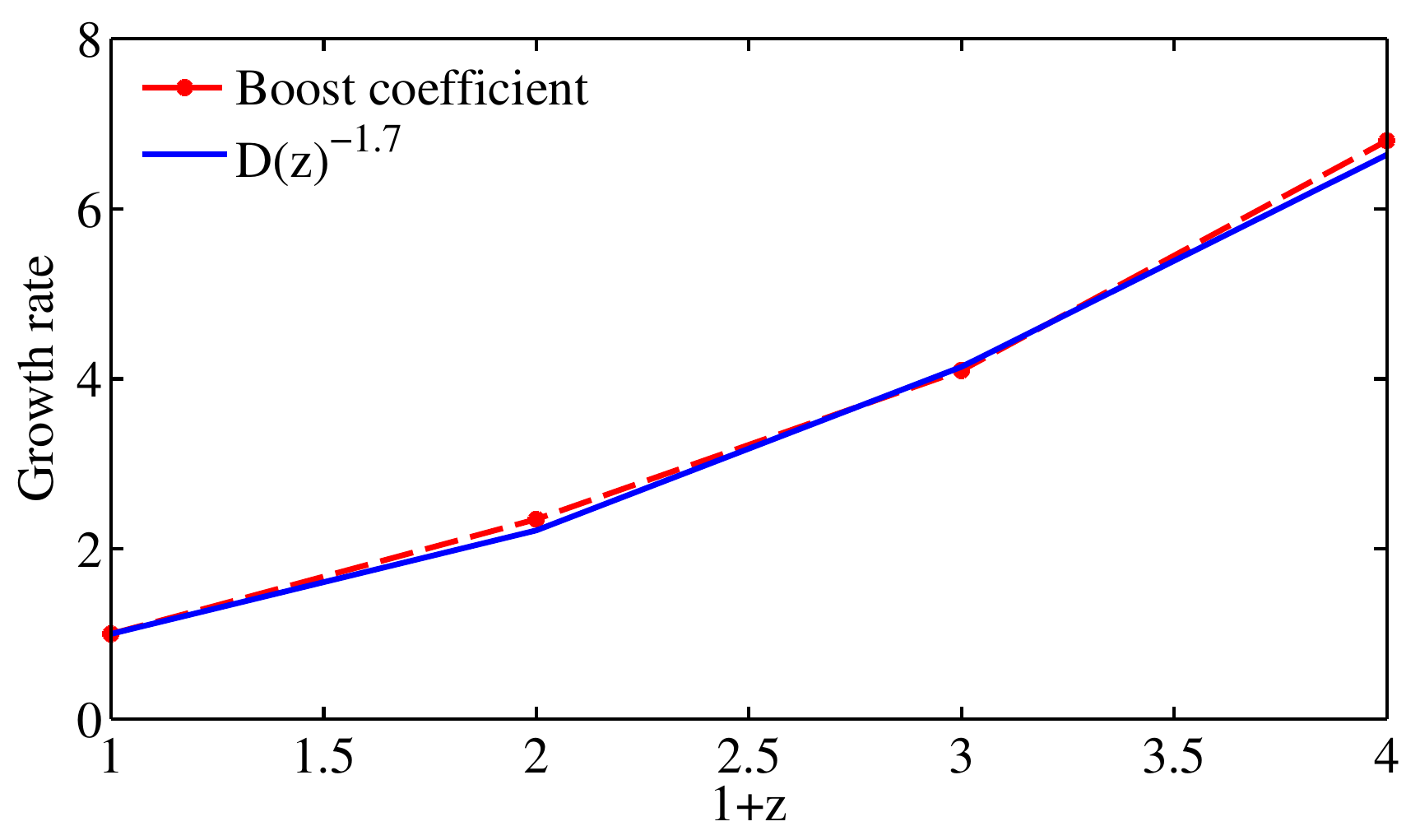} 
\caption{Best-fit boost coefficient to simulations for the two-halo term compared to $D(z)^{-1.7}$. }
\label{boost}
\end{center}
\end{figure}

 As a first simple method to improve the agreement between the model and the simulations, we increase the contribution of the two-halo term at higher redshifts in order to compensate for the deficit. We find that a `boosted' two-halo term can provide a much better fit to numerical simulations for redshifts $z>0$ by multiplying the existing two-halo term by the heuristic factor $D(z)^{-1.7}$.
We determine this `best-fit' factor by computing the total correlator $\mathcal{T}$ of Eq.~(\ref{totalcor}) between the model and the simulations separately at each redshift, and then obtaining the scaling law by maximising the correlator $\mathcal{T}$. 
We show in Fig.~\ref{boost} 
 that the function $D(z)^{-1.7}$ describes well the numerical values found over the relevant redshift range.
This simple method solves the power deficit in  the intermediate regime but we discussed previously how the halo model already has an excess of power as $k \to 0$, driven by the combination of one- and two-halo terms (for $z>0$). Therefore, there is a quantitative problem with simply boosting the two-halo term because it increases the excess on very large scales.  In Sec.~\ref{sec:testhalo}, we will make direct comparisons with the standard halo and other models.

\subsection{Two-shape time-shift model}

In Ref.~\cite{Schmittfull2013} using tree-level and constant bispectrum shapes it was already recognised that simple phenomenological models of the bispectrum could be constructed; this was motivated by explaining the different growth rates of primordial non-Gaussian shapes in terms of an initial time offset.   This time-shift model relies on the fact that in the nonlinear regime the matter bispectrum can be approximated by the constant bispectrum using the following ansatz (consistent with Eq.~\ref{separable}):
\begin{equation}
B_{\text{const}}\left(k_1,k_2,k_3\right)=c_1 D(z)^{n_h}K^{\nu} \, ,
\end{equation}
with two free parameters, an amplitude $c_1$ and a growth rate $n_h$ determined from simulations, plus a scale-dependence $\nu\approx -1.7$ for equilateral configurations in the one-halo model \cite{valageas2, 1475-7516-2012-08-036}. 
This two-shape model was further improved by replacing the tree-level bispectrum (Eq.~\ref{stree}) with the nonlinear tree-level bispectrum (Eq.~\ref{streeNL}), i.e. the tree-level bispectrum calculated with the nonlinear power spectrum from simulations:
\begin{equation}
B_{T\text{-shift}}\left(k_1,k_2,k_3\right)=c_1 D(z)^{n_h}K^{\nu} + S^{\text{treeNL}}\left(k_1,k_2,k_3\right)
\end{equation}
While this model produced a reasonable description of the matter bispectrum in terms of the shape correlation $\mathcal{S}$ (see Ref.~\cite{Schmittfull2013}), our more detailed analysis here with the binned shape correlator $\mathcal{S}^S$ has revealed the possibility of further improvement on intermediate scales by extending the model with the additional squeezed shape of Eq.~(\ref{squeezed}).

\subsection{Three-shape bispectrum model}
\label{sec:three-shape}
Based on the three shapes we identified in the halo model in Sec.~\ref{hms}, we propose a more general benchmark model that incorporates the physical behaviour of all these components, but with rescaled growth factors to provide an improved quantitative fit to simulations. As shown in Fig.~\ref{comparatietot},  the one-, two- and three-halo terms have a high shape correlation with the constant, squeezed and tree-level shapes respectively on slices of constant $K \equiv k_1+k_2+k_3$.   Since these shapes also describe the measured matter bispectrum (see Fig.~\ref{bissh}),  we can construct it as a scale-dependent sum of three templates:
\begin{multline}
B_\text{3-shape}(k_1,k_2,k_3)= \sum _{i=1}^3 f_i(K)\, S^{i}(k_1,k_2,k_3) \\
=f_{1h}(K) \, S^{\text{const}}(k_1,k_2,k_3) 
+  f_{2h}(K) \, S^{\text{squeez}}(k_1,k_2,k_3) \\
 +   f_{3h}(K) \, S^{\text{treeNL}}(k_1,k_2,k_3)  \, ,
\label{shapes}
\end{multline}
where the nonlinear tree-level, squeezed, and constant shapes $S^{\text{treeNL}}, S^{\text{squeez}}, S^{\text{const}}$ are defined in Eqs.~(\ref{constantsh}), (\ref{squeezed}) and (\ref{streeNL}) respectively, and the amplitudes $f_{1h}, f_{2h}, f_{3h}$ were discussed in Sec.~\ref{hms} in the context of the halo model.\footnote{An even simpler three-shape model can be obtained by substituting the linear tree level (Eq.~\ref{stree}) for the flattened three-halo shape;  it provides a satisfactory fit to the simulations. In this simple scenario, the fitting functions $f_{1h}$ (Eq.~\ref{eq:f1h}) and $f_{3h}=1$ are given by the standard halo model, while for the two-halo term we allow an improved fit and growth scaling (Eq.~\ref{2ht}) with coefficients $C=240 \, D(z)^{-1}$ and $D=2.35 \,h\,\text{Mpc}^{-1}\, D(z)^{-1}$.  However, the three-shape model of Eq.~(\ref{shapes}) above provides an improved fit in the flattened limit in the quasi-nonlinear regime.}

\begin{figure}[!tb]
\begin{center}
\includegraphics[width=0.9\linewidth]{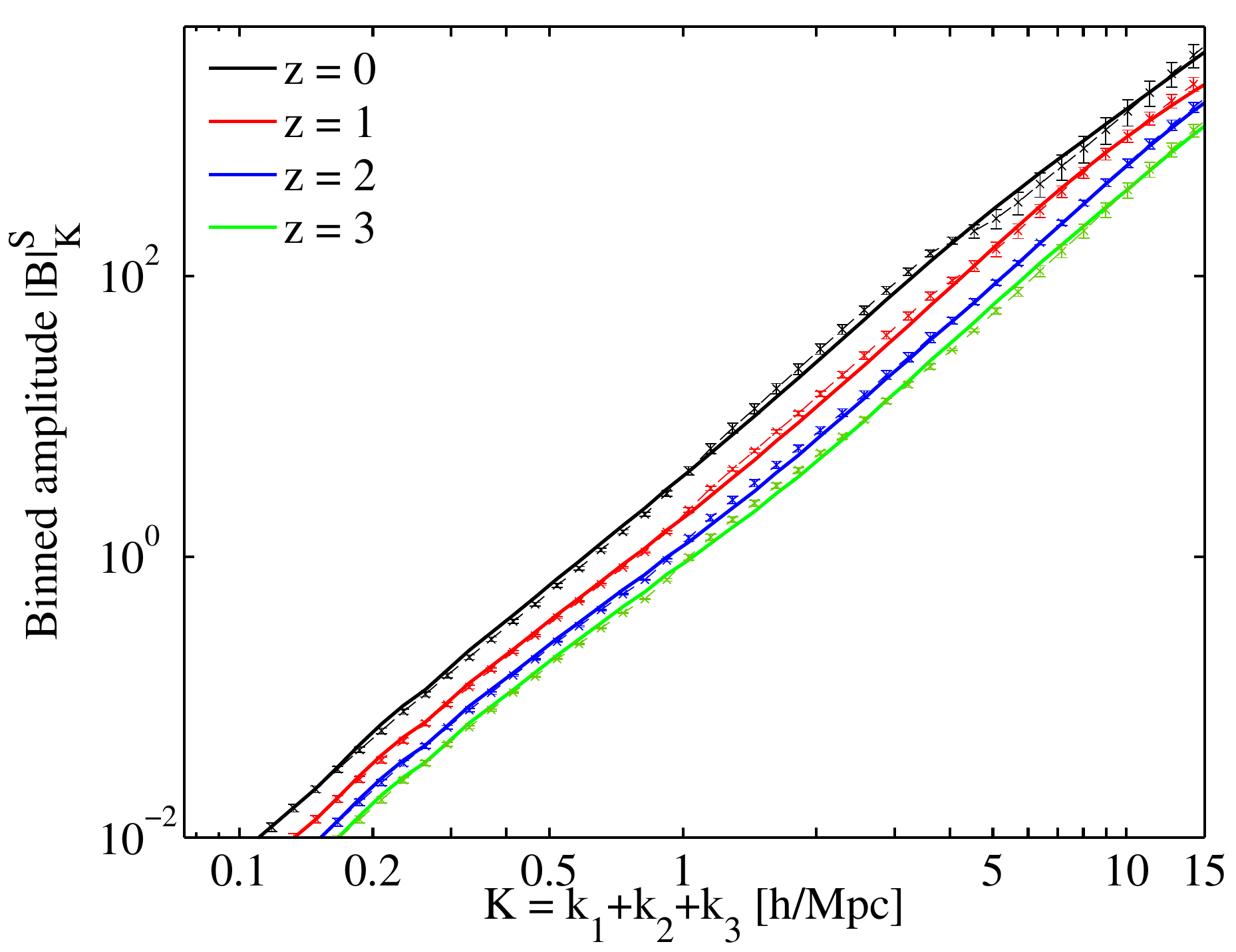} \\
\includegraphics[width=0.9\linewidth]{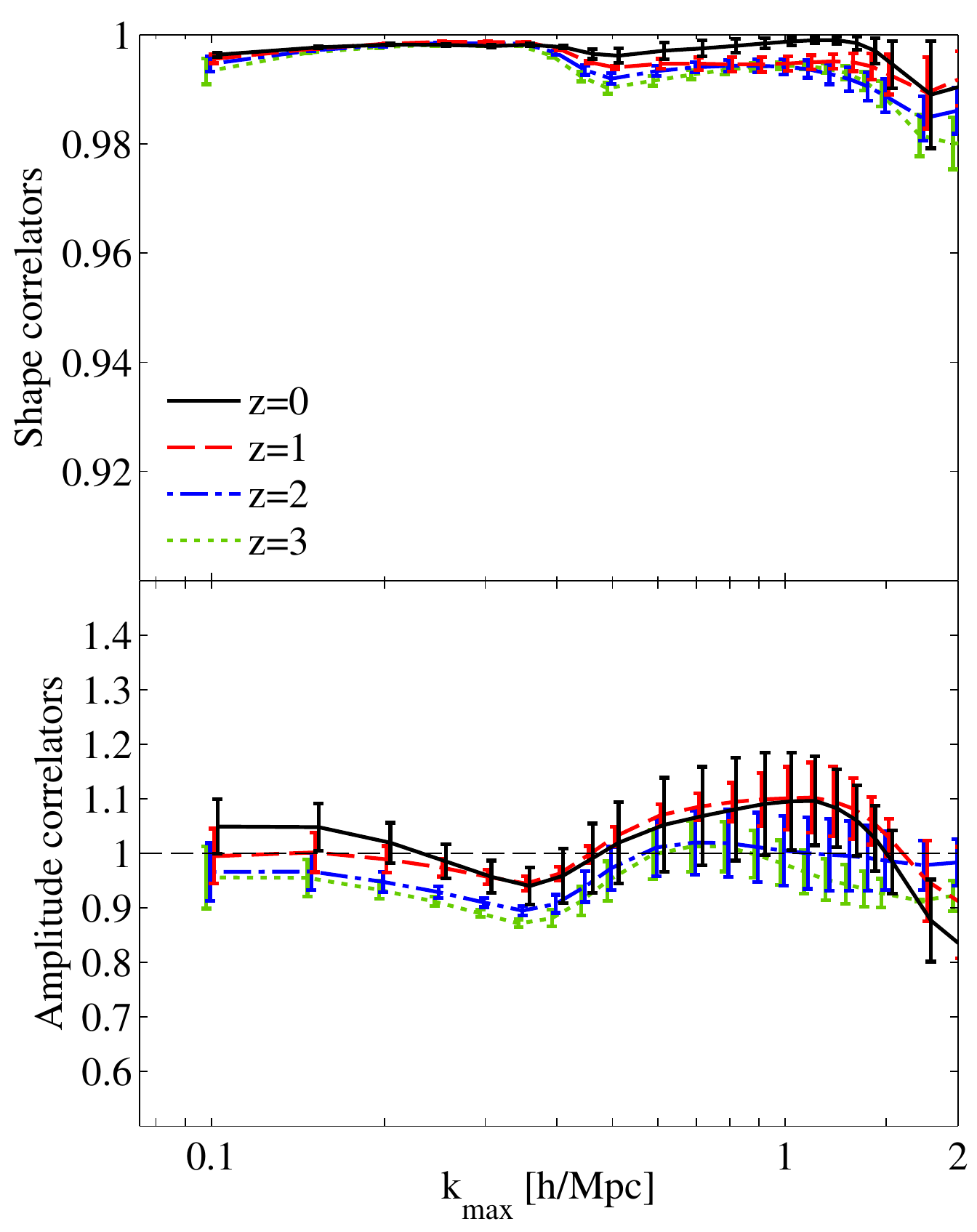} 
\caption{Comparison between the measured $N$-body matter bispectrum and the three-shape model. The top panel shows the binned amplitude $|B|^S(K)$ from the simulations (points and dashed lines) and from the fitted three-shape model (solid lines) at redshifts $z = \{0, 1, 2, 3\}$.  The middle and bottom panels show a relative comparison  between the simulations and the benchmark model, using the binned shape and amplitude correlators, $\mathcal{S}^S$ and $\mathcal{A}^S$. These results demonstrate that the three-shape model exhibits a high shape correlation on all scales and describes the simulated data well.}
\label{absb}
\end{center}
\end{figure}

\begin{figure}[tb]
\begin{center}
\includegraphics[width=2.5in]{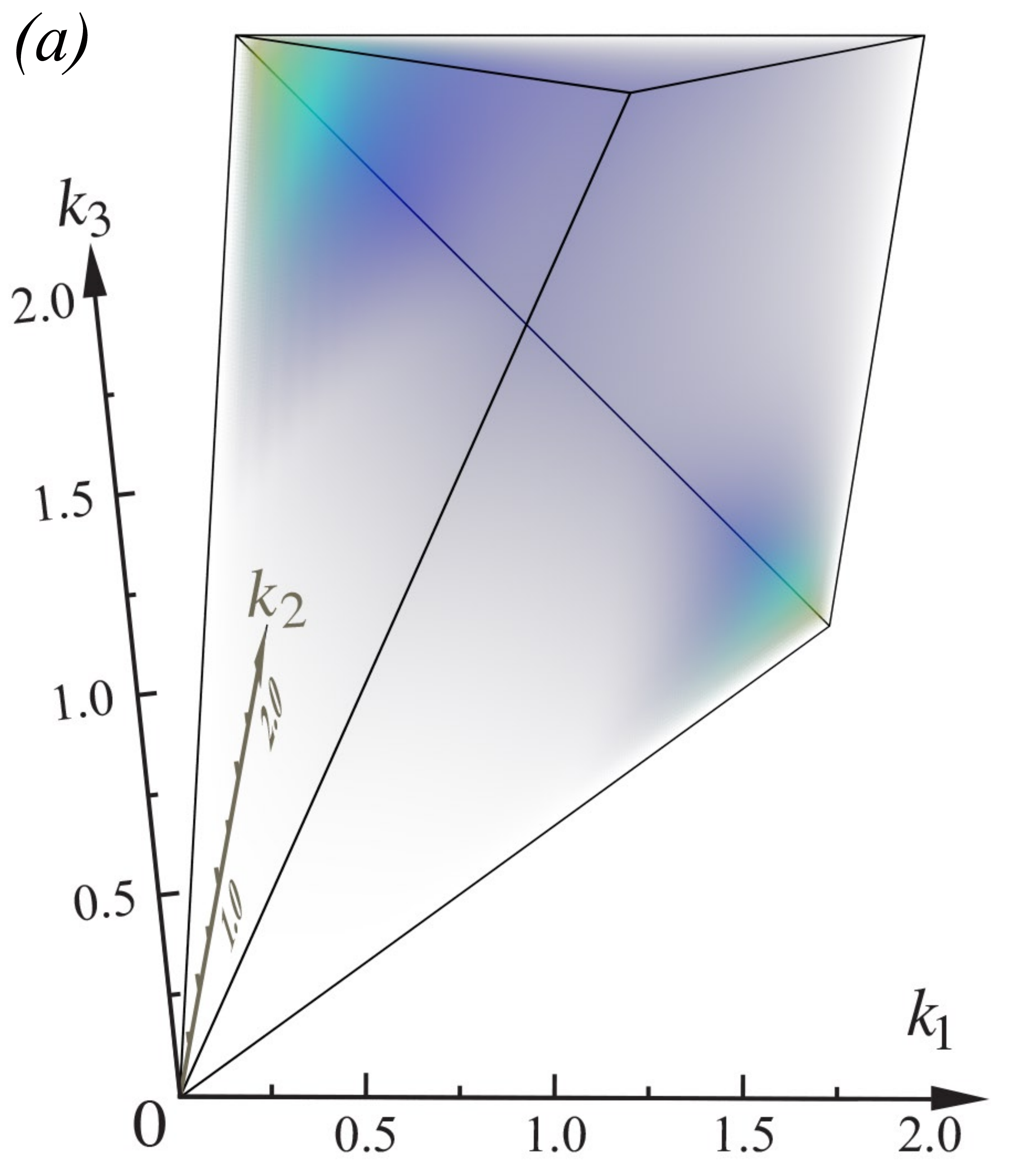}\\
\includegraphics[width=2.5in]{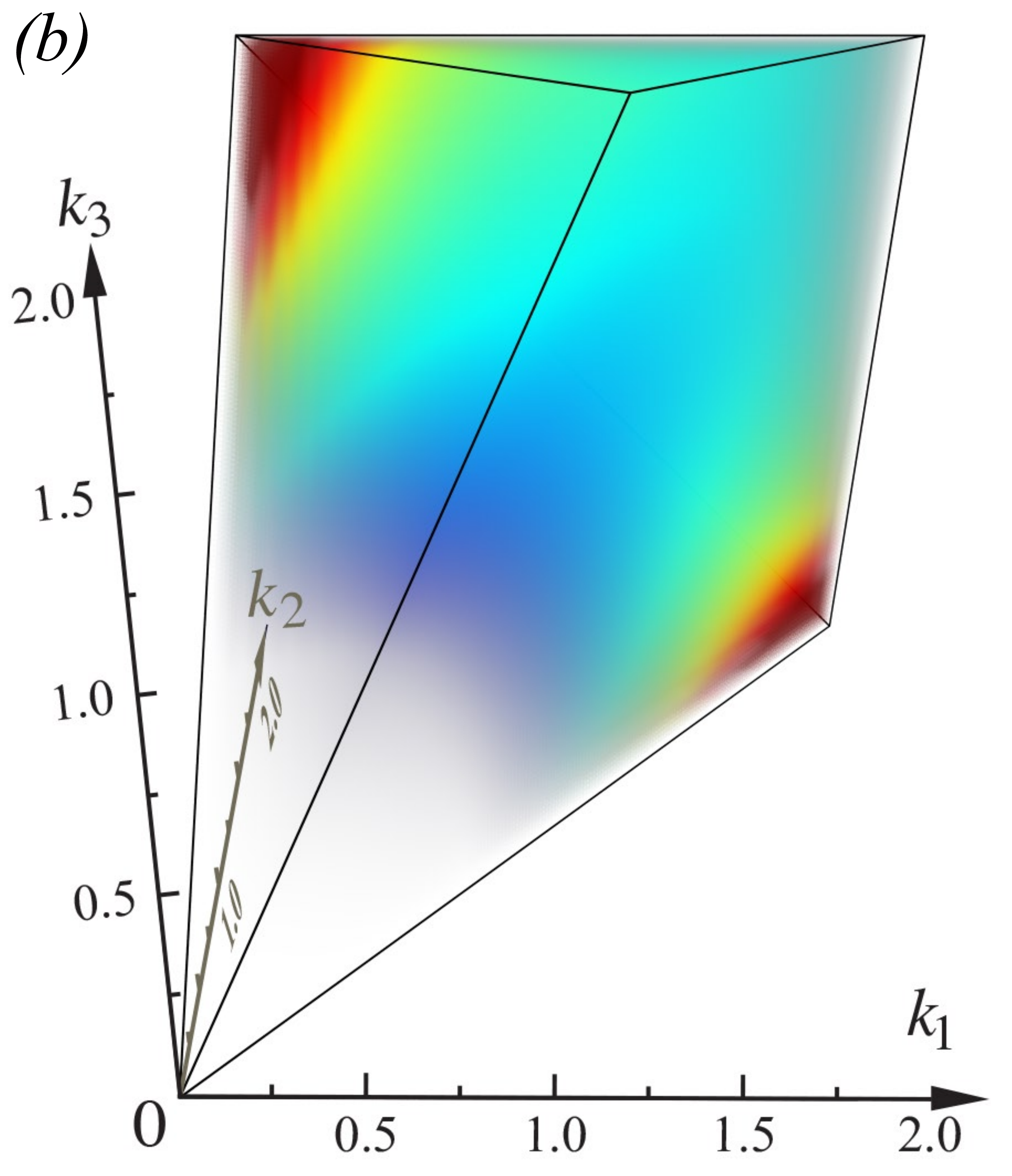}\\
\includegraphics[width=2.5in]{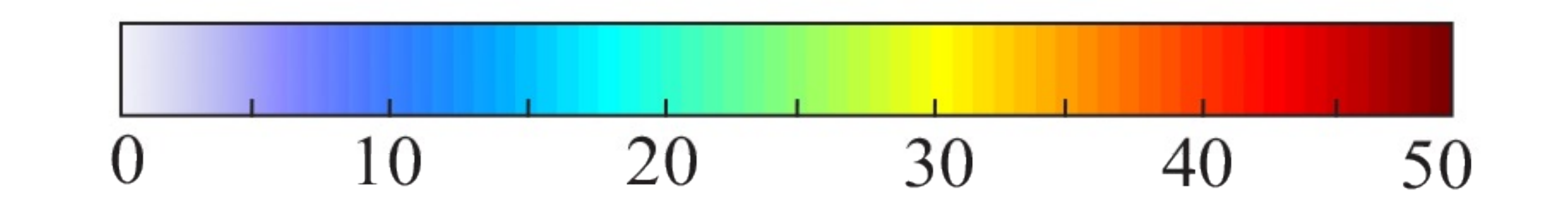}
\caption{The SN-weighted two-halo bispectrum (Eq.~\ref{b2h}) (upper panel) at $z=2$ compared to the best-fit two-halo squeezed shape ansatz (Eq.~\ref{b2hbis}); this allows the `three-shape' benchmark model to accurately match the simulation data shown in Fig.~\ref{bispectrum_halo}.  The two-halo model clearly exhibits a large deficit and does not describe squeezed contributions adequately at higher redshift.}
\label{two-halo_comp}
\end{center}
\end{figure}

We know that the one-halo term provides an adequate description of the matter bispectrum on small scales, so we fix the amplitude $f_{1h}$ to the simple functional fit of Eq.~(\ref{eq:f1h}) for the one-halo model presented in Sec.~\ref{hms}. On the largest scales, where the three-halo term is dominant, we know that the tree-level shape (Eq.~\ref{b3hbis}) provides an excellent fit to simulations.  However, on intermediates scales, while the shape correlation remains good beyond the strictly perturbative regime (see Fig.~\ref{bissh}), its amplitude is insufficient, as can be seen by comparing Fig.~\ref{bispectrum_tree} with  Fig.~\ref{bispectrum_halo}).  For this reason, we have chosen the nonlinear tree-level form (Eq.~\ref{streeNL}) instead because of its higher amplitude and the fact that it is a better approximation to one-loop perturbative expansions.  Nevertheless, it is well known that introducing the nonlinear power spectra into halo models generically causes excess power at low redshifts $z\approx 0$, so we need a prescription for cutting off the flattened shape in nonlinear regions (see, for example, the discussion about the combined halo-PT model in Sec.~\ref{sec:ihm} or the discussion of halo exclusion in Ref.~\cite{PhysRevD.83.043526}).  In order to keep this three-halo suppression as simple as possible we take an exponential form: 
\begin{equation}
f_{3h}=\exp(-K/E)\,,
\label{f3hnew}
\end{equation}
where we fit $E$ to simulations at several redshifts to obtain an appropriate amplitude and growth rate; in principle, it should be linked to the nonlinear scale $k_{\rm NL}$ satisfying $k^3 P_{\text{lin}}(k,z)=2 \pi^2$.  Finally, for the squeezed shape scaling $f_{2h}$ we do not use the two-halo model amplitude, but instead the prescription of Eq.~(\ref{2ht}) with the two free parameters $C$ and $D$ obtained from simulations (see discussion in Sec.~\ref{hms}).    By matching $f_{2h}$ to the excess in the measured bispectrum at redshifts $z=\{0,1,2,3\}$, together with the cutoff scale in $f_{3h}$, we obtain the following approximate fit for the coefficients $C$, $D$ and $E$:
\begin{align}
\label{three-shape-fit}
C&=140 \, D(z)^{-5/4} \nonumber \\
D&=1.9 \,h\,\text{Mpc}^{-1}\, D(z)^{-3/2} \\
E&=7.5 \, k_{\text{NL}}(z) \, .\nonumber 
\end{align}
We emphasise that this is different from the previous two-halo fits of Eqs.~(\ref{2htf1}-\ref{2htf2}), because these were obtained by fitting to the two-halo model predictions, which underestimate power for $z>0$.  This is illustrated starkly at $z=2$ in Fig.~\ref{two-halo_comp}, where we compare the standard two-halo model prediction with the squeezed shape of Eq.~(\ref{2ht}) with best fit simulation parameters of Eq.~(\ref{three-shape-fit}).  We also note that for redshifts $z>1$, the lengthscale $E$ moves rapidly to large $K\gg 1 \, h/$Mpc, so the exponential suppression $f_{3h}$ term (Eq.~\ref{f3hnew}) acts primarily to reduce power in the $z=0$ bispectrum and is less relevant elsewhere.

In Fig.~\ref{absb} we plot the value of the binned amplitude $|B|^S(K)$ for the three-shape model of Eq.~(\ref{shapes}), which we compare directly to the measured bispectrum from simulations; we also show the binned shape $\mathcal{S}^S$ and amplitude $\mathcal{A}^S$  correlators between the model and $N$-body bispectrum.    The plots show a good fit using the three-scale model across all scales $k>0.1 \,h/$Mpc and all redshifts.   The shape correlations in this range are approximately 99\% or higher and the amplitude correlator is within 10\% of the measured bispectrum (consistent given present simulation uncertainties). These correlation results are in line with expectations for a good fit for an $n_{\rm max}=50$ eigenfunction decomposition (Eq.~\ref{eigenfns}) (see validation discussions in Ref.~\cite{Schmittfull2013}).  We note that given the high shape correlations, we could introduce additional degrees of freedom in $f_{1h}, f_{3h}$ to improve this quantitative fit further, but our purpose first is to demonstrate the efficacy of this simple approach.   

Employing this new three-shape model as a benchmark has several advantages over using the simulated bispectra directly, though we will use both in subsequent discussions.   First, it smooths out any systematic discontinuities appearing where the simulations are joined together. Secondly, it allows direct comparisons with theoretical models without performing eigenfunction decompositions on the latter, so residual offsets do not have to be subtracted.  And finally the model is simple, capturing the most important features of the halo model without requiring computationally costly re-evaluations at all wavenumber combinations $(k_1,k_2,k_3)$, and thus it can be seen as an initial step towards a full \textsc{Halofit}-style phenomenological model of the matter bispectrum.

\subsection{Directions for further improvement}

The three-shape benchmark model achieves a high degree of correlation with the full bispectrum from $N$-body simulations, however undoubtedly further improvements of this model can be achieved in future, not least by deriving some key results from first principles, such as the modified two-halo growth rates. In principle, showing that the matter bispectrum is well approximated by the  separable form of Eq.~(\ref{shapes}) should considerably simplify mathematical modelling.

One improvement that can be incorporated into the model is to replace the nonlinear tree-level shape (Eq.~\ref{streeNL}) with specific one- and two-loop perturbative expansions.  However, while this approach could extend the tree-level shape further into the nonlinear regime, it requires prescriptions for suppressing the two- and one-halo terms more strongly to avoid over-prediction. This is similar in spirit to the suppression of the perturbative bispectrum contribution in the halo-PT model by Ref.~\cite{valageas2}; but it is clear that an exponential cut-off where the perturbative expansion breaks down is likely too aggressive, since Fig.~\ref{bissh} shows that the tree-level shape is present up to relatively high, $k \sim 1 \, h/$Mpc.

Clearly further improvement  of the three-shape model can be achieved through more extensive comparisons with higher-resolution $N$-body simulations, over a finer grid of scales and redshifts. The quality of fits obtained in the squeezed and flattened limits are constrained in accuracy by the restricted ansatzes chosen, allowing only three redshift-dependent parameters.  The likely outcome is a finer tuning of a larger number of phenomenological free parameters, again in the spirit of the \textsc{Halofit} method, with extensive surveys required to uncover dependencies on cosmological parameters.   

A final point of interest is the question whether the three-shape model we introduced satisfies well-known constraints in the squeezed limit. For example, Ref.~\cite{Chiang2014} derived a consistency relation between the integrated squeezed-limit bispectrum and a response function derived from the power spectrum.  In the case of our three-shape model (Eq.~\ref{shapes}), the tree-level shape term satisfies the consistency relation automatically, as was demonstrated by Ref.~\cite{Chiang2014} for tree-level SPT.
We know that the squeezed- and constant-shape terms of the benchmark model are similar to two- and one-halo terms of the standard halo model; furthermore, as we show in Sec.~\ref{sec:comparison} below, our model performs well compared with the $N$-body simulations in the squeezed limit over the configurations we have tested, so that it is unlikely that there is any large inconsistency. However, a more quantitative test of the consistency relation would require a full numerical evaluation of the integrated bispectrum, which we leave for future investigation.

 \begin{figure*}[tb]
\begin{center}
\includegraphics[width=2.25in]{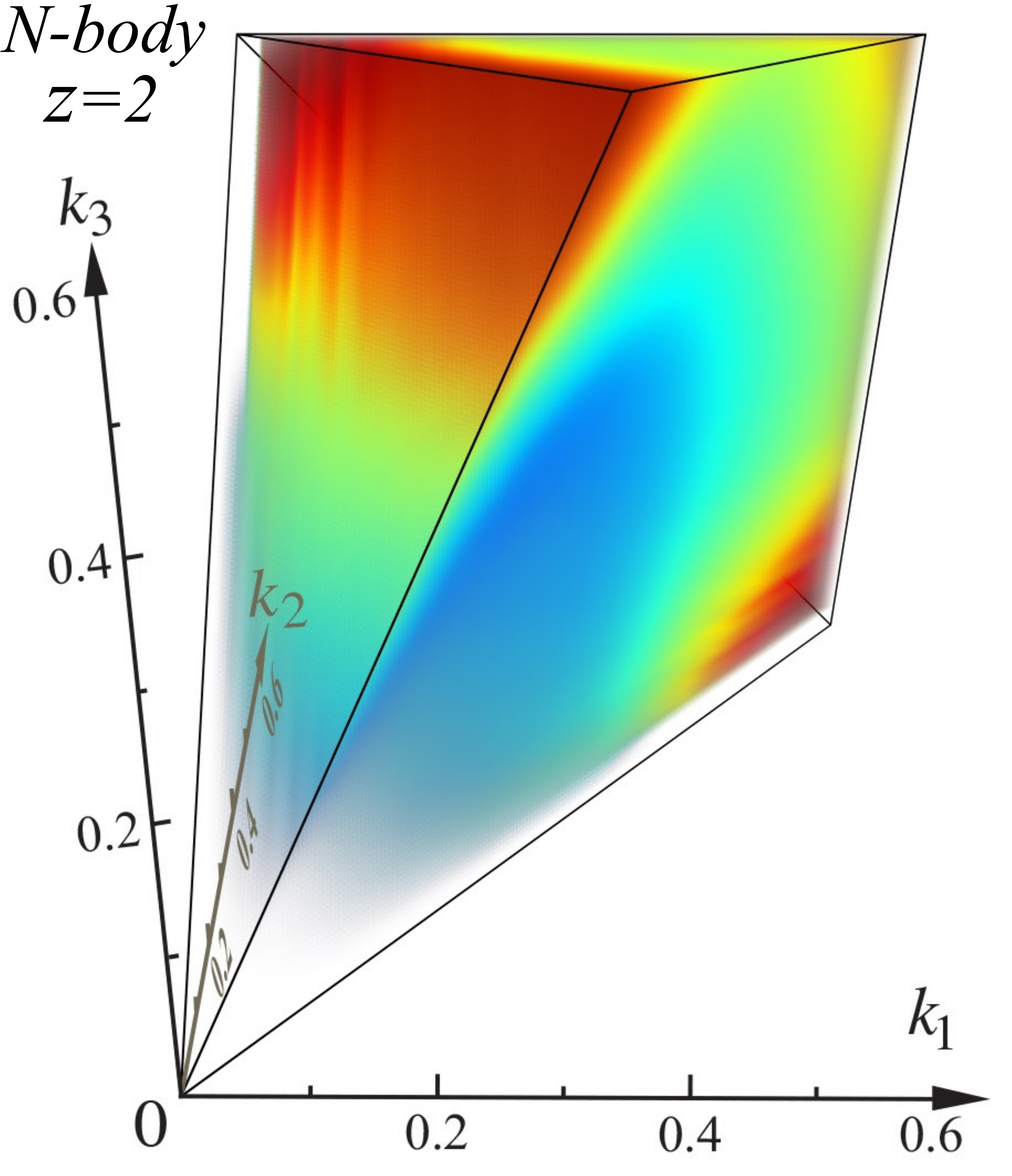}\includegraphics[width=2.25in]{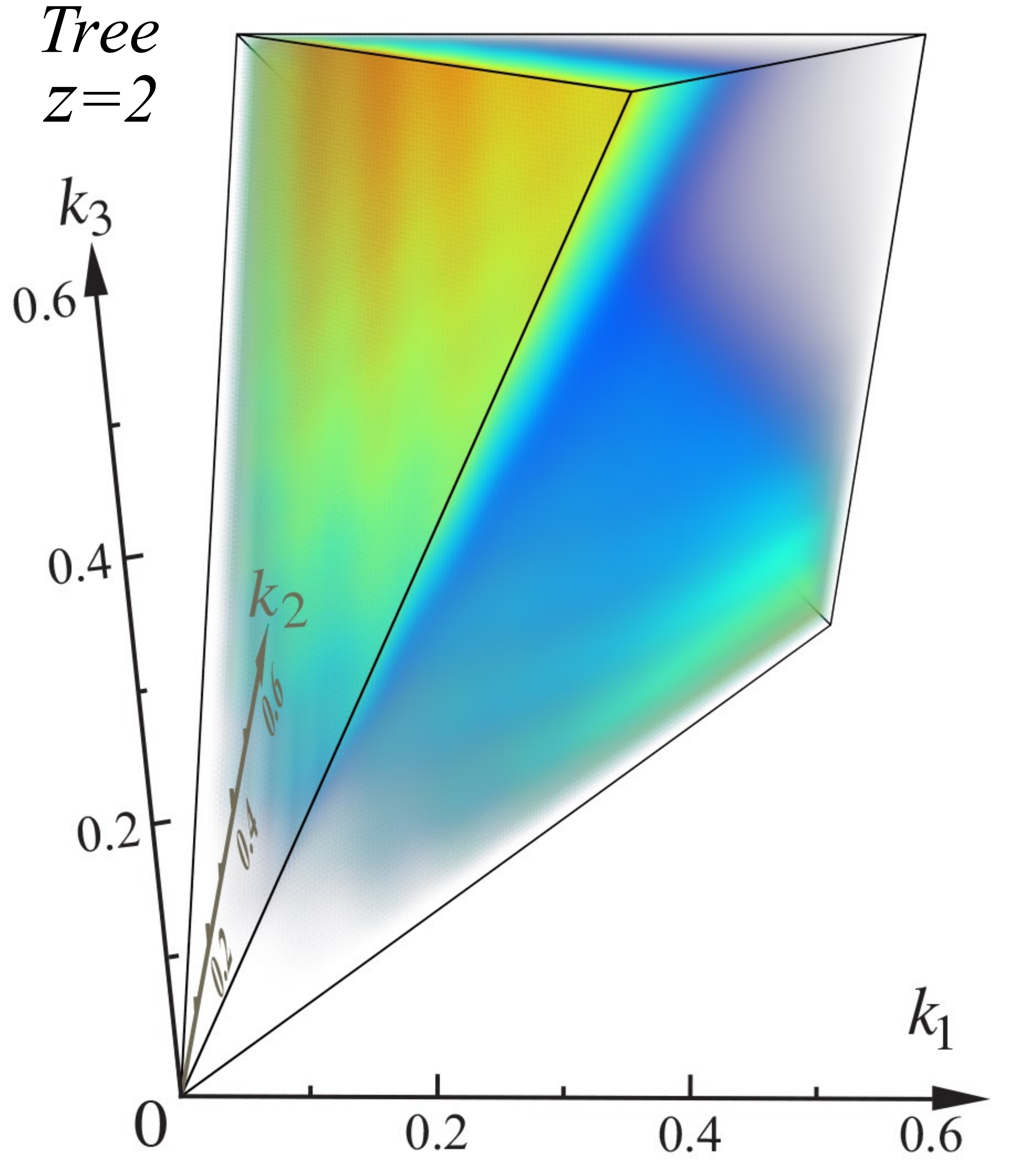}\includegraphics[width=2.25in]{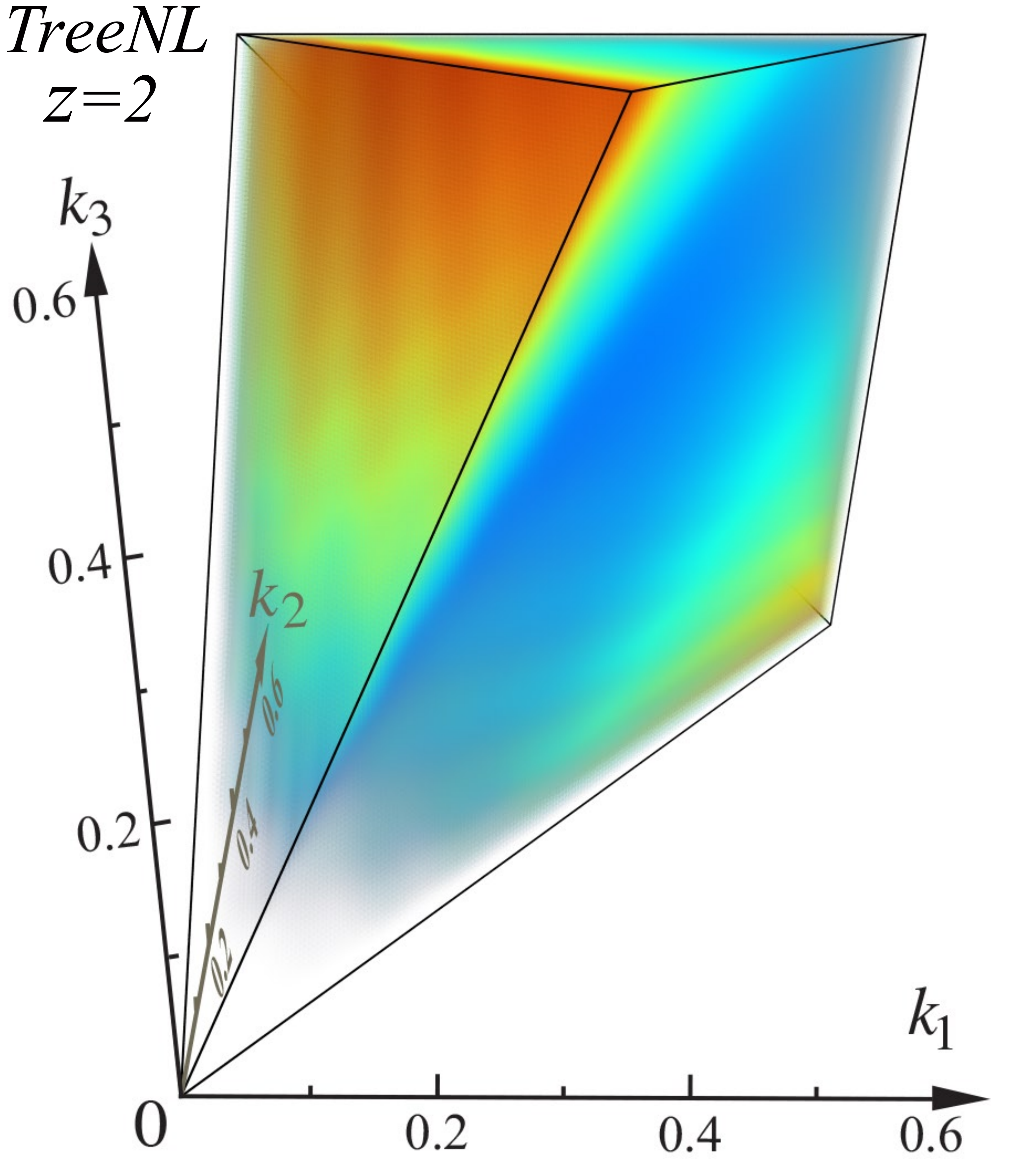}\\
\bigskip
\includegraphics[width=2.25in]{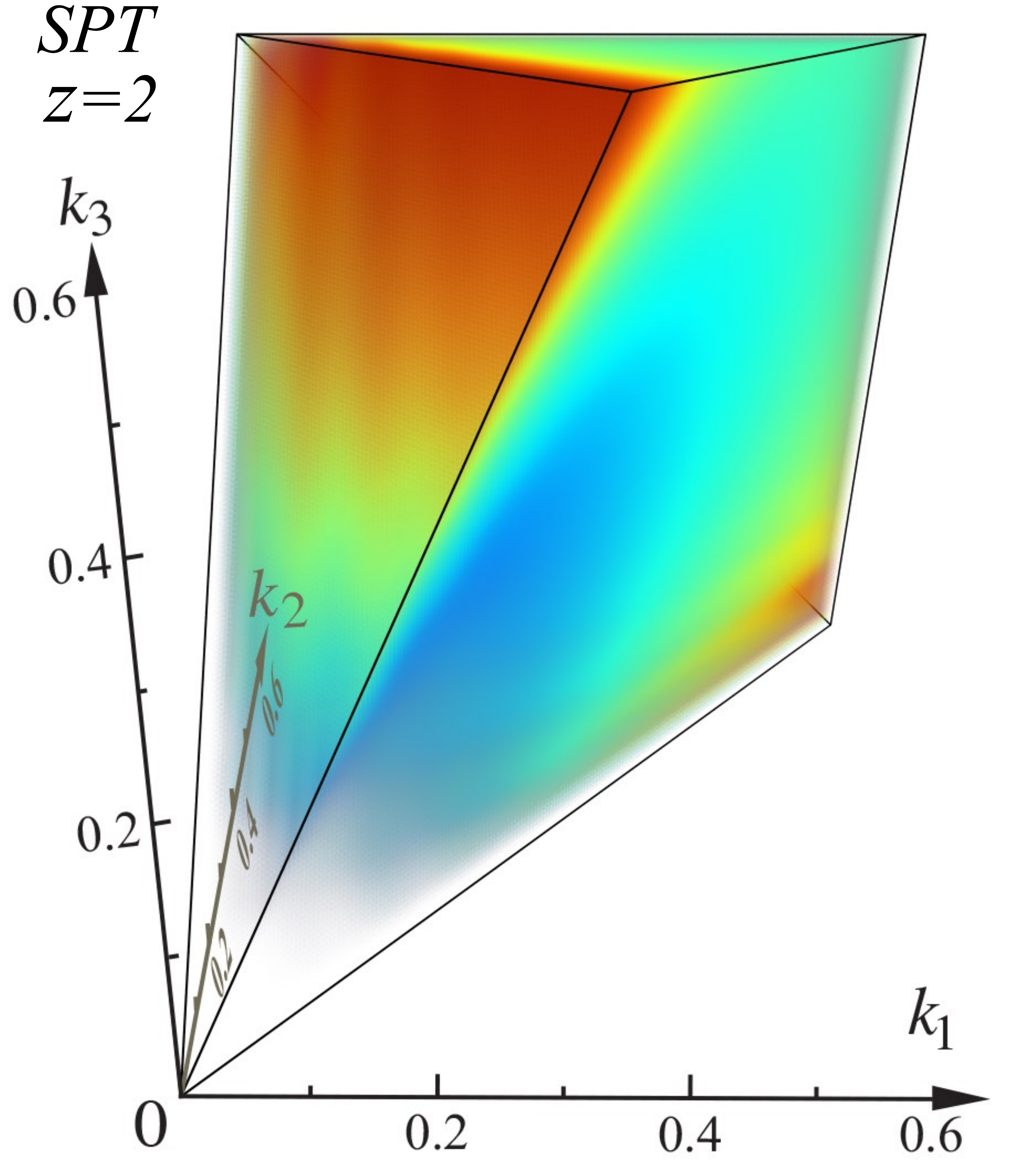}\includegraphics[width=2.25in]{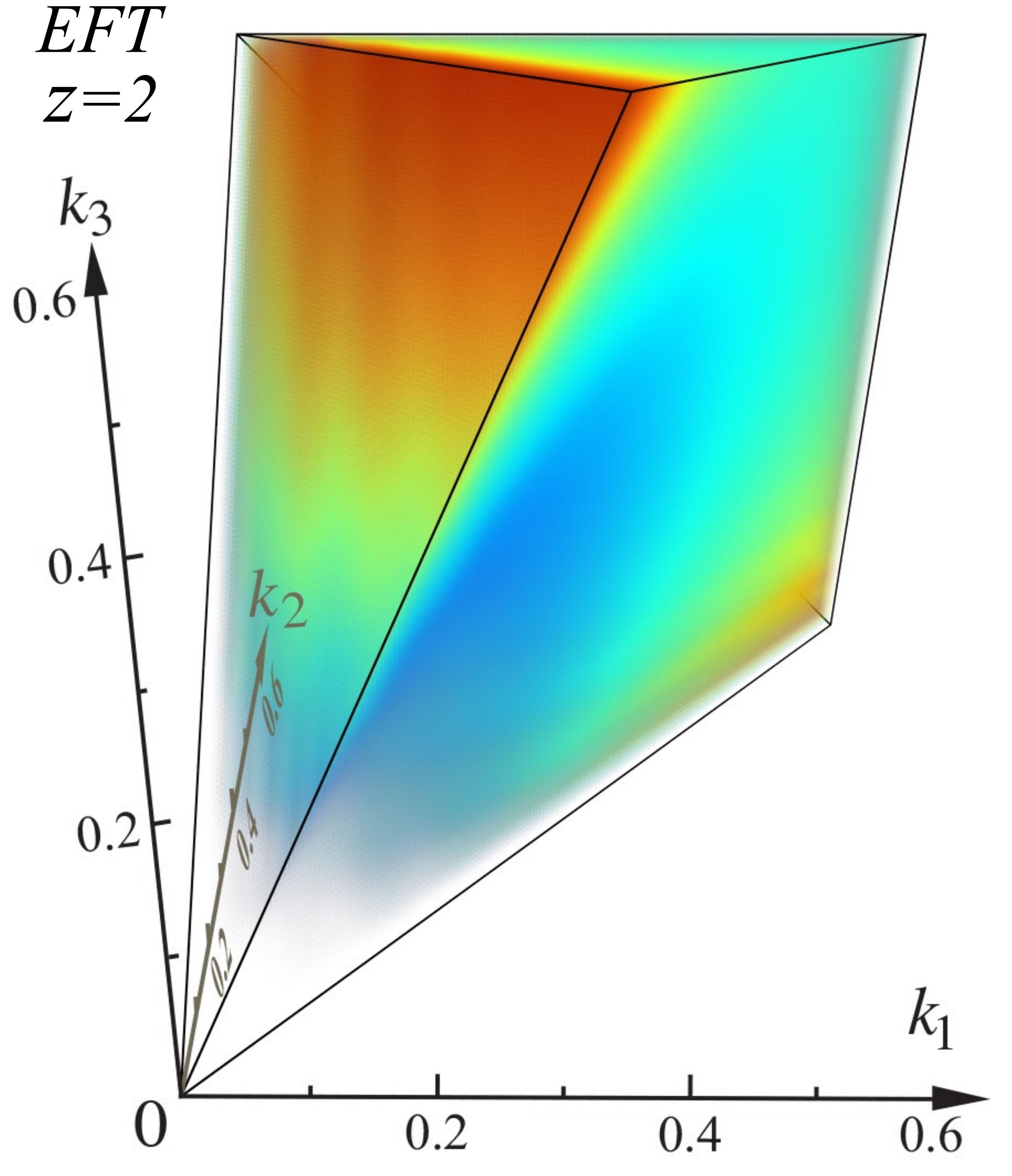}\includegraphics[width=2.25in]{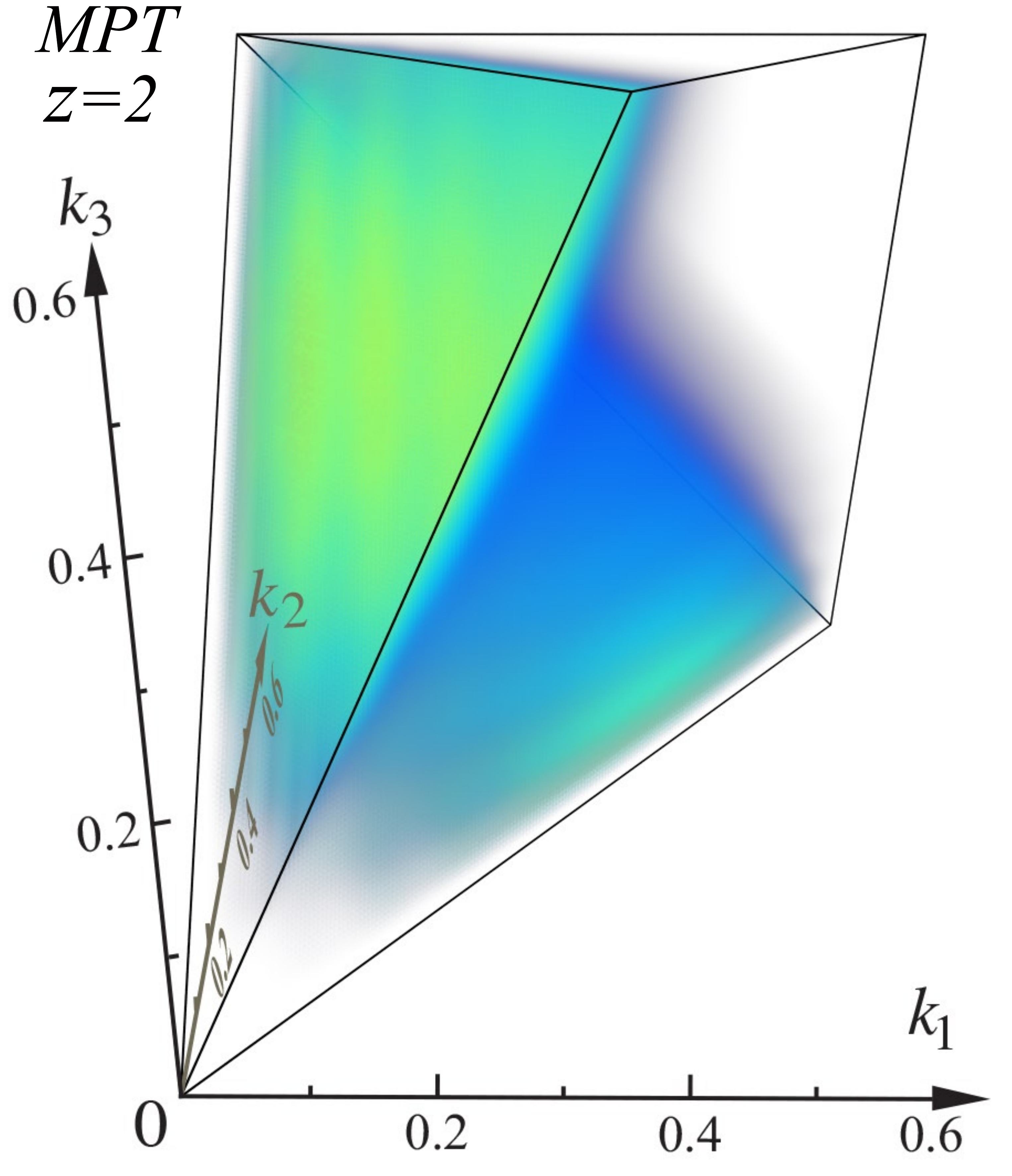}\\
\includegraphics[width=2.25in]{scale20.pdf}
\caption{Comparison at redshift $z=2$ of the SN-weighted bispectrum for perturbative models with the simulation data (top left):   the perturbative models are respectively tree-level bispectrum (top centre), nonlinear tree-level (top right), standard one-loop perturbation theory SPT (bottom left), one-loop effective field theory EFT (bottom centre)  and renormalised perturbation theory MPT (bottom right);  RLPT is not plotted as it appears  very similar to MPT.   Note that all perturbation theories have signal concentrated at flattened triangles (front left face), and so are highly correlated with the tree-level bispectrum shape of Eq.~(\ref{stree}), when using the binned shape correlator  (Eq.~\ref{shapecorbin}).  The $N$-body bispectrum also exhibits a squeezed signal for $k\gtrsim 0.4 \, h/$Mpc. We have chosen $z=2$ so that the PT models decay at higher $k$ and there is more signal to display, but the general behaviour is similar at lower $z$.}
\label{bispectrum_pert}
\end{center}
\end{figure*}

\section{Bispectrum model comparison with simulations}
\label{sec:comparison}

We next use the $N$-body simulations to compare the accuracy of the different theoretical bispectrum models described in the previous sections, both perturbative and non-perturbative.
We present this model comparison in two ways: we first directly compare the simulated and theoretical bispectra over a range of representative triangular configurations (equilateral, squeezed, and flattened), and we then use the full three-dimensional amplitude and shape correlators presented in Sec.~\ref{sec:bisp}.

 At high redshift, all models are expected to perform well over an extended range of scales, as the fluctuations are nearly linear, the power spectrum is linear and the bispectrum can be described by the tree-level expression. At lower redshifts, nonlinearities become more important and significant differences appear between the models. In the comparisons, we concentrate on redshifts $z = \{0, 1, 2\}$, as these span the observable redshift range of most current and future observations from galaxy surveys. We investigate the perturbative methods and the halo models separately, because the perturbative methods decay quickly in the nonlinear regime and therefore their predictions for high $k$ are of no interest; we present the comparison of PT models on scales $k \le 0.4 \, h$/Mpc only.
On the other hand, the phenomenological models, which are either based on or at least inspired by halo models, are expected to perform well even in the fully nonlinear regime; in this case we extend the model comparison up to the smallest scales accessible to the present simulations, i.e. $k \le 7.8 \, h/\text{Mpc}$.

\subsection{Testing alternative perturbative approaches}
\label{sec:testpert}

We first qualitatively compare perturbative bispectrum predictions with the matter bispectrum measured from simulations.  In Fig.~\ref{bispectrum_pert} we plot most of these predictions at redshift $z=2$ in three dimensions for wavenumbers $0.02 \, h/$Mpc $ < k< 0.6\, h/$Mpc, together with the actual $N$-body bispectrum (upper left).
We choose $z=2$ so that the perturbative models decay at higher $k$, and more of the signal is visible, but the overall behaviour is comparable at lower $z$.
  The $N$-body bispectrum shows a dominant flattened signal over the whole domain, which grows in amplitude as  $k$ increases.  Qualitatively, this measured  signal matches well the one-loop SPT and EFT models, as well as the nonlinear tree-level bispectrum, in regions well beyond the strictly perturbative regime.  However, the tree-level and MPTbreeze predictions are appreciably lower for large $k\approx 1\,h/\text{Mpc}$ (with the latter exponentially suppressed for large $k$ by prescription).      

\begin{figure*}[t]
\begin{center}
\includegraphics[height=0.29\textheight]{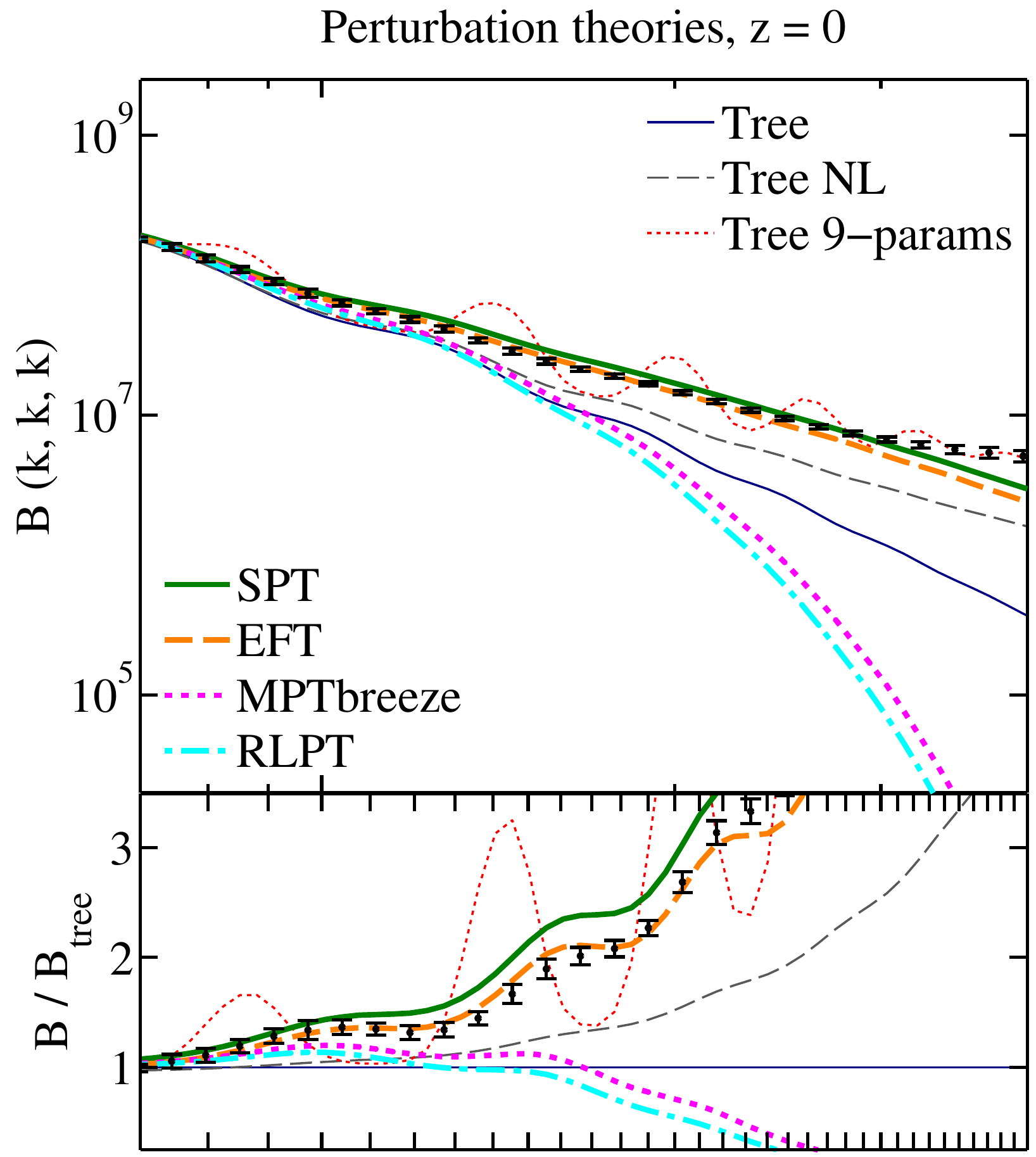}
\includegraphics[height=0.29\textheight]{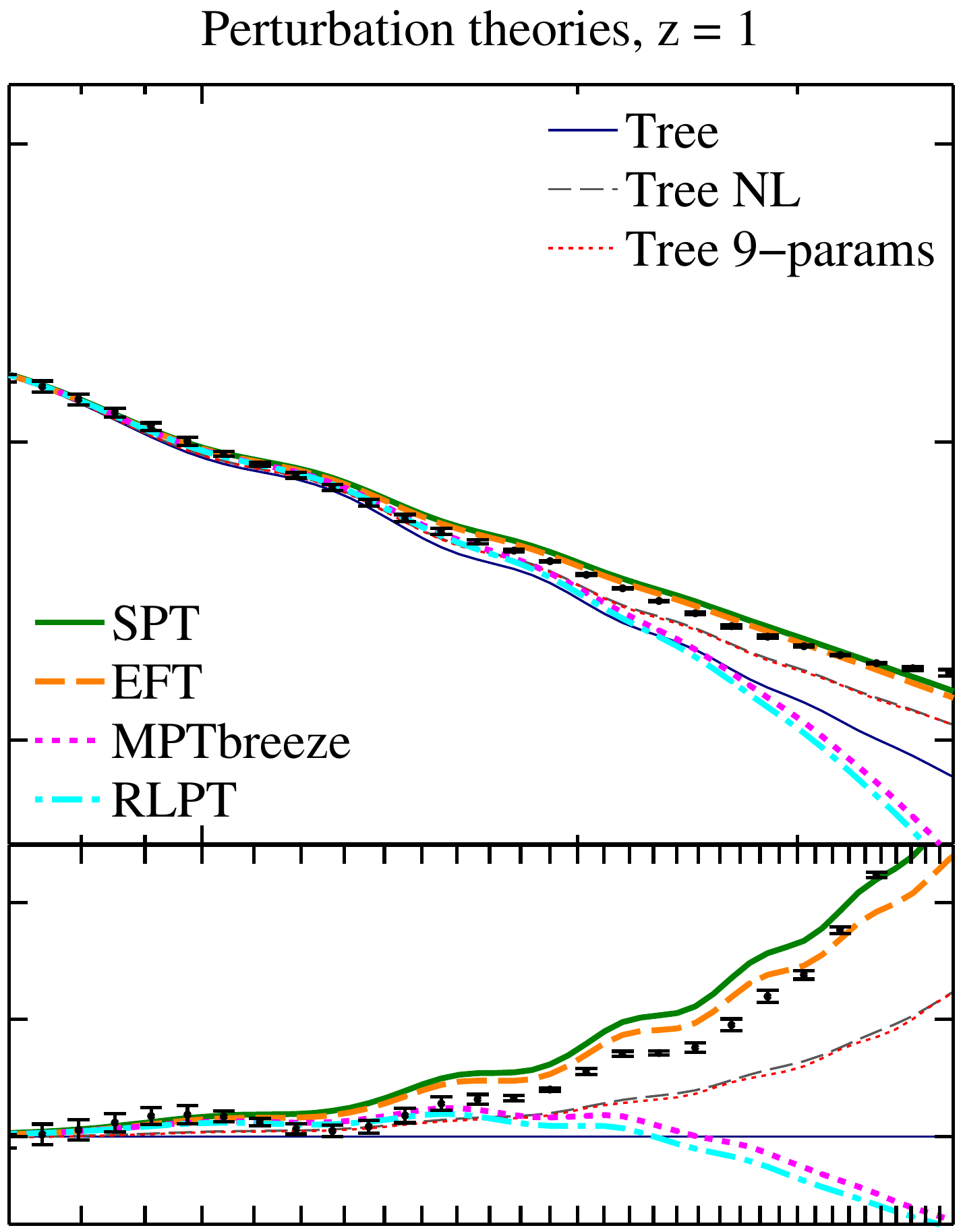}
\includegraphics[height=0.29\textheight]{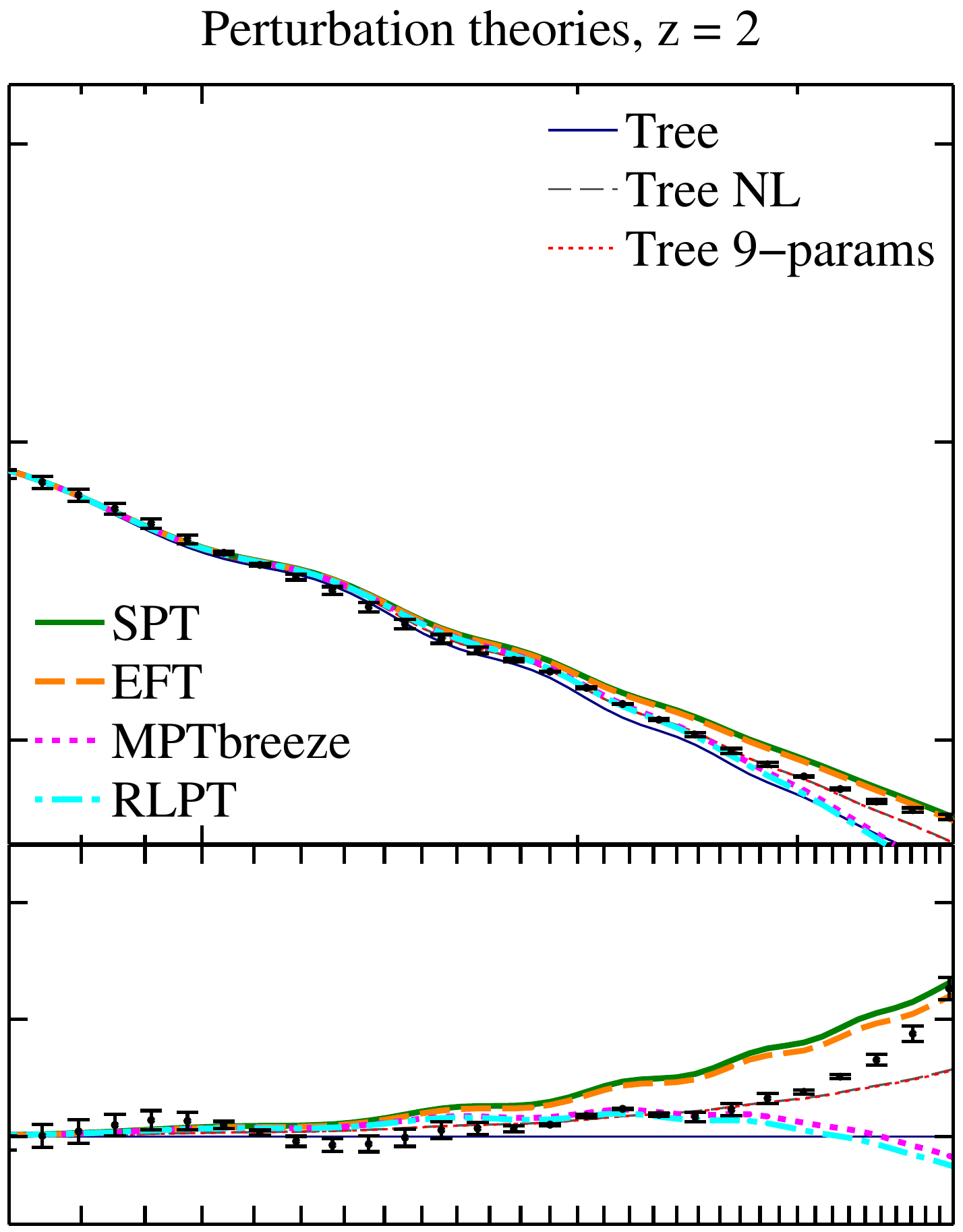}
\includegraphics[height=0.277\textheight]{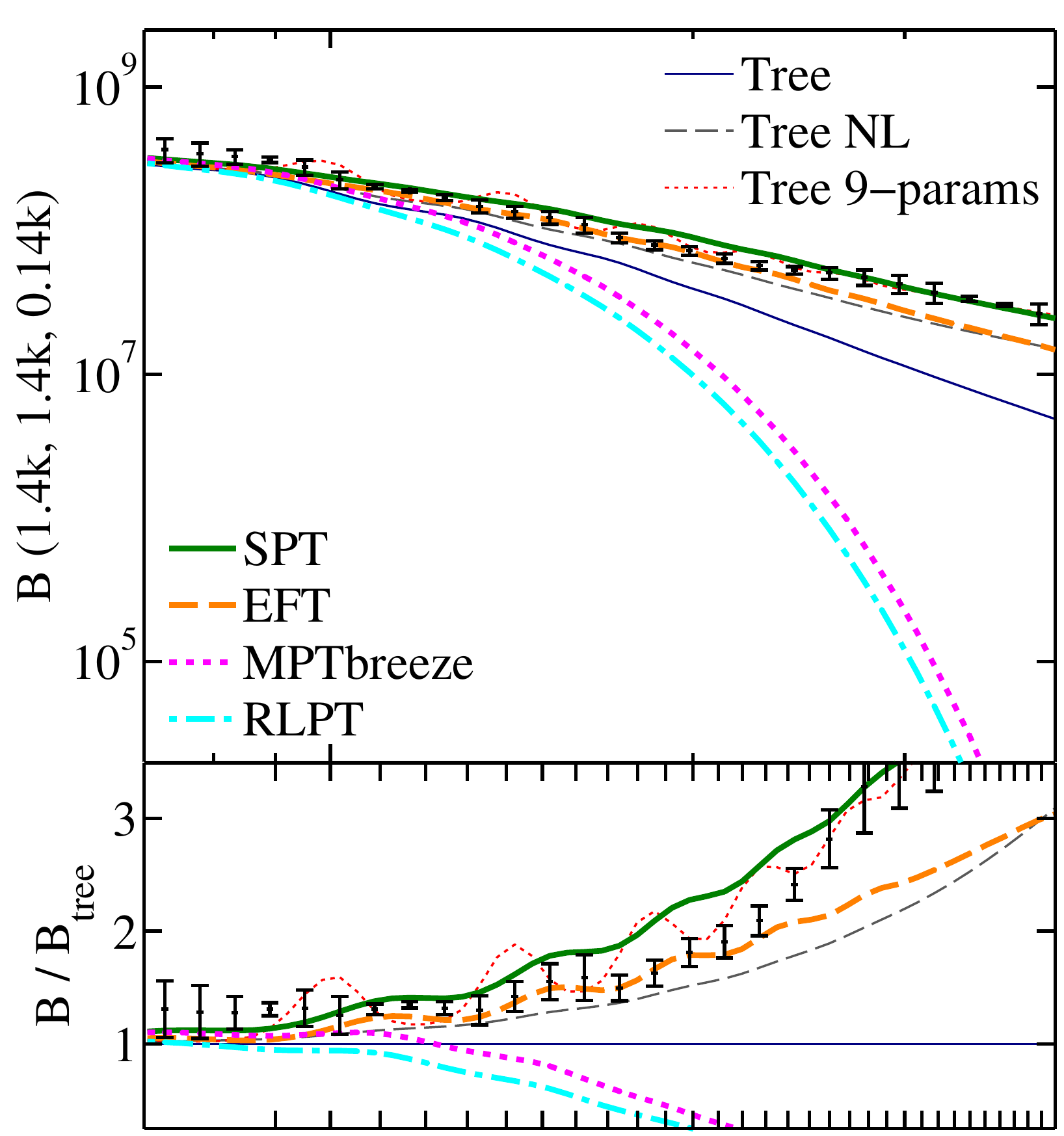}
\includegraphics[height=0.277\textheight]{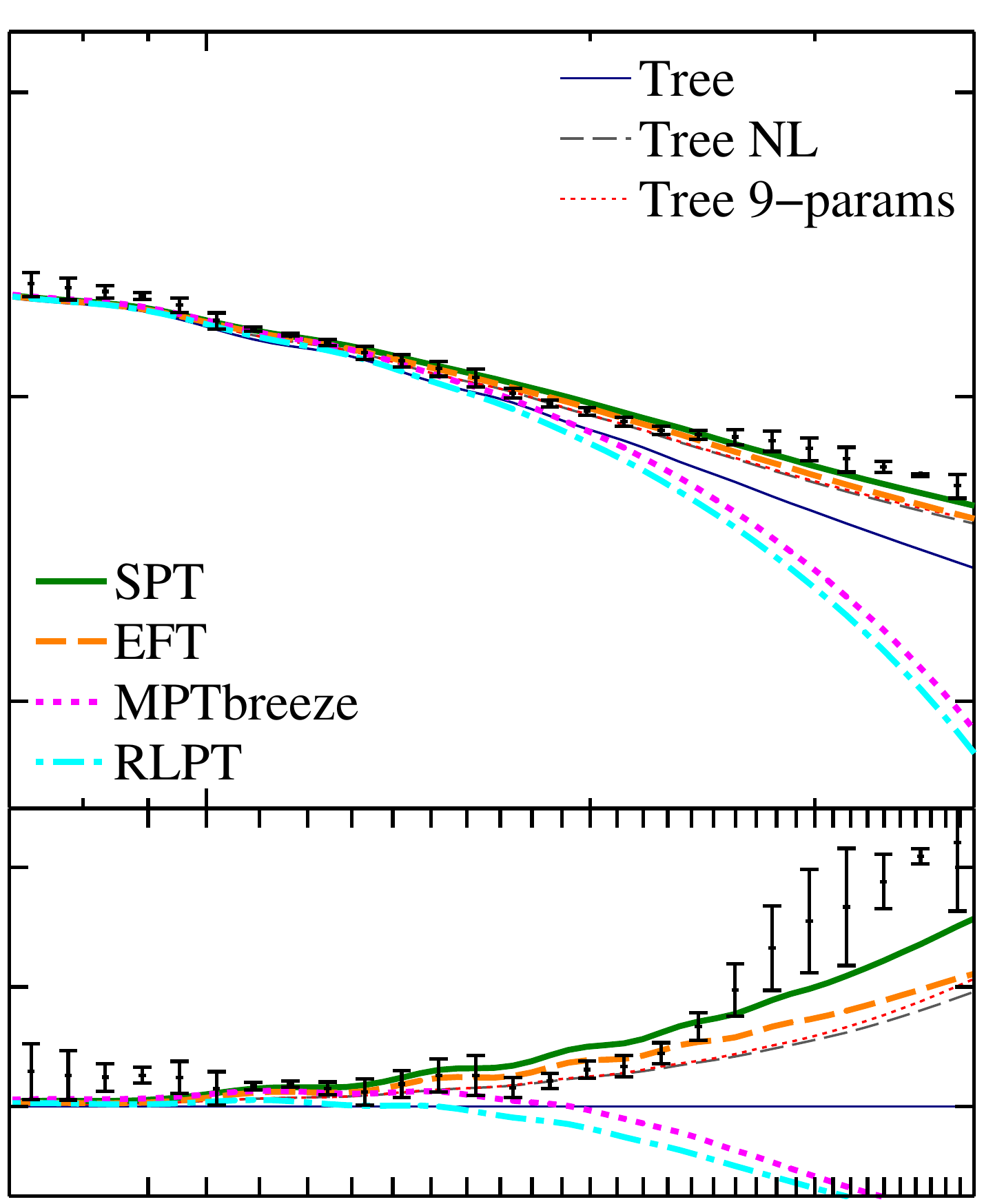}
\includegraphics[height=0.277\textheight]{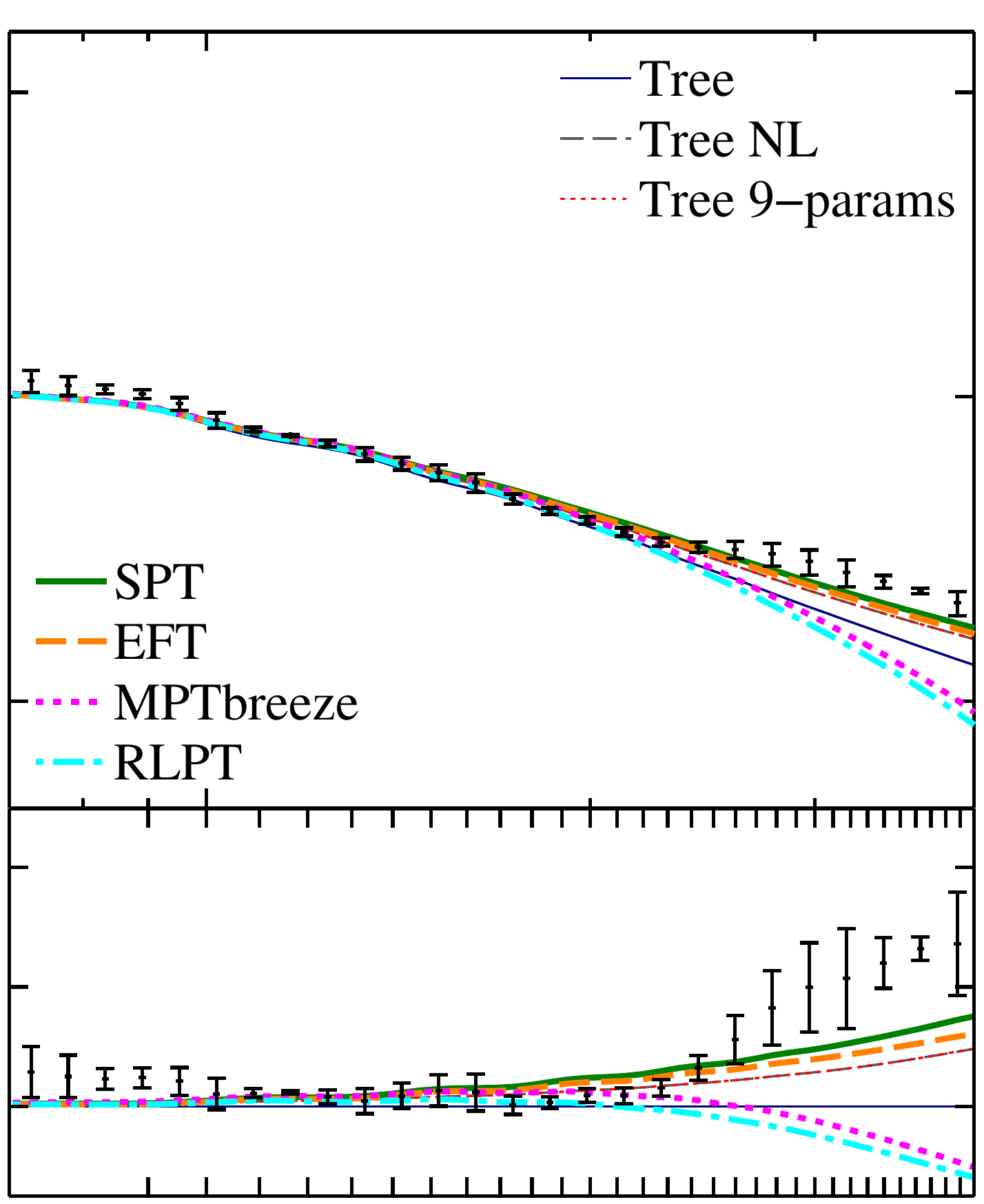}
\includegraphics[height=0.303\textheight]{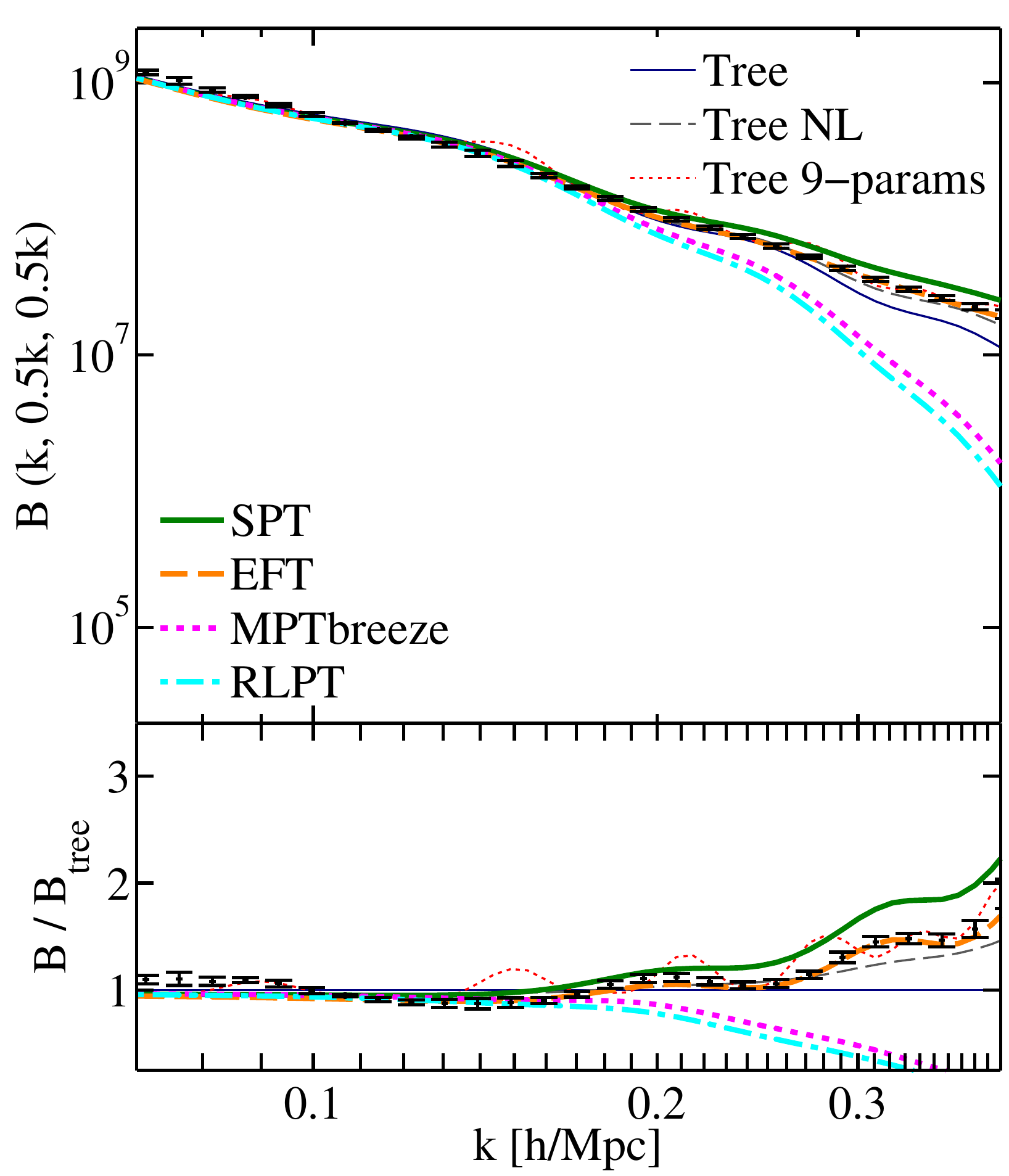}
\includegraphics[height=0.303\textheight]{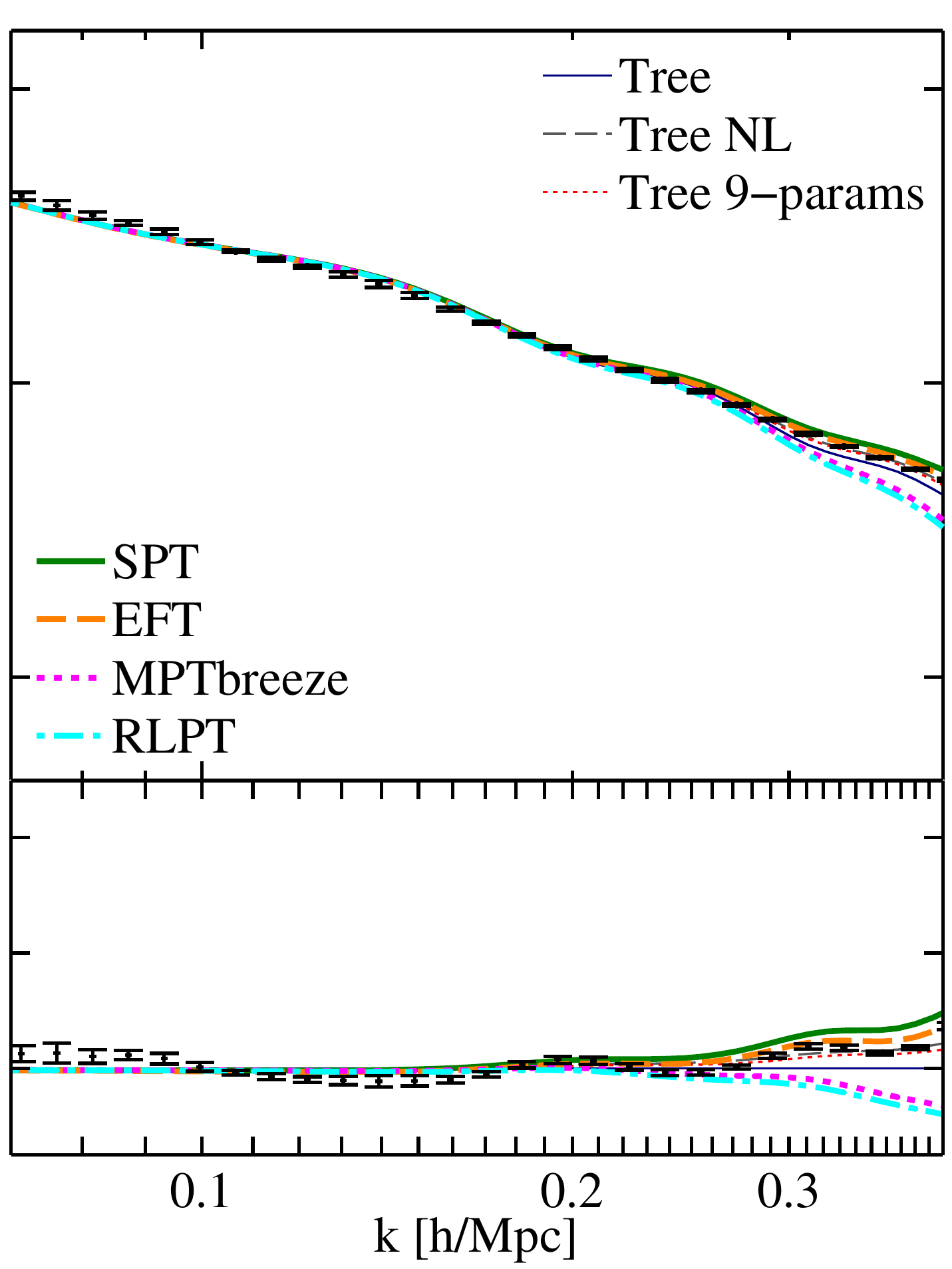}
\includegraphics[height=0.303\textheight]{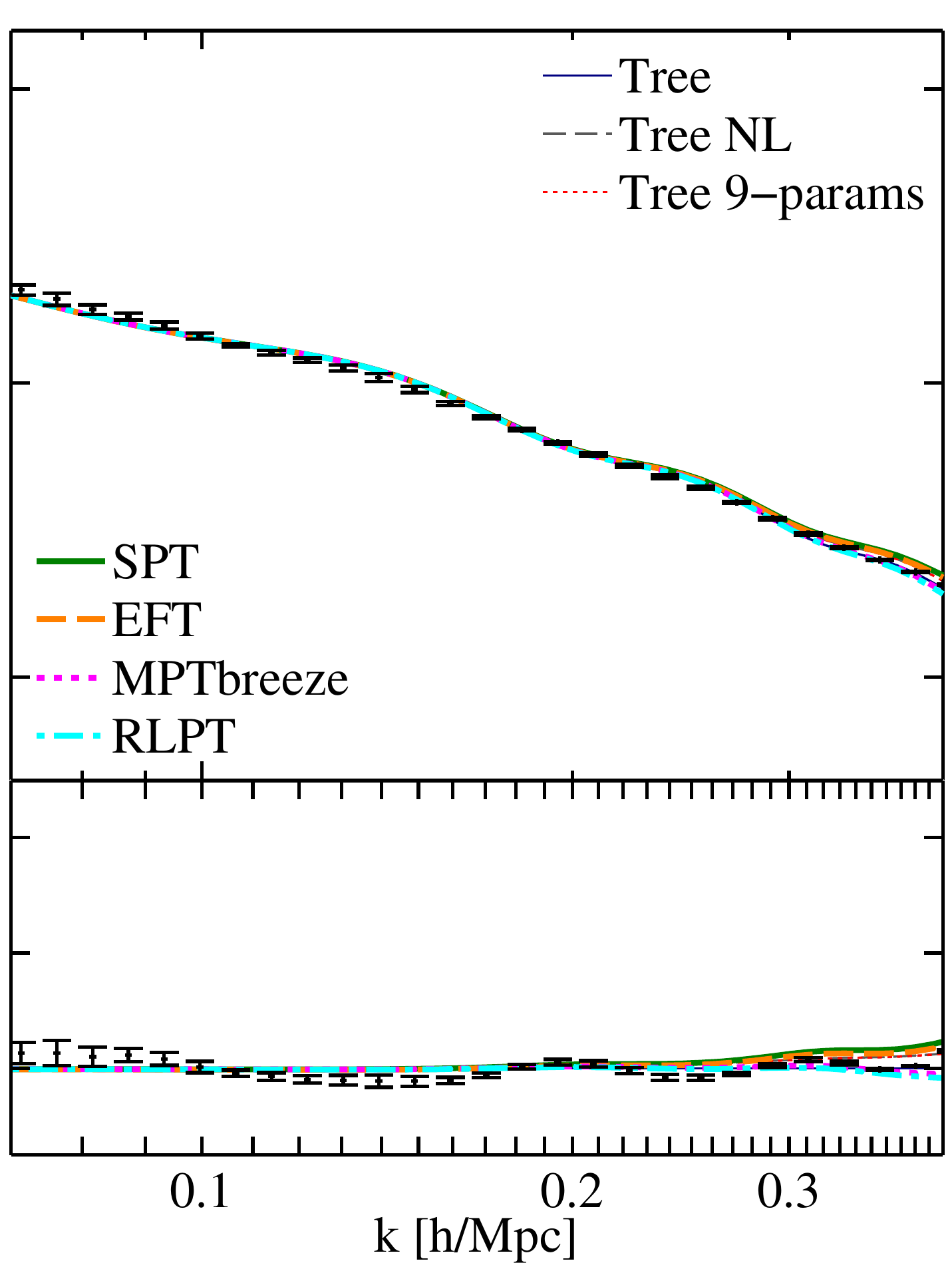}
\caption{Comparison of perturbation theory models of the matter bispectrum with $N$-body simulations, at redshifts 0, 1, 2 (left to right), for the equilateral, squeezed, and flattened configurations (top to bottom). The lower panels show the residuals with respect to the tree-level model.}
\label{fig:slices_PT}
\end{center}
\end{figure*}

We confirm these observations for three specific limiting cases in Fig.~\ref{fig:slices_PT} with a comparison of PT bispectrra amplitudes with measured values:  from top to bottom we show the  equilateral, squeezed, and flattened triangle configurations. In addition to the bispectrum amplitudes, in each case we also plot residuals with respect to the tree-level model.
Figure~\ref{fig:slices_PT} demonstrates that all models converge to the tree level for $k \lesssim 0.1 \, h /$Mpc at $z=0$ in agreement with simulations; the range of validity of the tree-level theory increases for higher redshift and for flatter shapes.   For the phenomenological models, we note that both the simple nonlinear tree-level model and the nine-parameter tree-level fit both increase the range over which there is agreement with simulations.   For $z\ge 1$ these two cases are nearly indistinguishable and both show a similar deficit in power for larger $k$.   In principle the nine-parameter model does provide a better match to the $z=0$ bispectrum, however, it also exhibits large oscillations which originate through the slope parameter $n$ for a power spectrum with BAO features, as noted and circumvented in Ref.~\cite{GilMarin2012}. While it is possible to remove these unwanted oscillations of $n$ with a spline smoothing, we do not apply this extra processing step here for simplicity.

\begin{figure*}[t!]
\begin{center}
\includegraphics[height=0.25\textheight, trim={0 0 0.25cm 0},clip]{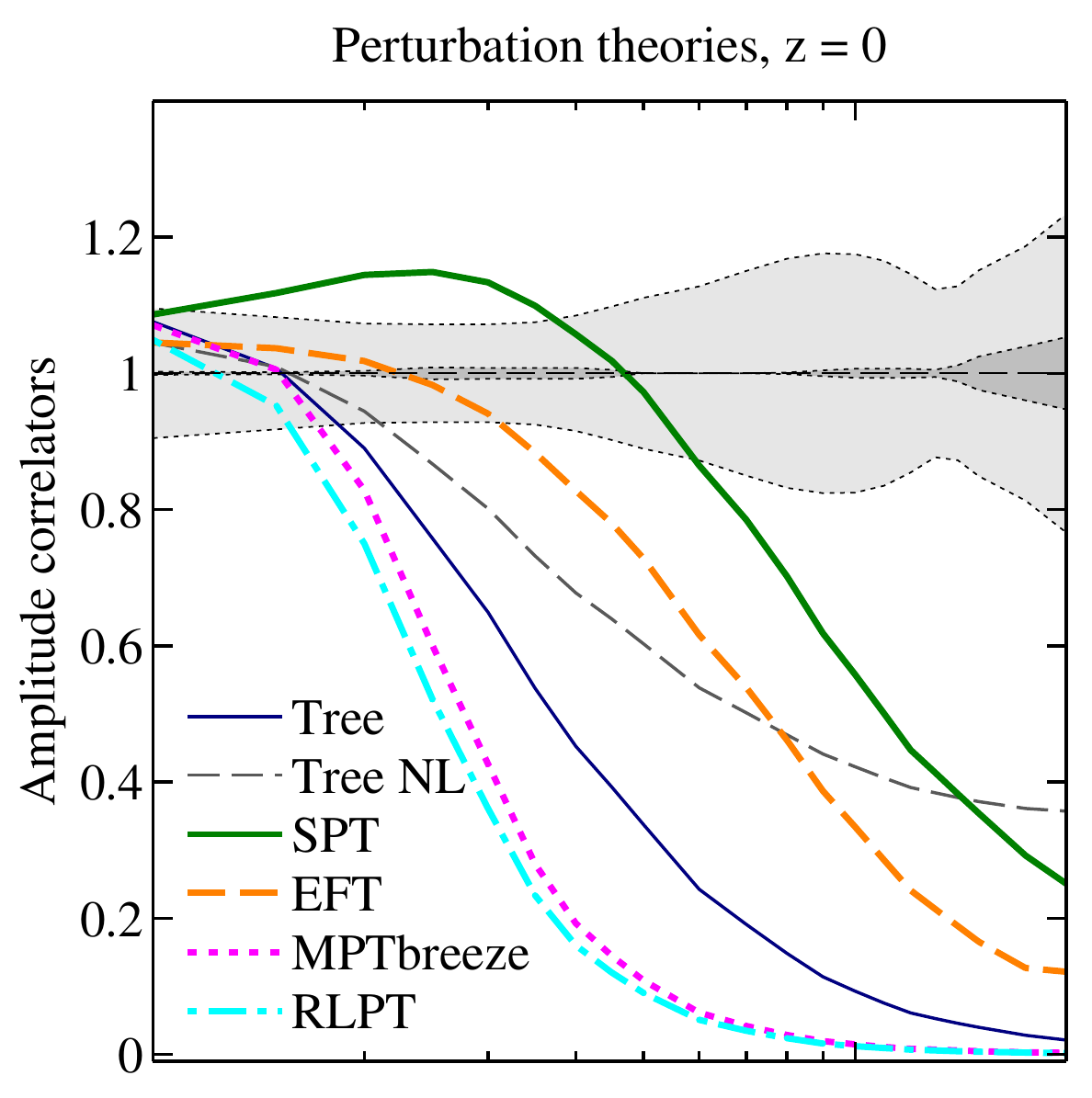}
\includegraphics[height=0.25\textheight, trim={-0.25cm 0 0.25cm 0},clip]{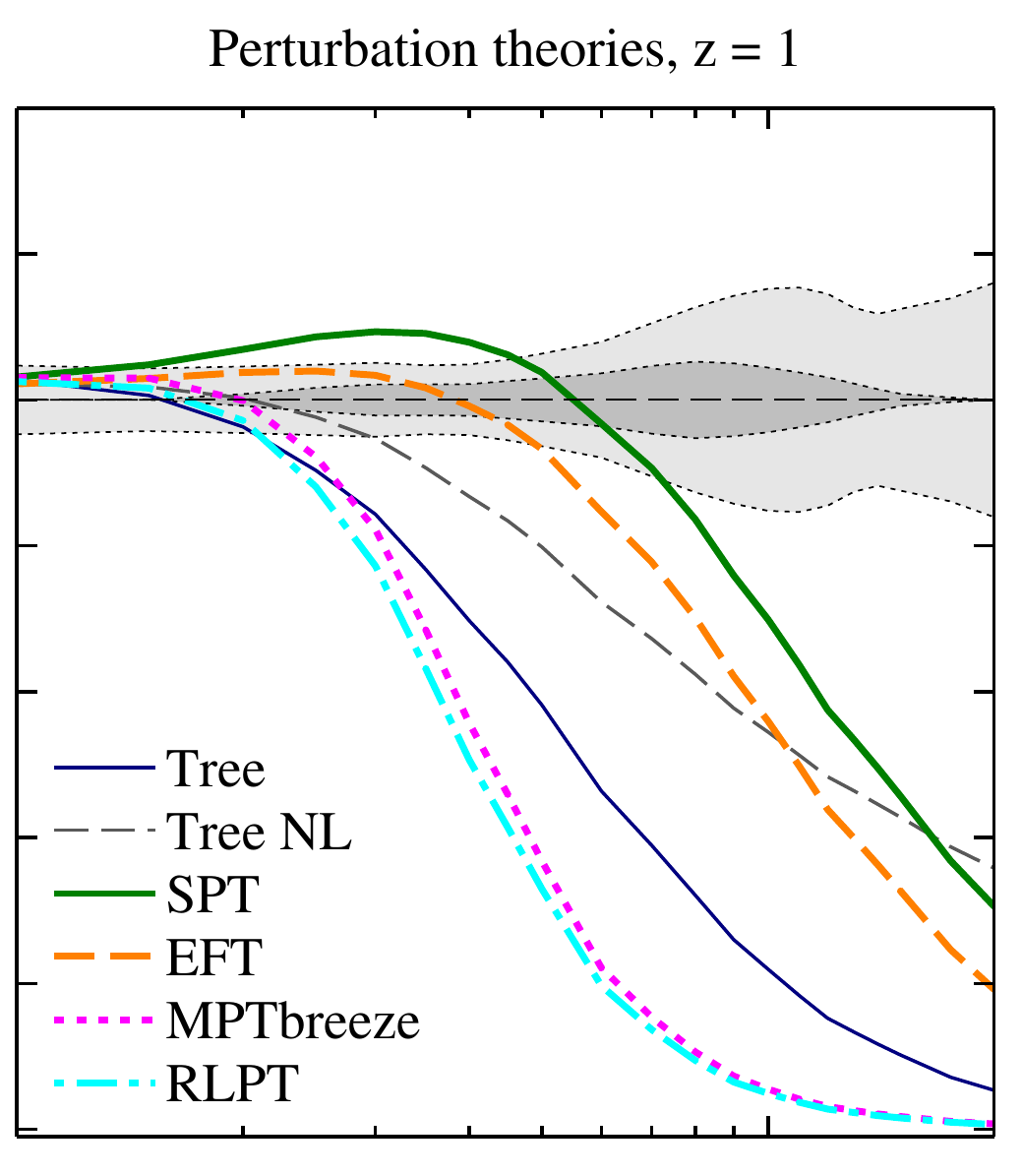} 
\includegraphics[height=0.25\textheight, trim={-0.25cm 0 0.25cm 0},clip]{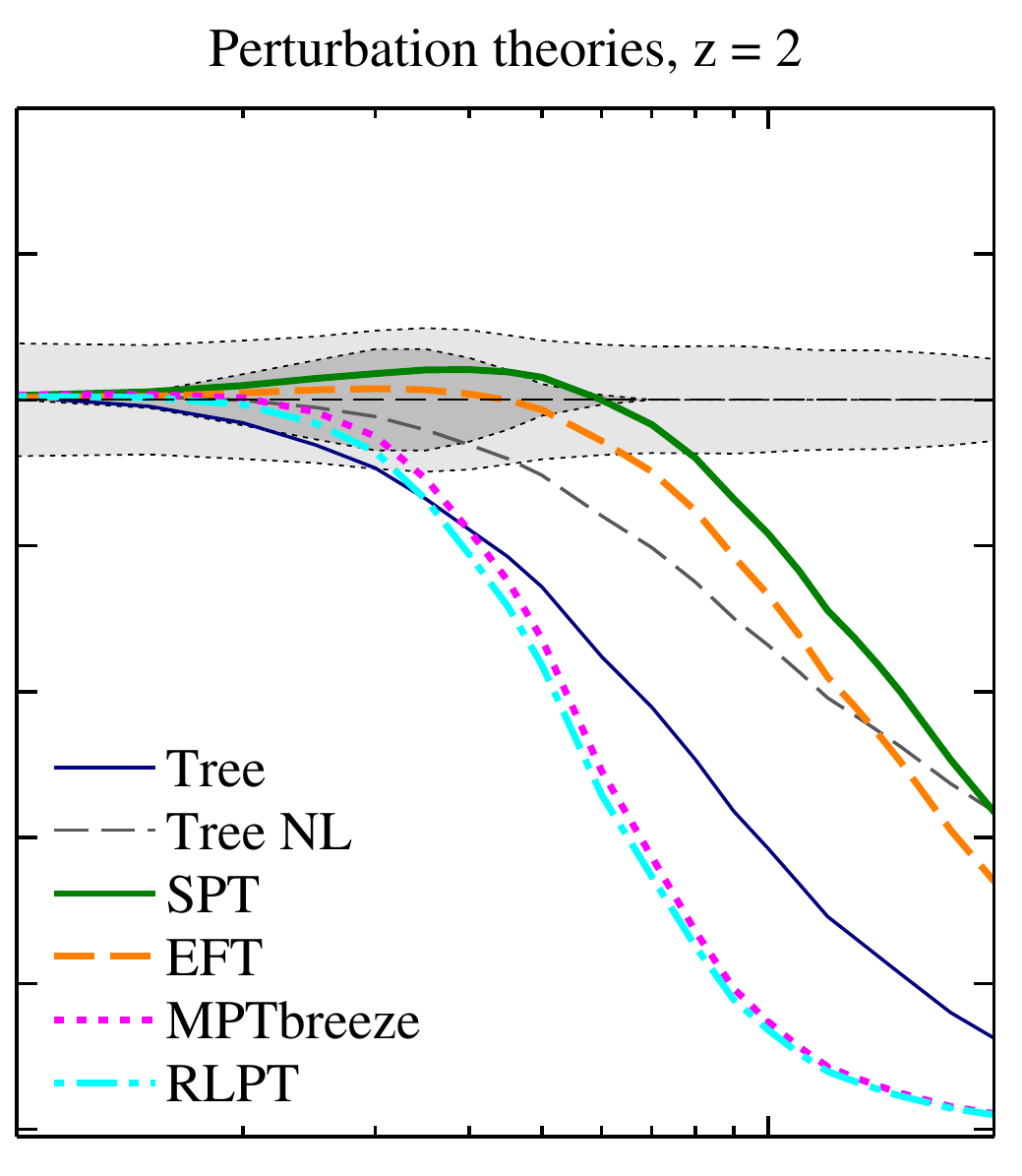} 
\includegraphics[height=0.255\textheight, trim={0 0 0.25cm 0.2cm},clip]{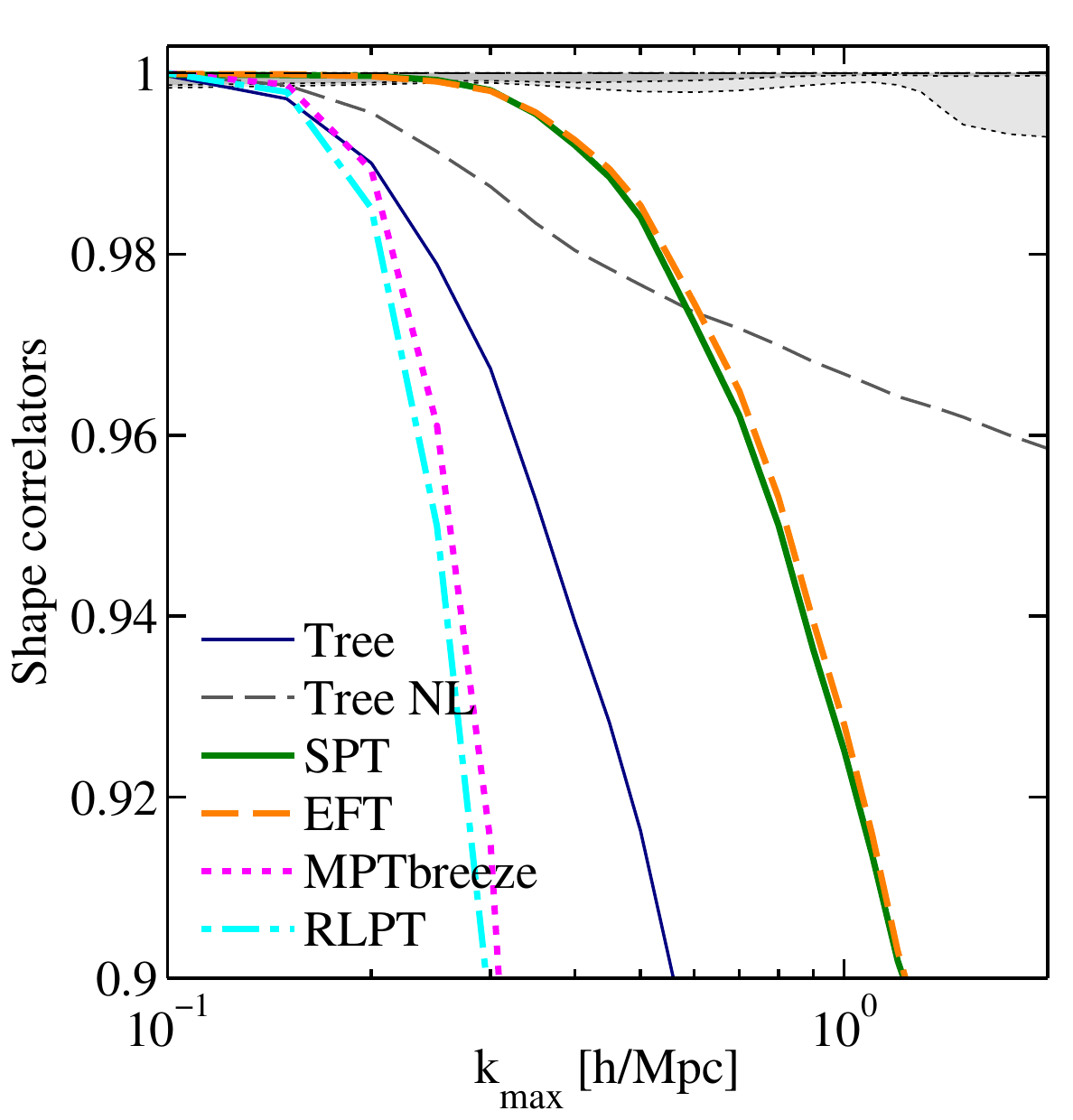} 
\includegraphics[height=0.255\textheight, trim={0 0 0.25cm 0.2cm},clip]{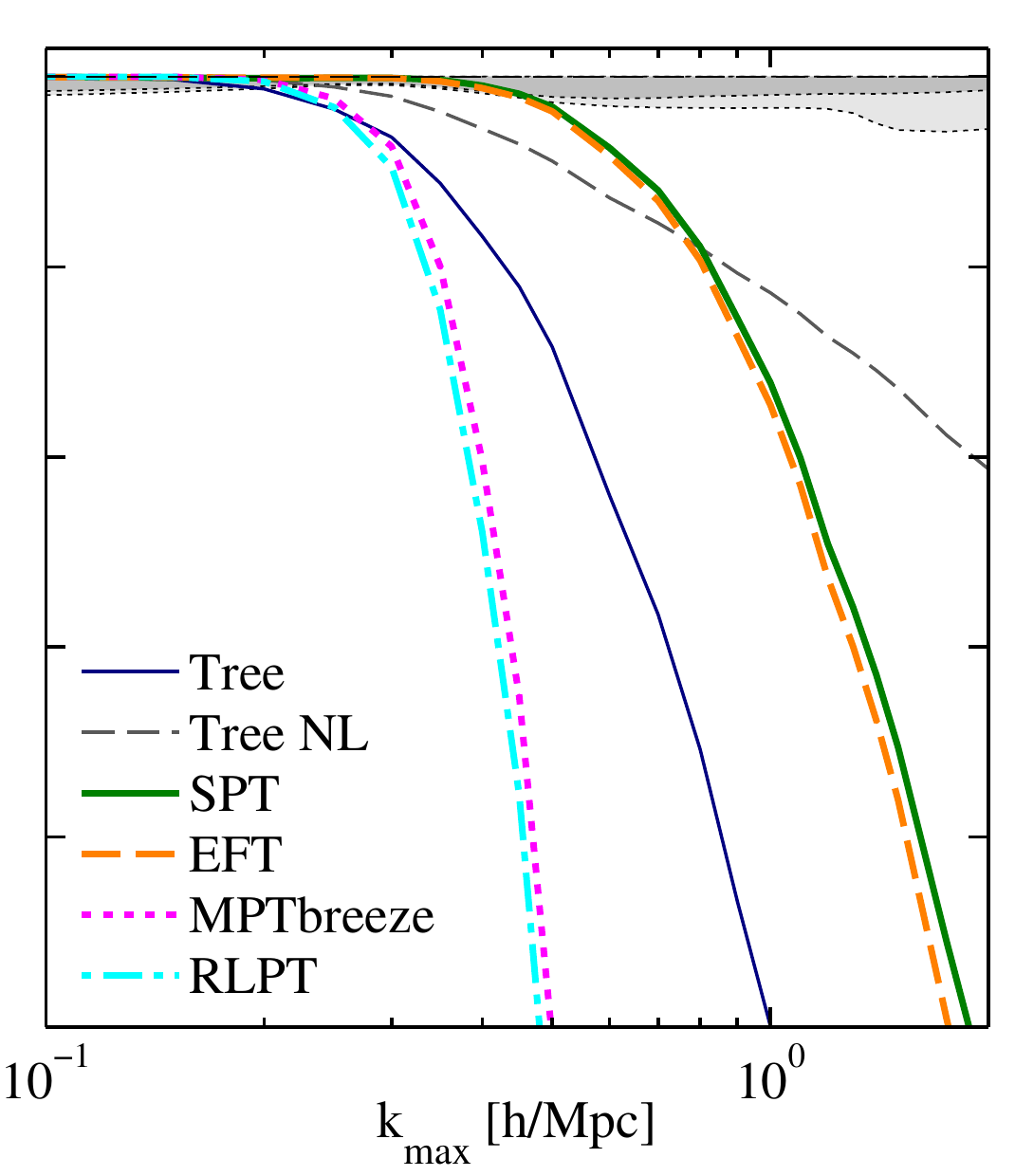} 
\includegraphics[height=0.255\textheight, trim={0 0 0.25cm 0.2cm},clip]{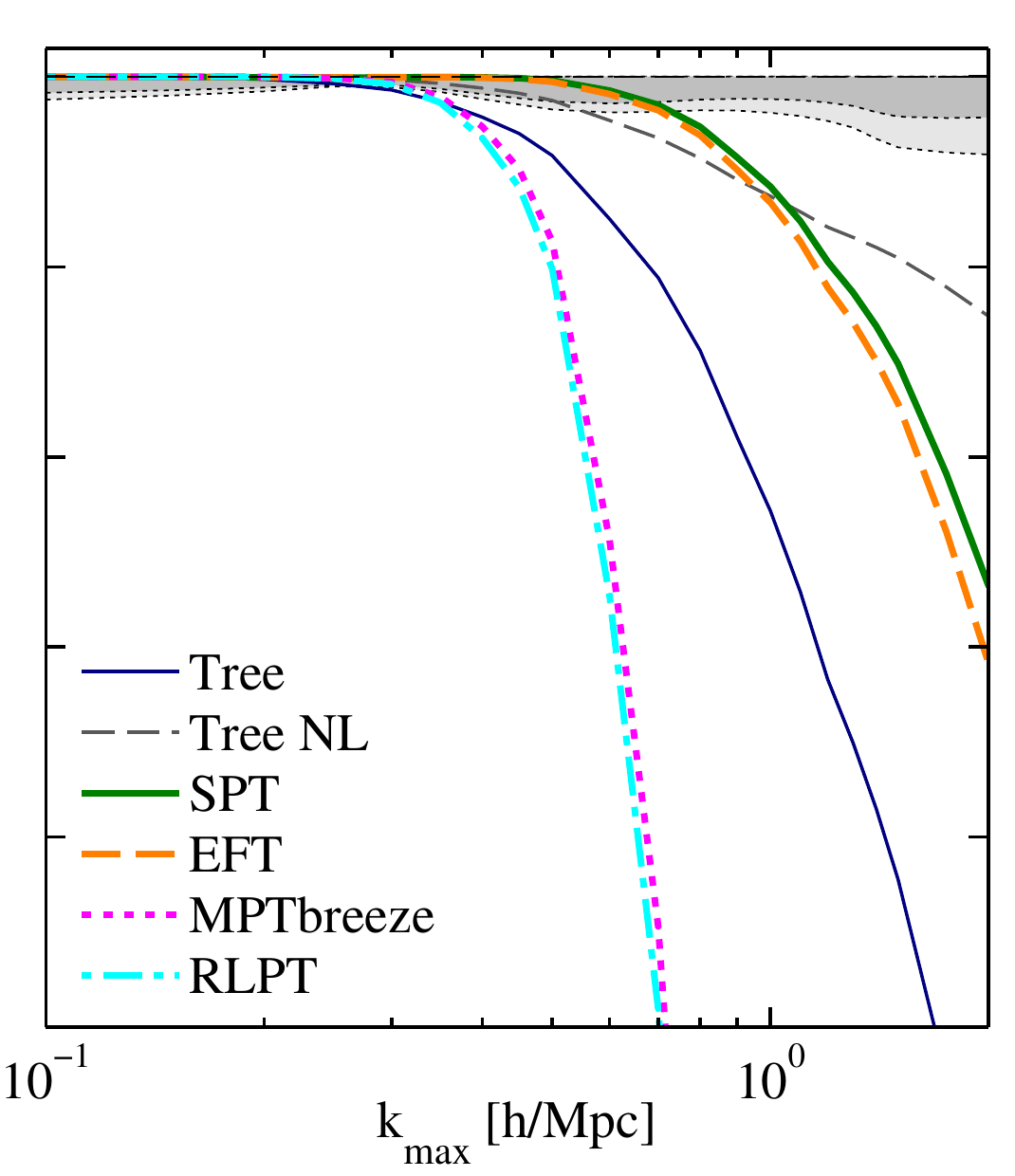} 
\caption{The amplitude $\mathcal{A}$ (top row) and shape $\mathcal{S}$ (bottom row) correlators at redshifts 0, 1, 2 for the perturbative methods, obtained by comparing with the benchmark model. The shaded areas represent error estimates between the benchmark model and the simulations and are explained in the main body of the paper.}
\label{combkmaxpt}
\end{center}
\end{figure*}

For the one-loop perturbative models plotted in Fig.~\ref{fig:slices_PT}, all approaches agree in the strictly perturbative regime at $z=0$.  However,  beyond this regime for larger $k,$ SPT generally overestimates the bispectrum, while the RLPT and RPT \textsc{MPTbreeze} models underestimate it.  The EFT approach lies in between the SPT and RLPT curves, and typically extends the range of agreement with simulations.   This trend is also apparent at higher redshift with the exception of the squeezed limit where even SPT falls below the measured bispectrum in the quasilinear regime. 

Given these interesting observations, we have undertaken a comprehensive quantitative comparative analysis using the integrated amplitude correlator  $\mathcal{A}$ (Eq.~\ref{ampcor1}) and the shape correlator  $\mathcal{S}$ (Eq.~\ref{shapecor}), the results of which are plotted in Fig.~\ref{combkmaxpt}. This corresponds to a signal-to-noise weighted integration over all triangular configurations up to a given resolution $k_\text{max}$, rather than the specific limiting configurations  Fig.~\ref{fig:slices_PT}.  Here, we directly compare the theoretical predictions $B_{\text{theory}}^j$ to the three-shape benchmark model $B_{\text{3-shape}}$ given in Eq.~(\ref{shapes}) with parameters given in Eq.~(\ref{three-shape-fit}), which provides an excellent fit and a smoother representation of the actual bispectrum from simulations (see Sec.~\ref{sec:three-shape}).

 We estimate the uncertainties on the correlators as follows.
From each simulation realisation $i$, we obtain the amplitude and shape correlators $\mathcal{A} (B_{\text{sim}}^i, B_{\text{3-shape}})$, $\mathcal{S}(B_{\text{sim}}^i, B_{\text{3-shape}})$ as a function of $k_{\max}$. For each value of $k_{\max}$, we can thus derive mean and standard deviation of the correlators: $\mu_{\mathcal{A}}$, $\sigma_{\mathcal{A}}$, and similarly for the shape. 
As we are comparing all theoretical models with the benchmark three-shape model, the total uncertainty $\sigma_{\text{tot}}$ on the correlators between each model $j$ and the benchmark, $\mathcal{A} (B_{\text{theory}}^j, B_{\text{3-shape}})$, will be larger than the variance $\sigma_{\mathcal{A}}$ obtained from the scatter of $\mathcal{A} (B_{\text{sim}}^i, B_{\text{3-shape}})$; this is because of the small $k$-dependent discrepancy that exists between the simulations and the smooth benchmark model.
The grey shaded areas in Fig.~\ref{combkmaxpt} represent two different estimates of $\sigma_{\text{tot}}$, as follows.
The light grey area represents a conservative error estimate obtained by adding the error bars of the simulation to the deviation from one of the mean of the correlator, i.e. assuming $\sigma_{\text{tot}} = | \mu_{\mathcal{A}} - 1 | + \sigma_{\mathcal{A}} $, while the darker grey area represents the part of the benchmark model outside the $1\sigma_{\mathcal{A}}$ error bars, i.e. assuming
\begin{equation}
\sigma_{\text{tot}} = \begin{cases}
  0 & \text{if } 1 \in [\mu_{\mathcal{A}} - \sigma_{\mathcal{A}}, \mu_{\mathcal{A}}  + \sigma_{\mathcal{A}}] \\
  \mu_{\mathcal{A}} - \sigma_{\mathcal{A}} -1 & \text{if } \mu_{\mathcal{A}} - \sigma_{\mathcal{A}}>1  \\
  1-\mu_{\mathcal{A}} - \sigma_{\mathcal{A}} & \text{if } \mu_{\mathcal{A}} + \sigma_{\mathcal{A}}<1 \, .
\end{cases}
\end{equation}
The same reasoning applies to the shape correlators $\mathcal{S}$, with the difference that $\mu_{\mathcal{S}} \le 1$.

From Fig.~\ref{combkmaxpt}, we note that that there are always high shape correlations well beyond the perturbative regime.   For example, at $z=0$ all theories have a shape correlation greater than 99\% up to $k<0.2 \, h/$Mpc, even when there are variations of $\mathcal{O}(\text{20\%})$ in the amplitude correlator. These remarkably high shape correlations imply that bispectrum estimators that measure the projection of the full bispectrum on these theoretical shapes (like in Ref.~\cite{Schmittfull2015}) should yield a high proportion of the total bispectrum signal-to-noise.  Since the shape correlator is not as discerning a tool for distinguishing between different perturbative models, we focus most attention on amplitude deviations. 

We also employ the total correlator $\mathcal{T}$, which combines the information of amplitude and shape, in order to directly estimate the range of validity of each model as a function of redshift. In Table~\ref{tablept} we give the maximum wavenumber $k_{\max}^*$ at which the total correlator between each model and the benchmark model deviates from unity by more than a fixed threshold of 10\% (and 5\%). While we show results at the three redshifts considered, $z = \{0, 1, 2 \}$, an important caveat is that the comparison at $z=0$ is more approximate, due to the less than perfect match between the simulations and the benchmark model; we therefore do not report the 5\% results at $z=0$, and choose to focus primarily on the results at $z = \{1, 2 \}$ in the following discussion. 
A striking feature of Table~\ref{tablept} is the wide range of wavenumbers for which there is good correspondence between theoretical predictions and the measured bispectrum, well beyond expectations for the limits of the perturbative regime estimated in Table \ref{check_tspt}. This shows that even where these theories are no longer expected to be accurate, they can nevertheless be successfully extrapolated into the  nonlinear regime for phenomenological modelling.

\begin{table}
\begin{center}
\caption{Wavenumber $k_{\max}^*$ where the total correlator $\mathcal{T}$ (Eq.~\ref{totalcor}) between the perturbative theory and the benchmark model deviates by more than 10\% (5\%) from unity. In the case of $z=0$, we only report the 10\% results, as the accuracy of the benchmark model is lower.}

\medskip
\begin{tabular}{  c c c c }
\toprule
\multicolumn{4}{ c }{Perturbation theories}\\
\colrule
  Threshold $10\%$ ($5\%$)          & \multicolumn{3}{c}{$k_{\max}^*\,[h/\text{Mpc}]$} \\ 
\colrule
 Theory     &  $z = 0 $  &  $z = 1 $  &  $z = 2 $ \\
\colrule
 Tree-level            & 0.13 & 0.22 (0.17)   & 0.27 (0.20)  \\
 NL tree-level & 0.17 & 0.30 (0.22)   & 0.42 (0.31) \\
 SPT                   & 0.11 & 0.37 (0.14)   & 0.66 (0.49) \\
 EFT                   & 0.29 & 0.45 (0.36)   & 0.60 (0.50) \\ 
 MPTbreeze             & 0.16 & 0.24 (0.21)   & 0.32 (0.28) \\ 
 RLPT                  & 0.15 & 0.22 (0.19)   & 0.30 (0.26) \\ 
\botrule
 \end{tabular}
\label{tablept}
\end{center}
\end{table}
 
 The tree-level (Eq.~\ref{stree}) and the nonlinear tree-level (Eq.~\ref{streeNL}) models are the simplest approximations to the matter bispectrum, and their range of validity can be verified from Fig.~\ref{combkmaxpt}: at $z=1$ we find $k_{\max}^* = 0.22\,h/\text{Mpc}$ for the tree level and $k_{\max}^* = 0.30\,h/\text{Mpc}$ for the nonlinear tree level (at 10\%). The nonlinear bispectrum improves faster than the linear one at higher redshifts: the tree-level increases by roughly $0.05 \, h/$Mpc at each redshift, while the nonlinear tree-level increases by $> 0.1 \, h/$Mpc.

The one-loop SPT bispectrum adds four extra terms to the tree-level shape. Two of them give positive contributions and the other two negative contributions. As seen in Fig.~\ref{combkmaxpt}, at low redshift the additional SPT contributions tend to overshoot the measured bispectrum, apparently lowering the value of $k_{\max}^*$ up to which predictions are accurate (see Table~\ref{tablept}). However, at $z=2$ the overshoot remains within bounds, extending the fit up as far as $k < 0.66 \,h/\text{Mpc}$ in the case of the 10\% threshold (almost accidentally at this specific redshift, possibly because of additional squeezed contributions in the measured bispectrum).  In general, 
SPT predicts an excess of power on quasi-linear scales, before finally decaying in the fully nonlinear regime. This overshoot phenomenon appears because the loop integrals involved require integrating momenta over an infinite range, a regime in which the basic assumption $\delta \ll 1$ is no longer valid.   Despite this problem, the shape correlation is excellent up to $k\sim 0.3\,h/$Mpc, improving significantly over the tree-level result. We also note that evidence for the amplitude overshoot is not very strong from our simulations because they have rather large uncertainty on $\mathcal{A}$, especially at $z=0$.

The one-loop EFT bispectrum includes one counterterm, which increases the accuracy of the model due to the one free parameter that is introduced and fitted at the level of the power spectrum. 
In Fig.~\ref{combkmaxpt} we observe that this method provides substantially improved agreement with the simulations, albeit at the cost of an extra parameter, which was calibrated on the power spectrum of $N$-body simulations, assuming a specific cosmological model.   This counterterm effectively removes excess power provided by SPT in the quasi-linear regime and the results that we obtain from the three-dimensional comparison are consistent with the improved agreement found in Ref.~\cite{Angulo:2014tfa}. The EFT method appears to work well up to $k_{\max}^* = 0.45 \, h/\text{Mpc}$ at $z = 1$ and  $k_{\max}^* = 0.60 \, h/\text{Mpc}$ at $z = 2$. However, we must proceed cautiously before using such projections because the detailed correspondence in the equilateral and squeezed limits shown in Fig.~\ref{fig:slices_PT} is not as encouraging.   (We also observe additional correlated squeezed signals emerging on these scales in the measured bispectrum which require more sophisticated joint fitting.)   At higher redshift, the contribution of the counterterm becomes less significant, because the growth rate of the term is $ \propto D^{7.1}(z) $ compared to $\propto D^6(z)$ from the one-loop SPT terms.
Although one can in principle add another three additional counterterms for the one-loop EFT bispectrum, we have found that the improvement in the accuracy is modest relative to the cost of introducing these further free parameters.

The RPT approach (\textsc{MPTbreeze} formalism) at one loop solves the SPT excess by cutting off terms appropriately with an exponential function, as can be seen in Fig.~\ref{combkmaxpt}.   Compared to SPT, all terms are positive to any number of loops, and so this is a convergent expansion.  With accuracy increasing with number of loops, the amplitude on all scales should always approach the measured bispectrum from below.  
We see in Table~\ref{tablept} that the RPT method appears to be accurate to 10\% at $k < 0.24\,h/\text{Mpc} $ at $z = 1$, improving to $k < 0.32 \,h/\text{Mpc} $ at $z = 2$. The main improvement of \textsc{MPTbreeze} compared with the other methods arises on large scales, before the exponential damping begins.    The disadvantage of this suppression is that it precludes any extrapolations into the nonlinear regime.  

The RLPT results we have obtained are similar to RPT, although the validity range is marginally smaller due to the increased power suppression; in this case we find  $k_{\max}^* = 0.22 \, h/\text{Mpc}$ at $z = 1$, and $k_{\max}^* = 0.30 \, h/\text{Mpc}$ at $z = 2$.

We conclude that all one-loop perturbative methods match simulations at present precision within the expected perturbative regime.   In terms of phenomenological extrapolation into the nonlinear regime, the EFT method goes furthest (once the counter-term coefficient has been appropriately fitted).  Both RLPT and RPT undershoot the measured bispectrum in this regime by construction, while SPT generically overshoots.  On the other hand, the nonlinear tree-level bispectrum Eq.~(\ref{streeNL}) provides a useful projection to larger $k$ which has the advantage of being much simpler to calculate.

\begin{figure*}[tb]
\begin{center}
\includegraphics[width=2.25in]{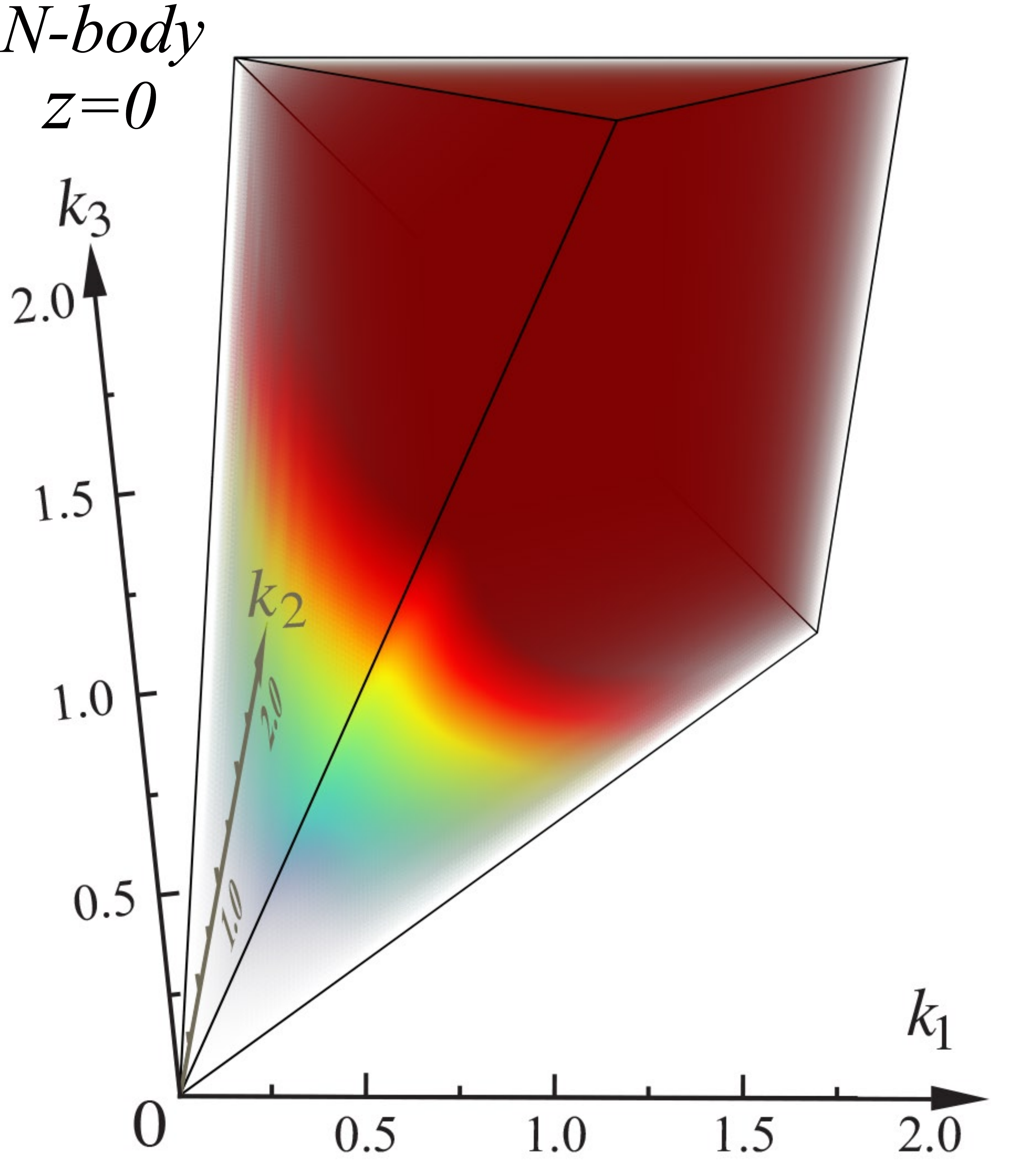}\includegraphics[width=2.25in]{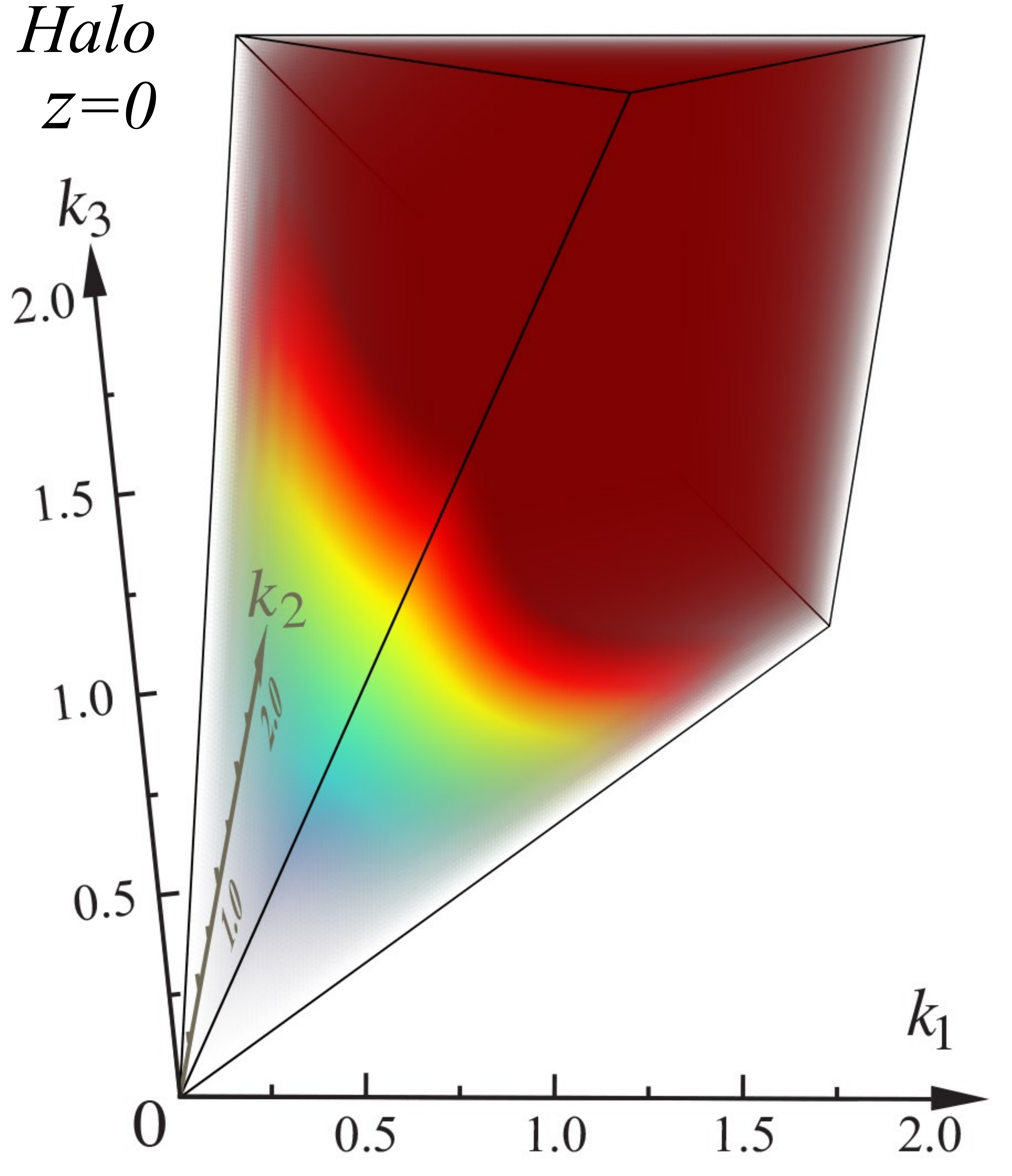}\includegraphics[width=2.25in]{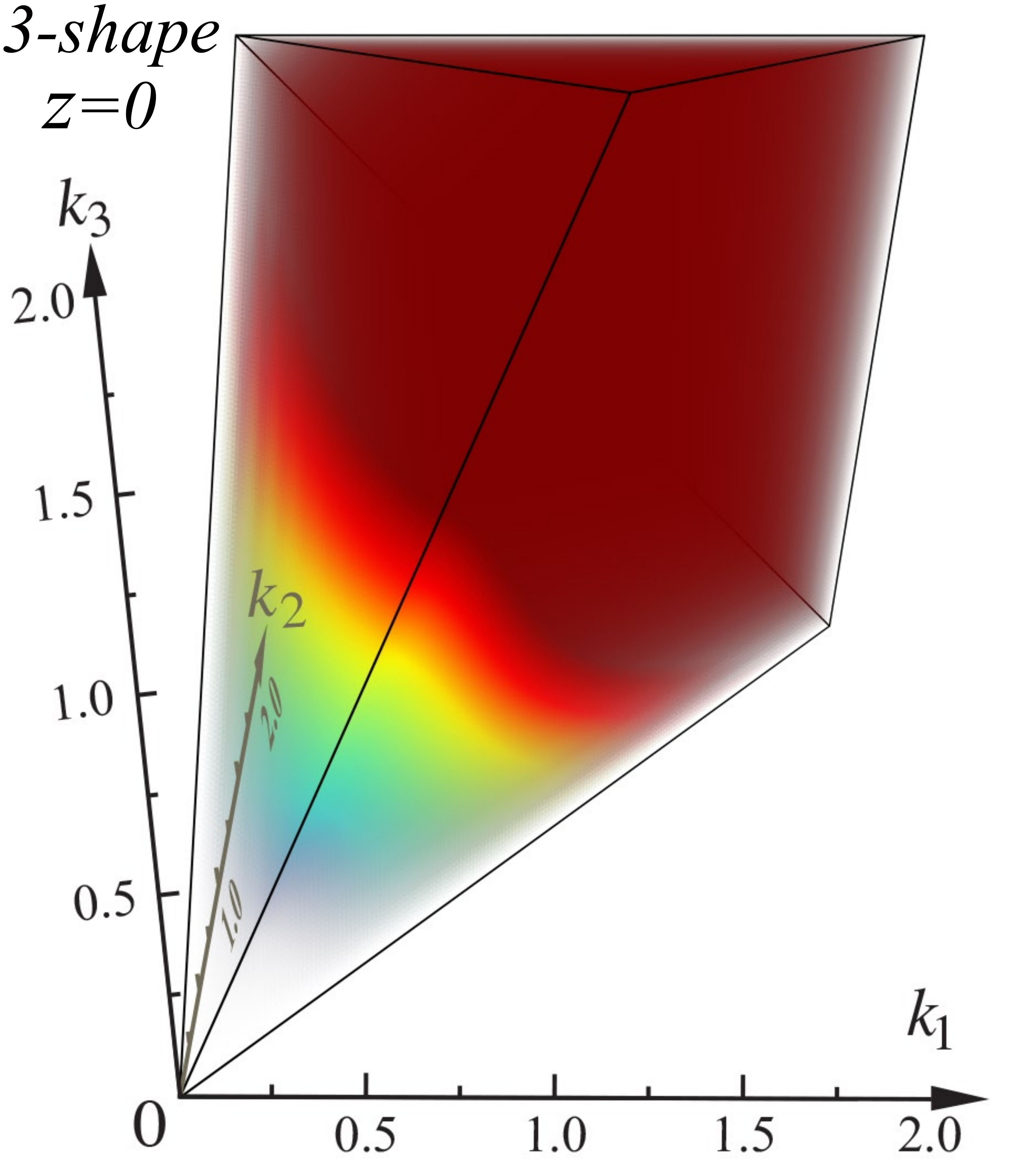}\\
\includegraphics[width=2.25in]{scale120.pdf}\\\medskip
\includegraphics[width=2.25in]{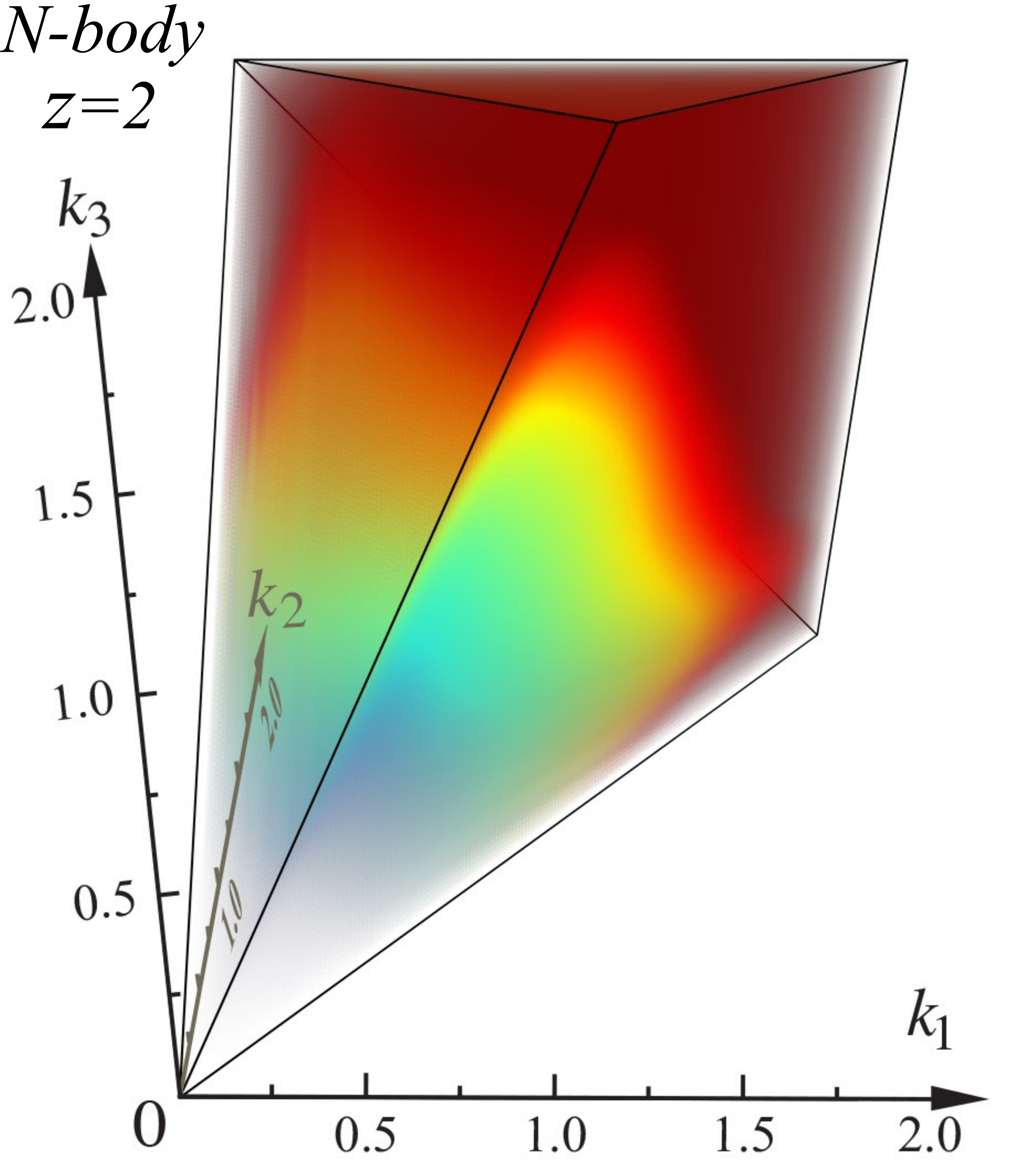}\includegraphics[width=2.25in]{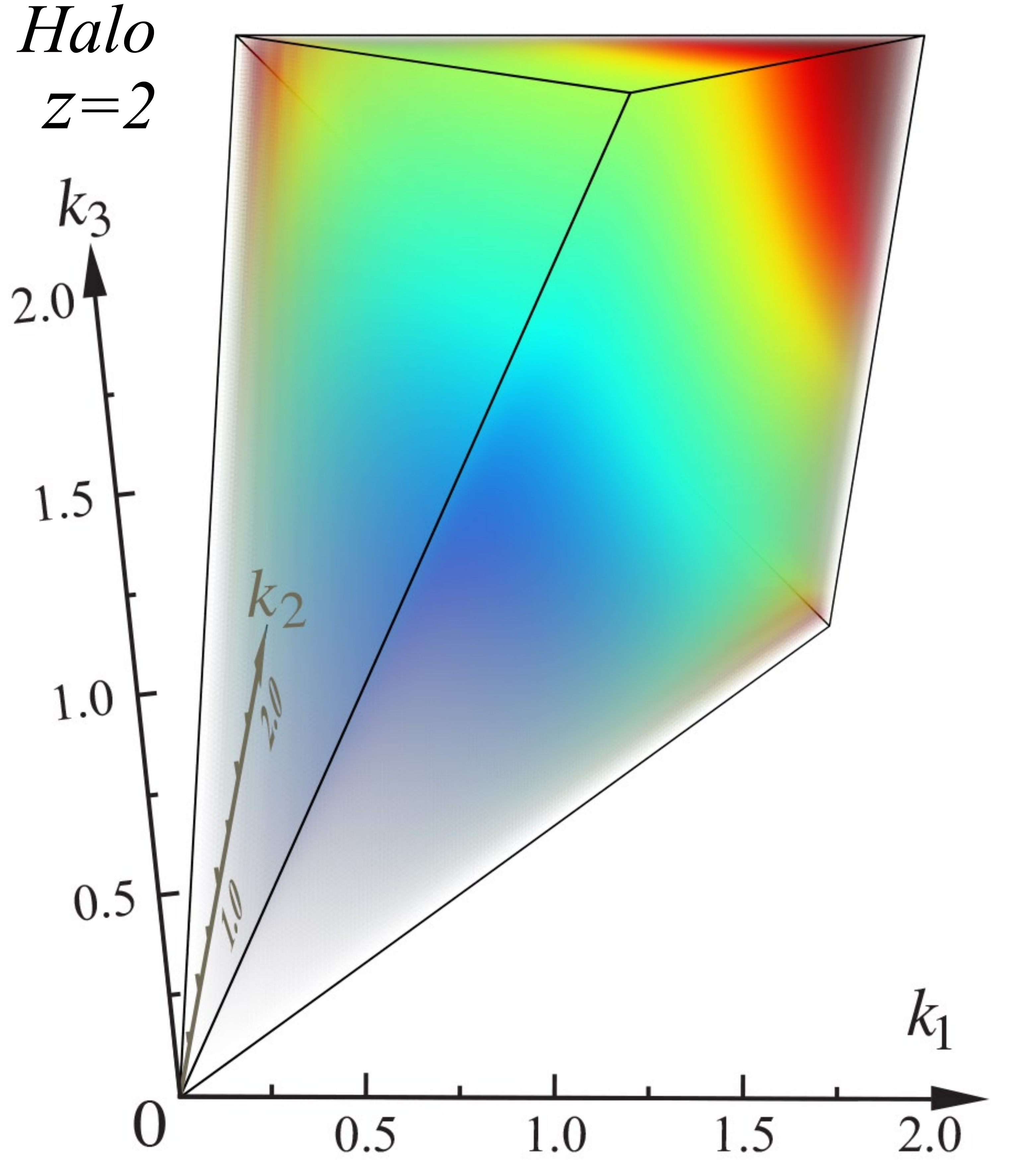}\includegraphics[width=2.25in]{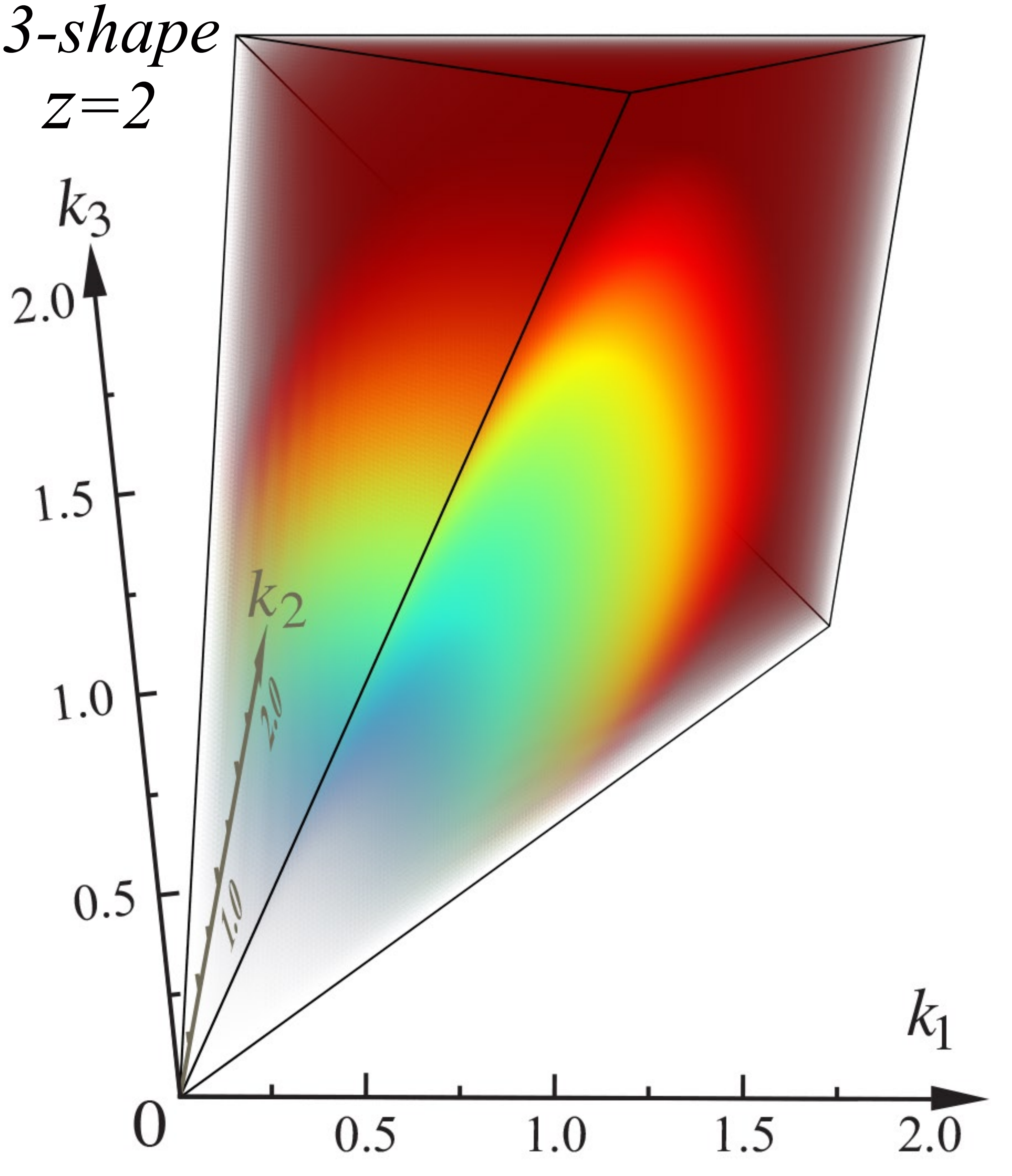}\\
\includegraphics[width=2.25in]{scale50.pdf}
\caption{ Comparison between between $N$-body simulation bispectrum (left panels) with the standard halo bispectrum model Eqs.~(\ref{b1h}-\ref{b3h}) (centre panels) and the `three-shape' benchmark model Eq.~(\ref{shapes}) (right panels) shown at two redshifts $z=1,2$.   The standard halo model is effectively normalised to fit the measured bispectrum at $z=0$, which is also achieved well by the phenomenological `three-shape' model (upper panels).  However, at higher redshift $z=2$ the halo model exhibits the wrong growth rates for the flattened three-halo and squeezed two-halo configurations, yielding a substantial deficit (lower panel centre); the measured bispectrum behaviour can be accommodated in the three-shape benchmark model (lower panel right).}
\label{bispectrum_phenohalos}
\end{center}
\end{figure*}

\begin{figure*}[t]
\begin{center}
\includegraphics[height=0.3\textheight]{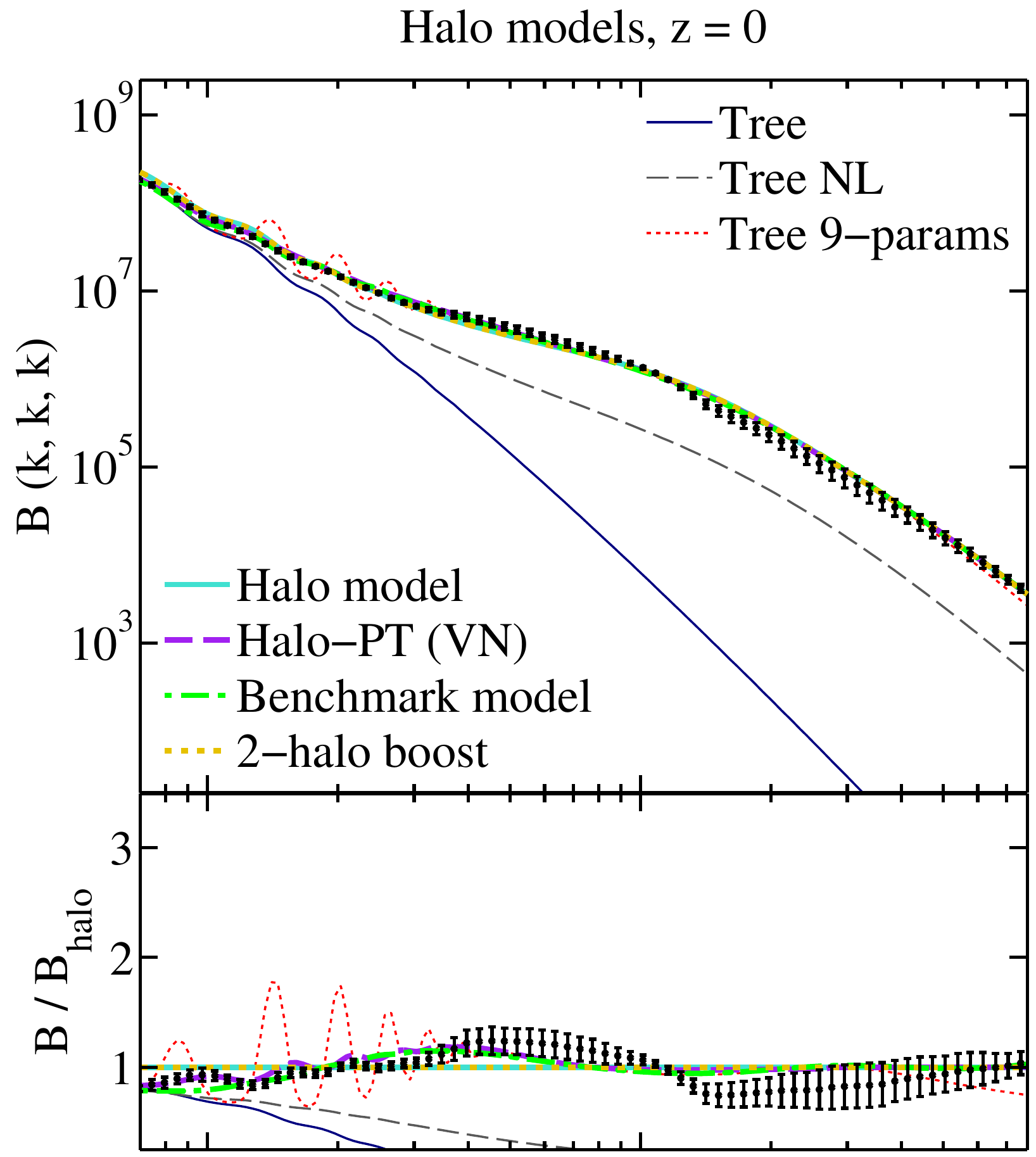}
\includegraphics[height=0.3\textheight]{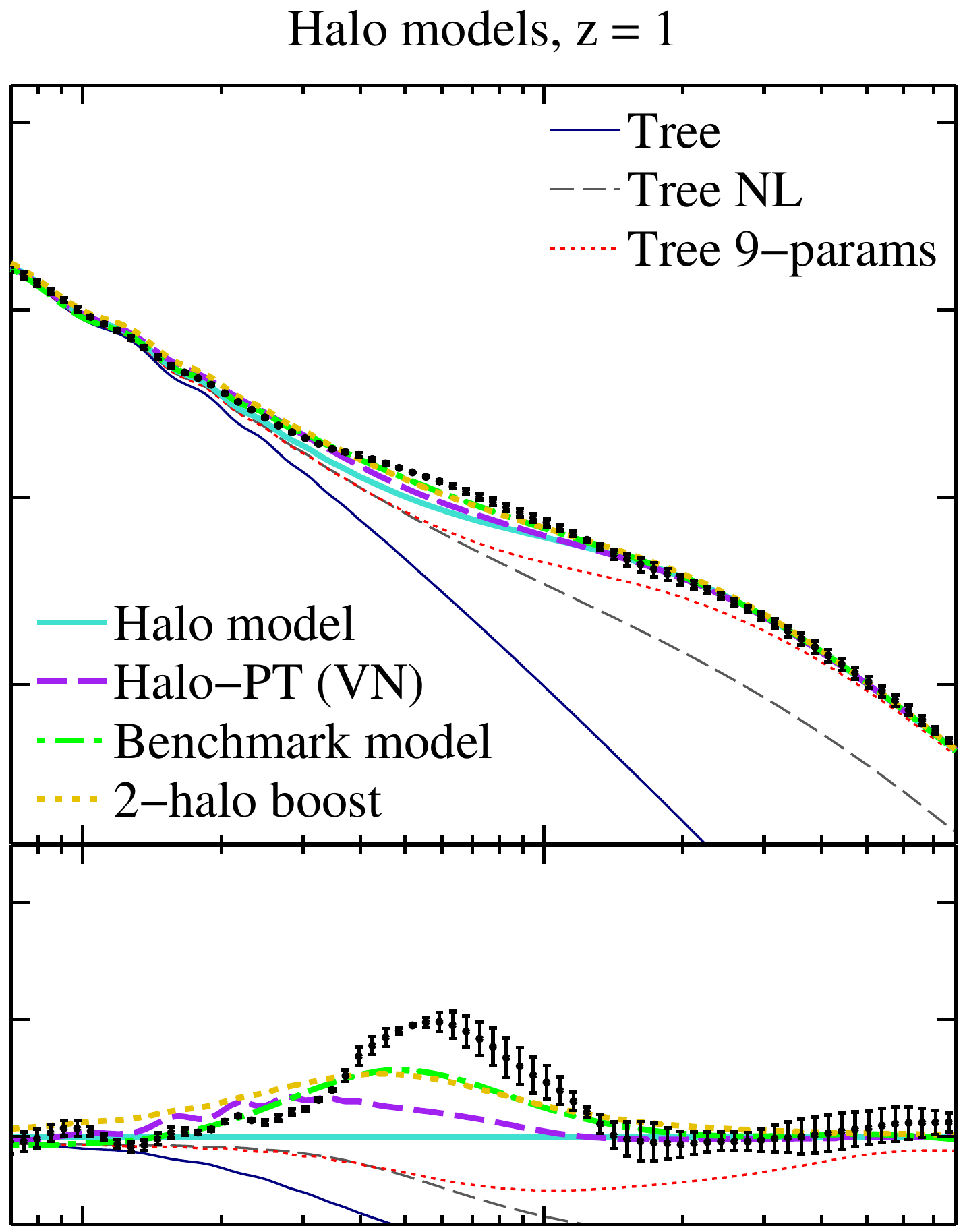}
\includegraphics[height=0.3\textheight]{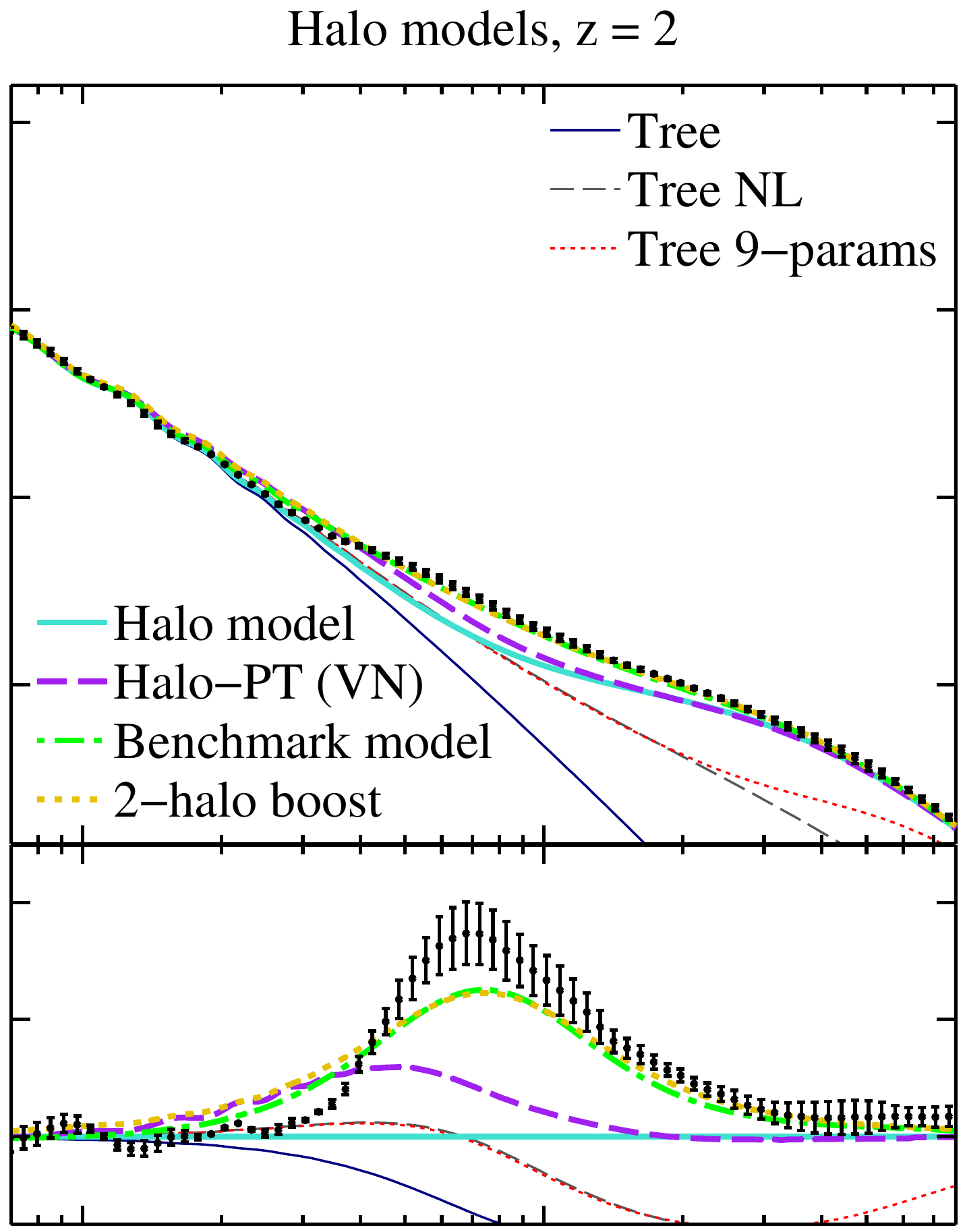}
\includegraphics[height=0.29\textheight]{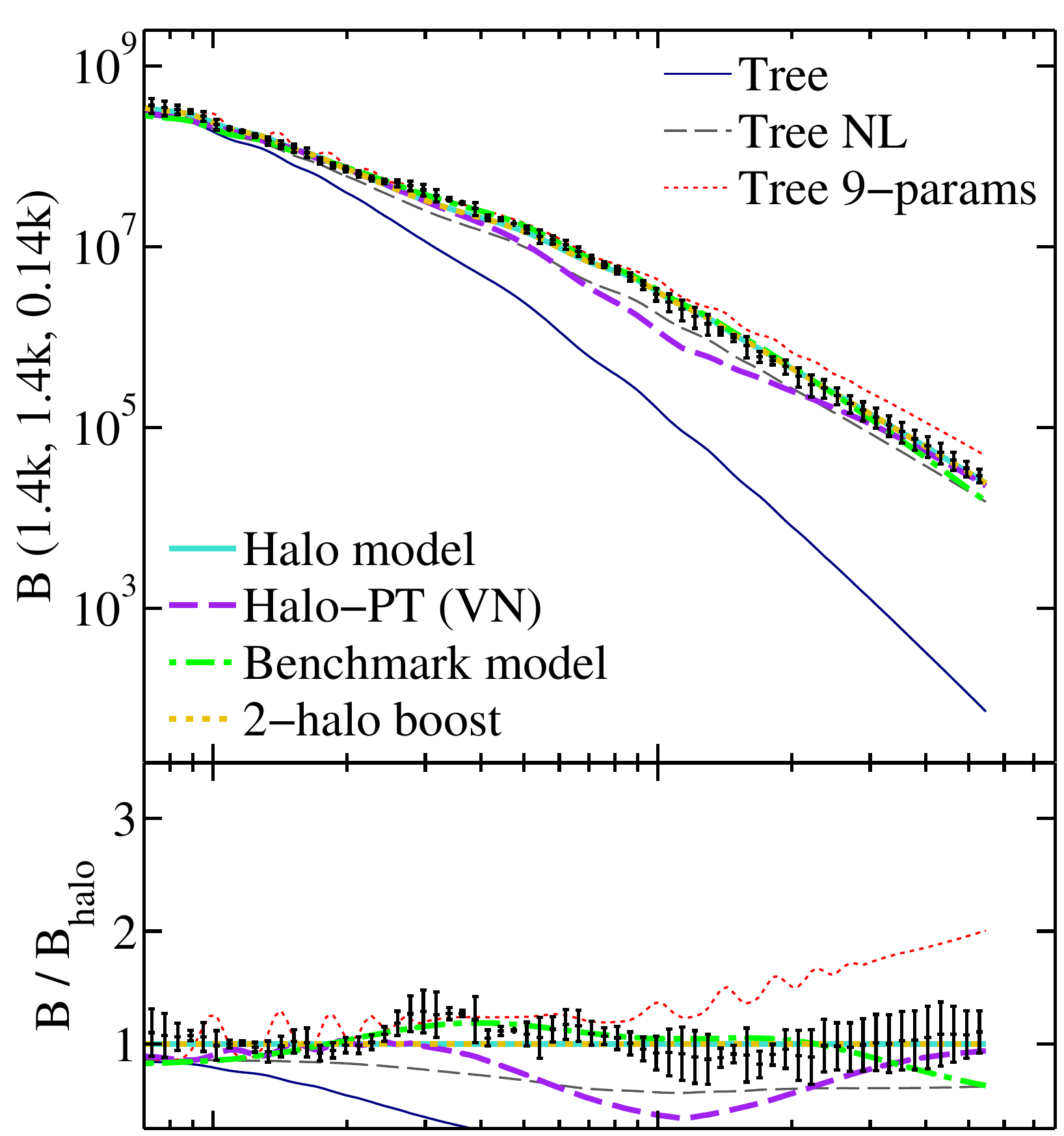}
\includegraphics[height=0.29\textheight]{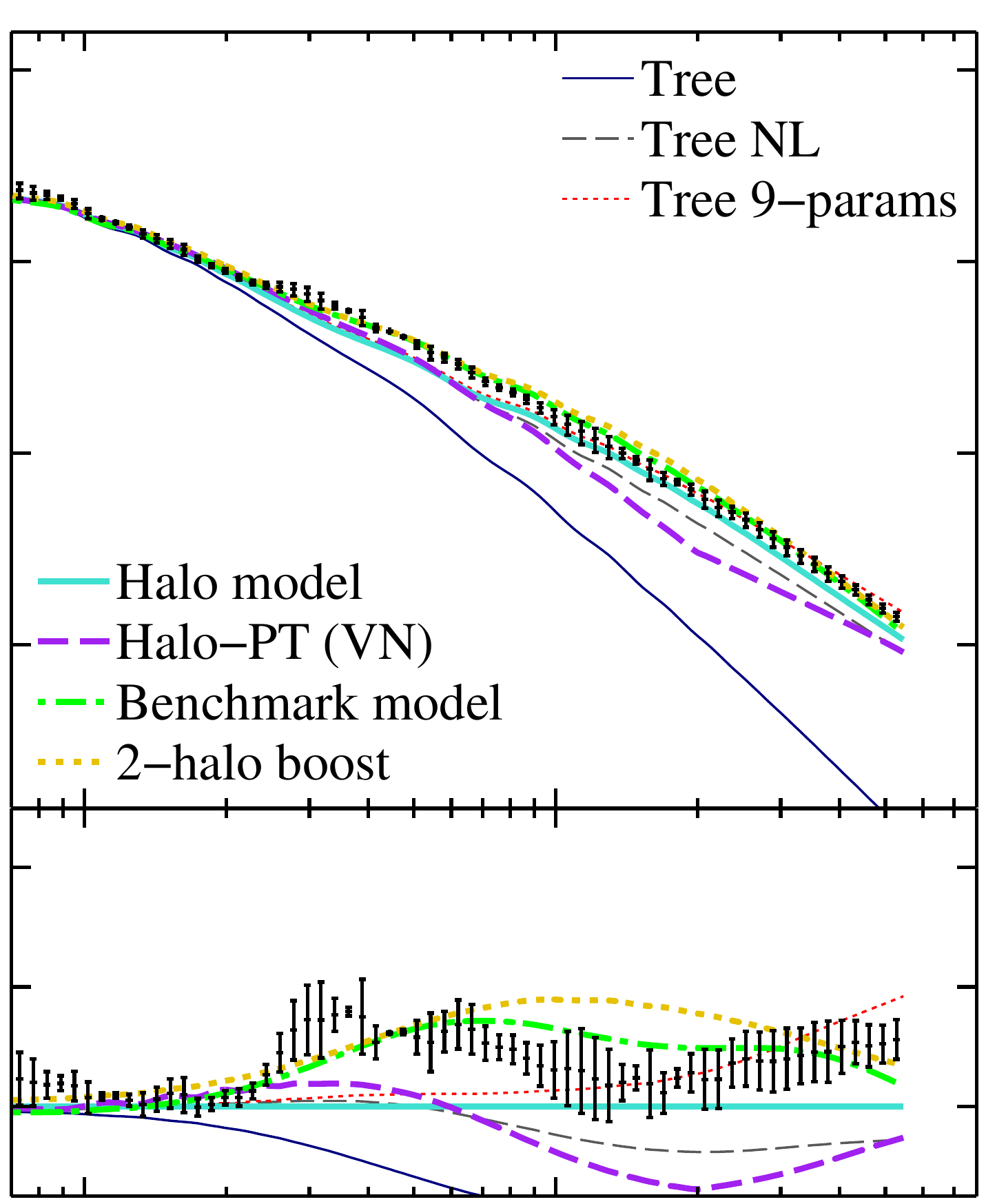}
\includegraphics[height=0.29\textheight]{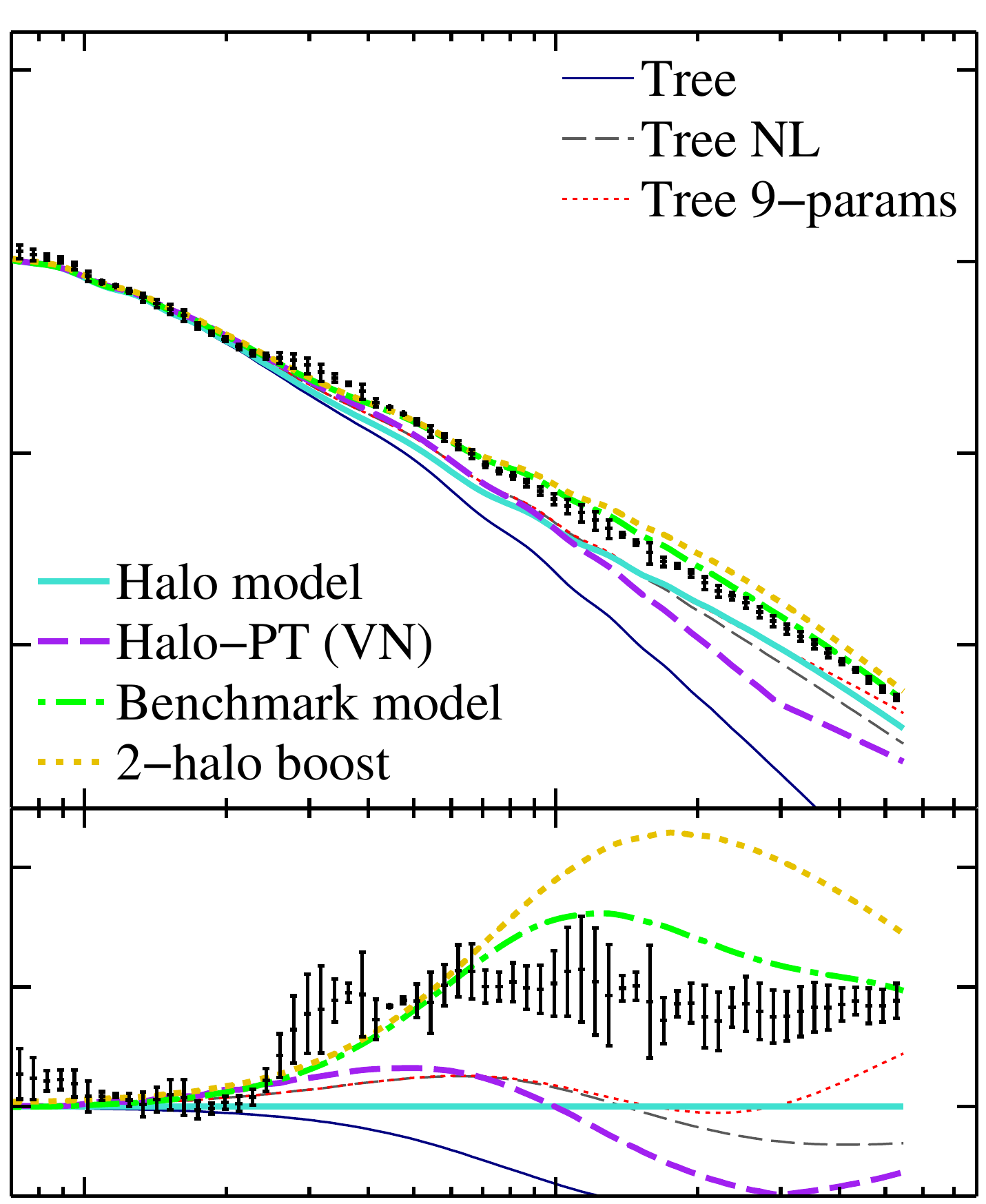}
\includegraphics[height=0.32\textheight]{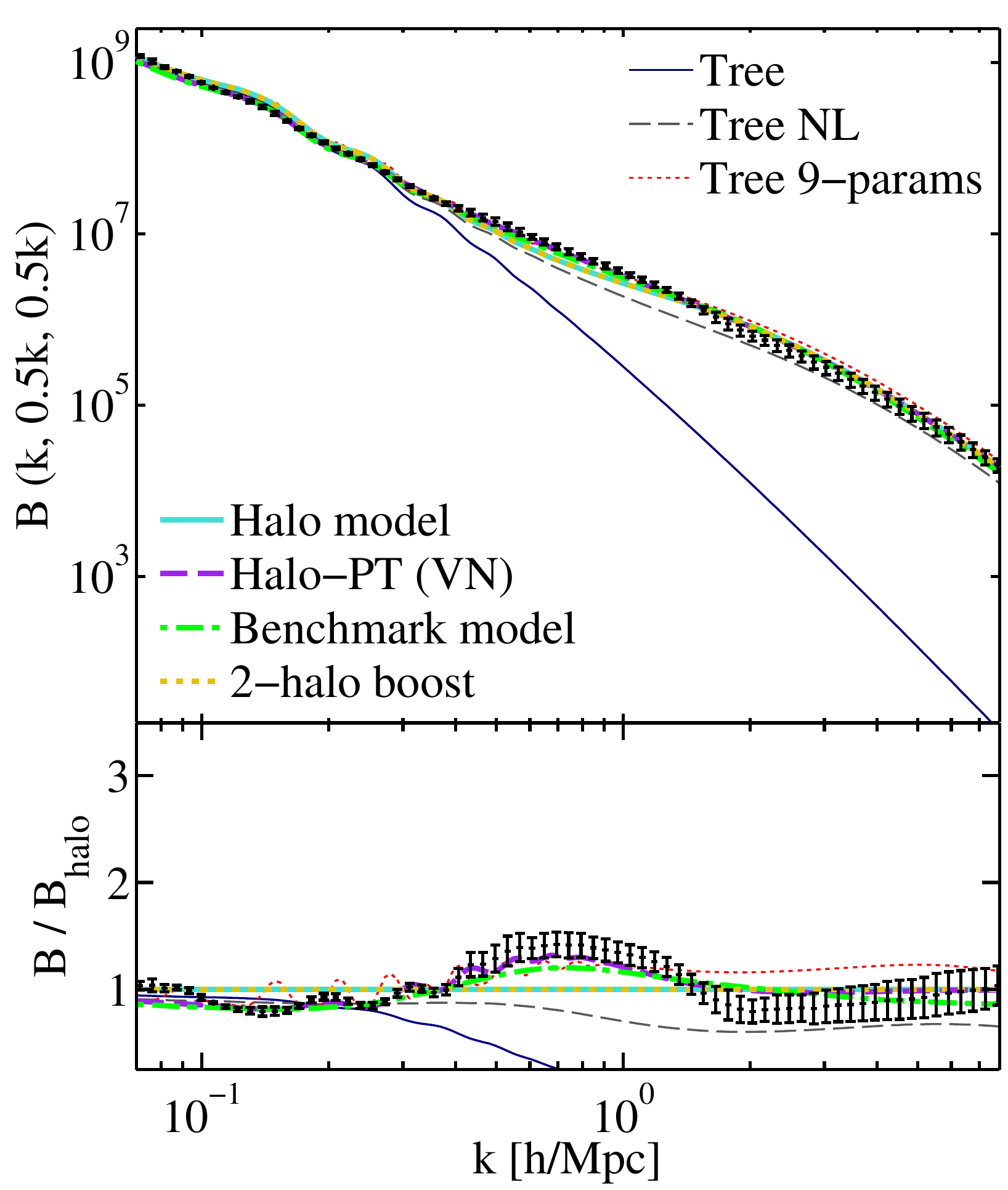}
\includegraphics[height=0.32\textheight]{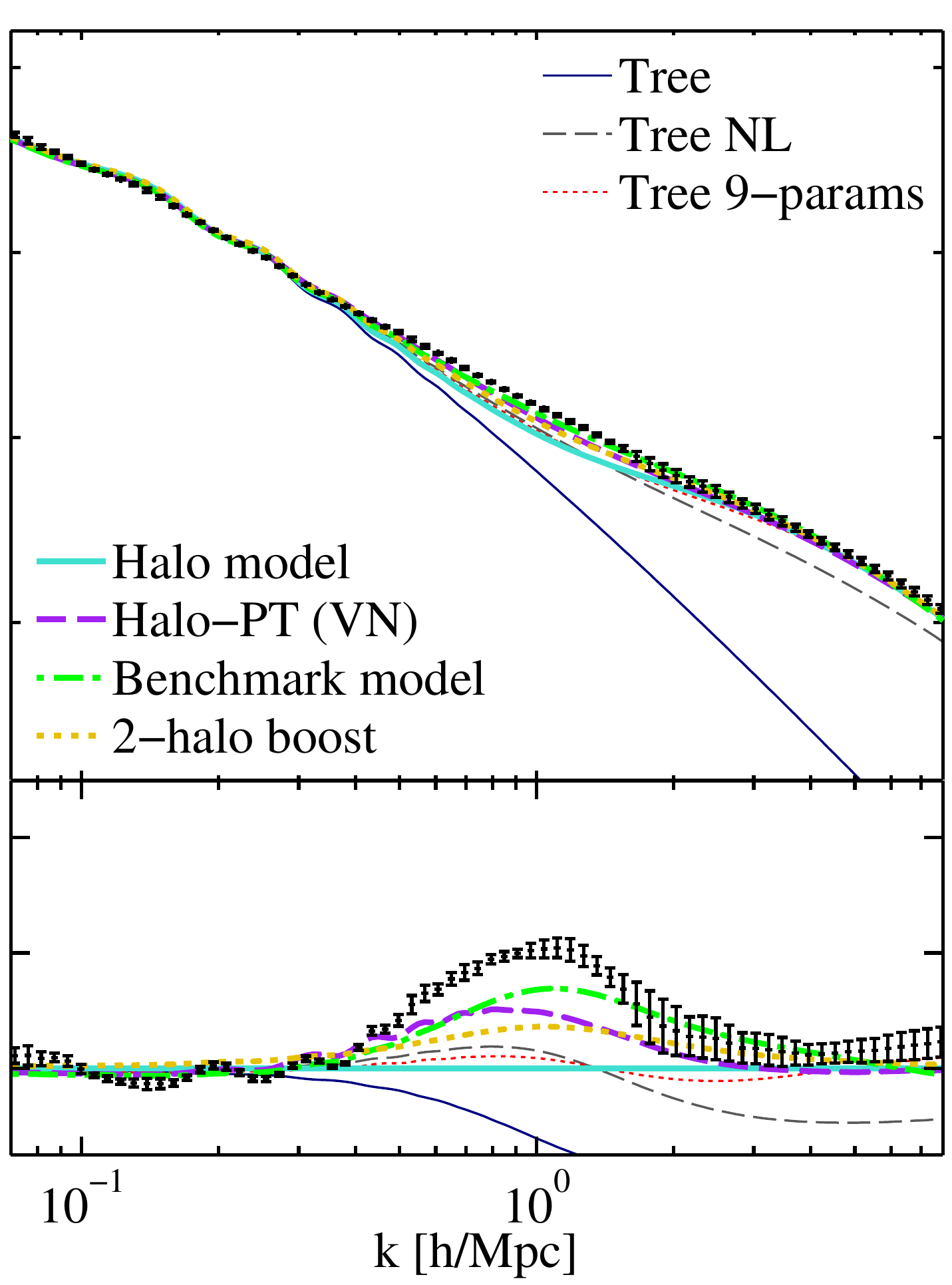}
\includegraphics[height=0.32\textheight]{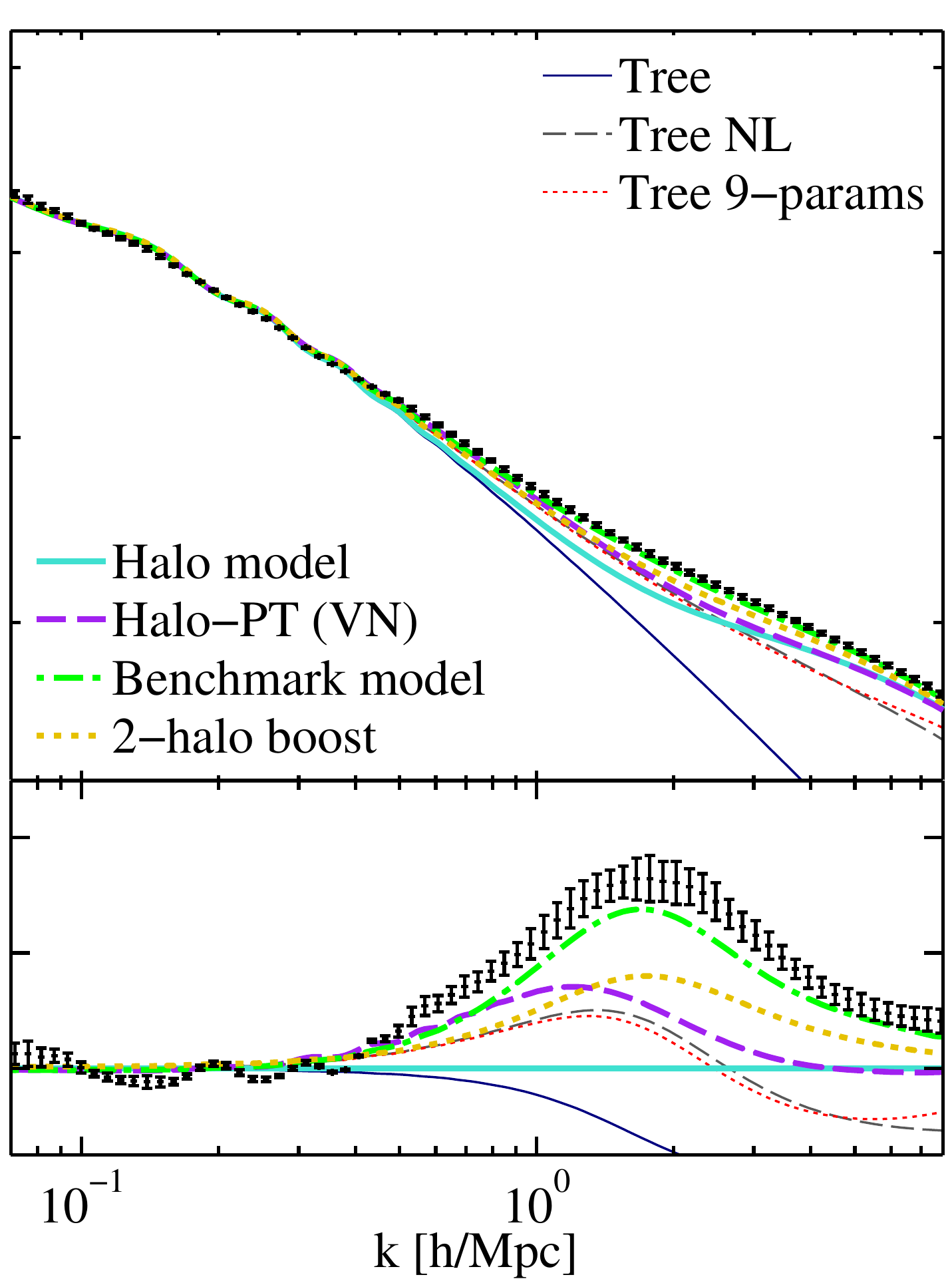}
\caption{Comparison of phenomenological nonlinear models of the matter bispectrum with $N$-body simulations, at redshifts 0, 1, 2 (left to right), for the equilateral, squeezed, and flattened configurations (top to bottom). The lower panels show the substantial residuals with respect to the standard halo model for $z>0$, demonstrating that the simple three-shape benchmark model provides a good fit to the $N$-body matter bispectrum for all three limits and redshifts.}
\label{fig:slices_halo}
\end{center}
\end{figure*}

\subsection{Testing phenomenological halo models}
\label{sec:testhalo}

By analogy with the discussion of the PT methods above, we first make qualitative comparisons of the phenomenological halo models with the measured bispectrum.   In Fig.~\ref{bispectrum_phenohalos}, we plot these bispectra in three dimensions at two redshifts $z=0,2$. While the standard halo model provides a reasonable fit at $z=0$, it reveals a large deficit on intermediate scales $k\sim 1 \, h/\text{Mpc}$. This is corrected in the three-shape model by using the nonlinear tree-level bispectrum and adopting a different growth rate for the squeezed signal at higher redshift.  In Fig.~\ref{fig:slices_halo} we offer a more detailed picture in the limiting equilateral, squeezed and flattened configurations,  also showing residuals relative to the standard halo model.
From Fig.~\ref{bispectrum_phenohalos}, we can see that for all configurations the standard halo model provides a good match to the $N$-body data on both linear and fully nonlinear scales, while a more significant mismatch appears in the transition regime at redshifts $z>0$.  The problem may be due in part to the approximate nature of the assumption in the halo model about all the matter in the Universe being in collapsed haloes, while other sources of inaccuracy are the spherical shapes of the haloes as well as neglecting their internal substructure; it is an issue acknowledged in the literature both for the power spectrum and the bispectrum \cite{PhysRevD.83.043526, Mead:2015yca, valageas1, valageas2}. 
 We confirm that this mismatch becomes more severe at higher redshift: for example, at $z = 2$ there is up to a factor of three mismatch on these intermediate scales. 
 
 The other phenomenological models we consider attempt to improve the behaviour in the transition region in different ways, and with varying degrees of success; they are also plotted in Fig.~\ref{fig:slices_halo}.  
 The combined halo-PT model provides some improvement at $z=0$ for flattened configurations, but it fails to significantly improve the situation at higher redshifts and especially in the squeezed limit.
The phenomenological two-halo boost and three-shape benchmark models improve the $N$-body results over a broader range of redshifts and configurations, largely by increasing the relative amplitude of the two-halo term at $z > 0$.  The three-shape benchmark, in particular, achieves a satisfactory fit in all limits and at all redshifts using only the restricted ansatz (Eq.~\ref{shapes}) by also increasing power in the flattened limit with the  nonlinear tree-level bispectrum.

\begin{figure*}[t!]
\begin{center}
\includegraphics[height=0.25\textheight, trim={0 0 0.25cm 0},clip]{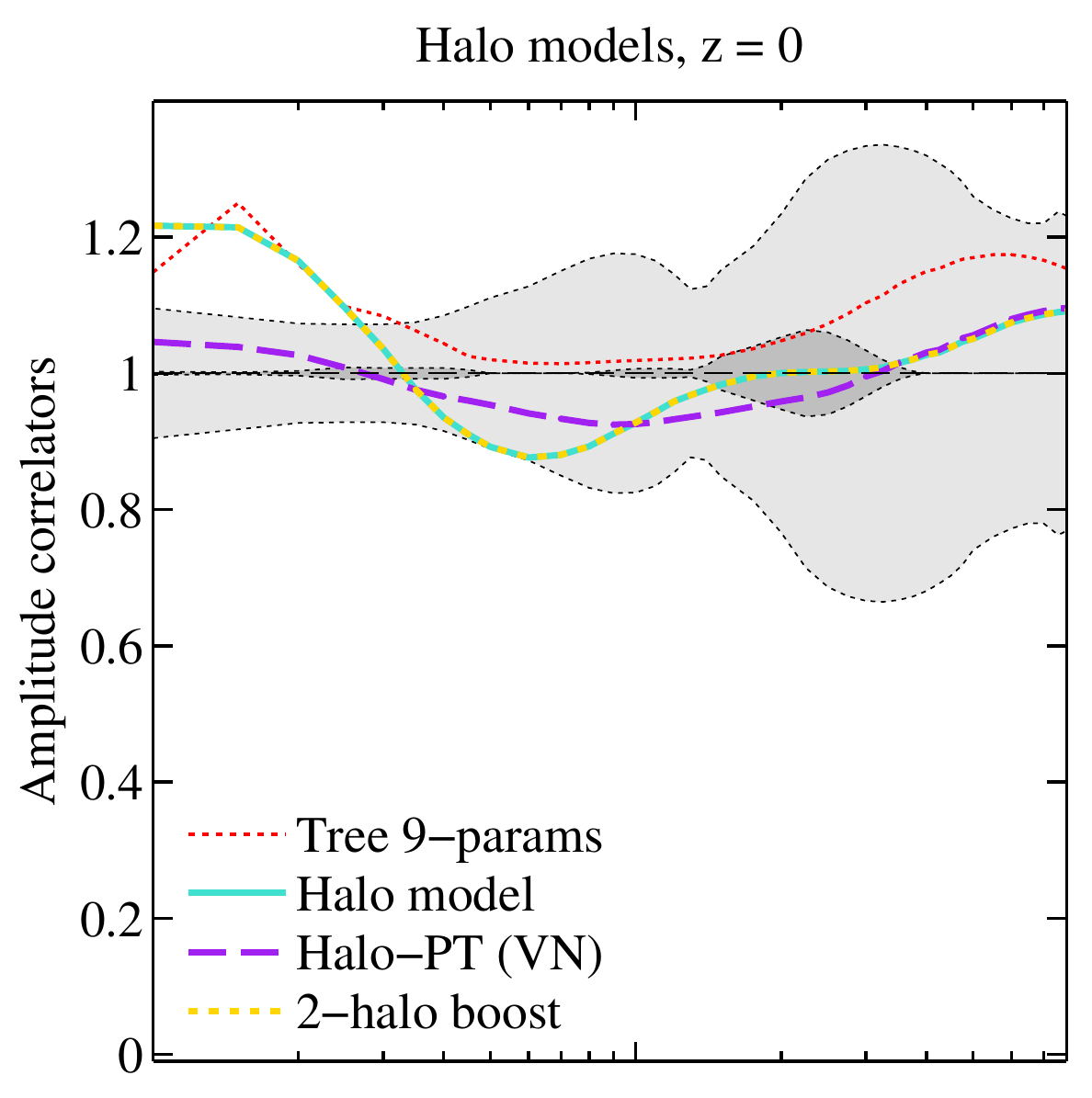} 
\includegraphics[height=0.25\textheight, trim={-0.25cm 0 0.25cm 0},clip]{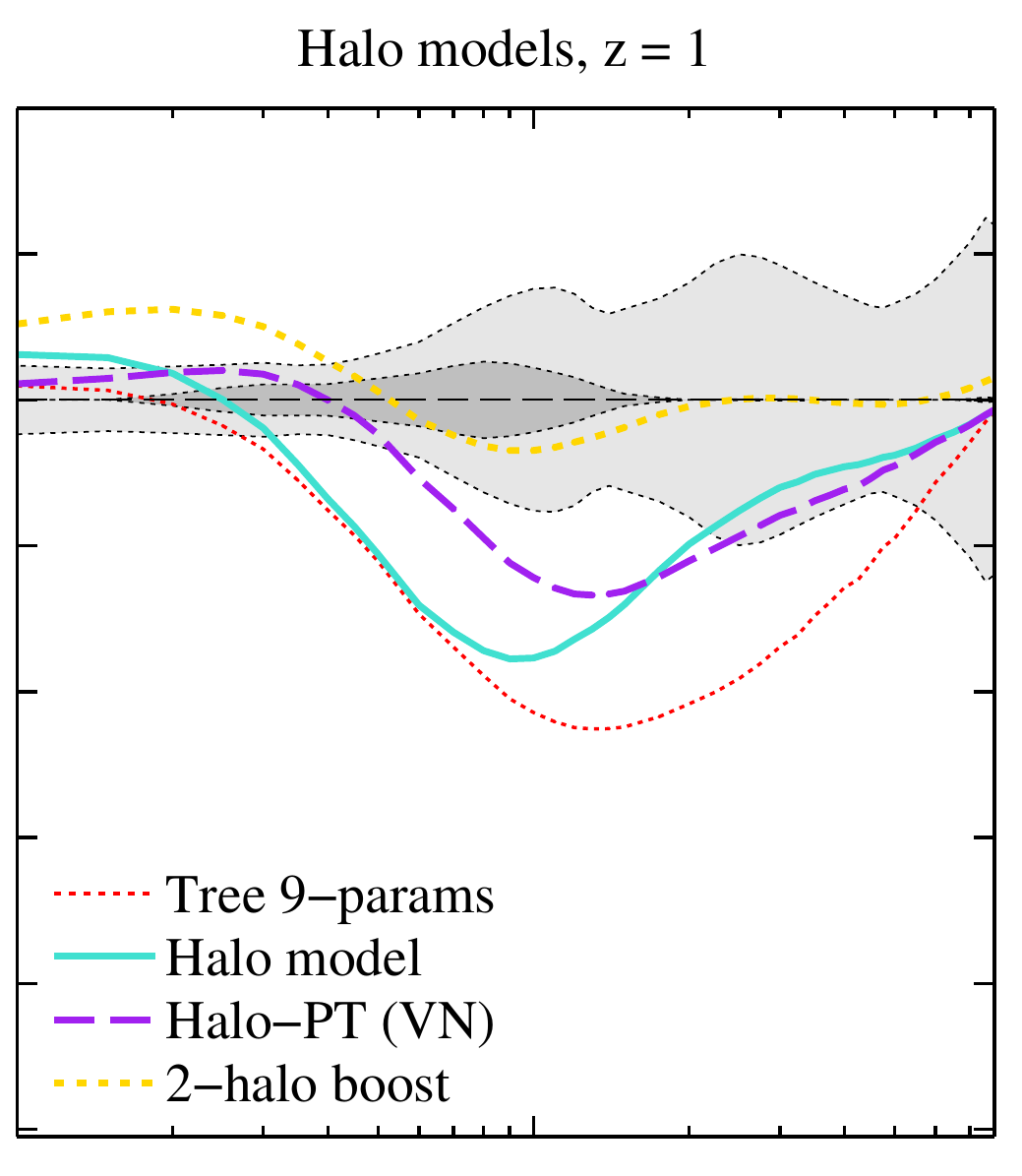} 
\includegraphics[height=0.25\textheight, trim={-0.25cm 0 0.25cm 0},clip]{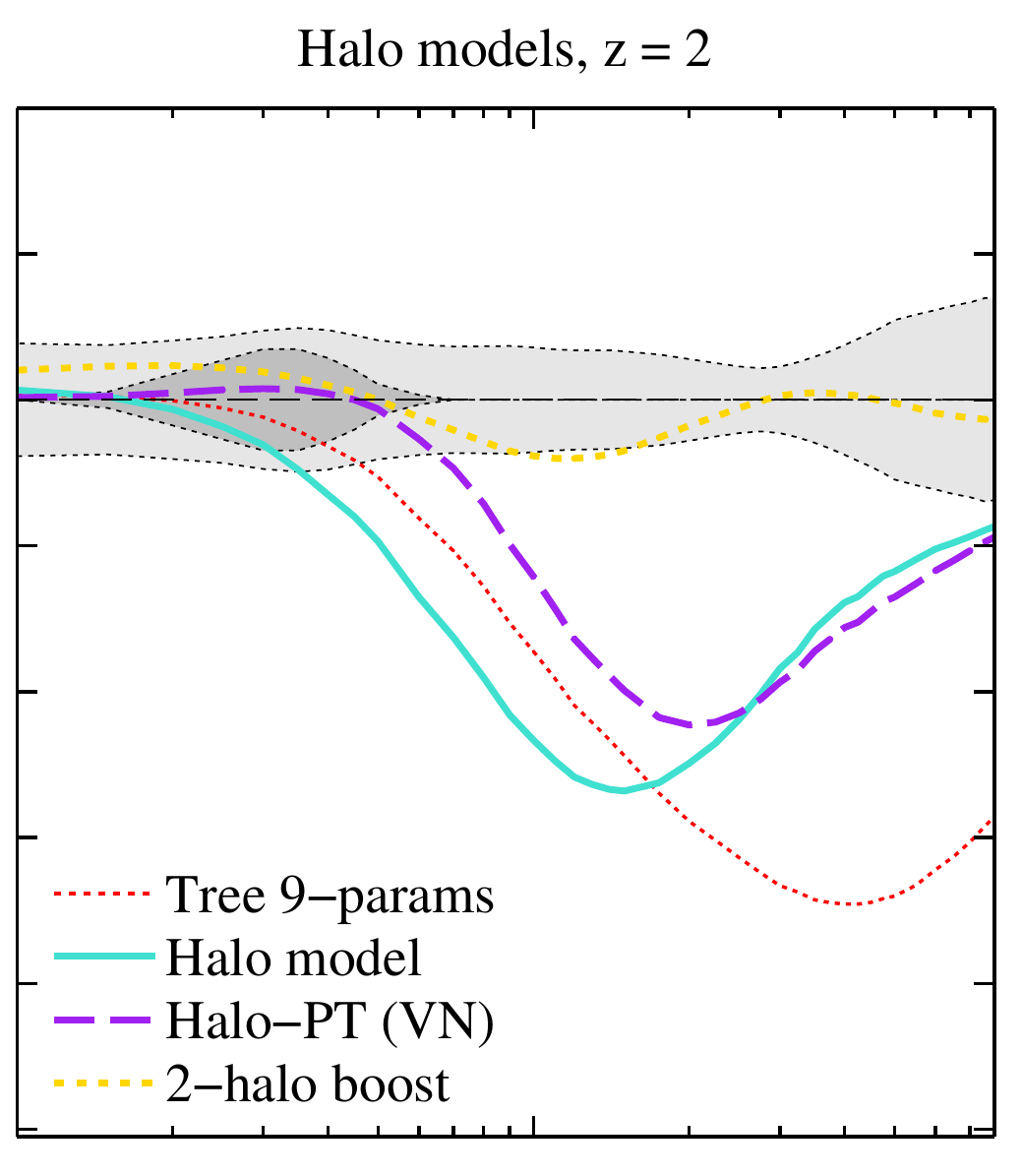} 
\includegraphics[height=0.255\textheight, trim={0 0 0.25cm 0.2cm},clip]{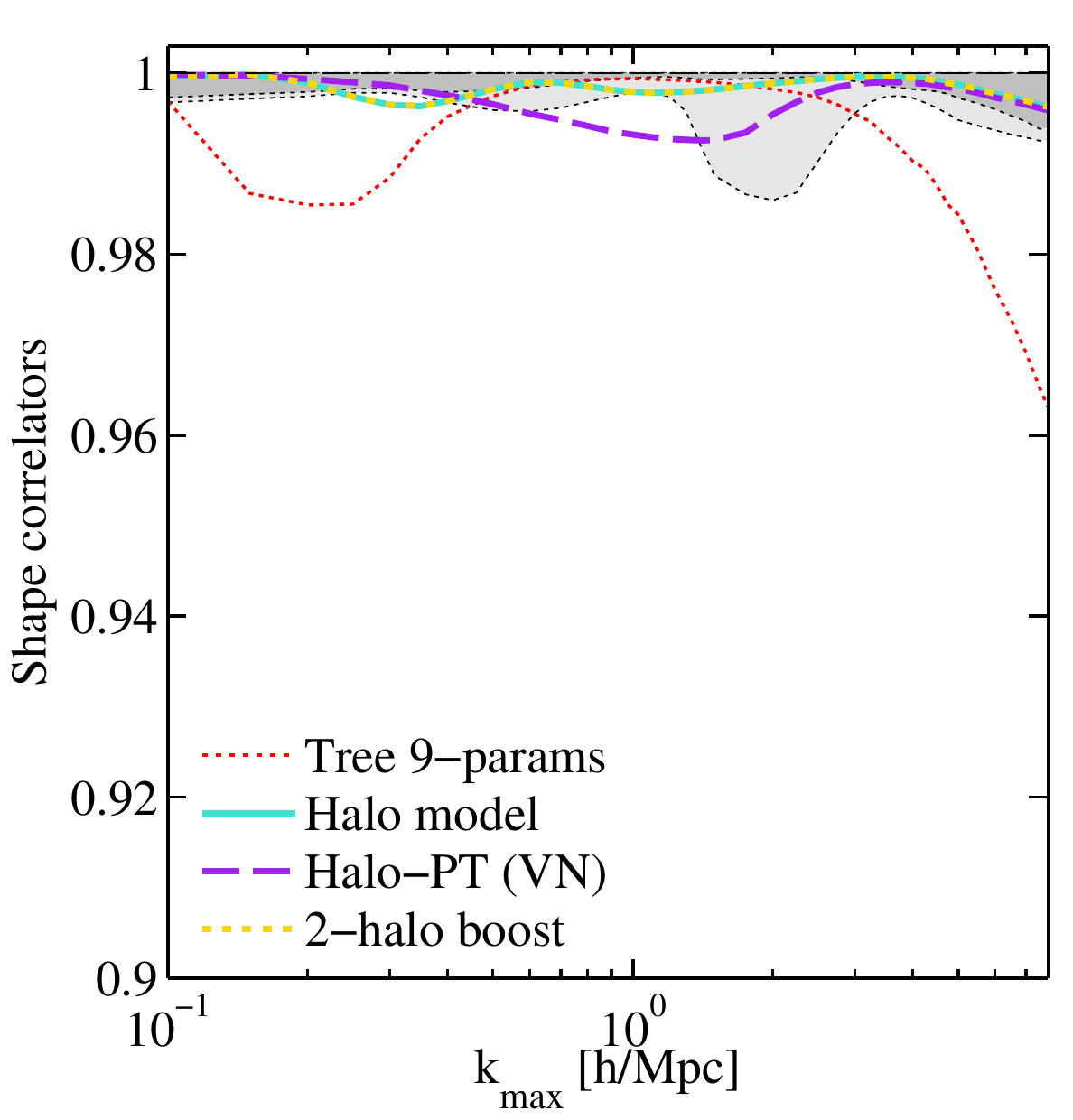} 
\includegraphics[height=0.255\textheight, trim={0 0 0.25cm 0.2cm},clip]{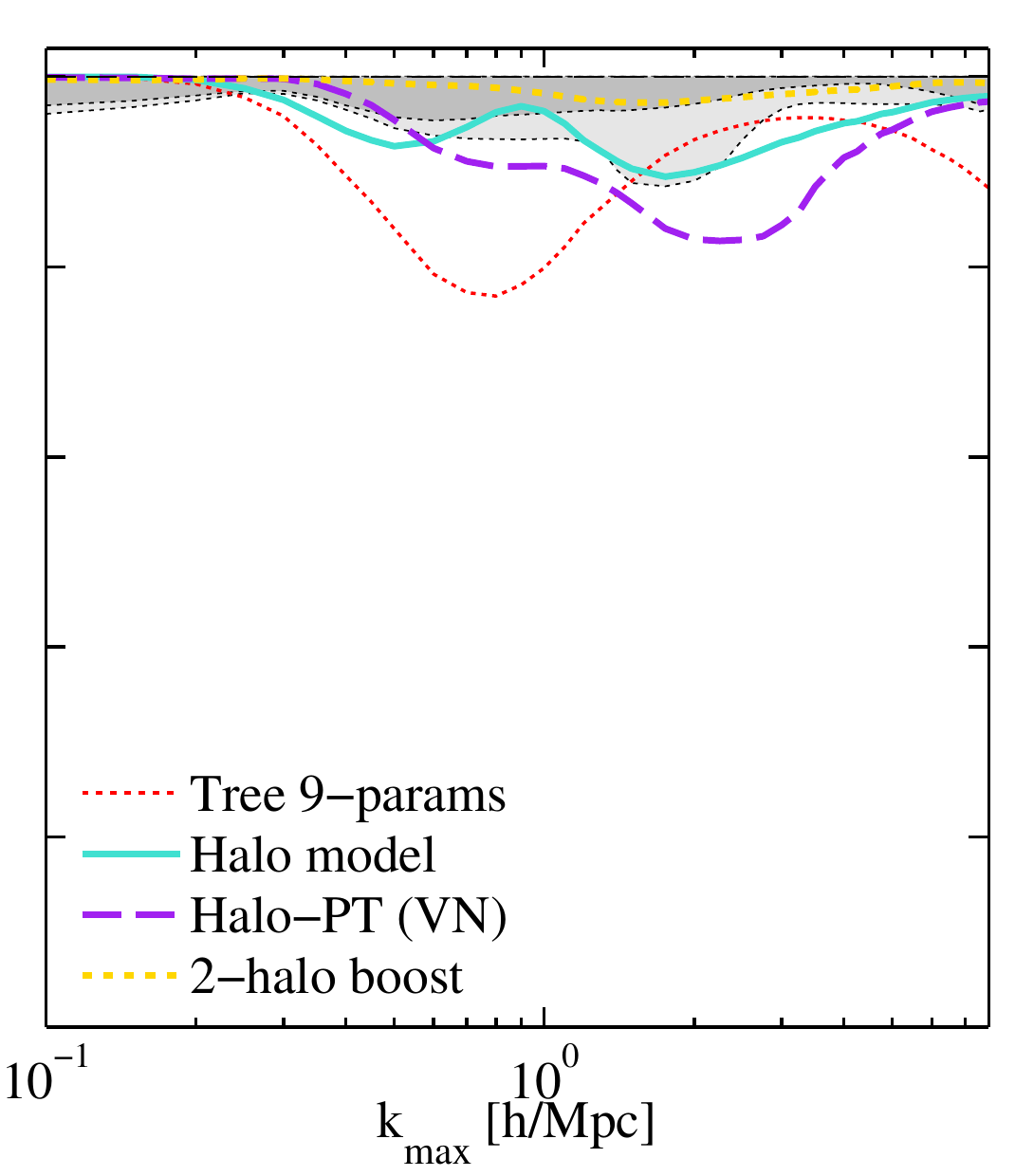}
\includegraphics[height=0.255\textheight, trim={0 0 0.25cm 0.2cm},clip]{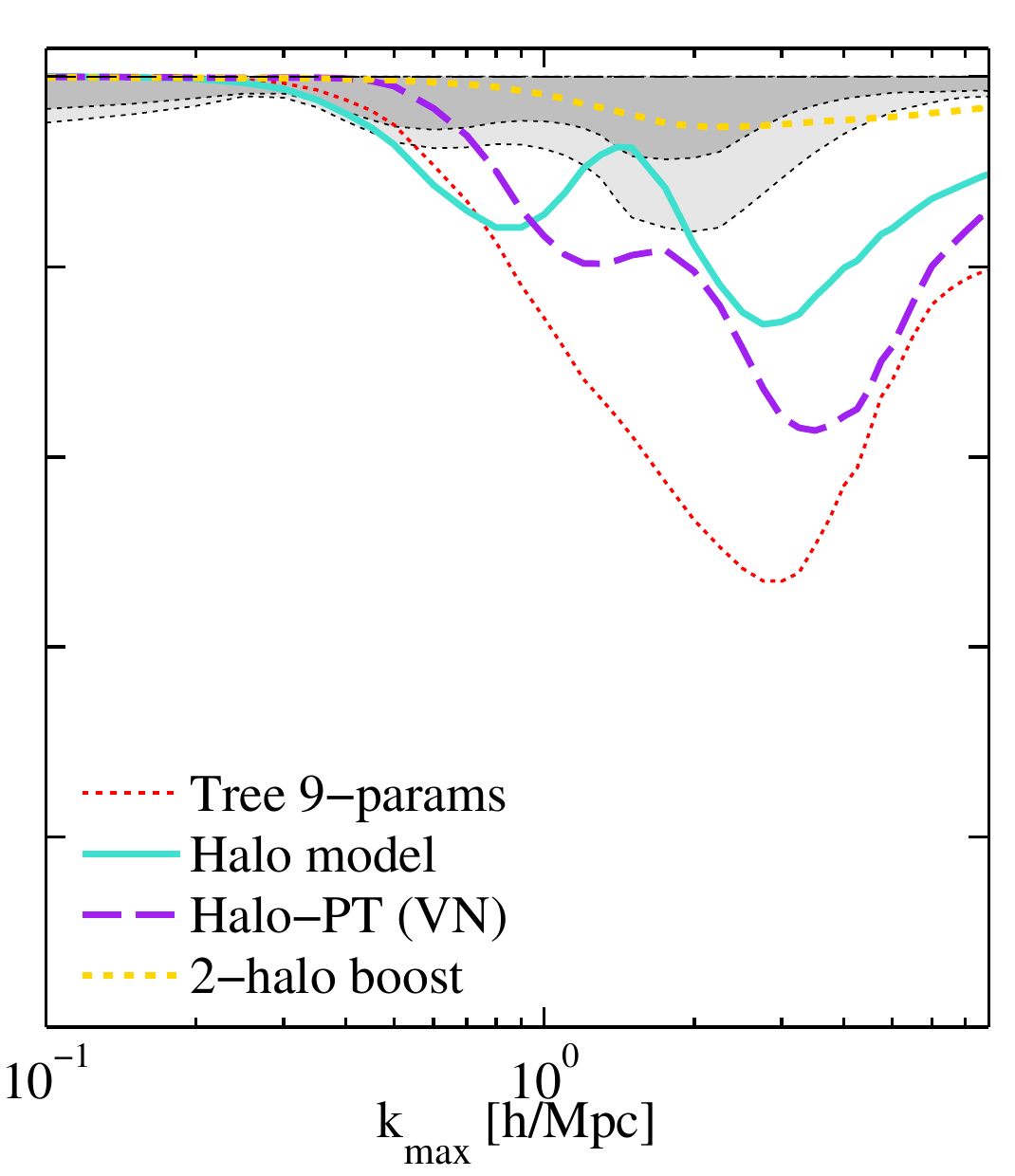} 
\caption{The amplitude $\mathcal{A}$ (top row) and shape $\mathcal{S}$ (bottom row) correlators at redshifts 0, 1, 2 for the phenomenological halo models, obtained by comparing with the three-shape benchmark model. The shaded areas represent error estimates between the three-shape benchmark model and the simulations and are explained in Sec.~\ref{sec:testpert}.}
\label{combkmaxh}
\end{center}
\end{figure*}

We now turn to a full three-dimensional analysis with the amplitude ($\mathcal{A}$) and shape ($\mathcal{S}$) correlators plotted in Fig.~\ref{combkmaxh} for redshifts $z=0,1,2$; as in the previous subsection, we again compare to the three-shape benchmark model with best-fit parameters of Eq.~(\ref{three-shape-fit}).   We also determine where the accuracy of different phenomenological models and fits break down in  Table~\ref{tablehalo}.

It is apparent from Fig.~\ref{combkmaxh} that the standard halo model offers an insightful description of the matter bispectrum in the nonlinear regime at redshift $z=0$; the shape correlation is above 99\% everywhere investigated and the amplitude deviates by less than 15\% from the measured simulation bispectrum over the range  $0.4 \, h/\text{Mpc} <k<8 \, h/\text{Mpc}$. Nevertheless, we observe some excess power on large scales, e.g. at $k_{\max}\sim 0.1 \, h/\text{Mpc}$, which is a well-known problem of the standard halo model, due to the one-halo term approaching a constant and the two-halo term not vanishing as $k \to 0$.
The large-scale excess is less important as the redshift is increased. However, we see in  Fig.~\ref{combkmaxh} that there is a new problem on intermediate scales where an amplitude deficit emerges, which increases significantly as a function of redshift: in the transition regime, the amplitude correlator decreases from 0.9 at $z = 0$ to 0.65 at $z=1$, and 0.45 at $z=2$. As discussed in previous sections, this is primarily due to an underprediction of the two-halo component (squeezed shape) in this $k$-range. Moreover, the lowest point in the transition regime shifts to higher $k$ at higher redshift, from $k \approx 0.5 \, h/\text{Mpc}$ at $z=0$ to $k \approx 1.5 \, h/\text{Mpc}$ at $z=2$ (see Fig.~\ref{h1eq} for an illustration of this in the equilateral configuration). In the strongly nonlinear regime, after the two-halo component has decayed and the one-halo term becomes dominant, the halo model again approaches the simulations. As we discussed above in Sec.~\ref{sec:sims}, a possible way of solving this problem is by boosting the two-halo component, which peaks exactly in the regime of interest; this leads to the two-halo boost model also shown in Fig.~\ref{combkmaxh}, which entails a minimal cost of introducing additional power on large scales.

The power excess produced on linear scales by the standard halo model is corrected in the combined halo-PT model of Sec.~\ref{sec:ihm}. As this model can use any perturbative theory on linear scales, we choose to use EFT, because we found it in the previous section to offer the most extended range of validity.   In this prescription, the two- and three-halo terms of the halo model are switched on as the perturbation theory is decaying. Hence, at $z=0$, this model provides the best fit across all scales considered; in the strongly nonlinear regime, the model converges towards the standard halo model result, because on small scales the improved two-halo and the improved one-halo terms are the same as their standard counterparts. Nonetheless, for $z>0$, the combined halo-PT model has the same problem as the standard halo model, as there is a deficit in the transition regime, though marginally weaker. In this model, the improved one- and especially the two-halo terms are heavily suppressed on large and intermediate scales. This is not visible in Fig.~\ref{combkmaxh}  because most of the signal comes from EFT on these scales, making it more challenging to solve the deficit by a simple boost of the improved two-halo term. 

The nine-parameter fit, which is based on the simple tree-level model, fitted to $k \le 0.4 \, h/\text{Mpc}$ and for $z \le 1.5$ is fairly accurate when extrapolated across the full domain at $z=0$.  (In principle, improvements  could be obtained by re-fitting the parameters to higher redshifts and further into the nonlinear regime, though the model does not naturally include the squeezed and constant shapes required.) Spurious peaks appearing at $z=0$ are produced by the BAO features of the power spectrum, as discussed previously.  However, at $z=1,2$ this model becomes increasingly inaccurate at large $k$ with its amplitude decreasing in a similar fashion to the nonlinear tree level bispectrum. Nevertheless, the nine-parameter model produces an accurate result up to $k_{\max}\sim 0.8 \, h/\text{Mpc}$ for all the redshifts considered.

As for perturbation theories,  in Table \ref{tablehalo} we present the maximum value of the wavenumber $k_{\max}^*$ for which the phenomenological halo models show good agreement, that is, by considering the point where the amplitude correlator deviates by more than 20\% from unity. The numerical results of the table confirm the general trends discussed above.  In contrast to the PT case, here the agreement between models and simulated data becomes worse at higher redshift, as the basic assumptions underlying the halo model become less valid. At higher redshifts, a secondary range of validity exists at high $k$ after the transition region, which is visible from Fig.~\ref{combkmaxh} but not reported in the table.

\begin{table}
\begin{center}
\caption{Wavenumber $k_{\max}^*$ where the amplitude deviation for phenomenological halo models is greater than 20\% when compared to the three-shape benchmark model matched to simulations.  (The small $k$ excess problem of the standard halo model  is ignored.) At $z=0$ all models agree within 20\% over the entire range of scales. }
\begin{tabular}{  c c c c }
\toprule
\multicolumn{4}{ c }{Phenomenological halo models}\\
\colrule
  Threshold $20\%$          & \multicolumn{3}{c}{$k^*_{\max}\,[h/\text{Mpc}]$} \\ 
\colrule
 Theory     &  $z = 0 $  &  $z = 1 $  &  $z = 2 $ \\
\colrule
Standard halo model    &  $>8$  & 0.47 & 0.51  \\
Combined halo-PT model & $>8$   & 0.48 & 0.68 \\
 9-parameter fit       & $>8$   & 0.82 & 0.90 \\
\botrule
 \end{tabular}
\label{tablehalo}
\end{center}
\end{table}

Among the alternative phenomenological models we tested, we conclude that the combined halo-PT model based on EFT is the most accurate, offering a physically well-motivated attempt to solve problems of the standard halo model.   Nevertheless, like the standard halo model, it also does not exhibit appropriate growth rates for the two-halo contribution at high redshift and, further, the prescription for transitioning between EFT and the other halo contributions deserves closer scrutiny.  From a phenomenological point of view there is a straightforward means to improve the theory by boosting the two-halo term at higher redshifts, as in the three-shape benchmark model.

\section{Conclusions}
\label{sec:concl}

The bispectrum of  large-scale structure has so far been a relatively neglected observable, due to the high cost of measuring it with most current sub-optimal estimators, and the relative complexity of its modelling and interpretation.
This is however bound to change in the current age of precision cosmology and ever-larger galaxy surveys, as the combination of two- and three-point statistics can improve the constraining power of the upcoming data, by breaking the existing degeneracies between cosmological and astrophysical parameters. The ultimate goal of large-scale structure bispectrum measurements is its potential to constrain models of the early universe via their non-Gaussian contribution to the primordial density perturbations, thus complementing and improving existing CMB constraints \cite{Sefusatti2007,Jeong2009}.  

Achieving these ambitious objectives will require efforts on multiple fronts. 
 A first issue shared with  power spectrum analysis is the endeavour to improve the theoretical modelling as far as possible into the nonlinear regime; other outstanding points include making the bispectrum estimation faster and more efficient, and developing a comprehensive method for comparing bispectrum predictions with observations.

 In this paper we have made progress on all these fronts.
Firstly, we studied how accurately different theoretical models for the matter bispectrum work on different scales, by comparing them with $N$-body simulations and introducing a new simplified phenomenological model based on three canonical bispectrum shapes. 
Secondly, we have used for our study the efficient modal bispectrum estimator by Ref.~\cite{Schmittfull2013}, which allowed us to reconstruct the full three-dimensional bispectrum information based on $\sim100$ modes only. Thirdly, we have introduced the amplitude, shape, and total correlators as instruments to estimate the overall goodness of match between a bispectrum model and measurements across its full three-dimensional domain, thus greatly simplifying the process of model comparison and parameter estimation.

The different bispectrum models we considered can be divided into two categories: methods based on perturbation techniques, and phenomenological models based on or inspired by the halo model. The perturbative methods assume a small departure from linear scales, when the density fluctuations are small, and therefore have limited range of validity. Multiple approaches exist for increasing the scales of validity of perturbative theories, such as effective field theories, and resummed perturbation theories. We have confirmed that such one-loop recipes manage to accurately model nonlinearities up to $k_{\max} \simeq 0.15\, h/\text{Mpc}$ at $z = 0$ for the matter bispectrum and further at higher redshift ($k_{\max} \simeq 0.4 \, h/\text{Mpc}$ at $z=2$).  This is already beyond the expectations for the strictly perturbative regime, but some methods appear to be amenable for even more ambitious extrapolations into the nonlinear regime, with effective field theory predictions apparently showing good agreement to $k \simeq 0.3 \, h/\text{Mpc}$ at $z = 0$, though at the cost of introducing free extra parameters calibrated to simulations.  The much simpler nonlinear tree-level bispectrum also offered useful nonlinear projections out to $k_{\max} \simeq 0.17\, h/\text{Mpc}$ at $z=0$.   

In addition, we have derived for the first time the expressions of the two-loop \textsc{MPTbreeze} bispectrum in an infrared-safe manner, demonstrating that it is analytically and numerically tractable, even if  computationally challenging. We have shown the improvement in the wavenumber range over the one-loop calculation for three triangle shape configurations. 

From a different perspective, the halo models  rely on models of matter collapse in order to describe nonlinearities from a phenomenological point of view. In that sense, they are valid much further beyond the scales that can be modelled by perturbation theories, and can match simulations reasonably well in the strongly nonlinear regime at $z = 0$. 
The combined halo-PT model \cite{valageas2} represents a compromise between the two approaches. It relies on a perturbative method on large scales, chosen here to be the EFT, where the halo model is not accurate, while relying on the halo model on nonlinear scales. For these reasons, we found that the 
halo-PT model gives the most accurate predictions on all scales at $z = 0$.   Nevertheless, at higher redshifts, a significant deficit appears at intermediate scales for all halo models.   

We have found that a simple way to solve this halo deficit problem is to increase the contribution of the squeezed or two-halo shape at $z>0$, which we have found dominates in the transition regime.
Driven by the observations from $N$-body simulations, we have generalised this idea, thus developing a simple phenomenological `three-shape'  model that fits the simulations well over the full range of scales and redshifts considered. This benchmark model is based on the fundamental shapes of the halo model  --- tree-level, squeezed and constant shapes, corresponding to the three-, two- and one-halo terms respectively. 
This model can be seen as a first step towards the development of an accurate phenomenological model calibrated on $N$-body simulations, translating the idea behind the \textsc{Halofit} method to the bispectrum domain.  This will be observationally relevant for weak gravitational lensing which is sensitive to the matter bispectrum.

Solving this two-halo deficit problem motivates our new benchmark model but it uncovers a more serious misconception in the standard halo approach built as it is on a hierarchical picture of structure formation.  The basic premise that nonlinear haloes form first and then using these to classify and calculate non-Gaussian structures may need to be carefully reconsidered.   This is clear already from the tree-level bispectrum, which is present at high redshifts $z>30$ long before any haloes form; fundamentally it is associated with the initial stage of gravitational collapse in the first dimension  which causes `pancake-like' structures to form.  The three-halo term accommodates this {\it a posteriori} by noting that the large-scale tree-level signal will be imprinted on the halo distribution.   In the same manner, there will be a squeezed signal from the formation of  filamentary structures (due to the onset of collapse in the second dimension), which again precedes haloes on any given lengthscale.   At present the two-halo model is flawed by assuming a hierarchical origin for this squeezed bispectrum contribution, and so it does not capture the appropriate growth rate at higher redshift.  Our investigations here present quantitative bispectrum data in the relevant intermediate regime, which shows clear pathways ahead for improving the halo model (see also Ref.~\cite{Valageas2013}), as well as mathematical simplifications due to the approximate separability of the underlying bispectrum.

Future developments of this work will on the one hand lead to a more comprehensive and accurate phenomenological model of the matter bispectrum, fitted on higher-resolution simulations, which will provide a bispectrum counterpart to the \textsc{Halofit} method.
On the other hand, we will extend the modelling and the comparison to the case of biased tracers, i.e.~dark matter haloes and galaxies, to bridge the gap between modelling and observations by galaxy surveys.
Finally, we plan to include the effects of primordial non-Gaussianity of different types to determine how it is amplified through gravitational collapse and how it can be optimally identified. 

\begin{acknowledgments}
We wish to thank Mart\'in Crocce, James Fergusson, Donough Regan, Daniel Baumann,  Cristiano Porciani, and Christian Wagner for useful discussions, and Eiichiro Komatsu and Emiliano Sefusatti for comments on an earlier version of this work.  A.L., T.G., and E.P.S.S. are supported by STFC Grants No. ST/L000636/1. T.G. also acknowledges support from the Kavli Foundation.
This work was undertaken on the COSMOS Shared Memory system at DAMTP, University of Cambridge, operated on behalf of the STFC DiRAC HPC Facility. This equipment is funded by BIS National E-infrastructure Capital Grant No. ST/J005673/1 and STFC Grants No. ST/H008586/1, No. ST/K00333X/1 and No. STM007065/1.
The 3D bispectrum visualisations used the OSPRay ray-tracing rendering engine and we are grateful for support from Johannes Guenther and Gregory Johnson, Intel Corporation. 
\end{acknowledgments}

\appendix

\section{Standard Eulerian perturbation theory}
\label{sec:SPT_full}

We summarise here the derivation of the SPT power spectrum and bispectrum, following Ref.~\cite{Bernardeau20021}.
The density contrast $\delta$ and the peculiar velocity $\textbf{u}$ are defined in terms of the average density and velocity $\textbf{v}$ as:
\begin{align}
\rho\left(\textbf{x},\tau\right)&=\bar\rho\left(\tau\right)\left[1+\delta\left(\textbf{x},\tau\right)\right] \\
\textbf{v}\left(\textbf{x},\tau\right)&=\mathcal{H}\textbf{x}+\textbf{u}\left(\tau\right) \, .
\end{align}
Then the cosmological gravitational potential $\Phi$ satisfies the Poisson equation:
\begin{equation}
\nabla^2\Phi\left(\textbf{x},\tau\right)=\frac{3}{2}\Omega_m\left(\tau\right)\mathcal{H}^2\left(\tau\right)\delta\left(\textbf{x},\tau\right) \, .
\end{equation}
If we define the momentum as $\textbf{p}=am\textbf{u}$, the particle number density in phase space $f\left(\textbf{x},\textbf{p},\tau\right)$ satisfies the Vlasov equation:
\begin{equation}
\frac{df}{d\tau}=\frac{\partial f}{\partial \tau}+\frac{\textbf{p}}{ma}\cdot \nabla f-am \nabla \Phi \cdot \frac{\partial f}{\partial \textbf{p}}=0 \, .
\label{vlasov}
\end{equation}

In order to obtain the spatial distribution of the particles, the moments of Eq.~(\ref{vlasov}) can be taken by appropriate integration in momentum space. The first 3 moments are of interest here:
\begin{align}
\label{ord0}
\int d^3\textbf{p}f\left(\textbf{x},\textbf{p},\tau\right)&=\rho\left(\textbf{x},\tau\right) \\
\label{ord1}
\int d^3\textbf{p}\frac{\textbf{p}}{am} f\left(\textbf{x},\textbf{p},\tau\right)&=\rho\left(\textbf{x},\tau\right)u\left(\textbf{x},\tau\right) \\
\label{ord2}
\int d^3\textbf{p}\frac{p_ip_j}{am} f\left(\textbf{x},\textbf{p},\tau\right)&= \nonumber \\ 
\rho\left(\textbf{x},\tau\right)u_i\left(\textbf{x},\tau\right)&u_j\left(\textbf{x},\tau\right)\sigma_{ij}\left(\textbf{x},\tau\right) \, .
\end{align}
Eq.~(\ref{ord0}) gives the continuity equation and Eqs.~(\ref{ord1}, \ref{ord2}) give the Euler equation, in analogy to fluid mechanics \cite{landau}:
\begin{align}
\label{partd}
\frac{\partial \delta\left(\textbf{x},\tau\right)}{\partial \tau}+\nabla \cdot \left[ \left(1+\delta\left(\textbf{x},\tau\right)\right)\textbf{u}\left(\textbf{x},\tau\right) \right]=0 \\
\label{partu}
\frac{\partial \textbf{u}\left(\textbf{x},\tau\right)}{\partial \tau}+\mathcal{H}\textbf{u}\left(\textbf{x},\tau\right)+\textbf{u}\left(\textbf{x},\tau\right) \cdot \nabla \textbf{u}\left(\textbf{x},\tau\right)= \nonumber \\
-\nabla \Phi\left(\textbf{x},\tau\right)-\frac{1}{\rho}\nabla_j\left(\rho\sigma_{ij}\right) \, .
\end{align}
On large scales, the Universe is expected to be smooth and hence Eqs.~(\ref{partd}, \ref{partu}) can be linearised. By defining the divergence and vorticity of the velocity field as:
\begin{align}
\theta\left(\textbf{x},\tau\right)&=\nabla \cdot \textbf{u}\left(\textbf{x},\tau\right) \\
\textbf{w}\left(\textbf{x},\tau\right)&=\nabla \times \textbf{u}\left(\textbf{x},\tau\right) \, ,
\end{align}
it can be shown that the vorticity decays quickly due to the expansion of the Universe, and hence it can be ignored. $\delta$ and $\theta$ satisfy the following equations in Fourier space:
\begin{align}
\label{deltan2}
\frac{\partial{\delta}\left(\textbf{k},\tau\right)}{\partial \tau}+{\theta}\left(\textbf{k},\tau\right)&= \nonumber \\
-\int d^3\textbf{k}_1 d^3\textbf{k}_2\delta_D\left(\textbf{k}-\textbf{k}_{12}\right)\alpha\left(\textbf{k}_1,\textbf{k}_2\right)&{\theta}\left(\textbf{k}_1,\tau\right){\delta}\left(\textbf{k}_2,\tau\right) \\
\frac{\partial {\theta}\left(\textbf{k},\tau\right)}{\partial \tau}+\mathcal{H}{\theta}\left(\textbf{k},\tau\right)+\frac{3}{2}\Omega_m\mathcal{H}^2{\delta}\left(\textbf{k},\tau\right)&= \nonumber \\
-\int d^3\textbf{k}_1 d^3\textbf{k}_2\delta_D\left(\textbf{k}-\textbf{k}_{12}\right)\beta\left(\textbf{k}_1,\textbf{k}_2\right)&{\theta}\left(\textbf{k}_1,\tau\right){\theta}\left(\textbf{k}_2,\tau\right) \, ,
\label{thentan2}
\end{align}
where $\textbf{k}_{12}=\textbf{k}_{1}+\textbf{k}_{2}$ and:
\begin{align}
\label{alpha}
\alpha\left(\textbf{k}_1,\textbf{k}_2\right)&=\frac{\textbf{k}_{12} \cdot \textbf{k}_{1}}{k_1^2} \\
\beta\left(\textbf{k}_1,\textbf{k}_2\right)&= \frac{k_{12} \left(\textbf{k}_{1} \cdot \textbf{k}_{2} \right)}{2k_1^2k_2^2} \, .
\label{beta}
\end{align}

In a $\Lambda$CDM universe, Eqs.~(\ref{deltan2}, \ref{thentan2}) can be solved with the expansions:
\begin{align}
\label{exp1}
{\delta} \left(\textbf{k},\tau\right)&=\sum_{n=1}^{\infty}D^n\left(a\right)\delta_n\left(\textbf{k}\right) \\
{\theta} \left(\textbf{k},\tau\right)&=-\mathcal{H}\sum_{n=1}^{\infty}D^n\left(a\right)\theta_n\left(\textbf{k}\right) \, ,
\label{exp2}
\end{align}
with $D\left(a\right)$ the linear growth factor and $\delta_n$ and $\theta_n$ given in terms of the expansions:
\begin{align}
\label{deltanf}
\delta_n\left(\textbf{k}\right)=&\int d^3\textbf{q}_1 \cdots \int d^3\textbf{q}_nF_n\left(\textbf{q}_1 \cdots \textbf{q}_n\right) \times \nonumber \\
&\delta_1\left(\textbf{q}_1\right) \cdots \delta_1\left(\textbf{q}_n\right) \delta_D(\textbf{k}-\textbf{q}_1-\cdots \textbf{q}_n)\\
\theta_n\left(\textbf{k}\right)=&\int d^3\textbf{q}_1 \cdots \int d^3\textbf{q}_nG_n\left(\textbf{q}_1 \cdots \textbf{q}_n\right) \times \nonumber \\
&\delta_1\left(\textbf{q}_1\right) \cdots \delta_1\left(\textbf{q}_n\right)\delta_D(\textbf{k}-\textbf{q}_1-\cdots \textbf{q}_n) \, .
\end{align}
In what follows, $D\left(a\right)$ and $D\left(z\right)$ will be used interchangeably, using the relation between the scale factor and redshift $1+z=\frac{1}{a}$. The kernels $F_n$ and $G_n$ are  homogeneous functions of the wavevectors and are given in terms of $\alpha, \beta$ by the following recurrence relations:
\begin{widetext}
\begin{align}
\label{Fn}
F_n\left(\textbf{q}_1,\cdots,\textbf{q}_n\right)&=\sum_{m=1}^{n-1} \frac{G_m\left(\textbf{q}_1, \cdots, \textbf{q}_m\right)}{\left(2n+3\right)\left(n-1\right)} 
\left[\left(2n+1\right)\alpha\left(\textbf{k}_1,\textbf{k}_2\right)F_{n-m}\left(\textbf{q}_{m+1},\cdots,\textbf{q}_n\right)+ \right.
\left.2\beta\left(\textbf{k}_1,\textbf{k}_2\right)G_{n-m}\left(\textbf{q}_{m+1},\cdots,\textbf{q}_n\right) \right] \\
G_n\left(\textbf{q}_1,\cdots,\textbf{q}_n\right)&=\sum_{m=1}^{n-1} \frac{G_m\left(\textbf{q}_1, \cdots, \textbf{q}_m\right)}{\left(2n+3\right)\left(n-1\right)} 
\left[3\alpha\left(\textbf{k}_1,\textbf{k}_2\right)F_{n-m}\left(\textbf{q}_{m+1},\cdots,\textbf{q}_n\right)+ \right.
\left.2n\beta\left(\textbf{k}_1,\textbf{k}_2\right)G_{n-m}\left(\textbf{q}_{m+1},\cdots,\textbf{q}_n\right) \right] \, ,
\label{Gn}
\end{align}
\end{widetext}
where $F_1 = G_1 = 1$, $\textbf{k}_1=\textbf{q}_1+\cdots+\textbf{q}_m$, and $\textbf{k}_2=\textbf{q}_{m+1}+\cdots+\textbf{q}_n$. For the correlation functions, the symmetrised versions of these functions are required, denoted $F_n^{(s)}$ and $G_n^{(s)}$.
In up to one-loop calculations for the power and bispectrum, only the expressions up to $n=4$ for $F_n$ are required. The explicit expressions For $F_3$ and $F_4$ are given explicitly in Ref.~\cite{1986ApJ...311....6G}. For $F_2$, the expression is given in Eq.~(\ref{f2s}) and here we show how it can be derived. Considering a matter-only universe, with $\Omega_m=1$, $a=\tau^2$ and $\mathcal{H}=\frac{2}{\tau}$, Eqs.~(\ref{deltan2}, \ref{thentan2}) become:
\begin{align}
\label{deltasimp}
\delta'+\theta=-I_1[\delta,\theta] \\
\theta'+\frac{2}{\tau}\theta+\frac{6}{\tau^2}\delta=-I_2[\delta, \theta] \, ,
\label{thetasimp}
\end{align}
where $I_1$ and $I_2$ are the expressions on the r.h.s. of Eqs.~(\ref{deltan2}, \ref{thentan2}). Expanding $\delta$ and $\theta$ to second order, one obtains the following equations:
\begin{align}
\delta=\tau^2 \delta_1+\tau^4 \delta_2 \\
\delta'=2\tau \delta_1+4\tau^3 \delta_2 \\
\theta=-2\tau \theta_1-2\tau^3 \theta_2 \\
\delta=-2\theta_1-6\tau^2 \theta_2 \, .
\end{align}
For $n=1$, $I_1$ and $I_2$ are second-order quantities and hence Eqs.~(\ref{deltasimp}, \ref{thetasimp}) are solved by $\theta_1=\delta_1$.
For $n=2$, one has to use the first-order solutions for the integrals on the r.h.s. of the expressions, and the following equations are obtained:
\begin{align}
4\tau^3 \delta_2-2 \tau^3 \theta_2=-I_1[\tau^2 \delta_1,-2\tau \delta_1] \\
-10\tau^2 \theta_2+6 \tau^2 \delta_2=-I_2[\tau^2 \delta_1,-2\tau \delta_1] \, .
\end{align}
By solving the above equations for $\delta_2$ and substituting the integral expressions $I_1$ and $I_2$ and $\alpha$  (Eq.~\ref{alpha}) and $\beta$ (Eq.~\ref{beta}), one finds the integral expression:
\begin{multline}
\delta_2(\textbf{k})=\int d^3 \textbf{q}_1 \int d^3 \textbf{q}_2 \delta_D(\textbf{k}-\textbf{q}_1-\textbf{q}_2) \\ \times
 \left[\frac{5}{7}+\frac{2}{7} \frac{(\textbf{q}_1 \cdot \textbf{q}_2)^2}{q_1^2 q_2^2}+ \frac{\textbf{q}_1 \cdot \textbf{q}_2}{7} \left(\frac{6}{q_1^2}+\frac{1}{q_2^2}\right)\right] \, .
\end{multline}
The expression of $F_2^{(s)}$ from Eq.~(\ref{f2s}) is finally obtained by symmetrisation over the arguments $\textbf{q}_1$ and $\textbf{q}_2$.

We have defined in Sec.~\ref{sec:bisp} the matter power spectrum and bispectrum.
Their expressions in SPT can be obtained by inserting the expansions of Eqs.~(\ref{exp1}, \ref{deltanf}) into Eqs.~(\ref{ps}, \ref{bis}) respectively. The full expansion is then grouped according to the number of $\delta_1$'s involved. This loop expansion can be interpreted in analogy with the loop diagrams from quantum field theory and this represents an intuitive manner of determining all the contributions at each order in the expansion. In this diagrammatic expansion, the exterior lines represent the arguments of the correlation function, vertices where $n$ lines meet are the kernels $F_n$, and the interior lines represent wavevectors that are integrated over. As usual, the sum of wavevectors into any vertex should be 0 and numerical factors in front of each diagram represent its symmetry. The loop order represents the number of interior lines in each of its vertices.
Following this procedure yields the power spectrum and bispectrum expressions presented in Sec.~\ref{sec:SPT} (Eqs.~\ref{p0loop}-\ref{b411}).

The actual numerical evaluation of the integrals (Eqs.~\ref{p0loop}-\ref{b411}) is non-trivial because the kernels may diverge. It has been shown \cite{1475-7516-2013-05-031, Kehagias2013514} that the divergences exactly cancel each other when summing the whole contributions at each loop order together, both in the power spectrum and bispectrum, provided that the linear power spectrum grows slowly enough on very large scales. However, for the numerical evaluation, a method to remove the divergences should be used. For the power spectrum at one loop only, a convenient split of the integration regions has been used in Ref.~\cite{1994ApJ...431..495J} which solves the divergence problems. More recently, both the power spectrum and bispectrum divergences have been eliminated in Refs.~\cite{1475-7516-2014-07-056, 1475-7516-2014-07-057, baldauf, Angulo:2014tfa}. We briefly explain this last method in the next paragraphs.

By considering Eqs.~(\ref{p13})-(\ref{b411}), it can be easily seen that divergences appear at $q=0$ and $\textbf{q}=\pm \textbf{k}_i$. The basic idea of the method is to first perform a convenient change of variable in order to move the divergences to 0 and then, as the variable of integration spans all space, to do a symmetrisation in $\textbf{q} \leftrightarrow -\textbf{q}$. For the power spectrum, this method yields:
\begin{widetext}
\begin{align}
&P_{\text{1-loop}}^{\text{SPT}}\left(k,z\right)=D^4\left(z\right)\int \frac{d^3q}{\left(2\pi\right)^3} 
\left[6 P_{\text{lin}}\left(k\right)P_{\text{lin}}\left(q\right)F_3^{\left(s\right)}\left(\textbf{k},\textbf{q},-\textbf{q}\right)+ \right. \nonumber \\
2 P_{\text{lin}}\left(q\right)P_{\text{lin}}&\left(|\textbf{k}-\textbf{q}|\right)\left[F_2^{\left(s\right)}\left(\textbf{q},\textbf{k}-\textbf{q}\right)\right]^2\Theta\left(|\textbf{k}-\textbf{q}|-q\right)+ 
\left. 2 P_{\text{lin}}\left(q\right)P_{\text{lin}}\left(|\textbf{k}+\textbf{q}|\right)\left[F_2^{\left(s\right)}\left(-\textbf{q},\textbf{k}+\textbf{q}\right)\right]^2\Theta\left(|\textbf{k}+\textbf{q}|-q\right) \right] \, ,
\end{align}
\end{widetext}
where $\Theta$ is the Heaviside step function.
For the bispectrum, $B_{321}^{(II)}$ and $B_{411}$ only have divergences at 0 and hence do not need any change of variable. The integrand $b_{222}$ of $B_{222}$ needs to be re-expressed in the following manner \cite{baldauf, Angulo:2014tfa} (where we correct a typo in the original paper):
\begin{widetext}
\begin{align}
b_{222}^{(k_3>k_1)} =& 8 P_{\text{lin}}(q) P_{\text{lin}}(|\textbf{k}_2-\textbf{q}|) P_{\text{lin}}(|\textbf{k}_3+\textbf{q}|) 
F_2^{(s)}(-\textbf{q},\textbf{k}_3+\textbf{q}) F_2^{(s)}(\textbf{k}_3+\textbf{q},\textbf{k}_2-\textbf{q}) \times \nonumber \\
& F_2^{(s)}(\textbf{k}_2-\textbf{q},\textbf{q})\Theta(|\textbf{k}_2-\textbf{q}|-q) \Theta(|\textbf{k}_3+\textbf{q}|-q)  \nonumber \\
&+8P_{\text{lin}}(|\textbf{k}_3+\textbf{q}|) P_{\text{lin}}(|-\textbf{k}_1+\textbf{q}|) P_{\text{lin}}(q)F_2^{(s)}(\textbf{k}_3+\textbf{q},-\textbf{q}) 
 F_2^{(s)}(-\textbf{q},-\textbf{k}_1+\textbf{q}) \times \nonumber \\
&F_2^{(s)}(-\textbf{k}_1+\textbf{q},-\textbf{q}-\textbf{k}_3)\Theta(|-\textbf{k}_1+\textbf{q}|-|\textbf{k}_3+\textbf{q}|) \Theta(|\textbf{k}_3+\textbf{q}|-\textbf{q}) \nonumber \\ 
&+8P_{\text{lin}}(|\textbf{k}_2-\textbf{q}|) P_{\text{lin}}(q) P_{\text{lin}}(|\textbf{k}_1+\textbf{q}|) 
\times F_2^{(s)}(-\textbf{k}_2+\textbf{q},-\textbf{k}_1-\textbf{q})  \times \nonumber \\ 
&F_2^{(s)}(-\textbf{k}_1-\textbf{q},\textbf{q})F_2^{(s)}(\textbf{q},\textbf{k}_2-\textbf{q}) \Theta(|\textbf{k}_2-\textbf{q}|-q) \Theta(|\textbf{k}_1+\textbf{q}|-|\textbf{k}_2-\textbf{q}|) \nonumber\\ 
&+8P_{\text{lin}}(|\textbf{k}_2-\textbf{q}|) P_{\text{lin}}(\textbf{q}) P_{\text{lin}}(|\textbf{k}_1+\textbf{q}|) 
F_2^{(s)}(-\textbf{k}_2+\textbf{q},-\textbf{k}_1-\textbf{q})
F_2^{(s)}(-\textbf{k}_1-\textbf{q},\textbf{q}) \times \nonumber \\
&F_2^{(s)}(\textbf{q},\textbf{k}_2-\textbf{q})\Theta(|\textbf{k}_2-\textbf{q}|-q) \Theta(|\textbf{k}_2-\textbf{q}|-|\textbf{k}_1+\textbf{q}|) \\
b_{222}^{(k_3<k_1)} =& \left. b_{222}^{(k_3>k_1)}\right|_{\textbf{k}_1\leftrightarrow \textbf{k}_3}\  \, ,
\end{align}
\end{widetext}
with the note that this expression is only valid under the integral sign due to the various remappings. Similarly, $b_{321}^{I}$ becomes:
\begin{equation}
b_{321}^{I} \to 2b_{321}^{I}\Theta\left(|\textbf{k}_2-\textbf{q}|-q\right) \, .
\end{equation}
The sum of the 4 contributions is then calculated by performing the integrals directly. They can be calculated numerically fast using the multi-dimensional integrator \textsc{Cuba} \cite{Hahn200578}.

\section{Effective field theory}
\label{sec:EFT_full}

The equations governing this effective field theory are obtained by considering the collisionless Boltzmann equation in an expanding universe and smoothing it on a lengthscale $\Lambda^{-1}$. Hence, the theory is determined by the equations of motion of the long-wavelength modes, sourced by a stress-energy tensor. In the absence of the stress-energy tensor, the SPT Eqs.~(\ref{deltan2}-\ref{thentan2}) are recovered. This stress-energy tensor only modifies the Euler equation for the velocity, by adding a term $-\frac{1}{\rho_l}\nabla \tau$ on the r.h.s. of Eq.~(\ref{partu}), the continuity and Poisson equation remaining unchanged, but valid for the long-wavelengths modes only. 

Therefore, the equations of motion are modified and the expansions for the density (Eq.~\ref{exp1}) and velocity (Eq.~\ref{exp2}) perturbations are modified to:
\begin{align}
\delta\left(k,a\right)=\sum_{n=1}^{\infty} D^n(a)\delta_{n}(\textbf{k})+\epsilon \sum_{n=1}^{\infty} D^{n+\zeta}(a)\tilde{\delta}_{n}(\textbf{k}) \\
\theta\left(k,a\right)=\sum_{n=1}^{\infty} D^n(a)\theta_{n}(\textbf{k})+\epsilon\sum_{n=1}^{\infty} D^{n+\zeta}(a)\theta{\delta}_{n}(\textbf{k})  \, ,
\end{align}
where $\delta_{n}$ and $\theta_{n}$ can be expressed in terms of the kernels $F_n$ and $G_n$ (Eqs.~\ref{deltanf}-\ref{Gn}), while the tilded expressions can be expressed similarly in terms of $\tilde{F}_n$ and $\tilde{G}_n$. The tilded kernels satisfy however slightly more complicated recurrence relations, shown in Ref.~\cite{Angulo:2014tfa}. $\zeta$ is a constant fixed from the scaling of the power spectrum.

In the case of the power spectrum, the lowest level counterterm that appears is the two-point correlation function between $\delta_{1}$ and $\tilde{\delta}_{1}$. This can be expressed in terms of a single free parameter, the sound speed $c_{s(1)}^2$.  Hence, the term of Eq.~(\ref{eq:eftP}) 
 is added to the linear and SPT one-loop terms \cite{1475-7516-2014-07-057}.
The free parameter is fixed by fitting the one-loop EFT power spectrum with the nonlinear power spectrum at a low value of $k$, where the SPT result is still valid, while $\zeta$ is fixed by looking at the redshift evolution of the power spectrum, and a value of $\zeta=3.1$ is found to best fit simulations as well as scaling properties of the Universe.

For the bispectrum, we use the counterterm corresponding to the tree-level bispectrum. Only the $\tilde{F}_2^{(s)}$ kernel is required, which has the following expression:
\begin{widetext}
\begin{align}
& \tilde{F}_2^{(s)}(\textbf{k}_1,\textbf{k}_2) =
	-\frac{\bar{c}_1}{(1+\zeta)(7+2\zeta)} \left\{ \left( 5 + \frac{113\zeta}{14} + \frac{17\zeta^2}{7} \right) (k_1^2+k_2^2)
	+ \left( 7 + \frac{148\zeta}{7} + \frac{48\zeta^2}{7} \right) \textbf{k}_1\cdot\textbf{k}_2 \right. \nonumber \\
\label{eq:TildeF2}
&\qquad\qquad + \left( 2+\frac{59\zeta}{7}+\frac{18\zeta^2}{7} \right) \left( \frac{1}{k_1^2}+\frac{1}{k_2^2} \right) 
	(\textbf{k}_1\cdot\textbf{k}_2)^2 
+ \left( \frac{7}{2}+\frac{9\zeta}{2}+\zeta^2 \right) \left( \frac{k_1^2}{k_2^2} + \frac{k_2^2}{k_1^2} \right)
	\textbf{k}_1\cdot\textbf{k}_2  \nonumber \\
&\qquad\qquad \left. + \left( \frac{20\zeta}{7} + \frac{8\zeta^2}{7} \right) \frac{(\textbf{k}_1\cdot\textbf{k}_2)^3}{k_1^2 k_2^2} \right\} \, .
\end{align}
\end{widetext}

The bispectrum counterterm that follows has no extra free parameters in addition to those needed for the power spectrum, and can be expressed as given in Eq.~(\ref{eq:eftB}).

Four counterterms corresponding to the one-loop bispectrum can be added to the one above, which have three free parameters. Their expressions are shown in Ref.~\cite{Angulo:2014tfa}, but the improvement in the accuracy of the bispectrum is modest, and we will thus disregard them.

\section{Renormalised perturbation theory}
\label{sec:RPT_full}

Using the notation from SPT and defining $\eta \equiv \log a\left(\tau\right)$ and the following two-component vector,
\begin{equation}
\Psi\left(\textbf{k},\eta\right)=\left(\delta\left(\textbf{k},\eta\right),-\theta\left(\textbf{k},\eta\right)/\mathcal{H}\right) \, .
\end{equation}
Eqs.~(\ref{deltan2}, \ref{thentan2}) may be recast in a matrix notation:
\begin{multline}
\partial_\eta \Psi_a\left(\textbf{k},\eta\right)+\Omega_{ab}\left(\textbf{k},\eta\right)=  \\
\gamma_{abc}^{(s)}\left(\textbf{k},\textbf{k}_1,\textbf{k}_2\right)\Psi_b\left(\textbf{k}_1,\eta\right)\Psi_c\left(\textbf{k},\eta\right) \, ,
\end{multline}
where:
\begin{equation}
\Omega_{ab}=\left(\begin{array}{cc}
                   0 & -1/2\\
                   -3/2 & 1/2
                  \end{array}
\right) \, ,
\end{equation}
and $\gamma_{abc}^{(s)}$ is a symmetrised vertex matrix given in terms of the functions $\alpha$ (Eq.~\ref{alpha}) and $\beta$ (Eq.~\ref{beta}).
Finally, the solution to the perturbation equations can be given in terms of an inverse Laplace transform:
\begin{multline}
\label{eqpsi}
\Psi_a\left(\textbf{k},\eta\right)=g_{ab}\left(\eta\right)\phi\left(\textbf{k}\right)+\int_0^{\eta}d\eta'g_{ab}\left(\eta-\eta'\right) \\ \times 
\gamma_{bcd}^{(s)}\left(\textbf{k},\textbf{k}_1,\textbf{k}_2\right)\Psi_c\left(\textbf{k}_1,\eta'\right)\Psi_d\left(\textbf{k},\eta'\right) \, ,
\end{multline}
where $g_{ab}$ is the linear propagator, defined for positive $\eta$ as:
\begin{align}
g_{ab}\left(\eta\right)=\frac{e^{\eta}}{5}\left(\begin{array}{cc}
                   3 & 2\\
                   3 & 2
                  \end{array}\right)-\frac{e^{-3\eta/2}}{5}\left(\begin{array}{cc}
                   -2 & -2\\
                   3 & -3
                  \end{array}\right)
\end{align}
and $g_{ab}\left(\eta\right)=0$ for $\eta<0$. 
Analogously to SPT, Eq.~(\ref{eqpsi}) can be solved by a series expansion:
\begin{equation}
\Psi_a\left(\textbf{k},\eta\right)=\sum_{n=1}^{\infty}\Psi_a^{(n)}\left(\textbf{k},\eta\right) \, ,
\label{sumpsi}
\end{equation}
where
\begin{multline}
\Psi_a^{(n)}\left(\textbf{k},\eta\right)= \int \delta_D\left(\textbf{k}-\textbf{k}_{1 \cdots n}\right)\mathcal{F}^{(n)}_{aa_1 \cdots a_n}\left(\textbf{k}_1, \cdots, \textbf{k}_n; \eta\right)  \\
\times \phi(\textbf{k}_1) \cdots \phi(\textbf{k}_n)
\end{multline}
and $\textbf{k}_{1 \cdots n}=\textbf{k}_1+ \cdots \textbf{k}_n$. The kernel function $\mathcal{F}$ satisfies recurrence relations that are analogous to Eqs.~(\ref{Fn}, \ref{Gn}). In this fashion, the SPT solutions are obtained. However, this approach allows for a simplified formalism, because Feynman diagrams can be used. The basic rules are described in detail in Ref.~\cite{PhysRevD.73.063519}.

Non-linearities modify the linear propagator into a fully nonlinear one, defined as:
\begin{equation}
G_{ab}\left(k,\eta\right)\delta_D\left(\textbf{k}-\textbf{k}'\right)=\left\langle\frac{\delta\Psi_a(\textbf{k},\eta)}{\delta \phi_b(\textbf{k}')}\right\rangle \, .
\end{equation}
This represents the response of the final density and velocity fields to variations in initial conditions. Using the series expansion (Eq.~\ref{sumpsi}), it can be expressed in terms of the linear propagator:
\begin{equation}
G_{ab}\left(k,\eta\right)=g_{ab}\left(k,\eta\right)+\sum_{n=2}^{\infty}\left\langle\frac{\delta\Psi_a^{(n)}(\textbf{k},\eta)}{\delta \phi_b(\textbf{k}')}\right\rangle \, .
\end{equation}
Non-linearities also modify the vertex functions. Thus, the symmetric full vertex function $\Gamma$ is defined in terms of the fully nonlinear propagator, with the vertex function $\gamma$ becoming just the first term of a perturbative expansion:
\begin{multline}
\left\langle\frac{\delta^2\Psi_a(\textbf{k},\eta)}{\delta \phi_e(\textbf{k}_1) \delta \phi_f(\textbf{k}_2)}\right\rangle = 2 \int_0^{\eta}ds\int_0^s ds_1 \int_0^s ds_2 G_{ab}\left(\eta-s\right)  \\
\times \Gamma_{bcd}^{(s)}\left(\textbf{k},s;\textbf{k}_1,s_1;\textbf{k}_2,s_2\right)G_{ce}(s_1)G_{df}(s_2) \, .
\end{multline}
Switching again to the Feynman diagram formalism, the nonlinear propagator satisfies Dyson's formula:
\begin{multline}
G_{ab}\left(\textbf{k},\eta\right)=g_{ab}\left(\eta\right)+\int_0^{\eta}ds_1 \int_0^{s_1}ds_2 g_ac\left(\eta-s_1\right) \\ \times 
\Sigma_{cd}\left(\textbf{k},s_1,s_2\right)G_{db}\left(\textbf{k},s_2,\eta'\right) \, ,
\end{multline}
where $\Sigma$ represents the sum of the principal path irreducible diagrams (diagrams that cannot be split into disjoint pieces by removing a linear propagator from the principal path).

In the small-scale limit, the infinite series for the propagator can be resummed after a lengthy computation to \cite{PhysRevD.73.063520}:
\begin{equation}
G_{ab}(k,a)=g_{ab}(a)\exp\left(-\frac{k^2\sigma_d^2}{2}\right) \, ,
\end{equation}
where $\sigma_d^2=\frac{(a-1)^2}{3}\int \frac{d^3q}{2\pi^3}\frac{P_{\text{lin}}}{q^2}$.

This method permits the calculation of the \textit{n}-point correlation function in RPT for an arbitrary number of loops. Explicit expressions for the power spectrum and bispectrum are presented in Ref.~\cite{Bernardeau2008}. Compared to SPT, this method has the advantage that all the contributions involved are positive and the resummation of the propagator terms gives a well-defined perturbative expansion in the nonlinear regime. However, the expressions involved are complicated and the solutions are computationally demanding, requiring to solve numerically a set of integro-differential equations. Moreover, more than one loop is required to obtain an accurate result, even on mildly nonlinear scales. 

In order to solve these problems, Refs.~\cite{Bernardeau2008, Crocce2012} proposed a method that simplifies the calculation dramatically. The scheme is called \textsc{MPTbreeze} and in this formalism only the late-time propagator is calculated and hence no time integrations are required. First, the nonlinear propagator is generalised to an arbitrary number of points. The $(n+1)$-point propagator $\Gamma^{(p)}$ has been defined as:
\begin{multline}
\frac{1}{p!}\left\langle \frac{\delta \Psi_a^p\left(\textbf{k},a\right)}{\delta \phi_{b_1}(\textbf{k}_1) \cdots \delta \phi_{b_p}(\textbf{k}_p)}\right\rangle \\
= \delta_D\left(\textbf{k}-\textbf{k}_{1 \cdots p}\right) \Gamma_{ab_1 \cdots b_p}^{(p)}\left(\textbf{k}_1,\cdots,\textbf{k}_p,a\right) \, ,
\end{multline}
where $\textbf{k}_{1 \cdots p}=\textbf{k}_1+\cdots+\textbf{k}_p$. In this framework, the power spectrum can be expressed as:
\begin{multline}
P\left(k,z\right)=\sum_{r \geq 1} r! \int\delta_D\left(\textbf{k}-\textbf{q}_{1 \cdots r}\right)\left[\Gamma^{(r)}\left(\textbf{q}_1, \cdots, \textbf{q}_r,z \right)\right]^2  \\
\times P_{\text{lin}}(q_1) \cdots P_{\text{lin}}(q_r) d^3q_1 \cdots d^3q_r \, .
\end{multline}

If only the growing mode initial conditions are considered, the growing mode solution reduces to the following  simple expression:
\begin{multline}
\Gamma_\delta^{(n)}\left(\textbf{k}_1, \cdots, \textbf{k}_n;z\right) =  \\
D^n\left(z\right)F_n^{(s)}\left(\textbf{k}_1, \cdots, \textbf{k}_n\right) \exp\left[f(k)D^2(z)\right] \, ,
\label{gamman}
\end{multline}
where the function $f$ depends only on the linear power spectrum today:
\begin{multline} \label{eq:fk}
f\left(k\right)=\int \frac{d^3q}{(2\pi)^3}\frac{P_{\text{lin}}\left(q,z=0\right)}{504k^3q^5} \left[ 6k^7q-79k^5q^3 +50q^5k^3 \right.  \\
\left.- 21kq^7 + \frac{3}{4}\left(k^2-q^2\right)^3\left(2k^2+7q^2\right)\log \frac{|k-q|^2}{|k+q|^2} \right] \, .
\end{multline}

The numerical results obtained with this method agree well with the more exact method \textit{RegPT} \cite{PhysRevD.86.103528} over the relevant range of scales.
Using the simplified \textsc{MPTbreeze} formalism, the power spectrum up to one loop can be expressed as:
\begin{align}
\MoveEqLeft[3] P_{\text{tree}}^{\text{MPTbreeze}}\left(k,z\right)=\left[\Gamma^{(1)}\left(k;z\right)\right]^2P_{\text{lin}}\left(k\right) \\
\MoveEqLeft[3] P_{\text{1-loop}}^{\text{MPTbreeze}}\left(k,z\right)=2\int \frac{d^3q}{(2\pi)^3}\left[\Gamma^{(2)}\left(\textbf{k}-\textbf{q},\textbf{q};z\right)\right]^2  \nonumber \\
\times{}& P_{\text{lin}}\left(|\textbf{k}-\textbf{q}|\right)P_{\text{lin}}\left(q\right)  \, .
\end{align}
Using Eq.~(\ref{gamman}) and Eqs.~(\ref{p0loop}) and (\ref{p22}) from Sec.~\ref{sec:SPT} and Eq.~(97) from Ref.~\cite{1475-7516-2014-07-057}, the \textsc{MPTbreeze} contributions can be expressed in terms of their SPT counterparts as given in Eq.~(\ref{eq:RPT_P}).

The bispectrum contributions can be treated in a similar manner \cite{PhysRevD.85.123519}, and the result up to one loop is given  in Eq.~(\ref{eq:RPT_B}) in terms of the SPT one-loop contributions (Eqs.~\ref{b222}, \ref{b321i}).

This prescription allows an easy computation of the power spectrum and bispectrum for this method once the SPT counterparts have been determined, as only one integral function ($f$) needs to be evaluated, the other terms being calculated in SPT. Unfortunately, applying this theory up to one loop is only expected to give a reliable prediction for the bispectrum up to $k_{\max}=0.15 \, h/\text{Mpc}$ at $z=0$. Therefore, it is desirable to go to two loops in order to increase the range of validity of the model.

\section{Resummed Lagrangian perturbation theory}
\label{sec:RLPT_full}
Alternatively, perturbation theory can be derived as a function of the Lagrangian coordinates $\mathbf{q}$, which are related to their Eulerian counterparts $\mathbf{x}$ by the displacement field $\mathbf{\Psi}$:
\begin{equation}
\mathbf{x} (\mathbf{q}, t ) = \mathbf{q} + \mathbf{\Psi} (\mathbf{q}, t ) \, .  
\end{equation}
Under the assumption that the density perturbations at initial times are negligible, Eulerian and Lagrangian coordinates are related by the continuity equation: $\rho (\mathbf{x}) d^3 x = \bar \rho d^3 q $. Using the properties of the Dirac $\delta_D$ distribution, this leads to
\begin{equation}
\delta(\mathbf{x}) = \int d^3 q \, \delta_D \left[ \mathbf{x} - \mathbf{q} - \mathbf{\Psi}(\mathbf{q}) \right] - 1 \, , 
\end{equation}
whose Fourier transform is \cite{Taylor01101996}
\begin{equation}
\delta (\mathbf{k}) = \int d^3 q \, e^{-i \mathbf{k} \cdot \mathbf{q}} \left[ e^{-i \mathbf{k} \cdot \mathbf{\Psi} (\mathbf{q}) }  -1 \right] \, .
\end{equation}
This expression can be used to derive the observable power spectrum in Eulerian space \cite{Taylor01101996,Fisher1996,Matsubara2008}
\begin{equation}  \label{eq:RLPT_P1}
P(\mathbf{k}) = \int d^3 \mathbf{\Delta}_{12} \, e^{-i \mathbf{k} \cdot \mathbf{\Delta}_{12}} \left\{ \langle e^{-i \mathbf{k} \cdot \left[  \mathbf{\Psi}(\mathbf{q_1}) -  \mathbf{\Psi}(\mathbf{q_2}) \right] }  \rangle - 1 \right\} \, ,
\end{equation}
where $ \mathbf{\Delta}_{ij} \equiv  \mathbf{q_i} -  \mathbf{q_j} $, and the expectation value only depends on the separation $\mathbf{\Delta}_{12}$ due to homogeneity. Likewise, the bispectrum can be written as \cite{Rampf2012b}
\begin{multline} \label{eq:RLPT_B1}
B(k_1, k_2, k_3) = \int d^3 \mathbf{\Delta}_{12} \int d^3 \mathbf{\Delta}_{13} \, e^{-i \mathbf{k} \cdot \left(\mathbf{\Delta}_{12} + \mathbf{\Delta}_{13} \right) } \\ \times \left\{ \langle e^{-i \mathbf{k_2} \cdot \left[ \mathbf{\Psi}(\mathbf{q_1}) -  \mathbf{\Psi}(\mathbf{q_2})   \right]   -i \mathbf{k_3} \cdot \left[ \mathbf{\Psi}(\mathbf{q_1}) -  \mathbf{\Psi}(\mathbf{q_3})   \right] }  \rangle  - 1  \right\} \, ;
\end{multline}
also here the expectation value only depends on the separations $\mathbf{\Delta}_{12}, \mathbf{\Delta}_{13}$.
Eqs.~(\ref{eq:RLPT_P1}, \ref{eq:RLPT_B1}) relate the observable (Eulerian) density polyspectra to the displacement field $\mathbf{\Psi}$. We follow Ref.~\cite{Matsubara2008} and use the cumulant expansion theorem:
\begin{equation}
\langle e^{-iX} \rangle = \exp {\left[  \sum_{N=1}^{\infty}  \frac{(-i)^N}{N!} \langle X^N  \rangle_c  \right]} \, ,
\end{equation}
where $\langle X^N  \rangle_c$ represents the cumulant of the random variable $X$ \cite{Bernardeau20021}. By applying the cumulant expansion to Eqs.~(\ref{eq:RLPT_P1}, \ref{eq:RLPT_B1}), and expanding the powers of $N$ with the binomial theorem, two types of terms are obtained: those depending on $\mathbf{\Psi}$ at one point, and those depending on $\mathbf{\Psi}$ at two different points. Refs.~\cite{Matsubara2008,Rampf2012b} demonstrated that, if both sets of terms are expanded to the same order, the LPT results are identical to those obtained in SPT for both power spectrum and bispectrum. However Ref.~\cite{Matsubara2008} found that, for large separations, the terms depending on $\mathbf{\Psi}$ at one point are much larger than those depending on $\mathbf{\Psi}$ at two points, so that the first set of terms should be kept as it is, and only the second set should be expanded. This renormalised approach is called RLPT.

In order to derive explicit expressions for the matter power spectrum and bispectrum, we need to expand the displacement field as a function of the matter overdensity $\delta$.
The displacement field follows the equation of motion
\begin{equation}
\frac{d^2 \mathbf{\Psi}}{dt^2} + 2H \frac{d \mathbf{\Psi}}{dt} = -\nabla_x \phi [\mathbf{q} + \mathbf{\Psi}(\mathbf{q})] \, ,
\end{equation}
where $\phi$ is the gravitational potential. The polyspectra of $\mathbf{\Psi}$ can be calculated by expanding it as a series of the density field:
\begin{multline}
\mathbf{\Psi}^{(n)} (\mathbf{p}) = \frac{iD^n}{n!} \int \frac{d^3p_1}{(2 \pi)^3} ... \frac{d^3p_n}{(2 \pi)^3} \delta_D \left( \sum_{j=1}^n \mathbf{p}_j - \mathbf{p} \right) \\ \times \mathbf{L}^{(n)} (\mathbf{p}_1, ..., \mathbf{p}_n) \, \delta_1 (\mathbf{p}_1) ... \delta_1 (\mathbf{p}_n) \, ,
\end{multline}
where $\delta_1$ indicates the linear density perturbation at present times, and the perturbative kernels $\mathbf{L}^{(n)}$ are the analogues of the SPT kernels $F_n, G_n$, and are also obtained from a recursion relation \cite{Bernardeau20021}.

Using this expansion leads to the RLPT power spectrum of Eq.~(\ref{eq:RLPT_P2}) \cite{Matsubara2008}, and to the bispectrum of Eq.~(\ref{eq:RLPT_B2}) \cite{Rampf2012b}.

\begin{widetext}
\section{Going to two loops}
\label{app:twoloops}
We outline in this Appendix the two-loop calculations for the matter power spectrum and bispectrum in perturbation theory.

\subsection{Two-loop power spectrum in perturbation theories}
\subsubsection{SPT two-loop terms}
The two-loop power spectrum in SPT can be expressed as \cite{PhysRevD.80.043531, 1475-7516-2014-07-057}:
\begin{equation}
P_{\text{2-loop}}=P_{15}+P_{24}+P_{33}^{(I)}+P_{33}^{(II)} \, ,
\end{equation}
where the four terms are:

\begin{align}
\label{P2l1}
P_{51}(k)&=D^6(z)\int \frac{d^3p}{(2\pi)^3} \int \frac{d^3q}{(2\pi)^3} 30 F_5^{(s)}(\textbf{k},\textbf{q},-\textbf{q},\textbf{p},-\textbf{p})P_{\text{lin}}(k)P_{\text{lin}}(q)P_{\text{lin}}(p) \\
P_{42}(k)&=D^6(z)\int \frac{d^3p}{(2\pi)^3} \int \frac{d^3q}{(2\pi)^3} 24 F_2^{(s)}(\textbf{q},\textbf{k}-\textbf{q}) F_4^{(s)}(-\textbf{q},\textbf{q}-\textbf{k},\textbf{p},-\textbf{p})P_{\text{lin}}(q)P_{\text{lin}}(p)P_{\text{lin}}(|\textbf{k}-\textbf{q}|) \\
P_{33}^{(I)}(k)&=D^6(z)\int \frac{d^3p}{(2\pi)^3} \int \frac{d^3q}{(2\pi)^3} 9 F_3^{(s)}(-\textbf{k},\textbf{p},-\textbf{p}) F_3^{(s)}(\textbf{k},\textbf{q},-\textbf{q})P_{\text{lin}}(k)P_{\text{lin}}(q)P_{\text{lin}}(p) \\
\label{P2l4}
P_{33}^{(II)}(k)&=D^6(z)\int \frac{d^3p}{(2\pi)^3} \int \frac{d^3q}{(2\pi)^3} 6 F_3^{(s)}(\textbf{q},\textbf{p},\textbf{k}-\textbf{q}-\textbf{p}) F_3^{(s)}(-\textbf{q},-\textbf{p},-\textbf{k}+\textbf{q}+\textbf{p}) P_{\text{lin}}(q)P_{\text{lin}}(p)P_{\text{lin}}(|\textbf{k}-\textbf{q}-\textbf{p}|) \, .
\end{align}

\subsubsection{Two-loop RPT power spectrum}
In the case of RPT, the two-loop calculation is simplified considerably, because only one of the terms from Eqs.~(\ref{P2l1}-\ref{P2l4}) appears in this theory \cite{Crocce2012}. Thus, the two-loop power spectrum can be calculated by modifying the expression of $P_{33}^{(II)}$ as follows:

\begin{equation}
P_{\text{2-loop}}^{\text{MPTbreeze}}(k,z)=P_{33}^{(II)}(k,z)\exp\left[2f(k)D^2(z)\right]. 
\end{equation}
The IR-safe evaluation of this integral has been described in Ref.~\cite{1475-7516-2014-07-056}.

\subsection{Two-loop bispectrum in perturbation theories}
In SPT, the loop expansion is obtained by considering the expansion from Eq.~(\ref{exp1}) up to the relevant order, together with the integral expression (\ref{deltanf}) and then using Wick's theorem. For two-loops, there are terms up to $F_6^{(s)}$ in the kernels, which make the numerical evaluation cumbersome. In RPT, the number of terms is however drastically reduced.

\subsubsection{Two-loop RPT calculation}

The tree-level and one-loop bispectrum in this theory have been discussed above, and hence we proceed directly to the two-loop terms. The generating function for the RPT bispectrum is given by Eq.~(59) of Ref.~\cite{Bernardeau2008}. At two loops, using the notation from  Ref.~\cite{Bernardeau2008}, we need to take $r+s+t=4$. As only one of these numbers can be 0, there are only three choices for $r$, $s$ and $t$ (plus permutations) giving non-vanishing contributions, which we will treat in turn:

\begin{enumerate}[(a)]
  \item $r=3$, $s=1$, $t=0$ (+ 5 perms.)
  \item $r=2$, $s=2$, $t=0$ (+2 perms.)
  \item $r=1$, $s=1$, $t=2$ (+2 perms.) \, .
\end{enumerate}

In all these three cases, the expressions involved will depend on the functions $\Gamma^{(n)}$ only up to $\Gamma^{(4)}$, which in turn can be expressed in terms of the corresponding kernel functions $F_n^{s}$ through Eq.~(\ref{gamman}). Even though the expressions that we obtained for the \textsc{MPTbreeze} two-loop bispectra are relatively simple, they cannot be integrated directly because they have various poles where at least one of the arguments of $F_n^{(s)}$ vanishes. However, we know that the divergences between the various terms must cancel exactly after performing the integration, but numerically this is an issue because the divergent parts are expected to be much bigger than the finite result and thus the numerical result may not be reliable. In order to solve this problem, we will use the methods developed in Refs.~\cite{1475-7516-2014-07-056, baldauf, Angulo:2014tfa}. Compared to SPT, where some of the terms involve the kernel $F_6^{(s)}$, the expressions appearing in this method represent a significant simplification.

We note that from Eq.~(\ref{gamman}) all expressions will have a prefactor:
\begin{equation}
D^8(z)\exp\left[\left(f(k_1)+f(k_2)+f(k_3)\right)D^2(z) \right] \, ,
\end{equation}
and therefore in the following paragraphs we will omit this factor because it does not affect the calculation. We will denote the three integrals by $B_a$, $B_b$ and $B_c$ and the integrands with the corresponding lowercase letters. Then the final two-loop \textsc{MPTbreeze} bispectrum is:
\begin{equation}
B_{\text{2-loop}}^{\text{MPTbreeze}}(k_1,k_2,k_3,z)=D^8(z)\exp\left[\left(f(k_1)+f(k_2)+f(k_3)\right)D^2(z) \right] \left[B_a(k_1,k_2,k_3)+B_b(k_1,k_2,k_3)+B_c(k_1,k_2,k_3)\right] \, .
\end{equation}

The expressions for the three bispectra are as follows:
\begin{align}
\MoveEqLeft[3]  B_a(k_1,k_2,k_3)=24 \int \frac{d^3 q_1}{(2\pi)^3} \int \frac{d^3 q_2}{(2\pi)^3} F_4^{(s)}(\textbf{q}_1,\textbf{q}_2,-\textbf{k}_3-\textbf{q}_1-\textbf{q}_2,-\textbf{k}_2)F_3^{(s)}(-\textbf{q}_1,-\textbf{q}_2,\textbf{k}_3+\textbf{q}_1+\textbf{q}_2) \nonumber \\
\times{}& P_{\text{lin}}(q_1)P_{\text{lin}}(q_2)P_{\text{lin}}(k_3)P_{\text{lin}}(|\textbf{q}_1+\textbf{q}_2+\textbf{k}_3|) \\
\MoveEqLeft[3]  B_b(k_1,k_2,k_3)=24 \int \frac{d^3 q_1}{(2\pi)^3} \int \frac{d^3 q_2}{(2\pi)^3} F_4^{(s)}(\textbf{q}_1,\textbf{k}_3-\textbf{q}_1,\textbf{q}_2,-\textbf{k}_2-\textbf{q}_2)F_2^{(s)}(-\textbf{q}_2,\textbf{k}_2+\textbf{q}_2)F_2^{(s)}(-\textbf{q}_1,\textbf{k}_3+\textbf{q}_1) \nonumber \\
\times{}&  P_{\text{lin}}(q_1)P_{\text{lin}}(q_2)P_{\text{lin}}(|\textbf{q}_1+\textbf{k}_3|)P_{\text{lin}}(|\textbf{q}_2+\textbf{k}_2|) \\
\MoveEqLeft[3]  B_c(k_1,k_2,k_3)=36 \int \frac{d^3 q_1}{(2\pi)^3} \int \frac{d^3 q_2}{(2\pi)^3} F_3^{(s)}(\textbf{q}_1,\textbf{q}_2,\textbf{k}_1-\textbf{q}_1-\textbf{q}_2)F_2^{(s)}(-\textbf{k}_1+\textbf{q}_1+\textbf{q}_2,-\textbf{k}_3-\textbf{q}_1-\textbf{q}_2) \nonumber \\
\times{}&  F_3^{(s)}(\textbf{k}_3+\textbf{q}_1+\textbf{q}_2,-\textbf{q}_1,-\textbf{q}_2) P_{\text{lin}}(q_1)P_{\text{lin}}(q_2)P_{\text{lin}}(|\textbf{k}_1-\textbf{q}_1-\textbf{q}_2|)P_{\text{lin}}(|\textbf{k}_3+\textbf{q}_1+\textbf{q}_2|) \, .
\end{align}
We will treat each of them in turn and show how to remove the singularities before the integration.
\paragraph*{$B_a$}
The expression for $B_a$ has singularities when $q_1=0$, $q_2=0$ and $\textbf{q}_1+\textbf{q}_2=-\textbf{k}_3$. By considering the variable $\textbf{q}_3=-\textbf{q}_1-\textbf{q}_2-\textbf{k}_3$, $B_a$ can be re-expressed in terms of a triple integral by adding a Dirac-delta function:
\begin{multline}
B_a(k_1,k_2,k_3)=24 \int \frac{d^3 q_1}{(2\pi)^3} \int \frac{d^3 q_2}{(2\pi)^3} \int \frac{d^3 q_3}{(2\pi)^3} F_4^{(s)}(\textbf{q}_1,\textbf{q}_2,\textbf{q}_3)F_3^{(s)}(-\textbf{q}_1,-\textbf{q}_2,-\textbf{q}_3) \delta_D(\textbf{q}_1+\textbf{q}_2+\textbf{q}_3+\textbf{k}_3) \\
\times P_{\text{lin}}(q_1)P_{\text{lin}}(q_2)P_{\text{lin}}(k_3)P_{\text{lin}}(q_3)  \, .
\end{multline}
This expression is now completely symmetric in $\textbf{q}_1 \leftrightarrow \textbf{q}_2 \leftrightarrow \textbf{q}_3$ and hence all ordering of the magnitudes of these three wavevectors are equivalent after a suitable relabelling of the variables. As there are six possible permutations of $q_1$, $q_2$ and $q_3$,
\begin{equation}
B_a= \int \int \int \frac{d^3 q_1 d^3 q_2 d^3 q_3}{(2\pi)^9} b_b 6 \Theta(q_3-q_2) \Theta(q_2-q_1) \, .
\end{equation}
Hence, the delta function and $\textbf{q}_3$ can now be eliminated and the final expression is obtained:
\begin{multline}
B_a(k_1,k_2,k_3)=24 \int \frac{d^3 q_1}{(2\pi)^3} \int \frac{d^3 q_2}{(2\pi)^3} F_4^{(s)}(\textbf{q}_1,\textbf{q}_2,-\textbf{k}_3-\textbf{q}_1-\textbf{q}_2,-\textbf{k}_2)F_3^{(s)}(-\textbf{q}_1,-\textbf{q}_2,\textbf{k}_3+\textbf{q}_1+\textbf{q}_2) \\
\times P_{\text{lin}}(q_1)P_{\text{lin}}(q_2)P_{\text{lin}}(k_3)P_{\text{lin}}(|\textbf{q}_1+\textbf{q}_2+\textbf{k}_3|)\times 6 \Theta(|\textbf{q}_1+\textbf{q}_2+\textbf{k}_3|-q_2)\Theta(q_2-q_1) \, .
\label{bafinal}
\end{multline}
The expression of Eq.~(\ref{bafinal}) has a leading divergence when $q_1=q_2=0$ and a subleading divergence when $q_1=0$,  $\textbf{q}_2$ fixed. The two divergences corresponding to $\textbf{q}_1+\textbf{q}_2=-\textbf{k}_3$ and $q_2=0$, at fixed $\textbf{q}_1$, have disappeared because the Heaviside functions evaluate to 0 in those limits. In order to eliminate all divergences at the integrand level, we  can also symmetrise in $\textbf{q}_{1,2} \leftrightarrow -\textbf{q}_{1,2}$:
\begin{align}
b_a(\textbf{q}_1,\textbf{q}_2) \to \frac{1}{4} \left[ b_a(\textbf{q}_1,\textbf{q}_2)+b_a(-\textbf{q}_1,\textbf{q}_2) + b_a(\textbf{q}_1,-\textbf{q}_2)+b_a(-\textbf{q}_1,-\textbf{q}_2) \right] \, .
\end{align}
We will use this symmetrisation for the \textit{b} and \textit{c} terms as well. 

\paragraph*{$B_b$}
The $B_b$ term has divergences for $q_1$=0, $\textbf{q}_1=-\textbf{k}_3$, $q_2=0$ and $\textbf{q}_2=-\textbf{k}_1$. We note that $b_b$ is symmetric under the transformations $\textbf{q}_1 \leftrightarrow -\textbf{k}_3-\textbf{q}_1$ and $\textbf{q}_2 \leftrightarrow -\textbf{k}_2-\textbf{q}_2$. 
We can exploit the three symmetries that now appear in the integrand by restricting the integration region to $q_1<|\textbf{k}_3+\textbf{q}_1|$ and $q_2<|\textbf{k}_1+\textbf{q}_2|$ and introducing two Heaviside functions and a factor of $2^2$:
\begin{multline}
B_b(k_1,k_2,k_3)=24 \int \frac{d^3 q_1}{(2\pi)^3} \int \frac{d^3 q_2}{(2\pi)^3} F_4^{(s)}(\textbf{q}_1,\textbf{k}_3-\textbf{q}_1,\textbf{q}_2,-\textbf{k}_2-\textbf{q}_2)F_2^{(s)}(-\textbf{q}_2,\textbf{k}_2+\textbf{q}_2)F_2^{(s)}(-\textbf{q}_1,\textbf{k}_3+\textbf{q}_1)  \\
\times P_{\text{lin}}(q_1)P_{\text{lin}}(q_2)P_{\text{lin}}(|\textbf{q}_1+\textbf{k}_3|)P_{\text{lin}}(|\textbf{q}_2+\textbf{k}_2|) \times 4 \Theta(|\textbf{k}_2+\textbf{q}_2|-q_2) \Theta(|\textbf{k}_3+\textbf{q}_1|-q_1) \, .
\end{multline}
This expression is not symmetric in $\textbf{q}_1 \leftrightarrow \textbf{q}_2$, but we can symmetrise it by symmetrising the whole integrand (including the delta functions):
\begin{equation}
b_b(\textbf{q}_1,\textbf{q}_2) \to \frac{1}{2} \left[ b_b(\textbf{q}_1,\textbf{q}_2)+b_b(\textbf{q}_2,\textbf{q}_1) \right]  \, .
\end{equation}
After the symmetrisation, we aim to restrict the integration range to $q_1<q_2$, and we achieve this by adding an additional $\Theta$-function, thus obtaining the final answer:
\begin{align}
\MoveEqLeft[3]  B_b(k_1,k_2,k_3)=24 \int \frac{d^3 q_1}{(2\pi)^3} \int \frac{d^3 q_2}{(2\pi)^3} [F_4^{(s)}(\textbf{q}_1,\textbf{k}_3-\textbf{q}_1,\textbf{q}_2,-\textbf{k}_2-\textbf{q}_2)F_2^{(s)}(-\textbf{q}_2,\textbf{k}_2+\textbf{q}_2)F_2^{(s)}(-\textbf{q}_1,\textbf{k}_3+\textbf{q}_1) \nonumber \\
\times{}& P_{\text{lin}}(q_1)P_{\text{lin}}(q_2)P_{\text{lin}}(|\textbf{q}_1+\textbf{k}_3|)P_{\text{lin}}(|\textbf{q}_2+\textbf{k}_2|) \times 4 \Theta(|\textbf{k}_2+\textbf{q}_2|-q_2) \Theta(|\textbf{k}_3+\textbf{q}_1|-q_1) \nonumber \\
+{}& F_4^{(s)}(\textbf{q}_2,\textbf{k}_3-\textbf{q}_2,\textbf{q}_1,-\textbf{k}_2-\textbf{q}_1)F_2^{(s)}(-\textbf{q}_1,\textbf{k}_2+\textbf{q}_1)F_2^{(s)}(-\textbf{q}_2,\textbf{k}_3+\textbf{q}_2) \nonumber \\
\times{}& P_{\text{lin}}(q_1)P_{\text{lin}}(q_2)P_{\text{lin}}(|\textbf{q}_2+\textbf{k}_3|)P_{\text{lin}}(|\textbf{q}_1+\textbf{k}_2|) \times 4 \Theta(|\textbf{k}_2+\textbf{q}_1|-q_1) \Theta(|\textbf{k}_3+\textbf{q}_2|-q_2)] \Theta(q_2-q_1) \, .
\end{align}
Hence all the leading and subleading divergences have been moved to $q_1=q_2=0$ and $q_1=0$, at $\textbf{q}_2$ fixed. For all the other poles in the kernels, the Heaviside functions vanish.

\paragraph*{$B_c$}
The expression for $B_c$ has only one direct symmetry $\textbf{q}_1 \leftrightarrow \textbf{q}_2$, but this is not enough. Therefore we introduce the following notation:
\begin{align}
&\textbf{q}_3=\textbf{k}_1-\textbf{q}_1-\textbf{q}_2 \\
&\textbf{q}_4=\textbf{k}_1+\textbf{k}_2-\textbf{q}_1-\textbf{q}_2  \, ,
\end{align}
and we then introduce two additional integrations and two Dirac delta functions. The integral becomes:
\begin{multline}
B_c(k_1,k_2,k_3)=36 \int\int\int\int \frac{d^3 q_1 d^3 q_2 d^3 q_3 d^3 q_4}{(2\pi)^{12}} F_3^{(s)}(\textbf{q}_1,\textbf{q}_2,\textbf{q}_3)F_2^{(s)}(-\textbf{q}_3,\textbf{q}_4) F_3^{(s)}(-\textbf{q}_4,-\textbf{q}_1,-\textbf{q}_2)  \\
\times P_{\text{lin}}(q_1)P_{\text{lin}}(q_2)P_{\text{lin}}(q_3)P_{\text{lin}}(q_4) \, \delta_D(\textbf{k}_1-\textbf{q}_1-\textbf{q}_2-\textbf{q}_3)\delta_D(\textbf{k}_2+\textbf{q}_3-\textbf{q}_4) \, .
\end{multline}
This expression is already symmetric in $\textbf{q}_1 \leftrightarrow \textbf{q}_2$, and we symmetrise it in all the other variables, obtaining 12 possible permutations and a fully symmetric expression. We can now introduce an ordering of the four variables in terms of their magnitude (e.g. $q_4\ge q_3 \ge q_2 \ge q_1$), knowing that all the other orderings can be obtained by a suitable re-labelling of the variables. There are $4!=24$ permutations of the four variables and, keeping only one of the permutations, we need to multiply it by the following product of Heaviside functions:
\begin{equation}
24\Theta(q_4-q_3)\Theta(q_3-q_2)\Theta(q_2-q_1) \, .
\end{equation}
There are now 12 summands, and each of them involves integrals over $\textbf{q}_1$, $\textbf{q}_2$, $\textbf{q}_3$ and $\textbf{q}_4$, three Heaviside functions and two delta functions. The aim is now to perform two of the integrations, in order to eliminate the delta functions. In ten of the terms, it turns out that it is possible integrate over  $\textbf{q}_3$ and $\textbf{q}_4$. In the other two, $\textbf{q}_3$ and $\textbf{q}_4$ appear in the same combination in both delta functions. For those terms we integrate over $\textbf{q}_2$ and $\textbf{q}_4$, and then relabel $\textbf{q}_3 \to \textbf{q}_2$. The final expression that we thus obtain only has divergences for $q_1=q_2=0$ and $q_1=0$, with $\textbf{q}_2$ fixed as required: 
\begin{align}
\MoveEqLeft[3] b_c(k_1,k_2,k_3)= 36 (2 F_2^{(s)}(\textbf{k}_2-\textbf{q}_1,\textbf{q}_1) F_3^{(s)}(-\textbf{q}_1,-\textbf{q}_2,-\textbf{k}_1-\textbf{k}_2+\textbf{q}_1+\textbf{q}_2) F_3^{(s)}(\textbf{q}_1-\textbf{k}_2,\textbf{k}_1+\textbf{k}_2-\textbf{q}_1-\textbf{q}_2,\textbf{q}_2) \nonumber \\
&\times P_{\text{lin}}(q_1) P_{\text{lin}}(|\textbf{q}_1-\textbf{k}_2|) P_{\text{lin}}(|\textbf{k}_1+\textbf{k}_2-\textbf{q}_1-\textbf{q}_2|) P_{\text{lin}}(q_2) \nonumber \\
& \times \Theta(|\textbf{k}_1+\textbf{k}_2-\textbf{q}_1-\textbf{q}_2|-|\textbf{q}_1-\textbf{k}_2|) \Theta(|\textbf{q}_1-\textbf{k}_2|-q_2|) \Theta(q_2)-q_1) \nonumber \\
+&2 F_2^{(s)}(-\textbf{q}_1,\textbf{k}_2+\textbf{q}_1) F_3^{(s)}(-\textbf{k}_2-\textbf{q}_1,-\textbf{q}_2,-\textbf{k}_1+\textbf{q}_1+\textbf{q}_2) F_3^{(s)}(\textbf{q}_1,\textbf{k}_1-\textbf{q}_1-\textbf{q}_2,\textbf{q}_2) P_{\text{lin}}(q_1) P_{\text{lin}}(|\textbf{k}_2+\textbf{q}_1|) \nonumber \\
& \times P_{\text{lin}}(|\textbf{k}_1-\textbf{q}_1-\textbf{q}_2|) P_{\text{lin}}(q_2) \Theta(|\textbf{k}_1-\textbf{q}_1-\textbf{q}_2|)-|\textbf{k}_2+\textbf{q}_1|) \Theta(|\textbf{k}_2+\textbf{q}_1)|-q_2) \Theta(q_2-q_1) \nonumber \\
+&2F_2^{(s)}(-\textbf{q}_1,\textbf{k}_2+\textbf{q}_1) F_3^{(s)}(-\textbf{k}_2-\textbf{q}_1,-\textbf{q}_2,-\textbf{k}_1+\textbf{q}_1+\textbf{q}_2) F_3^{(s)}(\textbf{q}_1,\textbf{k}_1-\textbf{q}_1-\textbf{q}_2,\textbf{q}_2) \nonumber \\
&\times P_{\text{lin}}(q_1) P_{\text{lin}}(|\textbf{k}_2+\textbf{q}_1|) P_{\text{lin}}(|\textbf{k}_1-\textbf{q}_1-\textbf{q}_2|) P_{\text{lin}}(q_2) \Theta(|\textbf{k}_2+\textbf{q}_1|-|\textbf{k}_1-\textbf{q}_1-\textbf{q}_2|) \Theta(|\textbf{k}_1-\textbf{q}_1-\textbf{q}_2|-q_2) \Theta(q_2-q_1) \nonumber \\
+&2 F_2^{(s)}(\textbf{k}_1+\textbf{k}_2-\textbf{q}_1-\textbf{q}_2,-\textbf{k}_1+\textbf{q}_1+\textbf{q}_2) F_3^{(s)}(-\textbf{q}_1,-\textbf{q}_2,-\textbf{k}_1-\textbf{k}_2+\textbf{q}_1+\textbf{q}_2) F_3^{(s)}(\textbf{q}_1,\textbf{k}_1-\textbf{q}_1-\textbf{q}_2,\textbf{q}_2) \nonumber \\ 
&\times P_{\text{lin}}(q_1) P_{\text{lin}}(|\textbf{k}_1-\textbf{q}_1-\textbf{q}_2|) P_{\text{lin}}(|\textbf{k}_1+\textbf{k}_2-\textbf{q}_1-\textbf{q}_2|) P_{\text{lin}}(q_2) \nonumber \\
& \times \Theta(|\textbf{k}_1+\textbf{k}_2-\textbf{q}_1-\textbf{q}_2|-|\textbf{k}_1-\textbf{q}_1-\textbf{q}_2|) \Theta(|\textbf{k}_1-\textbf{q}_1-\textbf{q}_2|-q_2) \Theta(q_2-q_1) \nonumber \\
+&2 F_2^{(s)}(\textbf{k}_2-\textbf{q}_1,\textbf{q}_1) F_3^{(s)}(-\textbf{q}_1,-\textbf{q}_2,-\textbf{k}_1-\textbf{k}_2+\textbf{q}_1+\textbf{q}_2) F_3^{(s)}(\textbf{q}_1-\textbf{k}_2,\textbf{k}_1+\textbf{k}_2-\textbf{q}_1-\textbf{q}_2,\textbf{q}_2) \nonumber \\
& \times P_{\text{lin}}(q_1) P_{\text{lin}}(|\textbf{q}_1-\textbf{k}_2|) P_{\text{lin}}(|\textbf{k}_1+\textbf{k}_2-\textbf{q}_1-\textbf{q}_2|) P_{\text{lin}}(q_2)  \nonumber \\
& \times \Theta(|\textbf{q}_1-\textbf{k}_2|)-|\textbf{k}_1+\textbf{k}_2-\textbf{q}_1-\textbf{q}_2|) \Theta(|\textbf{k}_1+\textbf{k}_2-\textbf{q}_1-\textbf{q}_2|-q_2) \Theta(q_2-q_1) \nonumber \\
+&2 F_2^{(s)}(\textbf{k}_1+\textbf{k}_2-\textbf{q}_1-\textbf{q}_2,-\textbf{k}_1+\textbf{q}_1+\textbf{q}_2) F_3^{(s)}(-\textbf{q}_1,-\textbf{q}_2,-\textbf{k}_1-\textbf{k}_2+\textbf{q}_1+\textbf{q}_2) F_3^{(s)}(\textbf{q}_1,\textbf{k}_1-\textbf{q}_1-\textbf{q}_2,\textbf{q}_2) \nonumber \\
& \times P_{\text{lin}}(q_1) P_{\text{lin}}(|\textbf{k}_1-\textbf{q}_1-\textbf{q}_2|) P_{\text{lin}}(|\textbf{k}_1+\textbf{k}_2-\textbf{q}_1-\textbf{q}_2|) P_{\text{lin}}(q_2) \nonumber \\
& \times \Theta(|\textbf{k}_1-\textbf{q}_1-\textbf{q}_2|-|\textbf{k}_1+\textbf{k}_2-\textbf{q}_1-\textbf{q}_2|) \Theta(|\textbf{k}_1+\textbf{k}_2-\textbf{q}_1-\textbf{q}_2|-q_2) \Theta(q_2-q_1) \nonumber \\
+&2 F_2^{(s)}(\textbf{k}_2-\textbf{q}_2,\textbf{q}_2) F_3^{(s)}(-\textbf{q}_1,-\textbf{q}_2,-\textbf{k}_1-\textbf{k}_2+\textbf{q}_1+\textbf{q}_2) F_3^{(s)}(\textbf{q}_1,\textbf{k}_1+\textbf{k}_2-\textbf{q}_1-\textbf{q}_2,\textbf{q}_2-\textbf{k}_2) \nonumber \\
&\times P_{\text{lin}}(q_1) P_{\text{lin}}(|\textbf{k}_1+\textbf{k}_2-\textbf{q}_1-\textbf{q}_2|) P_{\text{lin}}(q_2) P_{\text{lin}}(|\textbf{q}_2-\textbf{k}_2|) \nonumber \\
& \times \Theta(|\textbf{k}_1+\textbf{k}_2-\textbf{q}_1-\textbf{q}_2|-q_2) \Theta(|\textbf{q}_2-\textbf{k}_2|-|\textbf{k}_1+\textbf{k}_2-\textbf{q}_1-\textbf{q}_2|) \Theta(q_2-q_1) \nonumber \\
+&2 F_2^{(s)}(\textbf{k}_2-\textbf{q}_2,\textbf{q}_2) F_3^{(s)}(-\textbf{q}_1,-\textbf{q}_2,-\textbf{k}_1-\textbf{k}_2+\textbf{q}_1+\textbf{q}_2) F_3^{(s)}(\textbf{q}_1,\textbf{k}_1+\textbf{k}_2-\textbf{q}_1-\textbf{q}_2,\textbf{q}_2-\textbf{k}_2) P_{\text{lin}}(q_1) \nonumber \\
&\times  P_{\text{lin}}(|\textbf{k}_1+\textbf{k}_2-\textbf{q}_1-\textbf{q}_2|) P_{\text{lin}}(q_2) P_{\text{lin}}(|\textbf{q}_2-\textbf{k}_2|) \Theta(|\textbf{k}_1+\textbf{k}_2-\textbf{q}_1-\textbf{q}_2|-|\textbf{q}_2-\textbf{k}_2|) \Theta(|\textbf{q}_2-\textbf{k}_2|-q_2) \Theta(q_2-q_1) \nonumber \\
+&2 F_2^{(s)}(-\textbf{q}_2,\textbf{k}_2+\textbf{q}_2) F_3^{(s)}(-\textbf{q}_1,-\textbf{k}_2-\textbf{q}_2,-\textbf{k}_1+\textbf{q}_1+\textbf{q}_2) F_3^{(s)}(\textbf{q}_1,\textbf{k}_1-\textbf{q}_1-\textbf{q}_2,\textbf{q}_2) P_{\text{lin}}(q_1)\nonumber \\
&\times  P_{\text{lin}}(|\textbf{k}_1-\textbf{q}_1-\textbf{q}_2|) P_{\text{lin}}(q_2) P_{\text{lin}}(|\textbf{k}_2+\textbf{q}_2|) \Theta(|\textbf{k}_1-\textbf{q}_1-\textbf{q}_2|-q_2) \Theta(|\textbf{k}_2+\textbf{q}_2|-|\textbf{k}_1-\textbf{q}_1-\textbf{q}_2|) \Theta(q_2-q_1) \nonumber \\
+&2 F_2^{(s)}(-\textbf{q}_2,\textbf{k}_2+\textbf{q}_2) F_3^{(s)}(-\textbf{q}_1,-\textbf{k}_2-\textbf{q}_2,-\textbf{k}_1+\textbf{q}_1+\textbf{q}_2) F_3^{(s)}(\textbf{q}_1,\textbf{k}_1-\textbf{q}_1-\textbf{q}_2,\textbf{q}_2) \nonumber \\
&\times P_{\text{lin}}(q_1) P_{\text{lin}}(|\textbf{k}_1-\textbf{q}_1-\textbf{q}_2|) P_{\text{lin}}(q_2) P_{\text{lin}}(|\textbf{k}_2+\textbf{q}_2|) \Theta(|\textbf{k}_1-\textbf{q}_1-\textbf{q}_2|-|\textbf{k}_2+\textbf{q}_2|) \Theta(|\textbf{k}_2+\textbf{q}_2|-q_2) \Theta(q_2-q_1) \nonumber \\
+& 2 F_2^{(s)}(\textbf{k}_2-\textbf{q}_1,\textbf{q}_1) F_3^{(s)}(-\textbf{q}_1,-\textbf{q}_2,-\textbf{k}_1-\textbf{k}_2+\textbf{q}_1+\textbf{q}_2) F_3^{(s)}(\textbf{q}_1-\textbf{k}_2,\textbf{k}_1+\textbf{k}_2-\textbf{q}_1-\textbf{q}_2,\textbf{q}_2)  P_{\text{lin}}(q_1) \nonumber \\
& \times P_{\text{lin}}(|\textbf{q}_1-\textbf{k}_2|) P_{\text{lin}}(|\textbf{k}_1+\textbf{k}_2-\textbf{q}_1-\textbf{q}_2|) P_{\text{lin}}(q_2) \Theta(|\textbf{q}_1-\textbf{k}_2|-q_1) \Theta(|\textbf{k}_1+\textbf{k}_2-\textbf{q}_1-\textbf{q}_2|-q_2) \Theta(q_2-|\textbf{q}_1-\textbf{k}_2|) \nonumber\\
+& 2 F_2^{(s)}(-\textbf{q}_1,\textbf{k}_2+\textbf{q}_1) F_3^{(s)}(-\textbf{k}_2-\textbf{q}_1,-\textbf{q}_2,-\textbf{k}_1+\textbf{q}_1+\textbf{q}_2) F_3^{(s)}(\textbf{q}_1,\textbf{k}_1-\textbf{q}_1-\textbf{q}_2,\textbf{q}_2) \nonumber \\
& \times P_{\text{lin}}(q_1) P_{\text{lin}}(|\textbf{k}_2+\textbf{q}_1|) P_{\text{lin}}(|\textbf{k}_1-\textbf{q}_1-\textbf{q}_2|) P_{\text{lin}}(q_2) \Theta(|\textbf{k}_2+\textbf{q}_1|-q_1) \Theta(|\textbf{k}_1-\textbf{q}_1-\textbf{q}_2|-q_2) \Theta(q_2-|\textbf{k}_2+\textbf{q}_1|))  \, .
\end{align}

The three long expressions can be added together with their corresponding permutations to obtain the final two-loop result, which is then free of any divergences before the integration. 
\end{widetext}

\section{Nine-parameter model} \label{sec:9params}

The tree-level prediction is the simplest model for the bispectrum. As its simpler counterpart, the linear power spectrum, it is only accurate for very low values of the wavenumber. A simple improvement over the tree level would be to substitute the linear with the nonlinear power spectrum in Eq.~(\ref{btree}), e.g. as calculated with the \textsc{Halofit} method  \citep{2003MNRAS.341.1311S, 2012ApJ...761..152T}; this result can then be tuned further by modifying the kernel $F_2^{(s)}$ in order to better fit simulations. This idea has been proposed in Ref.~\cite{Scoccimarro21082001} and here we discuss a more elaborate version of it,  which  fits $N$-body simulation better, introduced by Ref.~\cite{GilMarin2012}. Each of the 3 terms of the kernel (Eq.~\ref{f2s}) is modified by a multiplicative function, as follows:
\begin{align}
F_2^{\text{eff}}&\left(\textbf{q}_1,\textbf{q}_2\right) = \frac{5}{7} \, a(n_1,q_1) \, a(n_2,q_2) \nonumber \\
&+ \frac{1}{2}\frac{\textbf{q}_1 \cdot \textbf{q}_2}{q_1q_2}\left(\frac{q_1}{q_2}+\frac{q_2}{q_1}\right) \, b(n_1,q_1) \, b(n_2,q_2) \nonumber \\ 
&+ \frac{2}{7}\frac{\left(\textbf{q}_1 \cdot \textbf{q}_2\right)^2}{q_1^2q_2^2} \, c(n_1,q_1) \, c(n_2,q_2) \, ,
\label{f2eff}
\end{align}
such that the bispectrum can be expressed as:
\begin{align}
B(k_1,k_2,k_3)=F_2^{\text{eff}}&\left(\textbf{k}_1,\textbf{k}_2\right)P_{\text{Halofit}}(k_1) \, P_{\text{Halofit}}(k_2) \nonumber \\
&+ 2 \text{ perms.}
\end{align}

The functions $a(n,k)$, $b(n,k)$, $c(n,k)$ can be expressed in terms of nine coefficients that are determined numerically ($a_1, \cdots, a_n$) by fitting $N$-body simulations:
\begin{align}
a(n,k)&=\frac{1+\sigma_8^{a_6}(z) [0.7Q_3(n)]^{1/2}(qa_1)^{n+a_2}}{1+(qa_1)^{n+a_2}} \\
b(n,k)&=\frac{1+0.2a_3(n+3)(qa_7)^{n+3+a_8}}{1+(qa_7)^{n+3.5+a_8}} \\
c(n,k)&=\frac{1+4.5a_4/[1.5+(n+3)^4](qa_5)^{n+3+a_9}}{1+(qa_5)^{n+3.5+a_9}} \, .
\end{align}
The functions $n(k)$, $Q_3(n)$ and $q$ are defined as:
\begin{align}
n(k)&=\frac{d\log P_{\text{lin}}(k)}{d\log k} \\
Q_3(n)&=\frac{4-2^n}{1+2^{n+1}}  \\
q&=\frac{k}{k_{\text{NL}}} \, ,
\end{align}
where $k_{\text{NL}}$ is the nonlinear scale defined as the solution to the equation:
\begin{equation}
\frac{k_{\text{NL}}^3P_{\text{lin}}(k_{\text{NL}})}{2\pi^2}=1 \, .
\end{equation}
The parameters have been calibrated to give a maximum of $10 \%$ error in the matter bispectrum for $z \in [0, 1.5]$ and $k \le 0.4 h/Mpc$ and they are: $a_1=0.484$, $a_2=3.740$, $a_3=-0.849$, $a_4=0.392$, $a_5=1.013$, $a_6=-0.575$, $a_7=0.128$, $a_8=-0.722$ and $a_9=-0.926$.

\section{Ingredients of the halo model} \label{sec:haloZutaten}

We assume that dark matter haloes are virialised spheres of mass 
$  m=\frac{4\pi}{3} R_v^3 \Delta_v \bar{\rho} $,
where $R_v $ is the virial radius.
The virial overdensity is $\Delta_v = 18 \pi^2 \simeq 180 $ in matter domination and it
depends weakly on cosmology;
we fix $\Delta_v = 200$ in our model in order to match the assumptions of the numerical fits for the mass function and concentration.
The initial overdensity of spherically collapsed objects, extrapolated to the present time using linear theory, is $\delta_c = \frac{3}{5}\left(\frac{3\pi}{2}\right)^{2/3} \simeq 1.686$ \cite{1972ApJ...176....1G}.

\subsection{Halo profile}

We use the Navarro-Frenk-White (NFW) profile \cite{1996ApJ...462..563N}, which can be expressed in terms of two parameters, $\rho_s(m)$ and $r_s(m)$, describing the scaling radius and associated density, where the profile slope changes:
\begin{equation}
\rho (r|m) = \frac {\rho_s}{ {r}/{r_s} \,  \left( 1 + {r}/{r_s} \right)^2} \, .
\label{NFW}
\end{equation}
The NFW profile can be recast in terms of the concentration $c \equiv R_v / r_s$, which can be calibrated from $N$-body simulations \cite{Bullock2001}. Hence:
\begin{equation}
\rho_s = \frac{ \Delta_v \bar{\rho}}{3} \frac{c^3}{\log{\left(1+c\right)} - \frac{1}{1+c}} \, ; \:\:\:  r_s = \left(\frac{3m}{4\pi c^3 \Delta_v\bar{\rho}}\right)^{1/3}  \, .
\end{equation}
Substituting these definitions into Eq.~(\ref{NFW}) and applying a Fourier transformation, we obtain \cite{0004-637X-546-1-20}:
\begin{multline}
u (k|m) = \frac{4\pi\rho_sr_s^3}{m} \left\{\sin (kr_s) \left[\mathrm{Si}\left((1+c)kr_s\right) - \mathrm{Si} (kr_s) \right] \right.  \\
- \left. \frac{\sin(kr_s)}{(1+c) kr_s} + \cos (kr_s) \left[\mathrm{Ci} \left((1+c)kr_s\right) - \mathrm{Ci} (kr_s) \right] \right\} \, ,
\end{multline}
where $\text{Si}\left(x\right)$ and $\text{Ci}\left(x\right)$ are the sine and cosine integral functions.
We use the fitting function to the concentration obtained from the Bolshoi simulation \cite{0004-637X-740-2-102}:
\begin{multline}
c (m, z) = 9.2 \kappa \left(z\right) D \left(z\right)^{1.3} \left( \frac{m}{10^{12}h^{-1}M} \right)^{-0.09}   \\ 
\times \left[ 1 + 0.013 \left( \frac{m}{10^{12} h^{-1} M} D (z)^{-\frac{1.3}{0.09}} \right)^{0.25}\right]
\end{multline}
with $\kappa\left(z\right)=1.26$ at $z=0$ and $\kappa\left(z\right)=0.96$ at $z \geq 1$, as in Ref.~\cite{Giannantonio01062012}.

\subsection{Halo mass function}

The number density of haloes of mass $m$ and redshift $z$ is given by $n(m,z)$, which can be written as \cite{1974ApJ...187..425P}:
\begin{equation}
\frac{m}{\bar{\rho}} \, n(m,z) \, dm = f (\nu) \,  d\nu \, .
\label{change_halo}
\end{equation}
Here the peak height $\nu \equiv \delta_c^2 / \sigma^2(m,z) $ is obtained from the variance of the linear density field filtered with a top-hat function in Fourier space $W_f(x) = (3/x^3) [\sin(x) - x \cos (x)]$ on the scale $R_f = [3m / (4 \pi \bar \rho)]^{1/3}$:
\begin{equation}
\sigma^2 (m, z ) = \frac{ D^2 (z) }{2\pi^2} \int_0^{\infty} dk \,  k^2 \,  W_f^2 (kR_f) \, P_{\text{lin}} (k) \, ,
\end{equation}
where $P_{\text{lin}}(k)$ is the linear matter power spectrum and $D(z)$ is the linear growth function.

Different choices for the mass function $f(\nu)$ are possible. The simplest form was derived by Press and Schechter \cite{1974ApJ...187..425P} analytically assuming spherical collapse, finding 
\begin{equation}
f^{\mathrm{PS}}(\nu) = \sqrt{2 \nu / \pi} \, e^{-\nu / 2} \, .
\end{equation}
This simple model only matches the results from $N$-body simulations within a factor of two (see e.g. Ref~\cite{Giannantonio2010}); more elaborated models calibrated on $N$-body simulations include those by Refs.~\cite{Sheth01091999, 0004-637X-688-2-709}. We use here the numerical fit by Tinker \emph{et al.} \cite{0004-637X-724-2-878}, where
\begin{equation}
f^{\mathrm{T}} (\nu) = \alpha \left[ 1 + (\beta \nu)^{-2 \phi} \right] \nu^{2\eta} \exp(-\gamma \nu^2 / 2) \, ;
\end{equation}
the coefficients $\beta$, $\gamma$, $\eta$ and $\phi$ have the following redshift dependence:
\begin{align}
\beta = \beta_0 (1+z)^{0.20} \,; &\:\:\:\:\:\:\:\:\:\: \gamma = \gamma_0 (1+z)^{-0.01} \nonumber \\
\eta = \eta_0 (1+z)^{0.27} \,; &\:\:\:\:\:\:\:\:\:\:   \phi = \phi_0 (1+z)^{-0.08} \,
\end{align}
with $\alpha=0.368$, $\beta_0=0.589$, $\gamma_0=0.864$, $\eta_0=-0.243$ and $\phi_0=-0.729$.

\subsection{Halo clustering}
Finally, we need a model for the clustering of the dark matter halo centres, i.e. $P_{h}$ and $B_{h}$.
Under the simplest assumption of local deterministic bias, the halo
overdensity in real space can be expressed as a Taylor expansion of the dark matter overdensity field, where the coefficients are the bias parameters \cite{Fry1993}:
assuming both fields have been smoothed on a relatively large scale $R$. This expression also holds in Fourier space, where the first constant term $b_0$ is relegated to $k=0$ and is thus irrelevant.

If we calculate the matter power spectra using SPT at tree level, we obtain \cite{Fry1993}:
\begin{equation}
  P_{h} (k | m_1, m_2) = b_1 (m_1) \, b_1 (m_2) \,  D^2 (z) \, P_{0} (k) \, , 
  \label{phh}
\end{equation}
\\[-1.2cm]
\begin{multline}
B_{h} (k_1,k_2,k_3 | m_1, m_2, m_3, z) =  \,  \\
  b_1 (m_1) \, b_1 (m_2) \, b_1 (m_3) \, D^6(z) \, B_0 (k_1, k_2, k_3) \nonumber \\
  + \left[ b_1 (m_1) \, b_1 (m_2) \, b_2 (m_3) \, D^4(z) \, P_{\text{lin}} (k_1) \, P_{\text{lin}} (k_2) + \mathrm { 2 \: cyc.}  \right] \, , \nonumber
\end{multline}
where we  only need the first two bias parameters $b_1, b_2$.
They can be derived from the halo mass function using the peak-background split technique  \cite{Cole1989, Mo1996, Sheth01091999, 0004-637X-546-1-20, Mo01011997}; this method consists of dividing the Lagrangian density perturbations into short and long wavelength modes,
and assuming that halo collapse happening on the short scales is enhanced or suppressed by the long-scale modulations in the dark-matter perturbations, which effectively alter the collapse threshold. 
After transformation from Lagrangian to Eulerian space assuming spherical collapse, the first two bias coefficients for the Press-Schechter mass function are \cite{Mo1996}:
\begin{align}
  b^{\mathrm{PS}}_1(\nu) &=  1 + \frac{\nu - 1}{\delta_c} \, , \\
  b^{\mathrm{PS}}_2(\nu) &= \frac{8}{21} \frac{\nu - 1}{\delta_c} +   \frac{\nu^4 - 3 \nu^2}{\delta_c^2} \, .
\end{align}
While using the Tinker \emph{et al.} mass function we find:
\begin{align}
\MoveEqLeft[3] b^{\mathrm{T}}_1(\nu) = \frac{2 \phi }{\delta_c  \left[ (\beta  \nu )^{2 \phi }+1\right]}+\frac{\gamma  \nu ^2+\delta_c -2 \eta -1}{\delta_c } \, , \\
\MoveEqLeft[3] b^{\mathrm{T}}_2(\nu) = \frac{2 \left(42 \gamma  \nu ^2 \phi +8 \delta_c  \phi -84 \eta  \phi +42 \phi ^2-21 \phi \right)}{21 \delta_c ^2 \left[(\beta  \nu )^{2 \phi }+1\right]} \nonumber \\ 
+{}& \frac{21 \gamma ^2 \nu ^4+8 \gamma  \delta_c  \nu ^2-84 \gamma  \eta  \nu ^2-63 \gamma  \nu ^2 }{21 \delta_c ^2}  \nonumber \\
+{}& \frac{-16 \delta_c  \eta -8 \delta_c +84 \eta ^2+42 \eta}{21 \delta_c ^2} \, .
\end{align}

In order to enforce consistency with the definition of matter overdensity, and to recover linear theory for $k \to 0$, we must finally impose the following conditions \cite{PhysRevD.66.043002}:
\begin{align}
\label{norm1}
\int_0^{\infty} dm \frac{m}{\bar{\rho}} \, n (m) &= 1 \\
\label{norm2}
\int_0^{\infty} dm \frac{m}{\bar{\rho}} \, n (m) \, b_1 (m) &= 1 \, , \\
\int_0^{\infty} dm \frac{m}{\bar{\rho}} \, n (m) \, b_i (m) &= 0 \, , \:\:\:\:\: \forall \, i > 1 \, .
\end{align}

\section{Combined halo-PT model}
\label{appendix:haloPT}

The derivation of this model can be summarised as follows.
The probability that a particle at Lagrangian position $\textbf{q}_1$ belongs to a halo with mass in $\left[m,m+dm \right]$ is
$ dF = f (\nu) \, d\nu $.
The probability that a particle at position $\textbf{q}_2$, at a distance $q=|\textbf{q}_2-\textbf{q}_1|$ is situated in the same halo can be expressed as:
\begin{equation}
 F_m(q) =   \frac{ (2q_m - q)^2 \, (4q_m + q)} {16 \, q_m^3}  \text{ , if } 0 \leq q \leq 2q_m \,  ,
\end{equation}
and $F_m(q) = 0$ if $q > q_m$. Then the probability that the pair $\left[\textbf{q}_1,\textbf{q}_2\right]$ belongs to one (or two) haloes is respectively:
\begin{align}
F_{1h} (q) &= \int_{\nu_{q/2}}^{\infty} d \nu f(\nu) \, F_m (q) \, , \\
F_{2h} (q) &= 1 - F_{1h} (q) \, . \label{eq:F2}
\end{align}
In order to derive an analytic expression in Eulerian space,  the function $F_{2h}(q)$ at perturbative level is further approximated by its value at $F_{2h}(q \sim 1/k)$.  

In terms of the
Eulerian particle positions $\textbf{x}\left(\textbf{q},t\right)$,
 the matter power spectrum can be expressed as \cite{1995MNRAS.273..475S, Taylor01101996}:
\begin{equation}
P (k) = \int d^3q \langle e^{i \textbf{k} \cdot \Delta \textbf{x}} - e^{i \textbf{k} \cdot \textbf{q}} \rangle \, ,
\end{equation}
where
$ \Delta \textbf{x} = \textbf{x}(\textbf{q}) - \textbf{x}(0) $
and $\langle \cdots \rangle$ represents statistical average. The term $ e^{i \textbf{k} \cdot \textbf{q}} $ is normally neglected in perturbation theory as it only produces a non-zero contribution at $k = 0$, but it is important in the halo model \cite{valageas1}.
The power spectrum can be split between the contributions coming from pairs in one (or two) haloes as:
\begin{equation}
\label{p1hv}
P_{1h \, (2h)} (k) = \int_0^{\infty} d^3q \, F_{1h \, (2h)}(q) \langle e^{i\textbf{k} \cdot \Delta \textbf{x}} - e^{i\textbf{k} \cdot \textbf{q}}\rangle_{1h \, (2h)} \, , 
\end{equation}
where in this case the averages are conditional on the set of pairs being in exactly one of the terms. The terms described in Eq.~(\ref{p1hv}) correspond to the one- and two-halo terms from the halo model of Sec.~\ref{sec:halo}.
 
We can then split the power spectra further between perturbative and non-perturbative regimes.
Considering the perturbative case and the expected physical behaviour of the two terms, it must hold $F_{1h} \equiv 0$ at all levels of perturbation theory. Hence $F_{2h} \equiv 1$, and the two-halo contribution is fully perturbative.
However, the two-halo power spectrum can be obtained more easily by replacing the conditional average of Eq.~(\ref{p1hv}) with the full average given by perturbation theory, and by weighting instead the results with $F_{2h}$ from Eq.~(\ref{eq:F2}). This yields the result of Eq.~(\ref{p2VN}).

For the one-halo contribution, it is assumed that the haloes are fully virialised, and hence:
\begin{equation}
\langle e^{i \textbf{k} \cdot \Delta \textbf{x}} \rangle_m = u^2 (k|m) \, .
\label{eikdx}
\end{equation}
Substituting back the result into Eq.~(\ref{p1hv}) and changing the order of integration yields the result of Eq.~(\ref{p1VN}).

\bibliography{Bibliografie}{}

\end{document}